\documentstyle[12pt,twoside,a42,epsfig,cite,axodraw,amstex,amssymb,verbatim,amsxtra,afterpage,slashed,float]{article}

 \newcommand{\SigmaP}{{\sf Sigma}}
 
 \newcommand{\D}{\displaystyle}

 \newcommand{\N}{\nonumber}
 \newcommand{\MS}{\overline{{\sf MS}}}
 \newcommand{\MOM}{\tiny{\mbox{MOM}}}
 \newcommand{\ep}{\varepsilon}
 \newcommand{\adag}{/\!\!\!\! }
 \newcommand{\si}{{\rm sign}}
 
 \newcommand{\GeV}{{\rm GeV}}
 
 \newcommand{\eV}{{\rm eV}}
 \newcommand{\bra}[1]{\langle#1\mid}
 \newcommand{\ket}[1]{\mid#1\rangle}
 
 \newcommand{\Li}{\mbox{Li}}

 \newcommand{\SN}{\mbox{S}}
 \newcommand\SH{\,\mbox{$\sqcup \! \sqcup$}\,}

 \newcommand{\gsim}{\raisebox{-0.07cm   }
 {$\, \stackrel{>}{{\scriptstyle\sim}}\, $}}

 \newcommand{\Ahathat}{\hat{\hspace*{-1mm}\hat{A}}}

 \newcommand{\Ctil}{\tilde{C}}

\newcommand {\de} [0] {\delta}

\newcommand{\bea}{\begin{eqnarray}}
\newcommand{\bq}{\begin{equation}}
\newcommand{\eea}{\end{eqnarray}}
\newcommand{\eq}{\end{equation}}

\newcommand{\lsim}{\raisebox{-0.07cm   }
{$\, \stackrel{<}{{\scriptstyle\sim}}\, $}}

\begin{document}
 \thispagestyle{empty}
 \begin{flushleft}
 \end{flushleft}

 \setcounter{page}{0}

 \begin{center}
 
  {\LARGE\bf\boldmath $O(\alpha_s^3 T_F^2 N_f )$ Contributions to the Heavy Flavor Wilson
 Coefficients of the Structure Function $F_2(x,Q^2)$ at $Q^2 \gg m^2$ \\
   }
  \vspace{3cm}

  \large 
   Diplomarbeit \\
  \vspace{2mm}
   zur Erlangung des wissenschaftlichen Grades \\
  \vspace{2mm}
   Diplom-Physiker \\
  \vspace{2mm}

  \vspace{3em}

   {\small vorgelegt von}

  \vspace{2em}
  \large
  Fabian Philipp Wi\ss brock $^{a}$ \\
  {\small geboren am 15.08.1983 in Hannover} 

  \vspace{8mm}
  Betreuer: PD Dr. habil. Johannes Bl\"umlein$^{b}$ \\
  \normalsize
  \vspace{4em}

   {\it$^a$Freie Universit\"at Berlin, Fakult\"at Physik}\\
   {\it    Arnimallee 14, D-14195 Berlin}\\
  \vspace{1mm}
   {\it$^b$Deutsches Elektronen--Synchrotron, DESY}\\
   {\it  Platanenallee 6, D--15738 Zeuthen}\\
 \end{center}
 \vspace*{\fill}
 

 \vspace*{\fill}
 \newpage
\vspace*{5cm}
 \begin{flushleft}
1. Fachgutachter: \hspace*{1.5cm} Prof. Dr. H. Kleinert\\
\vspace{1.5cm}
2. Fachgutachter: \hspace*{1.5cm} 
 \end{flushleft}
 \thispagestyle{empty}
 \setcounter{page}{0}

 \vspace{170mm}

  \vspace{5mm}

 
  \newpage
 \tableofcontents
 \listoftables
 \listoffigures
 \newpage
\mbox{             }
 \newpage
\section{\boldmath{Introduction} \label{INTR}}

\vspace*{2mm}
\noindent
During the 1950s an intense search 
for new elementary particles 
was performed
at various particle accelerators 
which led to the discovery of several hundreds of hadron states. 
Based on their spin hadrons were classified as mesons (spin~$=~0,1$) 
or baryons (spin~$=~1/2,3/2$). The richness and variety of the hadronic spectrum  
made it likely, that hadrons are not fundamental particles. First attempts to classify 
the hadronic spectrum  by properties as mass, charge, isospin and flavor were undertaken 
by W. Heisenberg, M. Gell-Mann and different Japanese groups. In 1961 M. Gell-Mann and 
Y. Ne'eman independently proposed 
hadron representations based on 
the flavor 
group $SU(3)$, \cite{Neeman:1961cd,GellMann:1961ky,GellMann:1964xy}. 
The mesons and spin 1/2-baryon states can be grouped into octets and the spin 
3/2-baryon states into decuplets of similar properties.  The prediction of the unknown 
baryon $\Omega^{-}$ with strangeness $-3$ and electric charge $-1$ was a triumphant success 
of this theory. A mathematical description of the hadron states was introduced in 1964, when 
M.~Gell-Mann~\cite{GellMann:1964nj}, and G.~Zweig \cite{Zweig:1964jf} proposed the static quark-model 
due to which hadrons are bound states out of three quarks in case of baryons and a quark-antiquark 
pair for mesons. Here the quarks are assumed to be spin 1/2-fermions. Quantum numbers of all hadron 
states known by that time could be described assuming three different quark flavors~: up ($u$), 
down ($d$), and strange($s$). The quarks had to carry fractional elementary charges 
$e_u = + 2/3, e_d = e_s = -1/3$. The $\Omega^{-}$--baryon, which is composed out of three $s$-quarks, 
was obtained as a fermionic state with a spin and flavor--symmetric wave function. Since 
quarks are fermions, this is not in accordance with the spin-statistics theorem 
\cite{Pauli:1940zz,Luders:1958zz,Duck:1998cp}. This 
contradiction 
was resolved, after some intermediary steps, by introducing the three-valued quantum number 
color~\cite{Greenberg:1964pe,Han:1965pf,Nambu:1966,Bardeen:1972xk} for quarks.
Since colored particles could not be observed experimentally, it was proposed, that all physical 
hadron states are color-singlets,~\cite{Nambu:1966,Han:1965pf,Fritzsch:1972jv}. 

Because of their anomalous magnetic moments, protons and neutrons were suspected to be composite states since the 1930s, cf.~\cite{Frisch:1933,Alvarez:1940zz}. In a series of lepton-nucleon scattering 
experiments Hofstadter and collaborators \cite{Hofstadter:1963,Hofstadter:1956,Mcallister:1956ng,Schopper:1961} 
have shown during the 1950s, that baryons possess extended charge distributions irrespective of their net-charge, 
which was a further clear sign of compositeness. 

In the late 1960s deeply inelastic lepton--nucleon scattering (DIS) experiments at the Stanford Linear 
Accelerator {\sf SLAC},  \cite{Mo:1965dv,Taylor:1967qv}, made it possible to study the substructure 
of nucleons at much higher spatial resolution. One may 
parametrize the differential scattering cross section in terms of different non--perturbative 
nucleonic structure functions $F_i$, 
\cite{Drell:1963ej,Derman:1978iz}. They describe the structure of the 
corresponding hadron at a given  energy-transfer $\nu$ in its rest frame and the virtual 4-momentum-transfer from the 
lepton to the nucleon $q^2=-Q^2$. The spatial resolution reached is $\Delta x \sim 
1/\sqrt{Q^2}$. Based on current algebra techniques J. Bjorken predicted, that in the limit, 
$Q^2,~\nu~\rightarrow~\infty$,~~~$Q^2/\nu=$~{\sf fixed}, 
the structure functions do not depend on $Q^2$ and $\nu$ independently but on the Bjorken variable $x=Q^2/2M\nu$, 
\cite{Bjorken:1968dy}. Here $M$ is the mass of the hadron and $0 <x \leq 1$. 
This behaviour has been discovered in the {\sf SLAC-MIT} experiments for
nucleons,~\cite{Coward:1967au,Panofsky:1968pb,Bloom:1969kc,Breidenbach:1969kd} cf. also 
\cite{Kendall:1991np,Taylor:1991ew,Friedman:1991nq}. R.~Feynman provided a phenomenological 
explanation of this behaviour by creating the parton model, 
\cite{Feynman:1969wa,Feynman:1969ej,Feynman:1973xc}.
The strong correlation between $Q^2$ and $\nu$ in the Bjorken limit can only be understood if the nucleon
is composed out of point-like constituents, Feynman's {\it partons}, at short distances. Partons appear 
as
basically massless quantum fluctuations which live long enough to be resolved during the short interaction 
times $\tau_{\rm int} \sim 1/q_0$, cf. \cite{Drell:1970yt}, implying strict conditions for the validity of 
the parton model. The hadronic structure functions can then be understood as incoherent weighted sums over 
individual parton distribution functions $f_i(x)$ at lowest order.
The discovery, that the longitudinal structure function $F_L$ vanishes 
in the Bjorken limit, known as Callan-Gross relation \cite{Callan:1969uq},
showed that the quark-partons possess spin $1/2$. 
Feynman's parton model and Gell-Mann's quark model were finally joint by Bjorken and Paschos~\cite{Bjorken:1969ja}, 
who identified the partons as the quarks of the group theoretic approach.

In the 1950s C.N.~Yang and R.L.~Mills, \cite{Yang:1954ek}, studied gauge field theories based on 
non-abelian gauge groups. 
The non-abelian nature implies three- and quadrilinear couplings between the gauge bosons contrary 
to the case of Abelian groups, like in Quantum Electrodynamics.  The renormalizibility of field theories of this 
type was proven by G.~'t~Hooft~\cite{'tHooft:1971fh} in the massless case.
In 1972/1973 M. Gell-Mann, H. Fritzsch and 
H. Leutwyler,~\cite{Fritzsch:1973pi}, cf. also \cite{Nambu:1966}, proposed to describe quark 
interactions by a $SU(3)_c$ Yang-Mills theory. Quantum Chromodynamics (QCD) has been introduced 
as the dynamical theory of quarks and color--octet vector gluons as gauge bosons.
In 1973, D.~Gross and F.~Wilczek, \cite{Gross:1973id}, and H.~Politzer,~\cite{Politzer:1973fx}, proved 
that Quantum Chromodynamics is asymptotically free, i.e. the strong coupling decreases toward 
shorter distances or larger momentum transfer. This allows to perform perturbative computations 
for scattering processes at large enough momentum scales. 

A useful tool to apply perturbative methods in deeply inelastic scattering processes 
is the light-cone expansion (LCE)~\cite{Wilson:1969zs,Zimmermann:1970,Frishman:1971qn,Brandt:1970kg}. 
Applying it to deeply inelastic scattering processes leads to a factorization theorem, which separates 
hadronic bound state effects from short distance effects allowing for a perturbative analysis of
the operators at the respective twist-level. 
The long-range bound state effects are collected in the parton distribution functions, 
which have to be determined from experimental data or computed using non-perturbative methods. 
The short-distance effects are associated with the Wilson-coefficients and allow for a perturbative 
analysis. 

In 1975 logarithmic scaling violations in the deep inelastic cross section were 
discovered,~\cite{Chang:1975sv,Watanabe:1975su}, which are due to the fact that QCD is not a free 
field theory neither is it conformally invariant, \cite{Ferrara:1973eg}. These scaling violations 
were predicted perturbatively in higher order calculations and 
constitute a great success for the Standard Model, \cite{Gross:1973juxGross:1974cs,Georgi:1951sr}.
Beyond the established light-quark picture a fourth quark, {\it charm}, has been 
proposed~\cite{Maki:1964ux,Hara:1963gw,Bjorken:1964gz}. It was discovered in 1974 almost simultaneously at {\sf SLAC} 
and at Brookhaven National Laboratory ({\sf BNL}) through the $J/\psi$ meson.
In 1977 the $\Upsilon$ resonance was observed, \cite{Herb:1977ek}, which was interpreted as a meson 
consisting of a new quark, 
the bottom quark ($b$). The search for quarks continued and in 1995 the top quark was discovered
at {\sf TEVATRON}, \cite{Abe:1994xtxAbe:1994stxAbe:1995hr,Abachi:1995iq}.

Today QCD is established as the theory of the strong interaction and together with the electroweak 
$SU_L(2) \times U_Y(1)$ sector it forms the Standard Model of elementary particle physics. The electroweak 
sector of the Standard Model has been introduced by 
S.~Glashow,~\cite{Glashow:1961tr} and
S. Weinberg in 1967,~\cite{Weinberg:1967tq}, cf. also \cite{Salam:1964ry,Salam:1968rm}. 
G.~'t~Hooft and M.~Veltman proved that this theory is renormalizable, \cite{'tHooft:1972ue},~see also 
\cite{Taylor:1971ff,Slavnov:1972fg,Lee:1972fjxLee:1973fn}.

Deep-inelastic scattering provides a clear
method to probe the short distance  substructure of hadrons in the space--like domain. Many DIS experiments have been 
performed through the last forty years, 
\cite{Diemoz:1986kt,Eisele:1986uz,Sloan:1988qj,Mishra:1989jc,Winter:1991ua,Schmitz:1997}. 
The proton substructure has been examined 
intensely at {\sf HERA} at {\sf DESY} \cite{:1981uka,Abt:1993wz,Derrick:1992nw,Ackerstaff:1998av,Hartouni:1995cf}.  
In case of unpolarized deep-inelastic scattering via single photon exchange the cross section is described 
in terms of the structure functions $F_2(x,Q^2)$ and $F_L(x,Q^2)$.  
While $F_2$ has been measured over a wide kinematic range, $F_L$ has mainly been measured 
at fixed targets. Experimental data from {\sf HERA} showed that the structure functions receive a
substantial part due to charm quark pair production, beyond the light flavor contributions which amounts
25--35~\% in the small $x$ region, cf. 
e.g.~\cite{Thompson:2007mx,Lipka:2006ny,Chekanov:2008yd,Blumlein:1996sc}. 

The scaling violations of the DIS structure functions are determined by the anomalous dimensions and the
Wilson coefficients both for the light partons and heavy flavors.
The anomalous dimensions have been computed at 
leading order (LO),~\cite{Gross:1973juxGross:1974cs,Georgi:1951sr}, and next-to-leading-order (NLO)
\cite{Floratos:1977auxFloratos:1977aue1,Floratos:1978ny,GonzalezArroyo:1979df,GonzalezArroyo:1979he,
Curci:1980uw,Furmanski:1980cm,Hamberg:1991qt,Ellis:1996nn}. At next-to-next-to-leading-order (NNLO) 
fixed moments were obtained in 
Refs. ~\cite{Larin:1996wd,Retey:2000nq,Blumlein:2004xt,Larin:1993vu} before the general result in the Mellin 
variable $N$ was determined  in 2004 ~\cite{Moch:2004pa,Vogt:2004mw} by Vermaseren et al. 
A first independent check on the NNLO moments of the unpolarized anomalous dimensions was performed in 
Ref. \cite{Bierenbaum:2009mv}. The massless Wilson coefficients for the structure functions $F_2$ and $F_L$, resp.
their moments, were computed at first order in~\cite{Zee:1974du,Bardeen:1978yd,Furmanski:1981cw}, at second order in
~\cite{Duke:1981ga,Devoto:1984wu,Kazakov:1987jk,
Kazakov:1990fu,SanchezGuillen:1990iq,vanNeerven:1991nnxZijlstra:1991qcxZijlstra:1992qd,
Kazakov:1992xj,Larin:1991fv,Moch:1999eb}, and at third order 
in~\cite{Larin:1993vu,Larin:1996wd,Retey:2000nq,Moch:2004xu,
Blumlein:2004xt,
Vermaseren:2005qc}. 

Due to the large size of the charm quark contribution to the DIS structure functions the precise computation of these 
terms is very important. The leading order massive Wilson coefficients have been computed completely in 
the late 1970s,~\cite{Witten:1975bh,Babcock:1977fi,Shifman:1977yb,Leveille:1978px,Gluck:1980cp}. 
The NLO--corrections are available in semi-analytic 
form,~\cite{Laenen:1992zkxLaenen:1992xs,Riemersma:1994hv}. A precise numerical implementation was given 
in~\cite{Alekhin:2003ev}. At three-loop order the computation of the heavy flavor Wilson coefficients over 
the whole kinematic range appears to be rather difficult at present. However, a very important region for 
deep-inelastic experiments at {\sf HERA} can be covered by computing the massive Wilson coefficients for the structure 
function $F_2$ in the limit of $Q^2 \gg m^2$. There the heavy flavor Wilson 
coefficients factorize, cf.~Ref. \cite{Buza:1995ie}, 
into the process dependent light flavor Wilson coefficients 
$C_{(q,g),(2,L)}(x,Q^2/\mu^2)$ and the process independent operator matrix elements (OMEs) $A_{ij}(x,\mu^2/m^2)$. 
The OMEs contain all mass dependence and are obtained as matrix elements of the leading twist local composite 
operators between partonic states $\mid j \rangle~(i,j=q,g)$. For the structure function $F_2(x,Q^2)$ this
representation becomes effective for $Q^2 \gsim 10~m^2$, cf. \cite{Buza:1995ie}.
In the case of the longitudinal structure 
function $F_L$ this factorization theorem applies 
for much larger momentum transfers, $Q^2 \gsim 800~m^2$, only, \cite{Buza:1995ie}, which lays outside  
the kinematic 
region probed at {\sf HERA}. The NNLO corrections for the longitudinal structure function 
have been computed in Ref. ~\cite{Blumlein:2006mh}.

At two-loop order the quarkonic OMEs  $A_{qj}$ have been calculated analytically in Ref.~\cite{Buza:1995ie} and 
confirmed, applying rather different methods, in Ref.~\cite{Bierenbaum:2007qe}. The contributions linear in the 
dimensional parameter $\ep$ were calculated in Ref.~\cite{Bierenbaum:2008yu}.
The remaining gluonic matrix elements $A_{gj}$ 
have been derived in Ref.~\cite{Buza:1996wv}. They were confirmed and extended to $O(\ep)$ 
in~\cite{Bierenbaum:2009zt}. The gluonic operator matrix elements are required to 
describe parton distribution functions in the variable flavor number scheme (VFNS). Furthermore they contribute to 
the NNLO quarkonic singlet OMEs through renormalization. The complete renormalization procedure for massive 
3-loop OMEs was developed in Ref.~\cite{Bierenbaum:2009mv}. There also the general structure of the NNLO OMEs 
was derived. In Ref.~\cite{Bierenbaum:2009mv} a large amount of Mellin moments for all contributing massive OMEs
were computed. All logarithmic terms $\propto \ln^k(Q^2/m^2),~~k=3,2,1$ are known in full analytic form 
by now.
The mathematical structure of these results is determined by nested harmonic sums 
\cite{Blumlein:1998if,Vermaseren:1998uu}.
Previous analyses of known results of different single--scale hard scattering processes have shown, that 
at least up to 
massless three-loop order calculations, single scale results are most simply expressed through nested 
harmonic sums, \cite{Blumlein:2004bb,Moch:2004pa,Vogt:2004mw,Vermaseren:2005qc,
Dittmar:2005ed,Blumlein:2005im,Blumlein:2006rr,Blumlein:2007dj}. 
They obey algebraic, 
\cite{Blumlein:2003gb}, and structural relations, 
\cite{Blumlein:2009ta,Blumlein:2009fz}. If one considers 
fixed moments only, one obtains representations in terms of multiple 
zeta values, \cite{Borwein:1999js,Blumlein:2009cf}.

In the present computation Feynman diagrams are evaluated by direct integration leading to a
representation
in terms of generalized hypergeometric functions~\cite{Slater,Bailey}, cf. also \cite{Blumlein:2009ta}.
In the case
of massless computations summation algorithms
as in Refs. \cite{Vermaseren:1998uu,Weinzierl:2002hv,Moch:2005uc} can be applied. In the massive
case, various additional infinite and finite sums occur which possess a much more involved structure.
These sums can be treated applying  modern summation technologies encoded in the
package~\SigmaP,~\cite{Refined,Schneider:2007,sigma1,sigma2}, 
written in ${\sf MATHEMATICA}$~\cite{MATHEMATICA}.

In Refs.~\cite{Buza:1995ie,Buza:1996wv,Blumlein:2006mh,
Bierenbaum:2007qe,Bierenbaum:2008yu,Bierenbaum:2009zt,Bierenbaum:2009mv}  all necessary calculations
to describe the massive Wilson coefficients in the asymptotic region to 2--loop 
order and to determine
all quantities which have to be known to renormalize these quantities at 3--loop order have been performed. Thereby all 
logarithmic contributions are known at general values of the Mellin variable $N$. For an essential 
piece~\footnote{For numerical studies of this aspect see~\cite{BBK10A}.} in the constant part of the 
renormalized OMEs the general $N$--dependence is not known yet. A series of Mellin-moments has been 
computed in Ref.~\cite{Bierenbaum:2009mv}. The general $N$ result is of numerical importance and 
has therefore to be computed exactly. This applies also to its small-$x$ behaviour which may cause
large effects.

In this thesis we perform a first step within this larger programme and compute the $O(a_s^3 T_F^2n_f)$ 
contribution to the massive operator 
matrix elements $A_{Qg}$, $A_{Qq}^{\sf PS}$, $A_{qq,Q}^{\sf PS}$, $A_{qq,Q}^{\sf NS}$ 
and $A_{qq,Q}^{\sf TR}$ for general values of the Mellin variable $N$. Due to the large fraction of the
heavy flavor contributions to the deep--inelastic structure functions the precise knowledge of the 
respective Wilson coefficients is of essential importance to consistently derive  the parton 
distribution functions at leading twist for the gluon, the valence quarks, and for the different
sea--quark species, along with a precision measurement of the strong coupling constant 
$\alpha_s(M_Z^2)$, cf. e.g. \cite{Alekhin:2009ni}. $\alpha_s(M_Z^2)$ in itself is one of the
fundamental quantities in nature and has to be known as precisely as possible.
Note that, despite of a large number of precision 
analyses using different high--energy observables, a final agreement on the value of $\alpha_s(M_Z^2)$,
being measurable with an accuracy of $\sim 1\%$ at present, could not be obtained yet, 
\cite{Bethke:2009jm}.
Needless to say that {\it all} measurements at the Large Hadron Collider {\sf LHC} at {\sf CERN} 
crucially rely on both the precision knowledge of the parton distribution functions 
and $\alpha_s(M_Z^2)$. Furthermore, $\alpha_s(M_Z^2)$ is an essential input-parameter in scenarios
of the potential unification of the fundamental forces of the strong--, weak--, and electromagnetic
interactions, cf.~\cite{Georgi:1974sy,Fritzsch:1974nn}. Its value is decisive for the question whether,
and in which theory, the fundamental forces of nature unify at high energy scales or not. This, in 
turn, touches the respective scenarios of the physics in the early universe, and is thus also 
connected 
to the major challenging problems in physics.

The outline of this thesis is as follows. In Section~2 we describe the basic high--energy 
process, deeply--inelastic lepton--nucleon scattering, to which the QCD corrections, which are 
calculated, belong. This includes the QCD-improved parton model at short distances, which is 
established through the light-cone expansion. The heavy flavor contributions to the deep--inelastic
structure functions can be viewed as a linear contribution in addition to the light parton 
contributions. The
leading order corrections in the strong coupling constant are re-calculated in Section~3.
We then discuss the general scenario which allows the {\it analytic} computation of the heavy flavor 
corrections to higher orders in the asymptotic region $Q^2 \gg m^2$ in terms of massive OMEs and 
the massless Wilson coefficients. The formalism is outlined to 3--loop orders in 
Section~4. In Section~5 the details of the renormalization
of the massive OMEs are summarized to 3--loop orders. The leading order massive operator matrix element
is then re-calculated in Section~6 as an introductory calculation, in which we discuss
the asymptotic factorization theorem for the heavy flavor Wilson coefficients following the formalism
of Section~4 and compare to the explicit calculation in Section~3.
We then turn to the computation of the analytic $O(a_s^3 T_F^2 N_f C_{F,A})$ contributions to
five heavy flavor OMEs at general values of the Mellin variable $N$ in Section~7. 
We describe the contributing Feynman diagrams, give details of their analytic evaluation, 
and illustrate the methods by different explicit examples. In Section~8
we present the analytic results for the OMEs 
$A_{Qg}^{(3)},
 A_{Qq}^{\sf PS, \rm(3)},
 A_{qq,Q}^{\sf PS, \rm(3)},
 A_{qq,Q}^{\sf NS, \rm(3)}$ and 
$A_{qq,Q}^{\sf NS, TR \rm(3)}$.
The new results are the constant parts of the respective unrenormalized 3--loop OMEs and a {\it
first} independent recalculation of the corresponding contributions to the 3--loop anomalous 
dimensions. 
This calculation generalizes results obtained for fixed integer moments in 
Ref.~\cite{Bierenbaum:2009mv} to general values of $N$, required by the experimental analyses, for the 
first time.
The final results are expressed in terms of nested harmonic sums, although intermediary results require
the treatment of generalizations thereof, which finally cancel for the class of graphs computed.
In Section~9 we summarize the main results. A series of technical details of the 
present computation is given in the Appendix. Basic conventions are summarized in 
Appendix~A. A consistent set of Feynman rules for QCD, including those of twist--2 composite 
operators, is given in Appendix~B. Relations for $D$--dimensional momentum integrals 
are summarized in Appendix~C. In Appendix~D the results for the individual 
diagrams, which were computed in this thesis, are given for reasons of documentation, to 
allow other groups for comparison, and to discuss their mathematical structure. Useful variable 
transformations are summarized in Appendix~E.
The present computation
relies on the use of special higher transcendental functions, which allow a particularly compact 
treatment. Main results and relations of these functions, as Euler-integrals, generalized 
hypergeometric functions, as well as harmonic sums, are given in Appendix~F. The final 
results of the present work were obtained summing multiply nested sums of the hypergeometric type 
and their extensions,
which 
are 
of a sophisticated nature. They could be uniquely solved using general modern summation technologies
encoded in C. Schneider's programme package {\sf SIGMA}~\cite{Refined,Schneider:2007,sigma1,sigma2}. 
As an illustration for several
thousands of sums which had to be computed we show  a few examples in 
Appendix~G. Partly they contain a large amount of also generalized harmonic sums. In 
Appendix~H reference values for the moments of the 3--loop anomalous dimensions 
and the various constant parts $a_{ij}^{(3)}$ of the massive OMEs, being computed in 
\cite{Bierenbaum:2009mv}, are summarized. These values were used to test the results of the present
calculation.

\newpage
\section{\boldmath Deeply inelastic scattering\label{DIS}}
Deep-inelastic scattering (DIS) denotes the scattering process of highly energetic leptons off hadrons, 
and provides a very precise method to probe the substructure of hadrons at short space-like distances. 
The 4-momentum transfer $q^2=-Q^2$ is at least of the order $Q^2 \geq 4 \GeV^2$, such that space-like 
distances of approximately $1/\sqrt{Q^2}$ can be resolved. Different deeply inelastic scattering experiments exploring charged and neutral current reactions allow to probe the flavor and the gluonic structure of the hadron. By performing polarized scattering experiments, also the spin-structure of hadrons can be investigated.
\begin{figure}[H]
\centering
\includegraphics[scale=0.9]{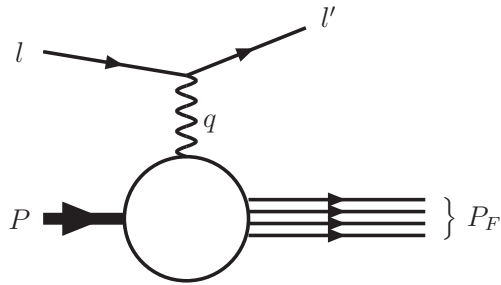}
\caption{\sf Schematic diagram of deeply inelastic scattering via single boson exchange}
\label{DIS-diag}
\end{figure}

\subsection{Kinematics}
The schematic diagram of deep-inelastic scattering at tree level is shown in Figure \ref{DIS-diag}. A lepton with momentum $l$ is scattered of a nucleon with momentum $P$ and mass $M$  via a virtual vector boson. In this process the nucleon state disintegrates and $F$ denotes a new linear combination of hadronic final states with allowed quantum numbers. The momenta of the outgoing lepton and hadronic states are denoted  by $l'$ and $P_F$, respectively. The 4-momentum $q$ of the virtual vector boson is space-like and the virtuality $Q^2$ is defined by
     \begin{eqnarray}
      Q^2&\equiv&-q^2~,\quad  q=l-l'~. \label{virtuality}
     \end{eqnarray}
Two more independent kinematic variables are sufficient to determine the scattering process:
     \begin{eqnarray}
      s  &\equiv&(P+l)^2~, \label{sdeep} \\
      W^2&\equiv&(P+q)^2=P_F^2~. \label{pdeepf}
     \end{eqnarray}
Here $s$ denotes the total center of momentum energy squared and $W$ is the invariant mass of the final hadron state $\bra{P_F}$. The kinematic variables can be measured from the final lepton or hadronic states, depending on the specific experiment, cf. e.g. \cite{Blumlein:1992we,Blumlein:1994ii,Arbuzov:1995id}. In the following analysis we will neglect the lepton mass and describe the process using the following Lorentz-invariant kinematic variables:
     \begin{eqnarray}
      \nu&\equiv&\frac{P.q}{M}~
             \,\,\,
             =~\frac{W^2+Q^2-M^2}{2M}~, 
               \label{nudef} \\
      x  &\equiv&\frac{-q^2}{2P. q}~
             =~\frac{Q^2}{2M\nu}~
             \,\,\,\,\,
             =~\frac{Q^2}{W^2+Q^2-M^2}~, 
               \label{Bjorkenx} \\
      y  &\equiv&\frac{P.q}{P.l}\hspace{3mm}
             =~\frac{2M\nu}{s-M^2}~
             =~\frac{W^2+Q^2-M^2}{s-M^2}~.
               \label{Bjorkeny}
     \end{eqnarray}
Here $\nu$ is the total energy transfer in the rest frame of the nucleon, $x$ is a Bjorken variable and 
$y$ is the inelasticity, cf.~\cite{Bjorken:1969mm}. In general, the momentum $q$ is transferred via 
the exchange of a $\gamma, Z, W^{\pm}$--boson. In this thesis we limit the investigation to photon 
exchange in unpolarized charged lepton-nucleon scattering. 
For not too large virtualities, i.e. $Q^2 \lsim 500~\GeV^2$, single photon exchange dominates this 
reaction, cf.~\cite{Blumlein:1987xk}. Thus from now on this region will be considered only and weak gauge boson effects caused by the exchange of a $Z$-Boson may be disregarded.
The invariant hadronic mass obeys the condition 
     \begin{eqnarray}
      W^2 \ge M^2~. \label{physreg}
     \end{eqnarray}
From (\ref{physreg}) one obtains
     \begin{eqnarray}
      &&W^2=(P+q)^2=M^2+Q^2\left(\frac{1}{x}-1\right) \ge M^2 ~.\label{xregion}
     \end{eqnarray}
Thereby $x$ is limited to the region
\begin{eqnarray}
0\leq x\leq 1~.
\end{eqnarray}
For $x=1$ the process is elastic, whereas $x<1$ describes the inelastic region \cite{Povh:1993jx}. By considering the nucleon's rest-frame and demanding a positive energy-transfer, we obtain further restrictions on the kinematic variables:
     \begin{eqnarray}
      \nu \ge 0~, \quad 0\le y \le 1~, s\ge M^2~. \label{physreg2}
     \end{eqnarray}
The cross section for $ep$-scattering is determined by the transition matrix element for the electromagnetic current. In Born approximation it is given by 
     \begin{eqnarray}
      M_{fi} 
            =e^2\overline{u}(l',\eta')\gamma^{\mu}u(l,\eta)
             \frac{1}{q^2}\bra{P_F}J^{em}_{\mu}(0)\ket{P,\sigma}~,\label{mfi}
     \end{eqnarray}
cf. e.g. \cite{Reya:1979zk,Roberts:1990ww,Muta:1998vi}. Here $\eta(\eta')$ and $\sigma$ denote the spin components of the leptons and gluons, respectively. The initial and final hadron states are denoted by $\ket{P,\sigma}$ and $\bra{P_F}$, respectively. $\gamma_{\mu}$ denotes the Dirac-matrices and $u (\overline{u})$ are the bi-spinors of the electron and its conjugate, respectively, see Appendix \ref{App-Con}. Furthermore $e$ denotes the electric charge and $J_{\mu}^{em}(\xi)$ is the quarkonic part of the electromagnetic current operator:
     \begin{eqnarray}
        J_{\mu}^{\dagger}(\xi)=J_{\mu}(\xi)~.\label{jself} 
     \end{eqnarray}
For QCD, the electromagnetic current is given by
     \begin{eqnarray}
      J^{em}_{\mu}(\xi)=\sum_{f,f'} \overline{\Psi}_f(\xi)\gamma_{\mu}
                        \lambda^{em}_{ff'}\Psi_{f'}(\xi)~, \label{current}
     \end{eqnarray}
where $\Psi_f(\xi)$ denotes the quark field of flavor $f$. $\lambda^{em}_{ff'}$ describes the electromagnetic charges 
of the different quark flavors. For three light quark flavors it is given by
     \begin{eqnarray}
      \lambda^{em}=\frac{1}{2}\Bigl(\lambda_{flavor}^3
                   +\frac{1}{\sqrt{3}}\lambda_{flavor}^8
                   \Bigr)~, \label{lambdaem} 
     \end{eqnarray}
where the $\lambda_{flavor}^i$ are the Gell-Mann matrices of the flavor group $SU(3)_{flavor}$, cf. \cite{Blumlein:1999sc,Yndurain:1999ui}. The unpolarized cross section is obtained by averaging over leptonic and hadronic spin degrees of freedom. The differential cross section, cf. ~\cite{Reya:1979zk,Field:1989uq,Roberts:1990ww,Muta:1998vi}, reads: 
     \begin{eqnarray}
      l_0'\frac{d\sigma}{d^3l'}=\frac{1}{32(2\pi)^3(l.P)}
                                \sum_{\eta',\eta ,\sigma ,F}
                                (2\pi)^4\delta^4(P_F+l'-P-l)
                                |M_{fi}|^2~. \label{scatcro}
     \end{eqnarray} 
Inserting the transition matrix element shows, that the cross section can be decomposed into a tensor $L_{\mu\nu}$ depending only on the leptonic states and a purely hadronic tensor $W_{\mu\nu}$ with
     \begin{eqnarray}
      L_{\mu \nu}(l,l')&=&\sum_{\eta',\eta}
                          \Bigl[\overline{u}(l',\eta')\gamma^{\mu}u(l,
                          \eta)\Bigr]^*
                          \Bigl[\overline{u}(l',\eta')\gamma^{\nu}u(l
                          ,\eta)\Bigr]~, \label{leptontens}\\
      W_{\mu\nu}(q,P)&=&\frac{1}{4\pi}\sum_{\sigma ,F}
                        (2\pi)^4\delta^4(P_F-q-P)
                        \bra{P,\sigma}J^{em}_{\mu}(0)\ket{P_F} 
                        \bra{P_F}J^{em}_{\nu}(0)\ket{P,\sigma}~. \N\\ 
                        \label{hadrontens}
     \end{eqnarray}
In terms of these quantities the cross section reads
     \begin{eqnarray}
      l_0'\frac{d\sigma}{d^3l'}&=&\frac{1}{4 P.l}
                                  \frac{\alpha^2}{Q^4}
                                  L^{\mu\nu}W_{\mu\nu}
                                =\frac{1}{2(s-M^2)}
                                  \frac{\alpha^2}{Q^4}
                                  L^{\mu\nu}W_{\mu\nu}~.\label{crosssec}
     \end{eqnarray}
Here $\alpha$ is the fine structure constant. The leptonic tensor can be computed easily by applying the conventions in Appendix \ref {App-Con}. One obtains      
\begin{eqnarray}
      L_{\mu \nu}(l,l')&=&Tr[l \hspace*{-1.3mm}/ \gamma^{\mu}
                        l' \hspace*{-1.7mm}/ \gamma^{\nu}]
                        =4\left(l_{\mu}l_{\nu}'+l_{\mu}'l_{\nu}-\frac{Q^2}{2}
                        g_{\mu \nu}\right)~. \label{leptontens2}
     \end{eqnarray}
The hadronic tensor cannot be evaluated purely perturbatively due to the non-perturbative nature of the matrix elements\footnote{Ab initio calculations would have to be based on lattice QCD methods. During the last years, an increased numerical precision has been achieved in this field,~cf.~e.g.~\cite{Dolgov:2002zm,Gockeler:2007qs,Baron:2007ti,Bietenholz:2008fe,Syritsyn:2009np,Renner:2010ks}}.
Using the integral representation of the $\delta$-distribution and applying elementary quantum mechanical identities, Eq. (\ref{hadrontens}) can be rewritten as, cf. \cite{Itzykson:1980rh,Muta:1998vi}, 
     \begin{eqnarray}
      W_{\mu\nu}(q,P)
                     &=&\frac{1}{4\pi}\sum_{\sigma}
                        \int d^4\xi\exp(iq\xi)
                        \bra{P}[J^{em}_{\mu}(\xi),
                        J^{em}_{\nu}(0)]\ket{P} \N\\
                     &=&\frac{1}{2\pi}
                        \int d^4\xi\exp(iq\xi)
                        \bra{P}[J^{em}_{\mu}(\xi),
                        J^{em}_{\nu}(0)]\ket{P}~. \label{hadrontens4}
     \end{eqnarray}
Here the bracket $[a,b]$ denotes the commutator of $a$ and $b$. The hadronic tensor obeys various symmetry and conservation laws, cf. \cite{Tangerman:1994eh}. These impose conditions on the Lorentz structure of the hadronic tensor and allow to parametrize it by different scalar structure functions. They contain all information about the structure of the proton. In the general case 14 independent structure functions exist,~\cite{Blumlein:1996vs,Blumlein:1998nv}, but in the case of unpolarized DIS via single photon exchange only two structure functions contribute. 
Here the leptonic tensor (\ref{leptontens}) is symmetric. Since any tensor of rank--2 can be decomposed 
into a 
symmetric and an anti-symmetric part, only the symmetric part of the hadronic tensor contributes. 
Thus the hadronic tensor must be a linear combination of the following tensors
     \begin{eqnarray}
      g_{\mu\nu}~,\quad q_{\mu}q_{\nu}~,\quad P_{\mu}P_{\nu}~,\quad
      q_{\mu}P_{\nu}+q_{\nu}P_{\mu}~. \label{Ansatz}
     \end{eqnarray} 
From the conservation of the electromagnetic current, 
\begin{eqnarray}
      \partial_{\mu}J^{em}_{\mu}(\xi)=0~,
     \end{eqnarray}
Lorentz- and time-reversal invariance it follows that 
     \begin{eqnarray}
      q_{\mu}W^{\mu\nu}=0~.
     \end{eqnarray}
Furthermore strong interactions preserve CP-invariance, cf. \cite{Tangerman:1994bb}.
Making a general ansatz in terms of (\ref{Ansatz}) and imposing gauge invariance leads to the following representation of $W_{\mu\nu}$, containing the two structure functions $F_L$ and $F_2$
     \begin{eqnarray}
      W_{\mu \nu}(q,P)=&&
                       \frac{1}{2x}\left(g_{\mu \nu}+\frac{q_{\mu}q_{\nu}}{Q^2}
                  \right)F_{L}(x,Q^2) \N\\
                  &+&\frac{2x}{Q^2}\left(
                   P_{\mu}P_{\nu}+\frac{q_{\mu}P_{\nu}+q_{\nu}P_{\mu}}{2x}
                   -\frac{Q^2}{4x^2}g_{\mu\nu}\right)F_{2}(x,Q^2)~.
      \label{hadrontens2}
     \end{eqnarray} 
Due to the hermiticity of the hadronic tensor, the structure functions $F_2(x,Q^2)$ and $F_L(x,Q^2)$ are real functions. Their arguments are Bjorken-$x$ and $Q^2$, whereas in the elastic case the cross section is determined by one kinematic variable, e.g. the total energy transfer, only. The differential cross section in terms of the structure function is obtained by inserting (\ref{hadrontens2}) into (\ref{crosssec}):
\begin{eqnarray}
      \frac{d\sigma}{dxdy}=\frac{2\pi\alpha^2}{xyQ^2}
                           \Bigg\{\Bigl[1+(1-y)^2\Bigr]F_2(x,Q^2)
                                  -y^2F_L(x,Q^2)\Biggr\}~.\label{crosssec1}
\end{eqnarray}
The two structure functions $F_2$ and $F_L$ can be extracted from (\ref{hadrontens2}) by applying the following projectors in $D$ dimensions:
 \begin{eqnarray}
      F_L(x,Q^2)&=&\frac{8x^3}{Q^2}P^{\mu}P^{\nu}W_{\mu\nu}(q,P)~,\N\\
      F_2(x,Q^2)&=&
                   \frac{2x}{D-2}\Biggl[
                   (D-1)\frac{4x^2}{Q^2}P^{\mu}P^{\nu}W_{\mu\nu}(q,P)
                   -g^{\mu\nu}W_{\mu\nu}(q,P)\Biggr]~.\label{project}
     \end{eqnarray}
Here and in the following we neglect target mass corrections and thereby set $P^2=0$. 

\subsection{The Parton Model}
In general, the structure functions in (\ref{hadrontens2}) depend on two kinematic variables $Q^2$ and $x$. However, in the Bjorken limit \cite{Bjorken:1968dy} ${Q^2,\nu}\rightarrow \inf, ~x~~\sf{fixed}$, the structure function depends on $x$ only,
  \begin{eqnarray}
      \lim_{\{Q^2,~\nu\}~\rightarrow~\infty,~x=const.} 
            F_{(2,L)}(x,Q^2)=F_{(2,L)}(x)~,
      \label{scaling}
     \end{eqnarray}
which is called Bjorken scaling. Experimental observations from electron--proton collisions performed at {\sf SLAC} in 1968,
\cite{Coward:1967au,Panofsky:1968pb,Bloom:1969kc,Breidenbach:1969kd}, confirmed the existence of an approximate scaling behaviour and  thereby supported Bjorken's predictions. Furthermore it was shown experimentally, that the cross section remained large at high momentum transfers $Q^2$. This behaviour indicates point-like particles in the target, as no further substructure could be probed with increasing 
momentum. The earlier picture according to which the size of the proton was about about $10^{-13}~$cm  
with a 
smooth charge distribution, \cite{Mcallister:1956ng,Schopper:1961,Hofstadter:1963}, which is valid at 
lower momentum transfer, was superseded. Feynman solved this evanescent contradiction by introducing 
the 
parton model, \cite{Feynman:1969wa,Feynman:1969ej}, cf. also 
\cite{Bjorken:1969ja,Feynman:1973xc,Roberts:1990ww,Reya:1979zk,Kogut:1972di,Yan:1976np}. At large enough scales the proton is a composite object, consisting of several point-like particles, the partons. During the interaction time with the virtual photon, the long-lived partons behave as quasi--free particles. The photon scatters elastically of a single parton, while the other partons act as ``spectators'' and do not interfere with the process. Thus the total cross section is given by the incoherent sum of the 
individual parton-photon cross sections, weighted by the probability to find a specific parton $i$ with a momentum fraction $z$ inside the proton. Feynman introduced this probability as parton distribution function (PDF), $f_i(z)$. From now on we will use the collinear parton model, according to which the momenta of the specific parton $p$ is taken to be collinear to the nucleon momentum $P$, 
\begin{eqnarray}
p=z P~.\label{MomFrac}
\end{eqnarray} 
Analogously to the scaling variable $x$ one may define a partonic scaling variable $\tau$, 
     \begin{eqnarray}
      \tau\equiv\frac{Q^2}{2 p.q}~. \label{taudef}
     \end{eqnarray}
Combined with (\ref{MomFrac}) one obtains
     \begin{eqnarray}
      \tau z=x~.
     \end{eqnarray}
Feynman's original approach, the naive parton model, neglected the radiative corrections. Its main issue consists in the strict correlation  
     \begin{eqnarray}
       \delta\left(\frac{q.p}{M}-\frac{Q^2}{2M}\right)~.\label{feyncor}
     \end{eqnarray}
This condition implies $z=x$. Furthermore, according to this model protons are always composed of two 
$u$ and one $d$ valence quarks. With the advent of QCD this model was modified and also virtual quark states and gluons were incorporated as additional partons. This more advanced model is known as QCD-improved parton model and can be derived by applying the  light-cone expansion.
Here the assumption is made, that the hadronic tensor factorizes into the parton distribution functions (PDFs) and a partonic tensor ${\cal W}_{\mu\nu}^i$~:
     \begin{eqnarray}
      W_{\mu\nu}(x,Q^2)=\frac{1}{4\pi}\sum_i \int_0^1 dz
                        \int_0^1 d\tau \left[f_i(z)+f_{\overline{i}}(z)
                        \right]{\cal W}_{\mu\nu}^i(\tau,Q^2)
                        \delta(x-z \tau)~. \label{hadrontens8}
     \end{eqnarray}
The partonic tensor is given by (\ref{hadrontens4}), where the hadronic states $\bra{P}$ are substituted with the corresponding partonic state $\bra{p}$ of the struck parton. $f_{\overline{i}}(z)$ denotes the PDF of the respective anti-parton to parton $i$. Assuming that the electromagnetic parton current takes the following form
     \begin{eqnarray} 
      \bra{i}j^i_{\mu}(\tau)\ket{i}=-ie_i\overline{u}^i\gamma_{\mu}u^i~,
     \end{eqnarray}
where $e_i$ is the charge of the respective parton $i$, one obtains
     \begin{eqnarray}
      {\cal W}^i_{\mu\nu}(\tau,Q^2)=\frac{2\pi e_i^2}{q. p^i}\delta(1-\tau)
                             \Bigl[2p^i_{\mu}p^i_{\nu}+p^i_{\mu}q_{\nu}
                               +p^i_{\nu}q_{\mu}
                             -g_{\mu\nu}q.p^i\Bigr]~.\label{partontensLO}
     \end{eqnarray}
Combining the $\delta$-distributions in (\ref{partontensLO}) and (\ref{hadrontens8}) leads to Feynman's assumption of the naive parton model: $z=x$.
Applying the projectors (\ref {project}) to the hadronic tensor (\ref{hadrontens8}) yields the following structure functions at lowest order~:
     \begin{eqnarray} 
      F_L(x,Q^2)&=&0~,\N\\
      F_2(x,Q^2)&=&x\sum_ie^2_i\left[f_i(x)+f_{\overline{i}}(x)\right]~.
      \label{resfeynLO}
     \end{eqnarray} 
The parton distribution functions are determined from DIS-world data analyses by different groups. 
Currently 
they are known to NNLO, i.e. at $O(\alpha_s^3)$, in the unpolarized case, cf. Refs.~\cite{
Martin:2009iq,Alekhin:2009ni,JimenezDelgado:2008hf}.

 \newpage
\section{\boldmath{Calculation of the Wilson coefficients 
${H_{(2,L),g}^{(1)}}$}\label{WilsonLO}}
In the following we calculate the leading order massive Wilson
coefficients contributing to the structure functions $F_2(x,Q^2)$ and $F_L(x,Q^2)$ for
pure photon exchange. It is given by the Bethe-Heitler fusion 
process of virtual photon-gluon scattering:
\begin{equation}
\gamma^* + g \rightarrow Q + \bar{Q}
\end{equation}
\begin{figure}[H]
\centering
\includegraphics[scale=0.7]{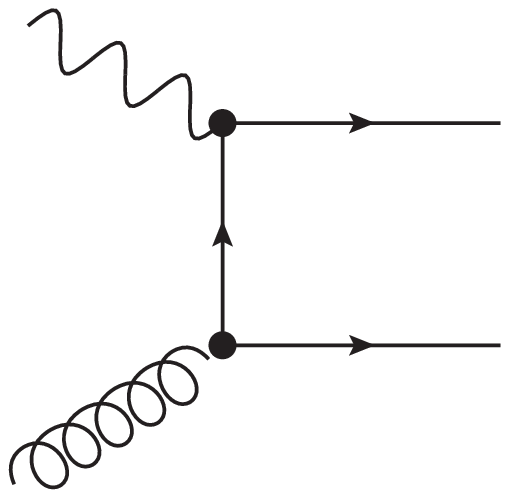}
\includegraphics[scale=0.6]{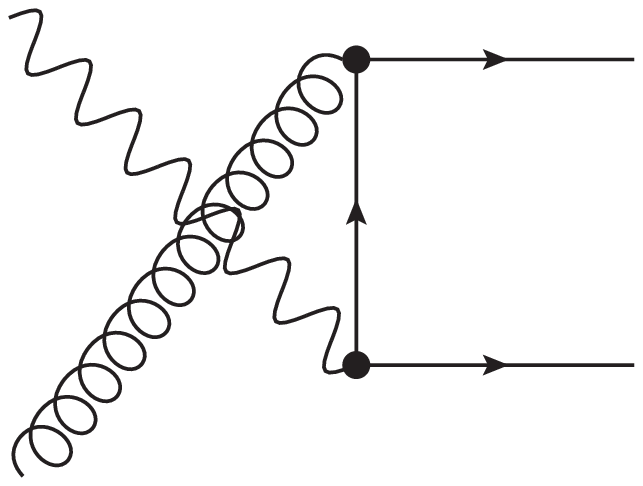}
\caption{\sf Feynman diagrams contributing to the leading order heavy flavor Wilson coefficient}
\label{Bethe-Heitler}
\end{figure}

{\noindent}The Feynman diagrams for the lowest order contributions to this process are shown in Figure~\ref {Bethe-Heitler}. 
The calculation is performed referring to the collinear parton model, i.e. the partons are assumed to act 
like free particles during the interaction and their momentum is parallel to the proton momentum,
\begin{equation}
k= z P~.
\end{equation}
Here $P$ denotes the proton momentum and $k$ is the gluon momentum.
The cms velocity $v$ of the outgoing heavy quarks of mass $m$ is then given by 
\begin{equation}
v = \left (1 - 4 \frac {m^2} {Q^2} \frac {\tau} {1-\tau} \right)^{1/2}\,,
\end{equation}
where $Q^2$ denotes the negative squared momentum of the space-like photon and $\tau$ is defined as the ratio of the 
Bjorken-variable $x$ and the momentum fraction $z$,
\begin{equation}
\tau:=\frac x z~.
\end{equation}
Applying the Feynman rules, cf. Appendix \ref{App-FeynRules}, to the diagrams in Figure \ref{Bethe-Heitler} 
yields the matrix element
\begin{equation}
M_{\mu\nu}=\bar{u}(p_2) i g_s \gamma_\mu i \frac {\slashed{p_{1}} - \slashed{q}+m} {\left(p_1-q\right)^2-m^2} 
i e_q \gamma_\nu v(p_1)+\bar{u}(p_2)i gs \gamma_\mu i \frac {\slashed{p_{1}} - \slashed{k}+m} 
{\left(p_1-k\right)^2-m^2} i e_q \gamma_\nu v(p_1) ~.
\end{equation}
The massive Wilson coefficients are obtained as projections of the squared matrix element 
\begin{equation}
H^{\mu\nu\rho\sigma}=M^{\mu\nu}M^{*\rho\sigma}~.
\end{equation}
Here the sum over the gluon polarization states is performed by contraction with $-g_{\mu\rho}$. 
The average over the spin directions gives an additional factor of
$1/(D-2)$, with $D$ the space-time dimension. The massive Wilson coefficients for $D=4$ read 

\begin{eqnarray}
H_{2,g}^{(1)}\left(\tau,\frac {Q^2} {m^2}, \frac {m^2} {\mu^2}\right)&=&\frac {T_F} {2} g_{\mu\rho} \int dR_2 
\left(g_{\nu\sigma} - 12 \frac {\tau^2} {Q^2} k_{\nu} k_{\sigma}\right) H^{\mu\nu\rho\sigma}~,
\\H_{L,g}^{(1)}\left(\tau,\frac {Q^2} {m^2}, \frac {m^2} {\mu^2}\right)&=&\frac {T_F} {2} g_{\mu\rho} \int dR_2 
\left( - 8 \frac {\tau^2} {Q^2} k_{\nu} k_{\sigma}\right) H^{\mu\nu\rho\sigma} ~.
\end{eqnarray}
Here $T_F$ is a color factor which is 
given by $1/2$ for $SU(N)$, see Appendix \ref{App-Con}. The phase space integral $\int dR_2$ in this case is given by, see e.g. \cite {Reya:1979zk},
\begin{equation}
\int dR_2 = \frac {1} {16 \pi} \frac {1} {\sqrt{s} p_{cm}}\int\limits_{-s-Q^2}^0 dt~,
\end{equation}
where 
\begin{eqnarray}
p_{cm}&=&\frac{1} {2 \sqrt{s}} \lambda^{1/2}\left(s,0,-Q^2\right)=\frac {1} {2 \sqrt{s}} \lambda^{1/2}\left(s,m^2,m^2\right)
\end{eqnarray}
denotes the modulus of the center-of-mass 3-momentum of the initial (final)
state particles and $s$, $t$ the are Mandelstam variables of the $2\rightarrow2$ process \cite{Byckling}, with
\begin{eqnarray}
\lambda(x,y,z)&=&(x-y-z)^2-4yz~.
\end{eqnarray}
Performing the integration yields :

\begin{align}
\begin{split}
H_{2,g}^{(1)}\left(\tau,\frac {Q^2} {m^2}, \frac {m^2} {\mu^2}\right) & = 8 T_F \left\{v \left[-\frac{1} {2} + 4 \tau 
- 4 \tau^2 + 2 \frac {m^2} {Q^2} (\tau^2-\tau) \right] \right.
\\ & \quad \left. +\left[-\frac 1 2 + \tau - \tau^2 + 2 \frac {m^2} {Q^2} (3\tau^2 - \tau) + 4 \frac {m^4} {Q^4}~,
\tau ^2\right] \ln \left(\frac {1-v} {1+v} \right)\right\} \label{H2g}
\end{split}
\\H_{L,g}^{(1)}\left(\tau,\frac {Q^2} {m^2}, \frac {m^2} {\mu^2}\right)&=16 \: T_F  \left[ v \: \tau (1-\tau) +
 2 \frac {m^2} {Q^2} \tau^2 \ln \left( \frac {1-v} {1+v} \right)
\right ]~.
\label{HLg}
\end{align}
The heavy quark coefficient functions (\ref{H2g}) and (\ref{HLg})
differ from the results of Ref. \cite{Gluck:1980cp} by a factor of $\tau/Q^2$, which
is due to different definitions of the structure functions $F_2$ and $F_L$. In the asymptotic limit $Q^2{\gg}~m^2$  the cms velocity $v$ is
written as an expansion in $(m^2/Q^2)$
\begin{eqnarray}
v=1-\frac{2 m^2} {Q^2} \frac {\tau} {1-\tau} + O\left(\frac {m^4}
  {Q^4} \right)~.
\end{eqnarray}
In this limit the logarithms in (\ref{H2g}) and (\ref{HLg}) become
\begin{eqnarray}
\ln\left(\frac {1-v} {1+v} \right) = \ln\left(\frac {m^2} {Q^2}\right)
+ \ln\left(\frac{\tau} {1-\tau}\right)+O\left(\frac{m^2}{Q^2}\right)~.
\end{eqnarray}
The expansion  of (\ref{H2g}) and (\ref{HLg}) yields
\begin{eqnarray}
     H^{(1)}_{2,g}\Bigg(\tau,\frac{Q^2}{m^2},\frac{m^2}{\mu^2}\Biggr)
                              &=&4T_F \cdot 
                                 \Biggl\{ 8\tau(1-\tau)-1+[\tau^2+(1-\tau)^2]
                                 \ln\Biggl(\frac{Q^2}{m^2}\Biggr)\N\\
                              &&+[\tau^2+(1-\tau)^2]\ln\Biggl(\frac{1-\tau}{\tau}
                                \Biggr)\Biggr\}\N\\
                              &&+4T_F\frac{m^2}{Q^2}\Biggl\{-10\tau^2-\tau
                                +4(3\tau^2-\tau)\Biggl[\ln\Biggl(\frac{\tau}{1-\tau}
                                \Biggr)   \N\\
                              &&+\ln\Biggl(\frac{m^2}
                                {Q^2}\Biggr)\Biggr]\Biggr\}
                                +O\Biggl(\frac{m^4}{Q^4}\Biggr)\label{H2h1}~,
\end{eqnarray}\begin{eqnarray}
     H^{(1)}_{L,g}\Biggl(\tau,\frac{Q^2}{m^2},\frac{m^2}{\mu^2}\Biggr)
                               &=&16T_F \cdot \tau(1-\tau)\N\\
                              &&+16T_F\tau^2
                                 \frac{m^2}{Q^2}\Biggr\{-1+\Biggl[
                                 \ln\Biggl(\frac{\tau}{1-\tau}\Biggr)+
                                 \ln\Biggl(\frac{m^2}{Q^2}\Biggr)\Biggr]
                                 \Biggr\} \N\\
                               &&+O\Biggl(\frac{m^4}{Q^4}\Biggr)
                                 \label{HLh1}~.
    \end{eqnarray}
In the limit $m^2/Q^2\rightarrow 0$ the Wilson coefficients obtain the following structure:
\begin{eqnarray}
     H^{(1)}_{2,g}\Bigg(\tau,\frac{Q^2}{m^2},\frac{m^2}{\mu^2}\Biggr)
       &\propto&
       4 T_F\Biggl\{[\tau^2+(1-\tau)^2]\ln\Biggl(\frac{Q^2}{m^2}\Biggr) +8\tau(1-\tau)-1 \Biggr\}~,
\label{H2gASYM}
\\
  H^{(1)}_{L,g}\Biggl(\tau,\frac{Q^2}{m^2},\frac{m^2}{\mu^2}\Biggr)
                       &\propto&16T_F\cdot \tau(1-\tau)      \label{HLh2}~.
\end{eqnarray}
Often one considers Wilson coefficients in Mellin-space by performing the integral transformation
\begin{eqnarray}
\tilde{H}_k(N) = \int_0^1 d\tau~~\tau^{N-1}~H(\tau)~.  
\end{eqnarray}
The factor in front of the logarithm denotes the leading order splitting function \cite{Gross:1973juxGross:1974cs,Georgi:1951sr}.
    \begin{eqnarray}
     \widehat{P}^{0}_{qg}(\tau)&=&8T_F[\tau^2+(1-\tau)^2]~, \label{Pqg0}
    \end{eqnarray}
which is a process-independent quantity. In Mellin-space it is given by
\begin{eqnarray}
P_{qg}^{(0)} = 8 T_F~~\frac{N^2 + N + 2}{N(N+1)(N+2)}~.
\end{eqnarray}
The logarithm in (\ref{H2gASYM}) indicates the presence of
a collinear singularity, if $m^2\rightarrow 0$. Both (\ref{H2gASYM}) and (\ref{HLh2}) were calculated in the so-called on mass-shell scheme for the outgoing quarks. While (\ref{HLh2}) is a scheme invariant quantity, redefinitions of the logarithmic contribution in (\ref{H2gASYM}) for $m^2\rightarrow0$ would yield different expressions, absorbing the divergent term and part of the constant contribution into the gluon distribution. 
The structure functions $F_{(2,L)}^{c\overline{c}}$ are obtained by a convolution with the gluon density 
$G(z,\mu^2)$:
    \begin{eqnarray}
    F^{Q\overline{Q}}_{(2,L)}\Biggl(x,\frac{Q^2}{m^2}\Biggr)=
         x\int\limits_{ax}^{1}\D \frac{dz}{z}
          H_{(2,L),g}\Biggl(\frac x z,\frac{Q^2}{m^2},\frac{m^2}{\mu^2} 
          \Biggr)G(z,\mu^2)~,\label{F2L}
   \end{eqnarray}
where $a=1+4 m^2/Q^2$.
Inserting (\ref{F2L}) into (\ref{crosssec1})  yields the cross section
\begin{eqnarray}
      \frac{d\sigma^{Q\overline{Q}}}{dxdy}=\frac{2\pi\alpha^2}{xyQ^2}
                           \Bigg\{\Bigl[1+(1-y)^2\Bigr]F_2^{Q\overline{Q}}(x,Q^2)
                                  -y^2F_L^{Q\overline{Q}}(x,Q^2)\Biggr\}~.\label{crosssecHQ}
\end{eqnarray}
 \newpage
\section{\boldmath{The Heavy Quark Coefficient Functions in the limit $Q^2\gg m^2$\label{HeavyQuarkCF}}}
We compute the inclusive DIS heavy flavor production cross section in the asymptotic region $Q^2~\gg~m^2$.
There the heavy flavor Wilson coefficients $H_{i,j}$ and $L_{i,j}$ factorize into
massive operator matrix elements $A_{ij}$ and the massless
coefficient functions $C_{i,k}$, as has been shown in Ref. \cite{Buza:1995ie}. 
Here $H_{i,j}$ are Wilson coefficients with a photon coupling to the heavy quark line, and 
$L_{i,j}$ are those with the coupling to a light quark line, see Eqs.~(\ref{eqWIL1}-\ref{eqWIL5}).
All process dependent quantities enter
only into the light flavor Wilson coefficients, whereas the complete
mass dependence is contained in the massive operator matrix
elements $A_{ij}$, which are process independent. The power corrections which are proportional to
$\left({m^2}/{Q^2} \right)^k$, $k\ge 1$, can be disregarded in this limit. A quantitative comparison
with the exact LO and NLO result in
Refs.
~\cite{Witten:1975bh,Babcock:1977fi,Shifman:1977yb,Leveille:1978px,Gluck:1980cp}
and \cite{Laenen:1992zkxLaenen:1992xs,Riemersma:1994hv}  shows, that
in the case of $F_2^{Q\overline{Q}}$ these power corrections can be
neglected for ${Q^2}/{m^2}~\gsim~ 10$, cf. \cite{Buza:1995ie}.

Applying the light cone expansion to the partonic tensor corresponding
to the inclusive Wilson coefficient ${\cal C}^{\sf S,PS,NS}_{i,j}$
yields the asymptotic factorization formula, \cite{Bierenbaum:2009mv} :
   \begin{eqnarray}
     {\cal C}^{{\sf S,PS,NS}, \small{{\sf \sf asymp}}}_{j,(2,L)}
          \Bigl(N,n_f+1,\frac{Q^2}{\mu^2},\frac{m^2}{\mu^2}\Bigr) 
        &=& \N\\ && \hspace{-55mm}
           \sum_{i} A^{\sf S,PS,NS}_{ij}\Bigl(N,n_f+1,\frac{m^2}{\mu^2}\Bigr)
                    C^{\sf S,PS,NS}_{i,(2,L)}
                      \Bigl(N,n_f+1,\frac{Q^2}{\mu^2}\Bigr)
           +O\Bigl(\frac{m^2}{Q^2}\Bigr)\label{CallFAC}~,
    \end{eqnarray}
where the quantifier $(n_f+1)$ denotes one heavy and $n_f$ light
flavors and $\mu$ is the factorization scale between the heavy and light
contributions in ${\cal {C}}_{j,A}$. The light flavor Wilson coefficients are
denoted by $C_{i,j}$ and taken at $(n_f + 1)$ flavors. 
In order to obtain the correct $Q^2$ behaviour, it is necessary to include all radiative corrections  
containing heavy quark loops into the heavy quark coefficient functions.

Let us consider first the unrenormalized OMEs~$\Ahathat_{ij}$.  They can be computed as projections of 
truncated Green's functions. During the computation of the Green's functions trace terms emerge. However they do not contribute, since the local operators are traceless. These terms are projected out from the beginning by contracting with
   \begin{eqnarray}
    J_N \equiv \Delta_{\mu_1}...\Delta_{\mu_N}~, \label{Jsource}
   \end{eqnarray}
where $\Delta_{\mu}$ is a light-like vector. 
The Green's functions used for the computation of the OMEs with external gluons are then given by, cf. \cite{Buza:1995ie},
   \begin{eqnarray} 
    \epsilon^\mu(p) G^{ab}_{Q,\mu\nu} \epsilon^\nu(p)&=&
    \epsilon^\mu(p)
    J_N \bra{A^a_{\mu}(p)} O_{Q;\mu_1 ... \mu_N} \ket{A^b_{\nu}(p)}
    \epsilon^\nu(p)~, 
        \label{GabmnQgdef} \\
    \epsilon^\mu(p) G^{ab}_{q,Q,\mu\nu} \epsilon^\nu(p)&=&
    \epsilon^\mu(p)
    J_N \bra{A^a_{\mu}(p)} O_{q;\mu_1 ... \mu_N} \ket{A^b_{\nu}(p)}_Q
    \epsilon^\nu(p)~.
     \label{GabmnqgQdef} 
   \end{eqnarray}
Here the the external gluon fields are denoted by $A_{\mu}^a$ with color index $a$ and Lorentz index 
$\mu$. The
polarization vectors of the external gluons with momentum $p$ are denoted by $\epsilon^{\mu}(p)$.
The indices $q,Q$ of the local operators $O$ label the operator coupled to a light or a heavy quark.
In the flavor non-singlet case the following Green's function contributes
   \begin{eqnarray} 
    \overline{u}(p,s) G^{ij, {\sf NS}}_{q,Q} \lambda_r u(p,s)&=&
    J_N
    \bra{\overline{\Psi}_i(p)}O_{q,r;\mu_1...\mu_N}^{\sf NS}\ket{\Psi^j(p)}_Q~
    \label{GijNS}~.
   \end{eqnarray}
Here $u(p,s)$ and $\overline{u}(p,s)$ denote the bi-spinors of the external massless quarks and 
antiquarks, respectively, and the corresponding fields are $\Psi$ 
and $\overline{\Psi}$.

Further OMEs are obtained from the following Green's functions with external quarks in the flavor 
singlet case
   \begin{eqnarray} 
    \overline{u}(p,s) G^{ij}_{Q}  u(p,s)&=&
    J_N\bra{\overline{\Psi}_i(p)} O_{Q,\mu_1 ... \mu_N}  \ket{\Psi^j(p)}~, 
      \label{GijQqPS} \\ 
    \overline{u}(p,s) G^{ij}_{q,Q}  u(p,s)&=&
    J_N\bra{\overline{\Psi}_i(p)}O_{q,\mu_1...\mu_N}  \ket{\Psi^j(p)}_Q
       \label{GijqqQPS} ~. 
   \end{eqnarray}

The OMEs $A^{\sf S,PS,NS}_{ij} \Bigl(N,n_f+1\Bigr)$ are obtained as expectation values of the following 
twist-2 operators, cf.~\cite{Geyer:1977gv,Blumlein:2009rg},
   \begin{eqnarray}
        \label{COMP1}
         O^{\sf NS}_{q,r;\mu_1, \ldots, \mu_N} &=& i^{N-1} {\bf S} 
                 [\overline{\psi}\gamma_{\mu_1} D_{\mu_2} \ldots D_{\mu_N} 
                  \frac{\lambda_r}{2}\psi] - {\rm trace~terms}~, \\
\label{COMP1a}
O_{{q,r};\mu, \mu_1, \ldots, \mu_N}^{\sf TR,NS}
&=& \frac{1}{2} i^{N-1}
 {\bf S} [\overline{\psi}\sigma_{\mu \mu_1}
D_{\mu_2} \ldots D_{\mu_N} \frac{\lambda_r}{2}
\psi ] - {\rm trace~terms}~, \\
        \label{COMP2}
         O^{\sf S}_{q;\mu_1, \ldots, \mu_N} &=& i^{N-1} {\bf S} 
                 [\overline{\psi}\gamma_{\mu_1} D_{\mu_2} \ldots D_{\mu_N}
                  \psi] - {\rm trace~terms}~, \\
        \label{COMP3}
         O^{\sf S}_{g;\mu_1, \ldots, \mu_N} &=& 2 i^{N-2} {\bf S} 
           {\rm \bf Sp}[F_{\mu_1 \alpha}^a D_{\mu_2} \ldots D_{\mu_{N-1}} 
            F_{\mu_N}^{\alpha,a}] - {\rm trace~terms}~,
       \end{eqnarray}
between corresponding on-shell partonic states $\bra{q}$ or $\bra{g}$ and
$\sigma^{\mu\nu} = (i/2)\left[\gamma^\mu \gamma^\nu -
\gamma^\nu \gamma^\mu \right]$. 
Here  ${\bf S}$ denotes the symmetrization operator of the Lorentz
indices $\mu_1,...,\mu_N$, $D_{\mu}$ is the covariant derivative,
$\Psi$ the quark field and $F_{\mu\nu}^a$ is the gluonic field-strength
tensor. The flavor matrix of $SU(n_f)$ is denoted
$\lambda_r$ and ${\rm \bf Sp}$ is the color trace. The operators are
classified as flavor singlet $(S)$ and non-singlet $(NS)$ with respect
to their symmetry properties under the flavor group $SU(n_f)$. 
{\noindent}

 For~$\Ahathat_{Qg}$ one obtains
\begin{equation}
\label{Projection}
\Ahathat_{Qg}\left(m^2/\mu^2,\ep\right)=\frac {1} {N_c^2-1} \frac 1 {D-2} (-g_{\mu\nu}) \delta_{ab} (\Delta.p)^{-N} G_{Q,\mu\nu}^{ab}~.
\end{equation}
The OMEs obey the perturbative expansion

 \begin{eqnarray}
     \label{eqAij}
      A_{ij}^{\sf S,NS}\Bigl(N,n_f+1,\frac{m^2}{\mu^2}\Bigr)
               = \langle j| O_i^{\sf S,NS}|j \rangle 
               =\delta_{ij}+\sum_{i=1}^{\infty}a_s^i A_{ij}^{(i),{\sf S,NS}}
                \label{pertomeren} ~.
    \end{eqnarray}
The singlet contribution has the following representation 
 \begin{eqnarray}
     A_{qq}^{\sf S} = A_{qq}^{\sf NS} + A_{qq}^{\sf PS} \label{splitS}
    \end{eqnarray}
in terms of the flavor non-singlet ({\sf NS}) and the pure-singlet ({\sf PS}) contribution.

Since any integral without scale vanishes in dimensional regularization due to the on-shell condition, all corresponding graphs but the  $O(a_s^0)$~ term do not contribute. Due to this we have to consider only those matrix elements with at least one heavy mass line.
For the singlet terms one has to distinguish the cases in which the operator is inserted on a light or a heavy quark line, respectively a vertex with a number of additional gluon lines, cf. Appendix \ref{App-FeynRules}. These contributions are referred to as $A_{qq,Q}^{\rm{PS}}$ and $A_{Qq}^{\rm{PS}}$ 
in the pure singlet case and by $A_{qg,Q}$ and $A_{Qg}$ in the gluonic case. 
In this thesis we compute contributions to 
$A_{qq,Q}^{\rm {NS}}$,
$A_{qq,Q}^{\rm {NS,TR}}$,
$A_{Qq}^{\rm {PS}}$, 
$A_{qq,Q}^{\rm {PS}}$,
and $A_{Qg}$, while the correction to $A_{qg,Q}$ will be given elsewhere \cite{ABKSW2010}.
Using the relation
    \begin{eqnarray}
      \label{eqLH}
{\cal C}^{\sf S,PS,NS}_{i,(2,L)}
               \left(\tau,n_f+1,\frac{Q^2}{\mu^2},\frac{m^2}{\mu^2}\right) 
      = 
       && C_{i,(2,L)}^{\sf S,PS,NS}\left(\tau,n_f,\frac{Q^2}{\mu^2}\right)
\N\\ &&
\hspace{-17mm}
      + H_{i,(2,L)}^{\sf S,PS}
               \left(\tau,n_f+1,\frac{Q^2}{\mu^2},\frac{m^2}{\mu^2}\right)
      + L_{i,(2,L)}^{\sf S,PS,NS}
             \left(\tau,n_f+1,\frac{Q^2}{\mu^2},\frac{m^2}{\mu^2}\right)\N\\
       \label{Callsplit}
     \end{eqnarray}
one may split Eq. (\ref{CallFAC}) into the following contributions, cf. \cite{Bierenbaum:2009mv},
\begin{eqnarray}
     C_{q,(2,L)}^{\sf PS}(n_f)
       +L_{q,(2,L)}^{\sf PS}
            (n_f+1)
     &=&
        \Bigl[
               A_{qq,Q}^{\sf NS}(n_f+1)
              +A_{qq,Q}^{\sf PS}(n_f+1)
              +A_{Qq}^{\sf PS}(n_f+1)
         \Bigr]
\N\\ &&
         \times
         n_f \tilde{C}_{q,(2,L)}^{\sf PS}(n_f+1)
        +A_{qq,Q}^{\sf PS}(n_f+1)
         C_{q,(2,L)}^{\sf NS}(n_f+1)
\N\\ &&        
+A_{gq,Q}(n_f+1)
         n_f \tilde{C}_{g,(2,L)}(n_f+1)~, \N\\ 
                 \label{LPSFAC} \\
      C_{g,(2,L)}(n_f)
     +L_{g,(2,L)}(n_f+1)
    &=&
           A_{gg,Q}(n_f+1)
           n_f \tilde{C}_{g,(2,L)}(n_f+1)
\N\\ &&
         + A_{qg,Q}(n_f+1)
           C_{q,(2,L)}^{\sf NS}(n_f+1)
\N\\ &&         
+\Bigl[
                A_{qg,Q}(n_f+1)
               +A_{Qg}(n_f+1)
          \Bigr]
        n_f\tilde{C}_{q,(2,L)}^{\sf PS}(n_f+1)~.\N\\
        \label{LgFAC}
    \end{eqnarray}
    \begin{eqnarray}
     H_{q,(2,L)}^{\sf PS}
          (n_f+1)
     &=&  
        A_{Qq}^{\sf PS}(n_f+1)
           \Bigl[ 
                 C_{q,(2,L)}^{\sf NS}(n_f+1)
                +\tilde C_{q,(2,L)}^{\sf PS}
                         (n_f+1)
          \Bigr]
\N\\ &&
          +\Bigl[ 
                A_{qq,Q}^{\sf NS}(n_f+1)
               +A_{qq,Q}^{\sf PS}(n_f+1)
         \Bigr]
        \tilde{C}_{q,(2,L)}^{\sf PS}(n_f+1)
\N\\ &&
       +A_{gq,Q}(n_f+1)
       \tilde{C}_{g,(2,L)}(n_f+1)~,         \label{HPSFAC} \\
     H_{g,(2,L)}(n_f+1)
      &=&
         A_{gg,Q}(n_f+1)
         \tilde{C}_{g,(2,L)}(n_f+1)
        +A_{qg,Q}(n_f+1)
          \tilde{C}_{q,(2,L)}^{\sf PS}(n_f+1)
\N\\ &&
        + A_{Qg}(n_f+1)
          \Bigl[ C_{q,(2,L)}^{\sf NS}(n_f+1)
            +\tilde{C}_{q,(2,L)}^{\sf PS}(n_f+1)
             \Bigr]~.         \label{HgFAC}
    \end{eqnarray}
Here we used the notation
\begin{eqnarray}
\tilde{f}(n_f)\equiv\frac{f(n_f)}{n_f}~.
\end{eqnarray}
The heavy flavor Wilson coefficients up to $O(a_s^3)$ can now be
obtained by expanding in $a_s$, \cite{Bierenbaum:2009mv},    
    \begin{eqnarray}
     \label{eqWIL1}
     L_{q,(2,L)}^{\sf NS}(n_f+1) &=& 
     a_s^2 \Bigl[A_{qq,Q}^{(2), {\sf NS}}(n_f+1)~\delta_2 +
     \hat{C}^{(2), {\sf NS}}_{q,(2,L)}(n_f)\Bigr]
     \N\\
     &+&
     a_s^3 \Bigl[A_{qq,Q}^{(3), {\sf NS}}(n_f+1)~\delta_2
     +  A_{qq,Q}^{(2), {\sf NS}}(n_f+1) C_{q,(2,L)}^{(1), {\sf NS}}(n_f+1)
       \N \\
     && \hspace*{5mm}
     + \hat{C}^{(3), {\sf NS}}_{q,(2,L)}(n_f)\Bigr]~,  \\
      \label{eqWIL2}
      L_{q,(2,L)}^{\sf PS}(n_f+1) &=& 
     a_s^3 \Bigl[~A_{qq,Q}^{(3), {\sf PS}}(n_f+1)~\delta_2
     +  A_{gq,Q}^{(2)}(n_f)~~n_f\Ctil_{g,(2,L)}^{(1)}(n_f+1) \N \\
     && \hspace*{5mm}
     + n_f \hat{\Ctil}^{(3), {\sf PS}}_{q,(2,L)}(n_f)\Bigr]~,
     \\
     \label{eqWIL3}
      L_{g,(2,L)}^{\sf S}(n_f+1) &=& 
     a_s^2 A_{gg,Q}^{(1)}(n_f+1)n_f \Ctil_{g,(2,L)}^{(1)}(n_f+1)
     \N\\ &+&
      a_s^3 \Bigl[~A_{qg,Q}^{(3)}(n_f+1)~\delta_2 
     +  A_{gg,Q}^{(1)}(n_f+1)~~n_f\Ctil_{g,(2,L)}^{(2)}(n_f+1)
     \N\\ && \hspace*{5mm}
     +  A_{gg,Q}^{(2)}(n_f+1)~~n_f\Ctil_{g,(2,L)}^{(1)}(n_f+1)
     \N\\ && \hspace*{5mm}
     +  ~A^{(1)}_{Qg}(n_f+1)~~n_f\Ctil_{q,(2,L)}^{(2), {\sf PS}}(n_f+1)
     + n_f \hat{\Ctil}^{(3)}_{g,(2,L)}(n_f)\Bigr]~,
 \\ \N \\
     H_{q,(2,L)}^{\sf PS}(n_f+1)
     &=& a_s^2 \Bigl[~A_{Qq}^{(2), {\sf PS}}(n_f+1)~\delta_2
     +~\Ctil_{q,(2,L)}^{(2), {\sf PS}}(n_f+1)\Bigr]
     \N\\
     &+& a_s^3 \Bigl[~A_{Qq}^{(3), {\sf PS}}(n_f+1)~\delta_2
     +~\Ctil_{q,(2,L)}^{(3), {\sf PS}}(n_f+1) \N
    \end{eqnarray}
    \begin{eqnarray}
 && \hspace*{-20mm}
     + A_{gq,Q}^{(2)}(n_f+1)~\Ctil_{g,(2,L)}^{(1)}(n_f+1) 
     + A_{Qq}^{(2), {\sf PS}}(n_f+1)~C_{q,(2,L)}^{(1), {\sf NS}}(n_f+1) 
        \Bigr]~,       \label{eqWIL4}
        \\ \N\\ 
     H_{g,(2,L)}^{\sf S}(n_f+1) &=& a_s \Bigl[~A_{Qg}^{(1)}(n_f+1)~\delta_2
     +~\Ctil^{(1)}_{g,(2,L)}(n_f+1) \Bigr] \N\\
     &+& a_s^2 \Bigl[~A_{Qg}^{(2)}(n_f+1)~\delta_2
     +~A_{Qg}^{(1)}(n_f+1)~C^{(1), {\sf NS}}_{q,(2,L)}(n_f+1)\N\\ && 
     \hspace*{5mm}
     +~A_{gg,Q}^{(1)}(n_f+1)~\Ctil^{(1)}_{g,(2,L)}(n_f+1) 
     +~\Ctil^{(2)}_{g,(2,L)}(n_f+1) \Bigr]
     \N\\ &+&
     a_s^3 \Bigl[~A_{Qg}^{(3)}(n_f+1)~\delta_2
     +~A_{Qg}^{(2)}(n_f+1)~C^{(1), {\sf NS}}_{q,(2,L)}(n_f+1)
     \N\\ &&
     \hspace*{5mm}
     +~A_{gg,Q}^{(2)}(n_f+1)~\Ctil^{(1)}_{g,(2,L)}(n_f+1)
     \N\\ && \hspace*{5mm}
     +~A_{Qg}^{(1)}(n_f+1)\Bigl\{
     C^{(2), {\sf NS}}_{q,(2,L)}(n_f+1)
     +~\Ctil^{(2), {\sf PS}}_{q,(2,L)}(n_f+1)\Bigr\}
     \N\\ && \hspace*{5mm}
     +~A_{gg,Q}^{(1)}(n_f+1)~\Ctil^{(2)}_{g,(2,L)}(n_f+1)
     +~\Ctil^{(3)}_{g,(2,L)}(n_f+1) \Bigr]~. \label{eqWIL5}
    \end{eqnarray}
Here $\de_2$ is given by $\de_2 = 1$ for $F_2$ and by $\de_2=0$ for
$F_L~.$ The relations (\ref{eqWIL1}-\ref{eqWIL5}) provide 
the basic scenario for the present work. To $3$--loop order they connect the various massive OMEs, 
$A_{ij}$, and the known massless Wilson coefficients, $C_{i(2,L)}^{(k)}$, cf. \cite{Vermaseren:2005qc}, to the {\it massive} Wilson coefficients in the asymptotic region $Q^2\gg m^2$. Henceforth, we will calculate the massive operator matrix elements.
 \newpage
\section{\boldmath{Renormalization}\label{Renorm}}
The renormalization of the massive OMEs proceeds in four  steps: {\it {i)}} mass renormalization, {\it {ii)}} renormalization of the strong coupling constant, {\it {iii)}} renormalization of the ultraviolet singularities of the composite local operators, and {\it {iv)}} subtraction of the collinear singularities. To 3-loop order this formalism has been developed in Ref. \cite{Bierenbaum:2009mv}. In the following we summarize the main steps.
  \subsection{\boldmath {Renormalization of the Mass}}
   \label{SubSec-RENMa}
  In order to renormalize the heavy quark mass, we will apply the on-shell renormalization
  scheme and define it as the pole
  mass. Hence the bare mass $\hat{m}$ is replaced by 

   \begin{eqnarray}
    \hat{m}&=&Z_m m
            =  m \Bigl[ 1 
                       + \hat{a}_s \Bigl(\frac{m^2}{\mu^2}\Bigr)^{\ep/2}
                                   \delta m_1 
                       + \hat{a}_s^2 \Bigl(\frac{m^2}{\mu^2}\Bigr)^{\ep}
                                     \delta m_2
                 \Bigr] + O(\hat{a}_s^3)~.
            \label{mren1}
   \end{eqnarray}
   The constants $\delta m_1$ and $\delta m_2$ are given by
   \begin{eqnarray}
    \delta m_1 &=&C_F
                  \Bigl[\frac{6}{\ep}-4+\Bigl(4+\frac{3}{4}\zeta_2\Bigr)\ep
                  \Bigr] \label{delm1}  \\
               &\equiv&  \frac{\delta m_1^{(-1)}}{\ep}
                        +\delta m_1^{(0)}
                        +\delta m_1^{(1)}\ep~, \label{delm1exp} \\
    \delta m_2 &=& C_F
                   \Biggl\{\frac{1}{\ep^2}\Bigl(18 C_F-22 C_A+8T_F(n_f+N_h)
                    \Bigr)
                  +\frac{1}{\ep}\Bigl(-\frac{45}{2}C_F+\frac{91}{2}C_A
 \N\\ &&
                   -14T_F(n_f+N_h)\Bigr)
                  +C_F\Bigl(\frac{199}{8}-\frac{51}{2}\zeta_2+48\ln(2)\zeta_2
                   -12\zeta_3\Bigr)
 \N\\ &&
                  +C_A\Bigl(-\frac{605}{8}
                  +\frac{5}{2}\zeta_2-24\ln(2)\zeta_2+6\zeta_3\Bigr)
 \N\\ &&
                  +T_F\Bigl[n_f\Bigl(\frac{45}{2}+10\zeta_2\Bigr)+N_h
                  \Bigl(\frac{69}{2}-14\zeta_2\Bigr)\Bigr]\Biggr\}
                  \label{delm2}  \\
               &\equiv&  \frac{\delta m_2^{(-2)}}{\ep^2}
                        +\frac{\delta m_2^{(-1)}}{\ep}
                        +\delta m_2^{(0)}~. \label{delm2exp}
   \end{eqnarray}
   Here $\zeta_k$ is the Riemann $\zeta$--function at integer values, cf.~(\ref{zetn}),  
   $n_f$ denotes the number of light flavors
   and $N_h$ the number of heavy flavors,
   which we will set equal to $N_h=1$ from now on.
   The pole contributions were given in 
   Refs.~\cite{Tarrach:1980up,Nachtmann:1981zg},
   and the constant term was derived 
   in Refs.~\cite{Gray:1990yh,Broadhurst:1991fy}, 
   cf. also \cite{Fleischer:1998dw}.

   After carrying out mass renormalization up to $O\left(\hat{a}_s^{3}\right)$ the OMEs
   have the following structure
   \begin{eqnarray}
    \Ahathat_{ij}\Bigl(\frac{m^2}{\mu^2},\ep,N\Bigr) 
                &=&\delta_{ij}+
                  \hat{a}_s~ 
                   \Ahathat_{ij}^{(1)}\Bigl(\frac{m^2}{\mu^2},\ep,N\Bigr) 
\N\\ 
    && \hspace{-25mm}
                         + \hat{a}_s^2 \left[~
                                        \Ahathat^{(2)}_{ij}
                                        \Bigl(\frac{m^2}{\mu^2},\ep,N\Bigr) 
                                      + \delta m_1 
                                        \Bigl(\frac{m^2}{\mu^2}\Bigr)^{\ep/2}
                                        \frac{md}{dm}~
                                                   \Ahathat_{ij}^{(1)}
                                           \Bigl(\frac{m^2}{\mu^2},\ep,N\Bigr) 
                                \right]
\N\\ && \hspace{-25mm}
                         + \hat{a}_s^3 \Biggl[~ 
                                         \Ahathat^{(3)}_{ij}
                                           \Bigl(\frac{m^2}{\mu^2},\ep,N\Bigr) 
                                        +\delta m_1 
                                         \Bigl(\frac{m^2}{\mu^2}\Bigr)^{\ep/2}
                                         \frac{md}{dm}~ 
                                                    \Ahathat_{ij}^{(2)}
                                           \Bigl(\frac{m^2}{\mu^2},\ep,N\Bigr)
\N\\ && \hspace{-15mm}
                                        + \delta m_2 
                                          \Bigl(\frac{m^2}{\mu^2}\Bigr)^{\ep}
                                          \frac{md}{dm}~ 
                                                    \Ahathat_{ij}^{(1)}
                                           \Bigl(\frac{m^2}{\mu^2},\ep,N\Bigr) 
                                        + \frac{\delta m_1^2}{2}
                                          \Bigl(\frac{m^2}{\mu^2}\Bigr)^{\ep}
                                                     \frac{m^2d^2}{dm^2}~
                                                     \Ahathat_{ij}^{(1)}
                                           \Bigl(\frac{m^2}{\mu^2},\ep,N\Bigr) 
    \Biggr]~. \N \\ \label{maren}
  \end{eqnarray}
Here and in the following we unify all renormalization and factorization scales to one scale $\mu^2$.
  \subsection{\boldmath Renormalization of the Coupling}
   \label{SubSec-RENCo}
     Considering only $n_f$ light flavors and no heavy flavors yields the
  following relation between the bare coupling constant $\hat{a}_s$ and the 
  renormalized coupling $a_s^{\MS}$
  \begin{eqnarray}
   \hat{a}_s             &=& {Z_g^{\MS}}^2(\ep,n_f) 
                             a^{\MS}_s(\mu^2) \N\\
                         &=& a^{\MS}_s(\mu^2)\left[
                                   1 
                                 + \delta a^{\MS}_{s, 1}(n_f) 
                                      a^{\MS}_s(\mu^2)
                                 + \delta a^{\MS}_{s, 2}(n_f) 
                                      {a^{\MS}_s}^2(\mu^2)    
                                     \right] + O({a^{\MS}_s}^3)~. 
                            \label{asrenMSb}
  \end{eqnarray}
  The coefficients in Eq.~(\ref{asrenMSb}) are given by 
  \cite{Khriplovich:1969aa,tHooft:unpub,Politzer:1973fx,Gross:1973id} and 
  \cite{Caswell:1974gg,Jones:1974mm}, 
  \begin{eqnarray}
    \delta a^{\MS}_{s, 1}(n_f) &=& \frac{2}{\ep} \beta_0(n_f)~,
                             \label{deltasMSb1} \\
    \delta a^{\MS}_{s, 2}(n_f) &=& \frac{4}{\ep^2} \beta_0^2(n_f)
                           + \frac{1}{\ep} \beta_1(n_f)~,
                             \label{deltasMSb2}
  \end{eqnarray}
  with 
  \begin{eqnarray}
   \beta_0(n_f)
                 &=& \frac{11}{3} C_A - \frac{4}{3} T_F n_f \label{beta0}~, \\
   \beta_1(n_f)
                 &=& \frac{34}{3} C_A^2 
               - 4 \left(\frac{5}{3} C_A + C_F\right) T_F n_f \label{beta1}~.
  \end{eqnarray}
  If one considers also heavy flavor contributions it is important to
  take into account that the factorization condition (\ref{CallFAC}) strictly requires the
  massless external particles to be on shell.
  This condition is violated by massive loop insertions to the gluon and
  ghost-propagators. Fortunately these corrections can be uniquely absorbed
  into the strong coupling constant, by applying the background field method, cf. \cite{Bierenbaum:2009mv}.
  The most direct way to do this, is to perform the renormalization first in 
  the $\sf{MOM}$--scheme and to transform afterwards into the $\overline{\sf
    MS}$--scheme.
 The following relations are obtained: 

  \begin{eqnarray}
   a_s^{\MOM}&=& a_s^{\MS}
                -\beta_{0,Q}\ln \Bigl(\frac{m^2}{\mu^2}\Bigr) {a_s^{\MS}}^2
\N \\ &&
                +\Biggl[ \beta^2_{0,Q}\ln^2 \Bigl(\frac{m^2}{\mu^2}\Bigr) 
                        -\beta_{1,Q}\ln \Bigl(\frac{m^2}{\mu^2}\Bigr) 
                        -\beta_{1,Q}^{(1)}
                 \Biggr] {a_s^{\MS}}^3
                         +O({a_s^{\MS}}^4)~, \label{asmoma}
  \end{eqnarray}
  or, 
  \begin{eqnarray}
   a_s^{\MS}&=&
               a_s^{\MOM}
              +{a_s^{\MOM}}^2\Biggl(
                          \delta a^{\MOM}_{s, 1}
                         -\delta a^{\MS}_{s, 1}(n_f+1)
                             \Biggr)
              +{a_s^{\MOM}}^{3}\Biggl(
                          \delta a^{\MOM}_{s, 2}
                         -\delta a^{\MS}_{s, 2}(n_f+1)
     \N\\ &&
                        -2\delta a^{\MS}_{s, 1}(n_f+1)\Bigl[
                             \delta a^{\MOM}_{s, 1}
                            -\delta a^{\MS}_{s, 1}(n_f+1)
                                                      \Bigr]
                             \Biggr)+O({a_s^{\MOM}}^4)~, \label{asmsa}
  \end{eqnarray}
  vice versa, where $a_s^{\sf \MS} = a_s^{\sf \MS}(n_f + 1)$. These identities are valid to all orders in $\ep$.
  In (\ref{asmoma}) the value for $\beta_{1,Q}^{(k)}$, \cite{Bierenbaum:2009mv}, are given by
    \begin{eqnarray}
   \beta_{1,Q} &=&\hat{\beta_1}(n_f)=
                  - 4 \left(\frac{5}{3} C_A + C_F \right) T_F~, \label{b1Q} \\
   \beta_{1,Q}^{(1)}&=&
                           -\frac{32}{9}T_FC_A
                           +15T_FC_F~, \label{b1Q1} \\
   \beta_{1,Q}^{(2)}&=&
                               -\frac{86}{27}T_FC_A
                               -\frac{31}{4}T_FC_F
                               -\zeta_2\left(\frac{5}{3}T_FC_A
                                        +T_FC_F\right)~. \label{b1Q2}
  \end{eqnarray}

  \subsection{\boldmath Operator Renormalization and Mass Factorization}
   \label{SubSec-RENOp}
   The ultraviolet singularities of the composite operators
   introduced in (\ref{COMP1}--\ref{COMP3}) is performed by introducing the corresponding
   $Z_{ij}$-factors:

 \begin{eqnarray}
        O^{\sf NS}_{q,r;\mu_1,...,\mu_N}&=&
                    Z^{\sf NS}(\mu^2)\hat{O}^{\sf NS}_{q,r;\mu_1,...,\mu_N}~,
                    \label{ZNSdef}\\
        O^{\sf S}_{i;\mu_1,...,\mu_N}&=&
                  Z^{\sf S}_{ij}(\mu^2)
                  \hat{O}^{\sf S}_{j;\mu_1,...,\mu_N}~,\quad~i=q,g~.
                  \label{ZSijdef}
       \end{eqnarray}
Due to their identical quantum numbers mixing occurs among the
different singlet operators, (\ref{ZSijdef}).
The anomalous dimensions of the operators are defined by

\begin{eqnarray}
        \gamma_{qq}^{\sf NS}&=&
                  \mu Z^{-1, {\sf NS}}(\mu^2)
                  \frac{\partial}{\partial \mu} Z^{\sf NS}(\mu^2)~,
                                               \label{gammazetNS}\\
        \gamma_{ij}^{\sf S}&=&
                           \mu Z^{-1, {\sf S}}_{il}(\mu^2)
                         \frac{\partial}{\partial \mu} Z_{lj}^{\sf S}(\mu^2)~.
                                               \label{gammazetS}
       \end{eqnarray}

The ${\sf NS}$ and ${\sf PS}$ contributions are split in the
following way 
   \begin{eqnarray}
    Z_{qq}^{-1}&=&Z_{qq}^{-1, {\sf PS}}+Z_{qq}^{-1, {\sf NS}}~,\\
    A_{qq}     &=&A_{qq}^{\sf PS}+A_{qq}^{\sf NS} \label{ZPSNS}~. 
   \end{eqnarray}
The anomalous dimensions can be expanded into a perturbative series in
${a}_s^{\sf \overline{MS}}$:

    \begin{eqnarray}
    \gamma_{ij}^{{\sf S,PS,NS}}(a_s^{\MS},n_f,N)
        &=&\sum_{l=1}^{\infty}{a^{\MS}_s}^l 
                      \gamma_{ij}^{(l), {\sf S,PS,NS}}(n_f,N)~.
            \label{pertgamma}
   \end{eqnarray}

The renormalization is performed in two steps, see \cite{Bierenbaum:2009mv}. First only $n_f$ light
flavors are considered and then the renormalization scheme is
extended to $n_f$ light and one heavy quark flavor.

In the first case, one finds up to $O(a_s^3)$
   up to $O({a_s^{\MS}}^3)$
   \begin{eqnarray}
    Z_{ij}(a^{\MS}_s,n_f) &=&
                            \delta_{ij}
                           +a^{\MS}_s \frac{\gamma_{ij}^{(0)}}{\ep}
                           +{a^{\MS}_s}^2 \Biggl\{
                                 \frac{1}{\ep^2} \Bigl(
                                     \frac{1}{2} \gamma_{il}^{(0)}
                                                 \gamma_{lj}^{(0)}
                                   + \beta_0 \gamma_{ij}^{(0)}
                                                 \Bigr)
                               + \frac{1}{2 \ep} \gamma_{ij}^{(1)}
                                   \Biggr\}
 \N \\ &&
                           + {a^{\MS}_s}^3 \Biggl\{
                                 \frac{1}{\ep^3} \Bigl(
                                     \frac{1}{6}\gamma_{il}^{(0)}
                                                \gamma_{lk}^{(0)}
                                                \gamma_{kj}^{(0)}
                                   + \beta_0 \gamma_{il}^{(0)} 
                                             \gamma_{lj}^{(0)}
                                   + \frac{4}{3} \beta_0^2 \gamma_{ij}^{(0)}
                                                  \Bigr)
\N\\ &&
                               + \frac{1}{\ep^2}  \Bigl(
                                     \frac{1}{6} \gamma_{il}^{(1)} 
                                                 \gamma_{lj}^{(0)}
                                   + \frac{1}{3} \gamma_{il}^{(0)} 
                                                 \gamma_{lj}^{(1)}
                                   + \frac{2}{3} \beta_0 \gamma_{ij}^{(1)} 
                                   + \frac{2}{3} \beta_1 \gamma_{ij}^{(0)}
                                                  \Bigr)
                              + \frac{\gamma_{ij}^{(2)}}{3 \ep}
                                   \Biggr\}~. \label{Zijnf}
  \end{eqnarray}
As a second step the additional heavy quark is implemented. In order to limit the investigation to the 
ultraviolet 
singularities for now, the external momentum is temporarily kept artificially off-shell. The 
corresponding 
$Z$-factors for the massive OMEs are then obtained by taking Eq.~(\ref{Zijnf}) at $(n_f+1)$ flavors and applying the scheme transformation (\ref{asmsa}). Up to $O\left({a_s^{\sf MOM}}^3\right)$ one has

  \begin{eqnarray}
   Z_{ij}^{-1}(a_s^{\MOM},n_f+1,\mu^2)&=&
      \delta_{ij}
     -a_s^{\MOM}\frac{\gamma_{ij}^{(0)}}{\ep}
     +{a^{\MOM}_s}^2\Biggl[
          \frac{1}{\ep}\Bigl(
                       -\frac{1}{2}\gamma_{ij}^{(1)}
                       -\delta a^{\MOM}_{s,1}\gamma_{ij}^{(0)}
                       \Bigr)
\N\\ &&
         +\frac{1}{\ep^2}\Bigl(
                        \frac{1}{2}\gamma_{il}^{(0)}\gamma_{lj}^{(0)}
                       +\beta_0\gamma_{ij}^{(0)}
                        \Bigr)
         \Biggr]
     +{a^{\MOM}_s}^3\Biggl[
          \frac{1}{\ep}\Bigl(
                       -\frac{1}{3}\gamma_{ij}^{(2)}
                       -\delta a^{\MOM}_{s,1}\gamma_{ij}^{(1)}
\N\\ &&
                       -\delta a^{\MOM}_{s,2}\gamma_{ij}^{(0)}
                       \Bigr)
         +\frac{1}{\ep^2}\Bigl(
                        \frac{4}{3}\beta_0\gamma_{ij}^{(1)}
                       +2\delta a^{\MOM}_{s,1}\beta_0\gamma_{ij}^{(0)}
                       +\frac{1}{3}\beta_1\gamma_{ij}^{(0)}
\N\\ &&
                       +\delta a^{\MOM}_{s,1}\gamma_{il}^{(0)}\gamma_{lj}^{(0)}
                       +\frac{1}{3}\gamma_{il}^{(1)}\gamma_{lj}^{(0)}
                       +\frac{1}{6}\gamma_{il}^{(0)}\gamma_{lj}^{(1)}
                        \Bigr)
         +\frac{1}{\ep^3}\Bigl(
                       -\frac{4}{3}\beta_0^{2}\gamma_{ij}^{(0)}
\N\\ &&
                       -\beta_0\gamma_{il}^{(0)}\gamma_{lj}^{(0)}
                       -\frac{1}{6}\gamma_{il}^{(0)}\gamma_{lk}^{(0)}
                                    \gamma_{kj}^{(0)} 
                     \Bigr)
       \Biggr]~. \label{ZijInfp1}
  \end{eqnarray}
The contributions $\propto \delta a_{s,k}^{\sf MOM}$ in (\ref{ZijInfp1}) stem from finite mass effects and cancel singularities due to virtual processes at $p^2 \rightarrow 0$ in real radiation. 
The OMEs are split into a purely light part~$\Ahathat_{ij}$ and a heavy flavor part~$\Ahathat_{ij}^{Q}$, 
which denotes any massive OME, we consider:
  \begin{eqnarray}
    {\Ahathat}_{ij}(p^2,m^2,\mu^2,a_s^{\MOM},n_f+1)&=&
                {\Ahathat}_{ij}\Bigl(\frac{-p^2}{\mu^2},a_s^{\MS},n_f\Bigr)
\N\\ &&
             +~{\Ahathat}^Q_{ij}(p^2,m^2,\mu^2,a_s^{\MOM},n_f+1)~.
   \label{splitNSHL1}
  \end{eqnarray}
Here the light flavor part $\Ahathat_{ij}$ depends on $a_s^{\MS}$ since the renormalization prescription for the strong coupling constant applies to the massive part only. The UV--renormalized expression is obtained by subtracting all terms that apply to the light flavors only :
  \begin{eqnarray}
   \bar{A}^Q_{ij}(p^2,m^2,\mu^2,a_s^{\MOM},n_f+1)&=&
               Z^{-1}_{il}(a_s^{\MOM},n_f+1,\mu^2) 
                  \hat{A}^Q_{ij}(p^2,m^2,\mu^2,a_s^{\MOM},n_f+1)
\N\\ &&
              +Z^{-1}_{il}(a_s^{\MOM},n_f+1,\mu^2) 
                   \hat{A}_{ij}\Bigl(\frac{-p^2}{\mu^2},a_s^{\MS},n_f\Bigr)
\N\\ &&
              -Z^{-1}_{il}(a_s^{\MS},n_f,\mu^2)
                   \hat{A}_{ij}\Bigl(\frac{-p^2}{\mu^2},a_s^{\MS},n_f\Bigr)~.
  \label{eqXX} 
  \end{eqnarray}
Here $Z_{ij}$ can be expressed as a series in $a_s$ by 
  \begin{eqnarray}
   Z_{ij}^{-1} = \delta_{ij} + \sum_{k=1}^\infty a_s^k Z_{ij}^{-1, (k)}~.
  \end{eqnarray}
Since in the dimensional regularization scheme all integrals without scale vanish, for the light flavor OMEs only the constant term $\delta_{ij}$ remains in the limit $p^2\rightarrow0$. Expanding in $a_s$ yields the following UV--finite OME:
  \begin{eqnarray}
  \bar{A}^Q_{ij}\Bigl(\frac{m^2}{\mu^2},a_s^{\MOM},n_f+1\Bigr) &=& 
             a_s^{\MOM}\Biggl( \hat{A}_{ij}^{(1),Q}
                              \Bigl(\frac{m^2}{\mu^2}\Bigr)
                     +Z^{-1,(1)}_{ij}(n_f+1,\mu^2)
                     -Z^{-1,(1)}_{ij}(n_f)
               \Biggr)
\N\\ && \hspace{-45mm} 
            + {a_s^{\MOM}}^2\Biggl( \hat{A}_{ij}^{(2),Q}
                                    \Bigl(\frac{m^2}{\mu^2}\Bigr)
                       +Z^{-1,(2)}_{ij}(n_f+1,\mu^2)
                       -Z^{-1,(2)}_{ij}(n_f)
\N\\ && \hspace{-45mm}   \phantom{{a_s^{\MOM}}^2\Biggl(}
     +Z^{-1,(1)}_{ik}(n_f+1,\mu^2)
                        \hat{A}_{kj}^{(1),Q}\Bigl(\frac{m^2}{\mu^2}\Bigr)
               \Biggr)
\N\\ && \hspace{-45mm} 
           +{a_s^{\MOM}}^3\Biggl( \hat{A}_{ij}^{(3),Q}
                                    \Bigl(\frac{m^2}{\mu^2}\Bigr)
                       +Z^{-1,(3)}_{ij}(n_f+1,\mu^2)
                       -Z^{-1,(3)}_{ij}(n_f)
  \N\\ && \hspace{-45mm}  \phantom{{a_s^{\MOM}}^3\Biggl(}
                       +Z^{-1,(1)}_{ik}(n_f+1,\mu^2)
                        \hat{A}_{kj}^{(2),Q}\Bigl(\frac{m^2}{\mu^2}\Bigr)
                       +Z^{-1,(2)}_{ik}(n_f+1,\mu^2)
                        \hat{A}_{kj}^{(1),Q}\Bigl(\frac{m^2}{\mu^2}\Bigr)
                        \Biggr)~. \label{GenRen1}
  \end{eqnarray}
In the on-shell limit $p^2\rightarrow0$ collinear singularities emerge. These are absorbed into the 
parton distribution functions and only occur in massless parts of the OMEs. Thus the renormalized OMEs 
are obtained by
  \begin{eqnarray}
   A^Q_{ij}\Bigl(\frac{m^2}{\mu^2},a_s^{\MOM},n_f+1\Bigr)&=&
     \bar{A}^Q_{il}\Bigl(\frac{m^2}{\mu^2},a_s^{\MOM},n_f+1\Bigr)
     \Gamma_{lj}^{-1}~. \label{genren1}
  \end{eqnarray}
The generic renormalization formula is given by 
  \begin{eqnarray}
   A_{ij}&=&Z^{-1}_{il} \hat{A}_{lk} \Gamma_{kj}^{-1}~. \label{genren}
  \end{eqnarray}
In case of massless quarks only the $\Gamma$-factors would be given by, cf. e.g. \cite{Buza:1995ie},
  \begin{eqnarray}
    \Gamma_{ij} = Z^{-1}_{ij}~. \label{GammaZ}
  \end{eqnarray}
Due to the fact that collinear singularities emerge in massless subgraphs, only the $\Gamma$-factors
have to be computed newly, cf. Ref.~\cite{Bierenbaum:2009mv}. 
Finally the renormalized operator matrix element reads~:
\begin{eqnarray}
   && A^Q_{ij}\Bigl(\frac{m^2}{\mu^2},a_s^{\MOM},n_f+1\Bigr)=
\N\\&&\phantom{+}
                a^{\MOM}_s~\Biggl(
                      \hat{A}_{ij}^{(1),Q}\Bigl(\frac{m^2}{\mu^2}\Bigr)
                     +Z^{-1,(1)}_{ij}(n_f+1)
                     -Z^{-1,(1)}_{ij}(n_f)
                           \Biggr)
\N\\&&
           +{a^{\MOM}_s}^2\Biggl( 
                        \hat{A}_{ij}^{(2),Q}\Bigl(\frac{m^2}{\mu^2}\Bigr)
                       +Z^{-1,(2)}_{ij}(n_f+1)
                       -Z^{-1,(2)}_{ij}(n_f)
                       +Z^{-1,(1)}_{ik}(n_f+1)\hat{A}_{kj}^{(1),Q}
                                              \Bigl(\frac{m^2}{\mu^2}\Bigr)
\N\\
&&\phantom{+{a^{\MOM}_s}^2\Biggl(}
                       +\Bigl[ \hat{A}_{il}^{(1),Q}
                               \Bigl(\frac{m^2}{\mu^2}\Bigr)
                              +Z^{-1,(1)}_{il}(n_f+1)
                              -Z^{-1,(1)}_{il}(n_f)
                        \Bigr] 
                             \Gamma^{-1,(1)}_{lj}(n_f)
                         \Biggr)
\N\\ 
&&
          +{a^{\MOM}_s}^3\Biggl( 
                        \hat{A}_{ij}^{(3),Q}\Bigl(\frac{m^2}{\mu^2}\Bigr)
                       +Z^{-1,(3)}_{ij}(n_f+1)
                       -Z^{-1,(3)}_{ij}(n_f)
                       +Z^{-1,(1)}_{ik}(n_f+1)\hat{A}_{kj}^{(2),Q}
                                              \Bigl(\frac{m^2}{\mu^2}\Bigr)
\N\N\\ 
&&\phantom{+{a^{\MOM}_s}^3\Biggl(}
                       +Z^{-1,(2)}_{ik}(n_f+1)\hat{A}_{kj}^{(1),Q}
                                              \Bigl(\frac{m^2}{\mu^2}\Bigr)
                       +\Bigl[ 
                               \hat{A}_{il}^{(1),Q}
                                 \Bigl(\frac{m^2}{\mu^2}\Bigr)
                              +Z^{-1,(1)}_{il}(n_f+1)
 \N\\ &&\phantom{+{a^{\MOM}_s}^3\Biggl(}
                              -Z^{-1,(1)}_{il}(n_f)
                        \Bigr]
                              \Gamma^{-1,(2)}_{lj}(n_f)
                       +\Bigl[ 
                               \hat{A}_{il}^{(2),Q}
                                 \Bigl(\frac{m^2}{\mu^2}\Bigr)
                              +Z^{-1,(2)}_{il}(n_f+1)
                              -Z^{-1,(2)}_{il}(n_f)
 \N\\ 
&&\phantom{+{a^{\MOM}_s}^3\Biggl(}
                              +Z^{-1,(1)}_{ik}(n_f+1)\hat{A}_{kl}^{(1),Q}
                                              \Bigl(\frac{m^2}{\mu^2}\Bigr)
                        \Bigr]
                              \Gamma^{-1,(1)}_{lj}(n_f)
                        \Biggr)+O({a_s^{\MOM}}^4)~. \label{GenRen3}
\end{eqnarray}
One notices that the order $O(\ep^2)$-terms of the leading order OMEs and the $O(\ep)$-terms of the NLO 
OMEs are required to renormalize the 3--loop OMEs, cf.~\cite{Buza:1995ie,Buza:1996wv,Bierenbaum:2007qe,Bierenbaum:2008yu,Bierenbaum:2009zt}. Transforming the coupling constant back into the $\overline{\sf MS}$--scheme and performing a series expansion in $a_s^{\sf MS}$ yields the final expression for the renormalized OME $A_{Qg}^{(3)}$, cf. \cite{Bierenbaum:2009mv},
  \begin{eqnarray}
   A_{Qg}^{(3), \MS}&=&
                  \frac{\hat{\gamma}_{qg}^{(0)}}{48}
                     \Biggl\{
                             (n_f+1)\gamma_{gq}^{(0)}\hat{\gamma}_{qg}^{(0)}
                            +\gamma_{gg}^{(0)}\Bigl(
                                      \gamma_{gg}^{(0)}
                                     -2\gamma_{qq}^{(0)}
                                     +6\beta_0
                                     +14\beta_{0,Q}
                                              \Bigr)
                            +\gamma_{qq}^{(0)}\Bigl(
                                      \gamma_{qq}^{(0)}
\N\\
 &&
                                     -6\beta_0
                                     -8\beta_{0,Q}
                                              \Bigr)
                            +8\beta_0^2
                            +28\beta_{0,Q}\beta_0
                            +24\beta_{0,Q}^2
                     \Biggr\}
                     \ln^3 \Bigl(\frac{m^2}{\mu^2}\Bigr)
                    +\frac{1}{8}\Biggl\{
                            \hat{\gamma}_{qg}^{(1)}
                                \Bigl(
                                        \gamma_{qq}^{(0)}
                                       -\gamma_{gg}^{(0)}
\N
\\ 
&&
                                       -4\beta_0
                                       -6\beta_{0,Q}
                                \Bigr)
                           +\hat{\gamma}_{qg}^{(0)}
                                \Bigl(
                                        \hat{\gamma}_{gg}^{(1)}
                                       -\gamma_{gg}^{(1)}
                                       +(1-n_f) \hat{\gamma}_{qq}^{(1), {\sf PS}}
                                       +\gamma_{qq}^{(1), {\sf NS}}
                                      +\hat{\gamma}_{qq}^{(1), {\sf NS}}
                                      -2\beta_1
\N\\ &&
                                      -2\beta_{1,Q}
                                \Bigr)
                     \Biggr\}
                     \ln^2 \Bigl(\frac{m^2}{\mu^2}\Bigr)
                    +\Biggl\{
                            \frac{\hat{\gamma}_{qg}^{(2)}}{2}
                           -n_f\frac{\hat{\tilde{\gamma}}_{qg}^{(2)}}{2}
                           +\frac{a_{Qg}^{(2)}}{2}
                                \Bigl(
                                        \gamma_{qq}^{(0)}
                                       -\gamma_{gg}^{(0)}
                                       -4\beta_0
                                       -4\beta_{0,Q}
                                \Bigr)
\N
\\ 
&&
                           +\frac{\hat{\gamma}_{qg}^{(0)}}{2}
                                \Bigl(
                                       a_{gg,Q}^{(2)}
                                      -n_fa_{Qq}^{(2), {\sf PS}}
                                \Bigr)
                           +\frac{\hat{\gamma}_{qg}^{(0)}\zeta_2}{16}
                                \Bigl(
                                       -(n_f+1)\gamma_{gq}^{(0)}
                                             \hat{\gamma}_{qg}^{(0)}
                                       +\gamma_{gg}^{(0)}\Bigl[
                                                2\gamma_{qq}^{(0)}
                                               -\gamma_{gg}^{(0)}
                                               -6\beta_0
\N
\end{eqnarray}
\begin{eqnarray}
&&
                                               -6\beta_{0,Q}
                                                         \Bigr]
                                       -4\beta_0[2\beta_0+3\beta_{0,Q}]
                                       +\gamma_{qq}^{(0)}\Bigl[
                                               -\gamma_{qq}^{(0)}
                                               +6\beta_0
                                               +4\beta_{0,Q}
                                                         \Bigr]
                                \Bigr)
                     \Biggr\}
                     \ln \Bigl(\frac{m^2}{\mu^2}\Bigr)
                           +\overline{a}_{Qg}^{(2)}
                                \Bigl(
                                        \gamma_{gg}^{(0)}
\N\\ 
&&
                                       -\gamma_{qq}^{(0)}
                                       +4\beta_0
                                       +4\beta_{0,Q}
                                \Bigr)
                           +\hat{\gamma}_{qg}^{(0)}\Bigl(
                                        n_f\overline{a}_{Qq}^{(2), {\sf PS}}
                                       -\overline{a}_{gg,Q}^{(2)}
                                                   \Bigr)
                           +\frac{\hat{\gamma}_{qg}^{(0)}\zeta_3}{48}
                                \Bigl(
                                        (n_f+1)\gamma_{gq}^{(0)}
                                                \hat{\gamma}_{qg}^{(0)}
\N\\ &&
                                       +\gamma_{gg}^{(0)}\Bigl[
                                                 \gamma_{gg}^{(0)}
                                               -2\gamma_{qq}^{(0)}
                                               +6\beta_0
                                               -2\beta_{0,Q}
                                                         \Bigr]
                                       +\gamma_{qq}^{(0)}\Bigl[
                                                \gamma_{qq}^{(0)}
                                               -6\beta_0
                                                         \Bigr]
                                       +8\beta_0^2
                                       -4\beta_0\beta_{0,Q}
\N\\ &&
                                       -24\beta_{0,Q}^2
                                \Bigr)
                           +\frac{\hat{\gamma}_{qg}^{(1)}\beta_{0,Q}\zeta_2}{8}
                           +\frac{\hat{\gamma}_{qg}^{(0)}\zeta_2}{16}
                                \Bigl(
                                        \gamma_{gg}^{(1)}
                                       -\hat{\gamma}_{qq}^{(1), {\sf NS}}
                                       -\gamma_{qq}^{(1), {\sf NS}}
                                       -\hat{\gamma}_{qq}^{(1),{\sf PS}}
                                       +2\beta_1
\N\\ &&
                                       +2\beta_{1,Q}
                                \Bigr)
                           +\frac{\delta m_1^{(-1)}}{8}
                                \Bigl(
                                       16 a_{Qg}^{(2)}
                                  +\hat{\gamma}_{qg}^{(0)}\Bigl[
                                           -24 \delta m_1^{(0)}
                                           -8 \delta m_1^{(1)}
                                           -\zeta_2\beta_0
                                           -9\zeta_2\beta_{0,Q}
                                                          \Bigr]
                                \Bigr)
\N\\ &&
                           +\frac{\delta m_1^{(0)}}{2}
                                \Bigl(
                                       2\hat{\gamma}_{qg}^{(1)}
                                      -\delta m_1^{(0)}
                                       \hat{\gamma}_{qg}^{(0)} 
                                \Bigr)
                           +\delta m_1^{(1)}\hat{\gamma}_{qg}^{(0)}
                                \Bigl(
                                        \gamma_{qq}^{(0)}
                                       -\gamma_{gg}^{(0)}
                                       -2\beta_0
                                       -4 \beta_{0,Q}
                                \Bigr)
\N\\ &&
                           +\delta m_2^{(0)}\hat{\gamma}_{qg}^{(0)}
                           +a_{Qg}^{(3)}~. \label{AQg3MSren}
\end{eqnarray}
Similar expressions are obtained for the other OMEs, cf. \cite{Bierenbaum:2009mv}. Both from the 
unrenormalized and the renormalized OMEs one may extract the information which is new at $O(a_s^3)$:
\begin{itemize}
\item  the constant parts $a_{ij}^{(3)}(N)$ of the unrenormalized OMEs \hspace*{1mm} $\Ahathat_{ij}$
\item  the corresponding contribution to the $3$--loop anomalous dimension 
$\hat{\gamma}_{ij}^{(2)}(N)~.$
\end{itemize}
Both quantities are computed in the following.
 \newpage
\section{\boldmath{Calculation of the Operator Matrix Element $ \Ahathat^{(1)}_{Qg}$}\label{OMEsLO}}
\begin{figure}[H]
\centering
\includegraphics[scale=1.0]{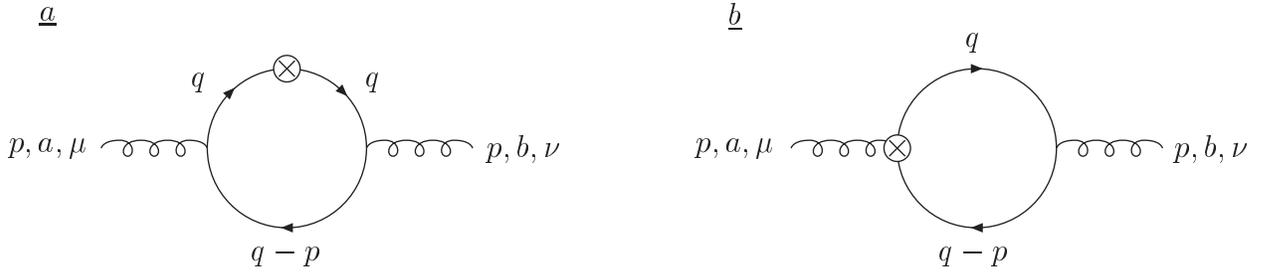}
\caption{\sf Feynman diagrams contributing to $\hat{A}_{Qg}^{(1)}$}
\label{FMG_AQg1}
\end{figure}

As an example to illustrate the method used in principle we recalculate the LO operator matrix element $\Ahathat_{Qg}^{(1)}$ in the following.
To the the lowest order in the coupling constant $\hat{a}_s=g_s^2/(4\pi)^2$ it receives contributions from the two 
Feynman diagrams given in Figure~\ref{FMG_AQg1}. The evaluation of the Feynman amplitudes
has been performed using the computer algebra system {\sf{FORM}} \cite{Vermaseren:2000nd}. One may split off an overall factor 
\begin{eqnarray}
\frac 1 2 \left[ 1+(-1)^N \right]
\end{eqnarray} 
from the OMEs, which we will do in the following. 
In the unpolarized case and for pure photon exchange only even moments contribute. This is a consequence of the current crossing relations, cf. e.g. 
\cite{Politzer:1974fr,Blumlein:1996vs,Blumlein:1998nv} 
and the local light cone expansion \cite{Wilson:1969zs,Brandt:1970kg,Zimmermann:1970,Frishman:1971qn}. One obtains

\begin{eqnarray}
\Ahathat_{Qg,(a)}^{(1)}&=&-8 \hat{a}_s T_F S_\ep {\left(\frac{m^2}{\mu^2}\right)}^{\ep/2} \frac 1 {(2+\ep)\ep} \exp 
\left(\sum_{l=2}^{\infty}\frac {\zeta_l} {l} {\left(\frac {\ep} {2} \right)}^l \right)
\N\\
&&
\frac {2 (N^2 + 3 N + 2) 
+ \ep (N^2 + N + 2)} {N (N+1) (N+2)}
\\
\Ahathat_{Qg,(b)}^{(1)}&=&32 \hat{a}_s T_F S_\ep {\left(\frac{m^2}{\mu^2}\right)}^{\ep/2} \frac 1 {(2+\ep)\ep} \exp 
\left(\sum_{l=2}^{\infty}\frac {\zeta_l} {l} {\left(\frac {\ep} {2} \right)}^l \right )\frac {1} {(N+1) (N+2)}~.
\end{eqnarray}
Here $\hat{a}_s$ denotes the bare coupling constant and $S_\ep$ is given by
\begin{eqnarray}
S_\ep=\exp\left[(\gamma_E-\ln(4\pi))~\frac {\ep} 2\right]~,
\end{eqnarray} 
with
\begin{eqnarray}
\frac {\Gamma(1-\ep/2)} {{(4 \pi)}^{(\ep/2)}}=S_\ep \exp\left(\sum_{l=2}^{\infty}\frac {\zeta_l} {l} 
{\left(\frac {\ep} {2} \right)}^l\right)~,
\end{eqnarray}
cf. Appendix \ref{App-SpeFun}. One obtains
\begin{eqnarray}
\tilde{A}_{Qg}^{(1)}&=&\frac {1} {\hat{a}_s} \left( \Ahathat_a^{Qg} + \Ahathat_b^{Qg}\right)
\\&=&-S_\ep T_F {\left(\frac{m^2}{\mu^2}\right)}^{\ep/2} \frac 1 {\ep} \exp\left[\sum_{l=2}^{\infty}\frac {\zeta_l} {l} 
{\left(\frac {\ep} {2} \right)}^l \right] \frac {8 (N^2 + N + 2)} {N (N+1) (N+2)} 
\\ &=& S_\ep T_F{\left(\frac{m^2}{\mu^2}\right)}^{\ep/2} \left ( - \frac {1} {\ep} - \frac {\zeta_2} {8} \ep - \frac {\zeta_3} {24} \ep^2 \right ) 
\frac {8 (N^2 + N + 2)} {N (N+1) (N+2)} + O(\ep^3)
\end{eqnarray}
The $N$-dependent term is identified as the Mellin transform of the LO splitting function $\hat{P}_{qg}^{(0)}(N)$, (\ref{Pqg0}).  
In $z$-space the unrenormalized leading order OME reads
\begin{equation}
\tilde{A}_{Qg}^{(1)}= S_\ep {\left(\frac{m^2}{\mu^2}\right)}^{\ep/2} \left[- \frac 1 \ep \hat{P}_{qg}^{(0)} (z) 
+ a_{Qg}^{(1)}+\ep \bar{a}_{Qg}^{(1)}+\ep^2 \bar{a}_{Qg}^{(2)} \right]~,
\end{equation}
with
\begin{eqnarray}
a_{Qg}^{(1)} &=& 0,\label{aQg1}
\\\bar{a}_{Qg}^{(1)} &=&-\frac {\zeta_2} 8 \hat{P}_{qg}^{(0)}(z)
\\\bar{a}_{Qg}^{(2)} &=&-\frac {\zeta_3} {24} \hat{P}_{qg}^{(0)}(z)~,
\end{eqnarray}
cf. Ref.\cite{Buza:1995ie}.
Expanding up to $O(\ep^2)$  yields

\begin{eqnarray}
\tilde{A}_{Qg}^{(1)} &=& S_\ep \hat{P}_{Qg}^{(0)}(z) \Biggl\{- \frac 1 \ep - \frac {1} {2} \ln \left(\frac {m^2} {\mu^2} \right) 
-\ep \Biggl[ \ln \left(\frac {1} {8} \frac {m^2} {\mu^2} \right)^2  + \frac{\zeta_2} 8 \Biggr]
\N\\
&& 
-\ep^2 \Biggl[\frac {1} {48}\ln \left(\frac {m^2} {\mu^2} \right)^3  +  \frac{\zeta_2} {16} \ln \left(\frac {m^2} {\mu^2} \right)+ \frac{\zeta_3} {24}\Biggr] \Biggr\}.
\end{eqnarray}
$\tilde{A}^{(1)}_{Qg}$ possesses a single pole, which is removed by operator renormalization later. The other contributions up to $O(\ep^2)$ all contribute to the renormalized OME at $3$-loop order.

Let us now discuss the results of Section~\ref{WilsonLO}, Eqs.~(\ref{H2gASYM},\ref{HLh2}) 
in the context 
of the asymptotic heavy flavor Wilson coefficient Eq.~(\ref{AQg3MSren}) 
to $O(a_s)$. In case of the 
structure function 
$F_L^{Q\overline{Q}}\left(x,Q^2\right)$, $H_{gL}^S$ 
is given by the massless result $\tilde{C}^{(1)}_{gL}$ since $\delta_2=0$. 
This applies to all schemes, cf. Ref. \cite{Furmanski:1981cw}. Eq.~(\ref{HLh2}) 
is actually the same in 
the heavy and light quark case after performing the limit 
$m^2/Q^2\rightarrow 0$. Comparing $H_{g2}^{(1)}$, Eq. (\ref{H2gASYM}), calculated using the on-mass-shell scheme in Section~\ref{WilsonLO} with the massless result obtained in the $\MS$--scheme, cf. \cite{Furmanski:1981cw}, one obtains the {\it same} result. This is in accordance with
\begin{eqnarray}
\label{EQ:aqg}
a_{Qg}^{(1)}=0~,
\end{eqnarray}
Eq.~(\ref{EQ:aqg}) determining the finite part of $A_{Qg}^{(1)}$ in Eq.~(\ref{AQg3MSren}). 
Corrections only occur in $O(a_s^3)$ and higher. This gives a first illustration of the formalism. 
The corresponding results at NLO 
are given in the literature, cf. 
Refs.~\cite{Buza:1995ie,Bierenbaum:2007qe,Bierenbaum:2008yu,Buza:1996wv,Bierenbaum:2009zt}.

\newpage
\section{\boldmath{Calculation of the ${O(a_s^3 T_F^2 N_fC_{F(A)}) }$ Contributions
  to the Massive Operator Matrix Elements}\label{OMEas3}}
In the following we describe the computation of the contributions to the massive 3-loop OMEs of 
$O(T_F^2n_f 
C_{F,A})$ being performed in this thesis. They concern $A_{Qg}$, $A_{Qg}^{PS}$, $A_{qq,Q}^{PS}$, $A_{qq,Q}^{NS}$ in the unpolarized case and $A_{qq,Q}^{NS,TR}$ for transversity. The results for $A_{qg,Q}$ will be presented in \cite{ABKSW2010}.
The unrenormalized OMEs are obtained by applying the following
projectors to the corresponding truncated Green's functions.
For external gluons the projector $P_g$, cf. \cite{Buza:1995ie},

 \begin{eqnarray}
    P_g \left[\hat{G}^{ab}_{l,(Q),\mu\nu}\right]
               &\equiv& 
            - \frac{\delta_{ab}}{N_c^2-1} \frac{g^{\mu\nu}}{D-2}
               (\Delta\cdot p)^{-N} \hat{G}^{ab}_{l,(Q),\mu\nu} ~,
               \label{projG1}
   \end{eqnarray}
defines the corresponding scalar contributions.
In Eq.~(\ref{projG1}) the
summation over $\mu$, $\nu$ includes unphysical
gluon-states, which have to be compensated by adding ghost-diagrams.
For the external quark contributions the Green's function is projected by
   \begin{eqnarray} 
    P_q \left[\hat{G}^{ij}_{l,(Q)}\right] &\equiv& 
              \frac{\delta^{ij}}{N_c} ( \Delta\cdot p)^{-N} 
               \frac{1}{4} {\sf Tr}[~\adag p~\hat{G}^{ij}_{l,(Q)}]~. 
               \label{projQ}
   \end{eqnarray}
Another projector is needed for the non-singlet transversity OME. It reads
   \begin{eqnarray} 
    P_q^{TR} \left[\hat{G}^{ij}_{l,(Q)}\right] &\equiv& 
              -i \frac{\delta^{ij}}{4 N_c (D-2)} ( \Delta\cdot
              p)^{-(N+1)} 
              \left\{ {\sf Tr}[~\adag{\Delta} ~ \adag{p} ~p^{\mu}
                ~\hat{G}^{ij,TR,NS}_{\mu,(Q)}]
\N\right.\\
&&\left.
               -\Delta{\cdot}p {\sf
                   Tr}[p^{\mu}~\hat{G}^{ij,TR,NS}_{\mu,(Q)}]
                 +i \Delta{\cdot}p {\sf Tr}[~ \sigma^{\mu \rho} ~ p^{\rho}
                ~\hat{G}^{ij,TR,NS}_{\mu,(Q)}]
\right\}
                ~, 
               \label{projQTR}
   \end{eqnarray}
cf. Ref. \cite{Blumlein:2009rg}.

\subsection{\boldmath Contributing diagrams}

\begin{figure}
\centering
\includegraphics[scale=1]{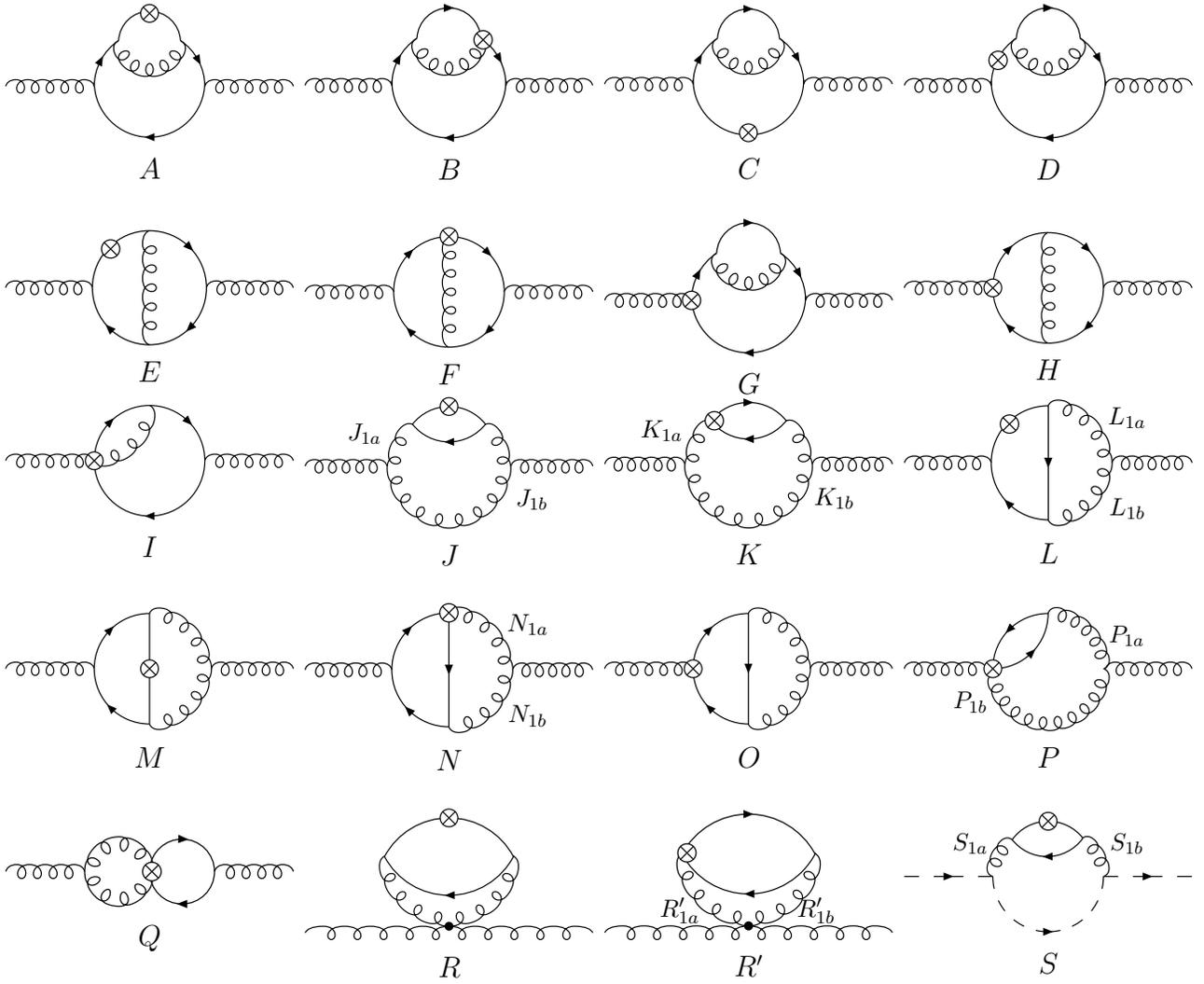}
\caption {\sf 2-loop diagrams, that were used to generate 3-loop diagrams by the insertion 
Eq.~(\ref{QuarkInsertion}) \label{graS1}}
\end{figure}
\begin{figure}
\centering
\includegraphics[scale=1]{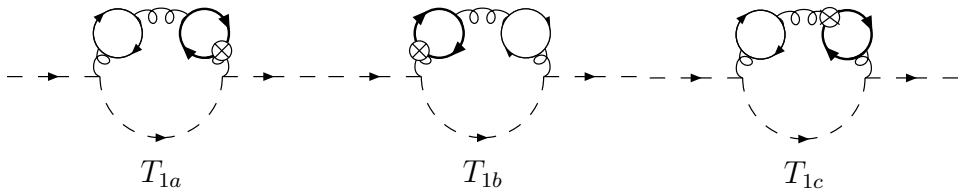}
\caption {\sf Additional ghost diagrams contributing to $A_{Qg}^{(3)}$\label{graGhosts}}
\end{figure}
\begin{figure}
\centering
\includegraphics[scale=1]{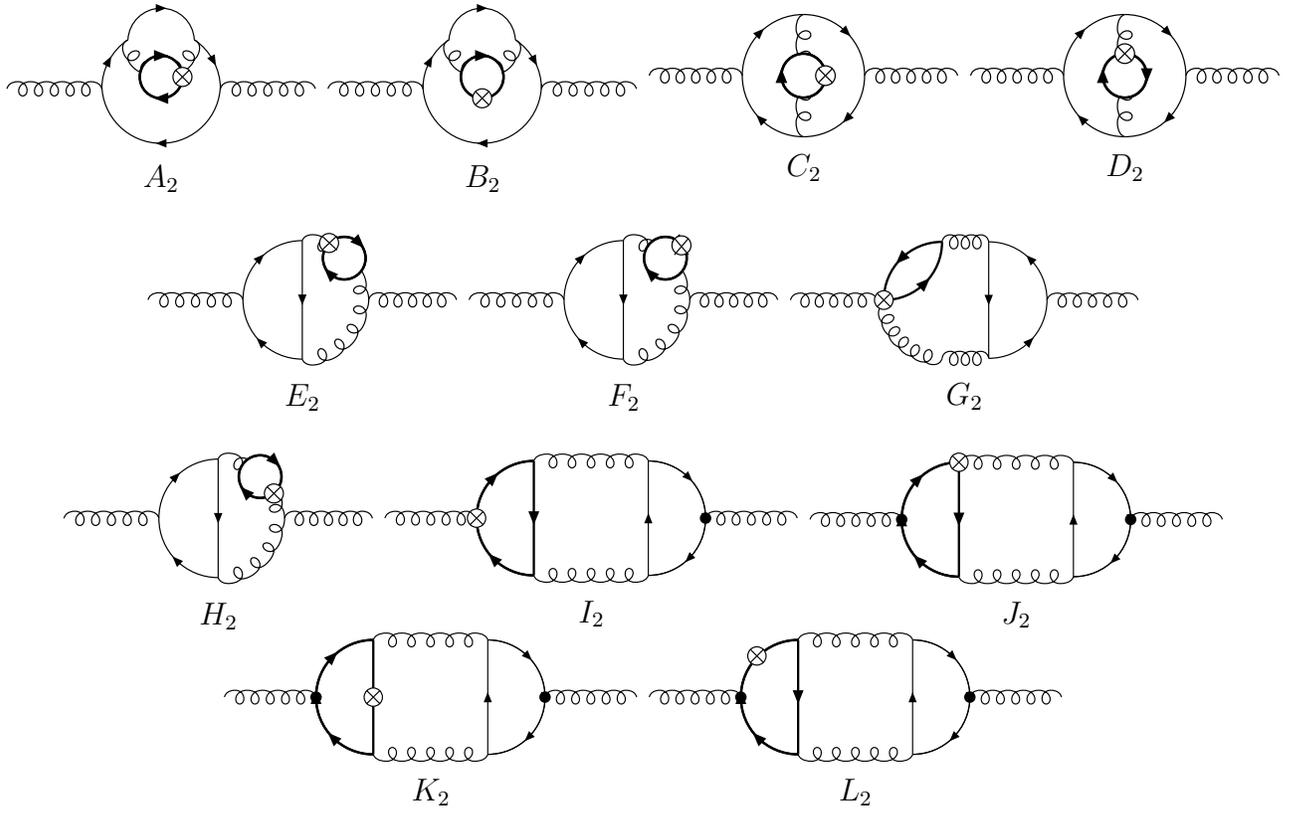}
\caption {\sf 3-loop diagrams, that are not generated from massive 2-loop graphs\label{graS2}}
\end{figure}

\begin{figure}
\centering
\includegraphics[scale=1]{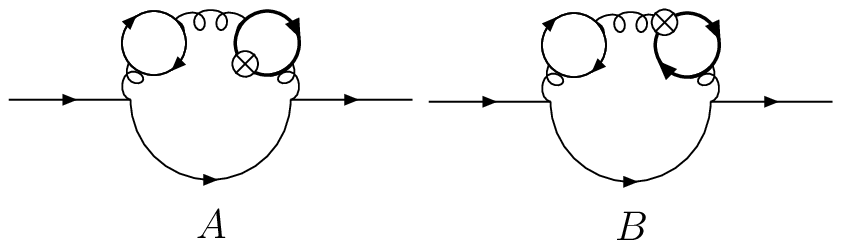}
\caption {\sf Diagrams contributing to $A_{Qq}^{PS}$\label{graPSQq}}
\end{figure}

\begin{figure}
\centering
\includegraphics[scale=1]{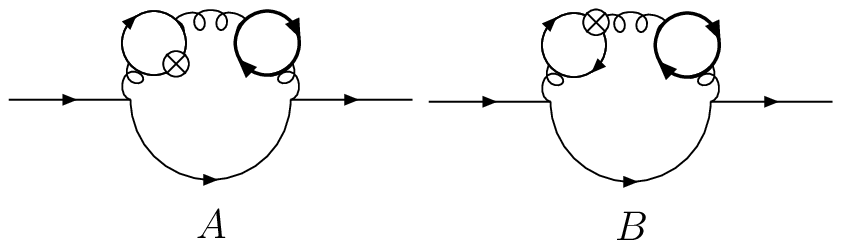}
\caption {\sf Diagrams contributing to $A_{qq,Q}^{PS}$\label{graPSqqQ}}
\end{figure}

\begin{figure}
\centering
\includegraphics[scale=1]{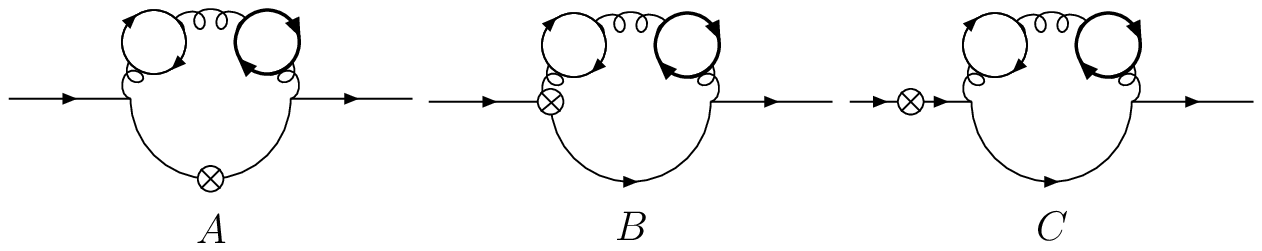}
\caption {\sf Diagrams contributing to $A_{qq,Q}^{NS}$\label{graNS}}
\end{figure}

The Feynman diagrams contributing to the respective Green's functions have been generated 
by a code \cite{Bierenbaum:2009mv,Klein:2009ig} allowing for local operator insertions in QCD-diagrams 
based on {\sf QGRAF}, \cite{Nogueira:1991ex}. 
The diagrams $A_{1}$ to $R'_{1b}$, see Figure~4, can be generated by including one
massless fermion loop into the gluon propagators of the corresponding
massive 2-loop diagrams \cite{Buza:1995ie,Bierenbaum:2007qe}. 
The color factors have been
evaluated using the ${\sf FORM}$-package ${\sf color}$, \cite{vanRitbergen:1998pn}. For
the diagrams that could be obtained from the 2-loop diagrams by
inserting a light quark loop, the color factor differs by $T_F n_f$.
This massless insertion can easily be integrated out and is given by the following expression for the extended
gluon propagator including the quark loop

\begin{eqnarray}
\Pi_{\mu\nu}^{ab}(p_{\mu})=\frac {8 i g_s^2 T_F n_f} {{4\pi}^{2+\ep/2}} B\left(2+ \frac{\ep}{2}, 
2+\frac{\ep} {2} \right)\, \Gamma\left(-\frac \ep 2 \right) 
\frac {\left\{-g_{\mu\nu}p^2 + p_{\mu}p_{\nu} \right\}} {(-p^2)^{2-\ep/2}} \label{QuarkInsertion}~,
\end{eqnarray}
with $g_s$ defined in $D = 4 +\ep$ dimensions.

The main differences to the Feynman rule for the gluon propagator, see Appendix \ref{App-FeynRules}, are slightly more complicated numerator
structures, and the $\ep$ dependence in the power
of the denominator. While the new numerator structure in most cases just increased
the size of the computation, the
occurrence of real exponents in many cases made it necessary to recompute the decorated $2$-loop diagrams completely, since new structures occurred.
In Figure~{\ref {graS1}} we show the $2$-loop diagrams of Refs. \cite{Buza:1995ie,Bierenbaum:2007qe} into which the massless fermion-bubble is inserted. This usually concerns more than one line. In some cases the insertions on different gluonic lines lead to different integrals, that cannot be mapped onto each other via a symmetry relation. We labeled this accordingly, e.g. $J_{1b}$, $J_{1b}$, etc. Figures {\ref {graS1}} and {\ref {graGhosts}} include the corresponding ghost diagrams which contribute, since the calculation is performed using a $R_{\xi}$-gauge. In Figure \ref{graS2} we show the other topologies which contribute to $A_{Qg}$ at $O(T_F^2 n_f C_{F,A})$, but cannot be generated by loop-insertions into known topologies. In this thesis the diagrams $A_2-H_2$ are computed. In the final result we will 
show the contributions to $I_2-L_2$ as well, which are calculated in Ref.~\cite{ABKSW2010}.\footnote{Results on the scalar integrals for diagrams $I_2-L_2$ are given in Ref.~\cite{BHKS} also.} 
Finally Figures~\ref{graPSQq}--\ref{graNS} show the corresponding diagrams contributing to the OMEs 
$A_{Qq}^{\sf PS}$, $A_{qq,Q}^{\sf PS}$, and $A_{qq,Q}^{\sf NS}$, $A_{qq,Q}^{\sf NS,TR}$. 
In Tables~1--4 we list the combinatorial multiplicities through which the different diagrams 
contribute. These were determined from the foregoing computation of the fixed moments for the respective 
OMEs in~\cite{Bierenbaum:2009mv,Klein:2009ig}.
\begin{table}[htb]
\caption{\sf Multiplicities of the individual diagrams contributing to $A_{Qg}^{\sf (3)}$}
\label{table:multiplicitiesQg}
      \begin{center}
       \renewcommand{\arraystretch}{1.1}
       \begin{tabular}{|l|c|l|c|l|c|}
        \hline \hline
        Diagram         & Multiplicity  & Diagram         &
        Multiplicity &     Diagram         & Multiplicity\\
        \hline \hline
        $A_{1}$		&	2  &    $L_{1a}$		&        4 &      $A_{2}$		&        4                         \\
        $B_{1}$		&	4  &     $L_{1b}$	&4 	&        $B_2$		&	4 \\        
        $C_{1}$		&	2  &    $M_{1}$		&	4 &      $C_{2}$		&	2   \\
        $D_{1}$		&	4  &    $N_{1a}$		&        4 &          $D_{2}$		&	2      \\      
        $E_{1}$		&	4  &    $N_{1b}$		&        4 &         $E_{2}$		&	4\\        
        $F_{1}$		&	2  &    $O_{1}$		&	4 &         $F_{2}$		&	8 \\
        $G_{1}$		&	4  &    $P_{1a}$		&        2 &       $G_{2}$		&	4  \\
        $H_{1}$		&	2  &    $P_{1b}$         &	2 &        $H_{2}$		&	4 \\
        $I_{1}$		&	4  &    $S_{1a}$         &2        &        $I_{2}$		&	4 \\
        $J_{1a}$		&	2  &    $S_{1b}$         &2&         $J_{2}$		&	8 \\
        $J_{1b}$		&	4  &    $T_{1a}$         &1&          $K_{2}$		&	4      \\
        $K_{1a}$		&       2  &    $T_{1b}$         &1&                 $L_{2}$		&	8 \\
        $K_{1b}$		&        4 &    $T_{1b}$         &2&&\\
  \hline\hline
       \end{tabular}

       \renewcommand{\arraystretch}{1.0}
      \end{center}
    \end{table}
\begin{table}[htb]
\caption{\sf Multiplicities of the individual diagrams contributing to $A_{Qq}^{\sf PS,(3)}$}
\label{table:multiplicitiesPSQq}
      \begin{center}
       \renewcommand{\arraystretch}{1.1}
       \begin{tabular}{|l|c|l|c|}
        \hline \hline
        Diagram         & Multiplicity  & Diagram         &
        Multiplicity  \\
        \hline \hline
        $A$		&	4  &    $B$		&        4 \\
  \hline\hline
       \end{tabular}
       \renewcommand{\arraystretch}{1.0}
      \end{center}
    \end{table}
\begin{table}[H]
\caption{\sf Multiplicities of the individual diagrams contributing to $A_{qq,Q}^{\sf PS,(3)}$}
\label{table:multiplicitiesPSqq}
      \begin{center}
       \renewcommand{\arraystretch}{1.1}
       \begin{tabular}{|l|c|l|c|}
        \hline \hline
        Diagram         & Multiplicity  & Diagram         &
        Multiplicity  \\
        \hline \hline
        $A$		&	4  &    $B$		&        4 \\
  \hline\hline
       \end{tabular}
       \renewcommand{\arraystretch}{1.0}
      \end{center}
    \end{table}
\begin{table}[H]
\caption{\sf Multiplicities of the individual diagrams contributing to $A_{qq}^{\sf NS,(3)}$ and $A_{qq}^{\sf NS,TR,(3)}$}
\label{table:multiplicitiesNS}
      \begin{center}
       \renewcommand{\arraystretch}{1.1}
       \begin{tabular}{|l|c|l|c|l|c|}
        \hline \hline
        Diagram         & Multiplicity  & Diagram         &
        Multiplicity  &Diagram         &        Multiplicity  \\
        \hline \hline
        $A$		&	2  &    $B$		&        4 & $C$& 2\\
  \hline\hline
       \end{tabular}
       \renewcommand{\arraystretch}{1.0}
      \end{center}
    \end{table}
\subsection{\boldmath Evaluation of the Feynman diagrams\label{Evaluation}}
\begin{figure}[H]
\centering
\includegraphics[scale=0.8]{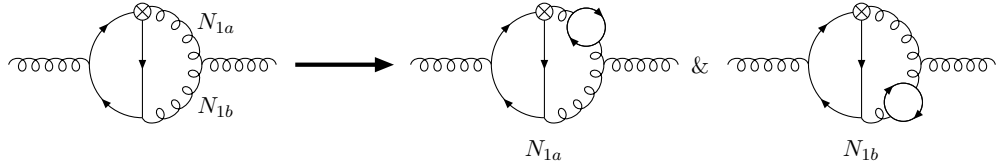}
\caption{\sf Replacement of the
gluon propagator by an extension including 1-loop massless fermion 
contributions\label{Replacement}}
\end{figure}

The two-loop massive operator matrix elements were calculated in 
\cite{Bierenbaum:2007qe,Bierenbaum:2008yu,SKdiploma}. Prototypes of graphs were computed in 
\cite{Bierenbaum:2007dm}. Many diagrams contributing to the $T_F^2n_f$-term of the three loop operator
matrix elements can be evaluated by considering the corresponding
2-loop diagrams, see Figure \ref{graS1}, and replacing one gluon
propagator by the extended gluon propagator containing the one loop self energy, cf. 
Figure~\ref{Replacement} for an example.

The first part of the evaluation of the individual Feynman diagrams has been performed
in a procedural way using the algebraic manipulation program ${\sf
  FORM}$, \cite{Vermaseren:2000nd}. The occurring fermionic traces of $\gamma$-matrices have been calculated using
the built-in functions, before the momenta were integrated. To integrate a respective momentum $k_i$ 
all denominators containing this variable were combined using Feynman parametrization, see Appendix~(\ref{FeynParam}). Here the order in which the momenta are 
integrated is not arbitrary. In many cases specific choices allow to avoid infinite sums emerging from 
generalized hypergeometric functions. As a next step the $\delta$-distributions are
integrated out and the momenta are shifted in order to symmetrize the $D$-dimensional momentum integral.
Due to this factors of the form $(\Delta.k_i+\Delta.l)^N$
occur, where $l$ denotes contributions from other momenta in the form 
\begin{eqnarray}
l&=&P_1(\{x_k\}) k_{j_1} +\cdots+P_n(\{x_n\}) k_{j_n}~,
\end{eqnarray}
 with polynomials in Feynman parameters $P_i$. These terms have to be expanded using the binomial theorem prior momentum integration~:
\begin{eqnarray}
(\Delta.k_i+\Delta.l)^N&=&\sum_{j=0}^N \binom {N} {j} (\Delta.l)^{N-j} (\Delta.k_{i})^j ~.
\end{eqnarray}
All but the first terms of this series can be dropped, as the symmetric momentum integrals over odd 
powers of $k_i$ vanish and the rules for $D$-dimensional integration, cf. Appendix \ref{sec:1}, lead to 
terms containing the contraction $\Delta.\Delta=0$ for higher powers of $\Delta.k_{i}$. In this 
computation a maximum 
of the first three terms had to be considered. The symmetric $D$-dimensional integral was evaluated by applying the rules in Appendix \ref{sec:1}. These steps were 
repeated for all internal momenta, which yields Feynman parameter integrals of various complexity. The application of simple 
algebraic transformations leads to representations of the following form~:
\begin{eqnarray}
I&=&\sum_{j=0}^{N-2} \int_{0}^1 \ldots \int_0^1~\prod_{i=1}^{n} dx_i 
\frac {\left(P_1(x_1, \cdots, x_n)\right)^{N-j+n_1} \left(P_2(x_1, \cdots, x_n)\right)^{j+n_2}} 
{(1-P(\{x_k\}))^{3/2 \ep + n_3}} 
\N\\
&&\times
P_3(x_1, \cdots, x_n,j)~\label{GenFormSum},
\end{eqnarray}
or 
\begin{eqnarray}
I&=&\int_0^1 \ldots \int_{0}^1 ~\prod_{i=1}^{n} dx_i \frac {\left(P_1(x_1, \cdots, x_n)\right)^{N+n_4}} 
{(1-P(\{x_k\}))^{3/2 \ep +n_5}} P_3(x_1, \cdots, x_n)~\label{GenFormNoSum}.
\end{eqnarray}
Here the $P_i$ are respective polynomials in the Feynman parameters, $n_i\in {\mathbb Z}$, and 
$P(\{x_k\})$ denotes a product in the Feynman parameters $\{x_k\}$. The physical sum $\sum_{j=0}^{N-2}$ 
emerges if the operator insertion is located at a vertex, cf. Appendix~\ref{App-FeynRules}. In general 
also diagrams with an operator insertion being represented by a physical double sum 
$\sum_{j=0}^{N-3}\sum_{l=j+1}^{N-2}$ have to be computed. For the contributions considered in this 
paper, these operator insertions where always located on an external vertex. Due to this one of the sums could be performed at 
the momentum level:  
\begin{eqnarray}
&&\hspace*{-2.5cm}
\sum_{0\leq j<l}(\Delta.k-\Delta.p)^{N-l-2}(-\Delta.p+\Delta.q)^{l-j-1}(\Delta.q)^j\N\\
&=&\frac {1} {\Delta.p} \sum_{j=0}^{N-2}\left\{(\Delta.k-\Delta.p)^{N-j-2} 
\left[(\Delta.q)^j-(\Delta.q-\Delta.p)^j \right]\right\}~, \label{doubleSum1}\\
&&\hspace*{-2.5cm}
\sum_{0\leq j<l}(\Delta.k-\Delta.p)^{N-l-2}(\Delta.k)^{l-j-1}(\Delta.q)^j\N\\
&=&\frac {1} {\Delta.p} \sum_{j=0}^{N-2}\left\{ \left[(\Delta.k)^j
-(\Delta.k-\Delta.p)^j \right] (\Delta.q)^{N-2-j} \right\}~. \label{doubleSum2}
\end{eqnarray}
In some cases a simpler structure is obtained, since variable transformations that map
that map $P(\{x_k\})$ to zero are applicable, cf. Appendix \ref{App-VarTrans}. 
If possible the physical sums $\sum_{j=0}^{N-2}$ were evaluated at the Feynman parameter level. All 
Feynman parameters that could be integrated through substitutions or in terms of Beta-functions were 
integrated at this point. In case the relations given in Appendix~\ref{App-VarTrans} were not 
applicable,  
and no further Feynman parameter could be integrated out directly, the binomial theorem was used thereby introducing additional finite sums. These steps have been repeated until all Feynman parameters, that do not contribute to the denominator are integrated out. 
The denominator in (\ref{GenFormSum}--\ref{GenFormNoSum}) and remaining Beta-like factors in the Feynman 
parameters $\{x_k\}$ were collected in terms of integral representations of generalized hypergeometric 
functions $_PF_Q$, \cite{Slater,Bailey}. Thus a representation in sums over generalized hypergeometric-- 
and $\Gamma$--functions has been obtained. The hypergeometric structure of the Feynman diagram is 
determined through the mass distribution of the Feynman graph and is widely independent of the operator 
insertion. We have implemented the relations being listed in Appendix \ref{App-SpeFunFPQ} into a {\sf 
FORM}--algorithm, which allowed to perform infinite sums stemming from generalized hypergeometric 
functions $_PF_Q$ on this level in many cases. 
 
The results are now given in terms of finite and infinite sums over $\Gamma$-functions depending on the
Mellin-variable $N$ and the dimensional parameter $\ep$. The respective expressions were simplified
and the integrals were expanded into a
Laurent series in $\ep$ using the computer
algebra system ${\sf MAPLE}$~\cite{MAPLE}. In some cases remaining sums could be evaluated
by applying known results from the literature, cf. Refs. 
\cite{Bierenbaum:2007qe,Bierenbaum:2007zz,Bierenbaum:2007zza,Bierenbaum:2008yu}. More complex sums
were evaluated using the {\sf MATHEMATICA}--based program \SigmaP,
\cite{sigma1,sigma2}. Some examples for typical sums that were obtained
during this work are given in Appendix~G.

For the cases of the diagrams $A_1$--$D_1$, $G_1$, $J_{1a}$--$K_{1b}$,
the insertion of the massless quark loop changed the structure of the
occurring Feynman parameter integrals moderately. The evaluation of
these diagrams could be performed in terms of a modification of the computation of the $2$-loop results
\cite{Bierenbaum:2007qe}. All other diagrams had to be computed newly. 

According to the complexity of the topology, between one and three 
sums had to be performed after the $\ep$-expansion. In the following we demonstrate details of the calculation considering some examples. 

\subsection{\boldmath Diagram $E_2$}

\begin{figure}[h]
\centering
\includegraphics[scale=1]{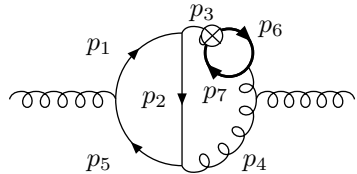}
\caption {\sf The Momentum flow for diagram $E_2$
\label{E2}}
\end{figure}
\noindent
A typical case for a  representation without any remaining sums prior the Laurent expansion 
is given by diagram $E_2$. After applying 
the projector (\ref{projG1}) the color and
$\gamma$-algebra  is performed. 
The momentum assignment is as follows: $p_1=k_1$, $p_2=k_1-k_2$, 
$p_3=k_2$, $p_4=k_2-p$, $p_5=k_1-p$, $p_6=k_3$, $p_7=k_3-k_2$, see Figure~\ref{E2}. 
One obtains the 
following 
representation~: 
\begin{eqnarray}
I_{E_{2}}&=&T_F^2 n_f C_A g^6_s
\frac {1} {(\Delta{\cdot}p)^{N}} 
\sum_{j=0}^{N-2}
\int \frac{d^D k_1}{(2 \pi)^D}
\int \frac{d^D k_2}{(2 \pi)^D}
\int \frac{d^D k_3}{(2 \pi)^D}
\N
\\
&&
\hspace{-21mm}
\times \frac {f(\Delta_{\mu},k_{1\mu},k_{2\mu},k_{3\mu},p_{\mu},m,\ep) 
(\Delta.k_3)^{j} (\Delta.k_3+\Delta.k_2)^{N-j-2}} {(-k_1^2)(-(k_1-p)^2) 
(-k_2^2)^2 (-(k_2-p)^2) (-(k_1-k_2)^2) (m^2-k_3^2) (m^2-(k_2-k_3)^2)} ~,
\label{E2Mom}
\end{eqnarray}
with $g_s$ the strong coupling constant, $\hat{a}_s = g_s^2 \cdot (\mu^2)^{\ep/2}/(4\pi)^2 $.

The factors $(\Delta.k_3)^{j} (\Delta.k_3+\Delta.k_2)^{N-j-2}$ in (\ref{E2Mom}) stem from the
operator insertion on a vertex. Applying the Feynman-parametrization (\ref{FeynParam}) to combine all
the denominators which contain the momentum $k_3$ and integrating out the $\delta$-distribution yields
\begin{eqnarray}
I_{E_{2}}&=&T_F^2 n_f C_A g^6_s
\frac {1} {(\Delta{\cdot}p)^{N}} 
\sum_{j=0}^{N-2}
\int_0^1 dx_0
\int \frac{d^D k_1}{(2 \pi)^D}
\int \frac{d^D k_2}{(2 \pi)^D}
\int \frac{d^D k_3}{(2 \pi)^D}
\N\\
&&
\times \frac {f(\Delta_{\mu},k_{1\mu},k_{2\mu},k_{3\mu},p_{\mu},m,\ep) (\Delta.k_3)^{j} (\Delta.k_3+\Delta.k_2)^{N-j-2}} {D(k_1,k_2,p,m) 
(m^2-k_3^2-2 x_0 k_3.k_2-x_0 k_2^2)^2} ~.
\end{eqnarray}
Here $D(k_1,k_2,p,m)$ denotes all the factors in the denominator of
(\ref{E2Mom}) which do not depend on the momentum $k_3$. In order to
symmetrize the momentum integral we shift the momentum $k_3
\rightarrow l_3 -x_0k_2$. This yields
\begin{eqnarray}
I_{E_{2}}&=&T_F^2 n_f C_A g^6_s
\frac {1} {(\Delta{\cdot}p)^{N}} 
\sum_{j=0}^{N-2}
\int_0^1 dx_0
\int \frac{d^D k_1}{(2 \pi)^D}
\int \frac{d^D k_2}{(2 \pi)^D}
\int \frac{d^D l_3}{(2 \pi)^D}
\N\\
&&
\hspace{-15mm}
\times \frac {f(\Delta_{\mu},k_{1\mu},k_{2\mu},l_{3\mu},p_{\mu},m,x_0,\ep) (\Delta.l_3-x_0 
\Delta.k_2)^{j} (\Delta.l_3+(1-x_0) \Delta.k_2)^{N-j-2}} {D(k_1,k_2,p,m) (m^2-k_3^2-x_0 (1-x_0) k_2^2)^2}
~. \end{eqnarray}
Now we perform a Wick rotation to obtain an Euclidean momentum integral, which can
be evaluated as described in Appendix \ref{sec:1}. The results depend 
strongly on the momentum structure in the numerator, and become rather lengthy in many cases.
During this computation many intermediary results amount to several {\tt Mbytes} of data. For this 
reason 
we consider a numerator of the form 
\begin{eqnarray}
f(\Delta_{\mu},k_{1\mu},l_{2\mu},p_{\mu},m,N,\ep,x_1,x_2)=m^4 \Delta.p^2
\end{eqnarray} 
to demonstrate further steps 
of the computation. In this case only the first term in both powers of the respective binomial 
expansions of the factors 
$(\Delta.l_3-x_0 \Delta.k_2)^{j} (\Delta.l_3+(1-x_0) \Delta.k_2)^{N-j-2}$ contribute. We thereby obtain the following
intermediate result after integrating out the momentum $l_3$ 

\begin{eqnarray}
I_{E_{2}}&=&-T_F^2 n_f C_A g^6_s
\frac {1} {(\Delta{\cdot}p)^{N} (4 \pi)^{D/2}} 
\frac {m^4 (\Delta.p)^2 B(D/2,2-D/2)} {(4 \pi)^{D/2} \Gamma(-D/2)}
\sum_{j=0}^{N-2} (-1)^j x_0^j (1-x_0)^{N-j-2}
\N\\
&&
\times
\int \frac{d^D k_1}{(2 \pi)^D}
\int \frac{d^D k_2}{(2 \pi)^D}
\frac {(\Delta.k_2)^{N-2}} {D(k_1,k_2,p,m) (m^2-x_0 (1-x_0) k_2^2)^{2-D/2}} ~.
\end{eqnarray}
The physical sum can now be evaluated in terms of a geometric sum
\begin{eqnarray}
\sum_{j=0}^{N-2} (-1)^j x_0^j (1-x_0)^{N-j-2}&=&(-1)^N x_0^{N-1}+(1-x_0)^{N-1} ~.
\end{eqnarray}
Thus the following representation is obtained
\begin{eqnarray}
I_{E_{2}}&=&-T_F^2 n_f C_A g^6_s
\frac {1} {(\Delta{\cdot}p)^{N}} 
\frac {m^4 (\Delta.p)^2 B(D/2,2-D/2)} {(4 \pi)^{D/2} \Gamma(-D/2)}
\N\\
&&
\times \int_{0}^{1} dx_0~~
x_0^{D/2-2} (1-x_0)^{D/2-2} \left[(-1)^N x_0^{N-1}+(1-x_0)^{N-1}\right]
\int \frac{d^D k_1}{(2 \pi)^D}
\int \frac{d^D k_2}{(2 \pi)^D}
\N\\
&&
\times 
\frac {(\Delta.k_2)^{N-2}} {D(k_2) (-k_1^2)(-(k_1-p)^2) (-(k_1-k_2)^2) (m^2/(x_0(1-x_0))-k_2^2)^{2-D/2}} 
~.
\end{eqnarray}
In the term, which contains the factor $(1-x_0)^{N-1}$, the variable shift $x_0\rightarrow (1-x_0)$ is 
performed. The same steps as above are applied to integrate over $k_1$. This yields
\begin{eqnarray}
I_{E_{2}}&=&-i T_F^2 n_f C_A g^6_s
\frac {1} {(\Delta{\cdot}p)^{N-2}} 
\frac {m^4 B(D/2,2-D/2) B(D/2,3-D/2)} {(4 \pi)^{D} \Gamma(-D/2)^2}
\N\\
&&
\times \int_{0}^{1} dx_0~~
x_0^{N+D/2-3} (1-x_0)^{D/2-2} \left(1+(-1)^N\right)
\\
&&
\times
\int_{0}^{1} dx_2
\int_{0}^{1} dx_3~~ \theta(1-x_2-x_3)
\int \frac{d^D k_2}{(2 \pi)^D}
\N\\
&&\N
\hspace{-23mm}
\times 
\frac {(\Delta.k_2)^{N-2}} {D(k_2) (-k_2^2)(-(k_2-p)^2) (-(x_2(1-x_2)k_2-x_2x_3 p)^2) (m^2/(x_0 (1-x_0))-k_2^2)^{2-D/2}} ~.
\end{eqnarray}
The transformation $x_3\rightarrow x_3 (1-x_2)$ maps the integral restricted by the $\theta$-function to 
the domain $[0,1]$. Finally, integrating over the momentum $k_2$, one obtains
\begin{eqnarray}
I_{E_{2}}&=& -T_F^2 n_f C_A g^6_s
\frac {m^4} {(4 \pi)^{3/2D}} \Gamma\left(8- \frac 3 2 D \right)
\N\\
&&
\times \int_{0}^{1} dx_0
x_0^{N+D/2-3} (1-x_0)^{D/2-2} \left[1+(-1)^N\right]
\N\\
&&
\times
\int_{0}^{1} dx_2
\int_{0}^{1} dx_3 x_2^{D/2-3} (1-x_2)^{D/2-2}
\int_{0}^{1} dx_7 x_7 (1-x_7)^{1-D/2}
\N\\
&&
\times
\int_{0}^{1} dx_5
\int_{0}^{1} dx_6~\theta(1-x_5-x_6) x_5^{2-D/2} (1-x_5-x_6)^{3-D/2} (x_6+x_3 x_5)^{N-2}
\N\\
&&
\times 
{\left[ \frac {x_0 (1-x_0)} {m^2 (1-x_7) (1-x_5-x_6)}\right]^{8-3/2 D}} ~,
\end{eqnarray}
with $D=4+\ep$. One notices, that after applying the transformation $x_6 \rightarrow (1-x_5) x_6$ the denominator factorizes into Beta-function like structures. No hypergeometric functions are required to represent this Feynman-integral. The integral over $x_3$ can be directly performed, yielding
\begin{eqnarray}
I_{E_{2}}&=& -T_F^2 n_f C_A g^6_s
\frac {(m^2)^{3/2 \ep}} {(4 \pi)^{3/2D}}  \Gamma \left( 8-\frac {3} {2} D\right)
\N\\
&&
\times \int_{0}^{1} dx_0
x_0^{5-D+N} (1-x_0)^{6-D} \left[1+(-1)^N\right]
\N\\
&&
\times
\int_{0}^{1} dx_2 ~~x_2^{D/2-3} (1-x_2)^{D/2-2}
\int_{0}^{1} dx_7~~ x_7 (1-x_7)^{-7+D}
\N\\
&&
\times
\int_{0}^{1} dx_5
\int_{0}^{1} dx_6 ~~x_5^{1-D/2} (1-x_5)^{D-4} (1-x_6)^{D-5} 
\N\\
&&
 \times \left\{\frac 1 {N-1} \left[(1-x_5)^{N-1} x_6^{N-1}-(x_5+x_6-x_5x_6)^{N-1}\right] \right\}~.
\end{eqnarray}
In the term, which contains the factor $(x_5+x_6-x_5x_6)^{N-1}$, the transformations 
\begin{eqnarray}
x_5\rightarrow x_5 x_6
\end{eqnarray}
 and 
\begin{eqnarray}
x_6\rightarrow\frac{x_5 (1-x_6)} {1-x_5 x_6}~,
\end{eqnarray}
cf. Appendix {\ref {App-VarTrans}}, are applied to obtain a representation which can be represented in 
terms of Beta-functions~:
\begin{eqnarray}
I_{E_{2}}&=& T_F^2 n_f C_A g^6_s
\frac {(m^2)^{3/2 \ep}} {(4 \pi)^{3/2D}} \Gamma\left(8 - \frac 3 2 \D \right)
\N\\
&&
\times \int_{0}^{1} dx_0
x_0^{5-D+N} (1-x_0)^{6-D} \left[1+(-1)^N\right]
\N\\&&
\times
\int_{0}^{1} dx_2 ~~x_2^{D/2-3} (1-x_2)^{D/2-2}
\int_{0}^{1} dx_7 ~~x_7 (1-x_7)^{-7+D}
\N\\
&&
\times
\int_{0}^{1} dx_5
\int_{0}^{1} dx_6~~ (1-x_5)^{1-D/2} (x_5)^{D-4} x_6^{D-5} 
\N\\
&&
 \times \frac 1 {N-1} \left[ x_5^{N+5-3/2 D} (1-x_5)^{-6+3/2 D} x_6^{6-3/2 D}- x_5^{N-1} (1-x_6)^{N-1} \right]~.
\end{eqnarray}

Performing the Feynman parameter integrals, and applying the conventions in Appendix \ref{App-Con}, yields the following representation in terms of $\Gamma$-functions:

\begin{eqnarray}
I_{E_{2}}&=& T_F^2 n_f C_A a^3_s
\left(\frac {m^2} {\mu^2}\right)^{3/2 \ep} \frac 1 {(4 \pi)^{3/2 \ep}} 
\frac {1} {N-1} 
\N\\
&&
\times
\Gamma\Biggl[\frac[0pt]{-\ep/2,\ep/2,1+\ep/2,\ep-2,2-3/2 \ep,3-\ep,2+N-\ep}{1+\ep,5+N-2\ep,N+\ep/2}\Biggr]
\N\\
&&
\times \Biggl\{\Gamma\Biggl[\frac[0pt] {N-\ep/2}{1-\ep/2}\Biggr]
-\Gamma\left({N}\right)
 \Biggr\}
\times \Biggl[1+(-1)^N\Biggr]~.\label{E2Allep}
\N\\
&&
\end{eqnarray}
In (\ref{E2Allep}) no more sums remain to be evaluated. Thus, the final result is obtained just by expanding in $\ep$. 

\subsection{\boldmath Diagram $L_{1a}$}

\begin{figure}[h]
\centering
\includegraphics[scale=1]{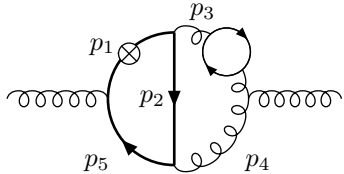}
\caption {\sf Momentum flow for diagram $L_{1a}$ \label{L1a}}
\end{figure}
\noindent
For this diagram the following representation is obtained after performing the color and
Dirac-algebra and integrating out the massless quark insertion~:

\begin{eqnarray}
I_{L_{1a}}&=&T_F^2 n_f C_A g_s^6 ~ \Gamma\Biggl[\frac[0pt]{-\ep/2,2+\ep/2,2+\ep/2}{4+\ep} \Biggr] \frac {1} {(4 \pi)^{D/2}}
\frac {1} {(\Delta{\cdot}p)^{N}} \int \frac{d^D k_1}{(2 \pi)^D}
\int \frac{d^D k_2}{(2 \pi)^D}
\N\\
&&
\hspace*{-6mm}
\times
\frac
{f(\Delta_{\mu},k_{1\mu},k_{2\mu},p_{\mu},m,\ep) (\Delta.k_1)^{N-1}} {(-k_2^2) (m^2-k_1^2)^2
  ((-k_2-p)^2)^{2-\ep/2} (m^2-(k_1-p)^2) (m^2-(k_1-k_2)^2)} ~.
\label{I1Mom}
\end{eqnarray}
Here the momentum flow is $p_1=k_1$, $p_2=k_1-k_2$, $p_3=k_2$, $p_4=k_2-p$, $p_5=k_1-p$, cf.~ 
Figure~\ref{L1a}.
The factor $(\Delta.k_1)^{N-1}$ in (\ref{I1Mom}) stems from the
operator insertion on a quark line. A Feynman-parametrization is applied to combine all
the denominators, which contain the momentum $k_2$. The emerging $\delta$-distribution is integrated out. This yields
\begin{eqnarray}
I_{L_{1a}}&=&T_F^2 n_f C_A g_s^6 ~ \Gamma\Biggl[\frac[0pt]{-\ep/2,2+\ep/2,2+\ep/2,3-\ep/2}{4+\ep,1-\ep/2} \Biggr] \frac {1} {(4 \pi)^{D/2}}
\frac {1} {(\Delta{\cdot}p)^{N}} 
\N\\
&&
\times
\int_0^1~ dx_1 \int_0^1~ dx_2 \theta(1-x_1-x_2) x_2^{-\ep/2} 
\int \frac{d^D k_1}{(2 \pi)^D}
\int \frac{d^D k_2}{(2 \pi)^D}
\N\\
&&
\times
\frac
{f(\Delta_{\mu},k_{1\mu},k_{2\mu},p_{\mu},m,N,\ep) (\Delta.k_1)^{N-1}}
{D(k_1,p,m) (-k_2^2+x_1 m^2+2 x_1 k_1.k_2+ 2 x_2 k_2.p - x_1
  k_1^2)^{4-\ep/2} }~.
\end{eqnarray}
Here $D(k_1,p,m)$ denotes all the factors in the denominator of
(\ref{I1Mom}) that do not depend on the momentum $k_2$. In order to
symmetrize the momentum integration we shift the momentum $k_2
\rightarrow l_2 +x_1 k_1+x_2 p$. This yields

\begin{eqnarray}
I_{L_{1a}}&=&T_F^2 n_f C_A g_s^6 ~ \Gamma\Biggl[\frac[0pt]{-\ep/2,2+\ep/2,2+\ep/2,3-\ep/2}{4+\ep,1-\ep/2} \Biggr] \frac {1} {(4 \pi)^{D/2}}
\frac {1} {(\Delta{\cdot}p)^{N}} 
\N\\
&&
\times
\int_0^1~ dx_1 \int_0^1~ dx_2 \theta(1-x_1-x_2) x_2^{-\ep/2} 
\int \frac{d^D k_1}{(2 \pi)^D}
\int \frac{d^D l_2}{(2 \pi)^D}
\N\\
&&
\times
\frac
{f(\Delta_{\mu},k_{1\mu},k_{2\mu},p_{\mu},m,N,\ep) (\Delta.k_1)^{N-1}}
{D(k_1,p,m) (-l_2^2+x_1 m^2- x_1 (1-x_1) k_1^2 + 2 x_1 x_2 k_1.p)^{4-\ep/2} }~.
\end{eqnarray}

Considering  only the contribution 
\begin{eqnarray}
f(\Delta_{\mu},k_{1\mu},l_{2\mu},p_{\mu},m,N,\ep,x_1,x_2)=(\Delta.p) m^4
\end{eqnarray}
for now and performing the $l_2$ integral gives
\begin{eqnarray}
I_{L_{1a}}&=& i T_F^2 n_f C_A g_s^6 \frac {1} {(4 \pi)^{D}} \frac {1} {(\Delta{\cdot}p)^{N}} 
\Gamma\Biggl[\frac[0pt]{-\ep/2,2+\ep/2,2+\ep/2,1-\ep}{1-\ep/2,4+\ep}\Biggr]
~
\N\\
&&
\times
\int_0^1~ dx_1 \int_0^1~ dx_2 {x_2^{-\ep/2}
  (1-x_1)^{1-\ep/2}}
\N\\
&&
\times
\int \frac{d^D k_1}{(2 \pi)^D}
\frac {(\Delta.p)~ m^4~(\Delta.k_1)^{N-1}} {D(k_1,p,m) (x_1 m^2 - x_1 (1-x_1) k_1^2 + 2 x_1 (1-x_1) x_2 k_1.p)^{1-\ep} }
\\
&=& i T_F^2 n_f C_A g_s^6 \frac {1} {(4 \pi)^{D}}   
\Gamma\Biggl[\frac[0pt]{-\ep/2,2+\ep/2,2+\ep/2,1-\ep}{1-\ep/2,4+\ep}\Biggr]
\N\\
&&
\times
\int_0^1~ dx_1 \int_0^1~ dx_2 ~x_1^{\ep-1} (1-x_1)^{\ep/2} x_2^{-\ep/2}
\N\\
&&
\times
\int \frac{d^D k_1}{(2 \pi)^D}
\frac
{\Delta.p~(\Delta.k_1)^{N-1}}
{(m^2-k_1^2) (m^2-(k_1-p)^2) (m^2/(1-x_1) - k_1^2 + 2 x_2 k_1.p)^{1-\ep} }
~.
\end{eqnarray}
Combining the remaining denominators inside the momentum integral and symmetrizing by performing the momentum shift $k_1\rightarrow l_1 +(x_5+x_2 x_6)p$ yields
\begin{eqnarray}
I_{L_{1a}}&=&i T_F^2 n_f C_A g_s^6 \frac {1} {(4 \pi)^{D}}  \Gamma\Biggl[\frac[0pt]{-\ep/2,2+\ep/2,2+\ep/2,4-\ep} {1-\ep/2,4+\ep}\Biggr] \frac {1} {(\Delta{\cdot}p)^{N}} 
\N\\
&&
\times
\int_0^1~ dx_1 \int_0^1~ dx_2 x_1^{\ep-1} (1-x_1)^{\ep/2} x_2^{-\ep/2}
\N\\
&&
\times
\int_0^1 dx_5 \int_0^1 dx_6 x_6^{-\ep} \theta(1-x_5-x_6) (1-x_5-x_6)
\N\\
&&
\times
\int \frac{d^D l_1}{(2 \pi)^D}
\frac
{(\Delta.p)~m^4(\Delta.l_1+(x_5+x_2 x_6) \Delta.p)^{N-1}}
{(m^2(1-x_6(1-1/(1-x_1))-l_1^2)^{4-\ep} }
~.
\end{eqnarray}
When expanding the factor $(\Delta.l_1+(x_5+x_2 x_6) \Delta.p)^{N-1}$ as described above, only the term $\Delta.p^{N-1} (x_5 + x_2 x_6)^{N-1}$ 
contributes. Applying the transformation $x_5\rightarrow x_5 (1-x_6)$ and performing the momentum integration yields
 \begin{eqnarray}
I_{L_{1a}}&=&T_F^2 n_f C_A g_s^6 \frac {1} {(4 \pi)^{3/2 D}}  \Gamma\Biggl[\frac[0pt]{-\ep/2,2+\ep/2,2+\ep/2,2-3/2 \ep} {1-\ep/2,4+\ep}\Biggr] \frac {1} {(\Delta{\cdot}p)^{N}} 
\N\\
&&
\times
\int_0^1~ dx_1 \int_0^1~ dx_2 x_1^{\ep-1} (1-x_1)^{\ep/2} x_2^{-\ep}
\N\\
&&
\times
\int_0^1 dx_5 \int_0^1 dx_6 (1-x_5) x_6^{-\ep} (1-x_6)^{2}
\N\\
&&
\times
\frac
{(\Delta.p) m^4~((x_5 (1-x_6) +x_2 x_6) \Delta.p)^{N-1}}
{(m^2(1-x_6(1-/(1-x_1)))^{2-3/2 \ep} }
~.
\end{eqnarray}
The factor $(1-x_5)$ is expanded. For reasons of brevity we will only consider the first term. The 
integral over $x_5$ can be performed directly. Shifting~$x_2 \rightarrow (1-x_2)$ gives
 \begin{eqnarray}
I_{L_{1a}}&=&
T_F^2 n_f C_A g_s^6 \frac {1} {(4 \pi)^{3/2 D}}  \Gamma\Biggl[\frac[0pt]{-\ep/2,2+\ep/2,2+\ep/2,2-3/2 \ep} {1-\ep/2,4+\ep}\Biggr]
\N\\
&&
\times
\int_0^1~ dx_1 \int_0^1~ dx_2 x_1^{\ep-1} (1-x_1)^{2-\ep} x_2^{-\ep/2}
\N\\
&&
\times
\int_0^1 dx_6 x_6^{-\ep} (1-x_6)
\frac
{1}
{(1-x_1+x_6 x_1)^{2-3/2 \ep} (m^2)^{3/2 \ep}}
\N\\
&&
\times
\frac {1} {N} \left[(1-x_6 x_2)^N-((1-x_2) x_6)^N \right]
~.
\end{eqnarray}
The factor $(1-x_6 x_2)^N$ cannot simplified further and a binomial expansion is applied

 \begin{eqnarray}
I_{L_{1a}}&=&
T_F^2 n_f C_A g_s^6 \frac {1} {(4 \pi)^{3/2 D}}  \Gamma\Biggl[\frac[0pt]{-\ep/2,2+\ep/2,2+\ep/2,2-3/2 \ep} {1-\ep/2,4+\ep}\Biggr]
\N\\
&&
\times
\int_0^1~ dx_1 \int_0^1~ dx_2 x_1^{\ep-1} (1-x_1)^{2-\ep} x_2^{-\ep/2}
\N\\
&&
\times
\int_0^1 dx_6 x_6^{-\ep} (1-x_6)
\frac
{(\Delta.p)^N}
{(1-x_1+x_6 x_1)^{2-3/2 \ep} (m^2)^{3/2 \ep}}
\N\\
&&
\times
\frac {1} {N} \left[\sum_{j_1=0}^{N} (-x_6 x_2)^{j_1}-(1-x_2)^N x_6^N \right]
~.
\end{eqnarray}
Finally the variable shift $x_6\rightarrow1-x_6$ is applied and the integration is performed. We obtain the following result:
\begin{eqnarray}
I_{L_{1a}}&=&T_F^2 n_f C_A \left(\frac{\mu^2} {\mu^2}\right)^{(3/2\ep} \frac {a_s^3} {(4 \pi)^{3/2 \ep}} \frac {1} {N}
    \sum_{j_1=1}^N \left[(-1)^{j_1}-\binom {N} {j_1}\delta_{j_{1}N}\right]
\N\\
&&
\times
\Gamma\Biggl[\frac[0pt]{-\ep/2,\ep,2+\ep/2,2+\ep/2,2-3/2\ep,3-\ep,1+j_1-\ep,1+j_1-\ep/2}{1-\ep/2,4+\ep,2+j_1-\ep/2,3+j_1-\ep}\Biggr]
\N\\
&&
\times
\empty_{3}F_2\Biggl[\frac[0pt]{2-3/2 \ep,2,\ep} {3+j_1-\ep,3} ;1\Biggr]~. \label{resL1}
\end{eqnarray}
As the generalized hypergeometric function in (\ref{resL1}) contains both one positive integer in the 
upper and lower indices, it can be summed by using the relations described in Appendix \ref{App-SpeFunFPQ}.

\subsection{\boldmath Diagram $I_1$}
\begin{figure}[h]
\centering
\includegraphics[scale=1]{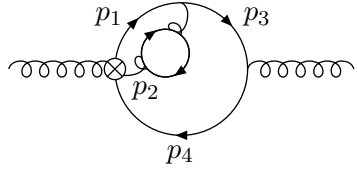}
\caption {\sf Momentum flow for diagram $I_{1}$
\label{I1a}}
\end{figure}
As a last example we consider diagram $I_1$. It has the following integral representation: 
\begin{eqnarray}
I_{I_{1}}&=&T_F^2 n_f g^6_s
\frac {1} {(\Delta{\cdot}p)^{N}} 
\sum_{j=0}^{N-3}\sum_{l=j+1}^{N-2}
\int \frac{d^D k_1}{(2 \pi)^D}
\int \frac{d^D k_2}{(2 \pi)^D}
\int \frac{d^D k_3}{(2 \pi)^D}
\\
&&
\hspace{-15mm}
\times \frac {f(\Delta_{\mu},k_{1\mu},k_{2\mu},k_{3\mu},p_{\mu},m,\ep) \left[C_1(\Delta.k_1)^{l-j-1} 
+C_2(\Delta.k_1+\Delta.k_2+\Delta.p)^{l-j-1}\right]} {(m^2-k_1^2)(m^2-(k_1-p)^2) (-k_2^2)^2 
(m^2-(k_1+k_2)^2) (-k_3^2) (-(k_2-k_3)^2)} ~.
\label{I2Mom}
\end{eqnarray}
The momentum flow is $p_1=k_1$, $p_2=k_2$, $p_3=k_1+k_2$, $p_4=k_1-p$, see Figure~\ref{I1a}. 
One of the sums is performed according to Eqs. (\ref{doubleSum1},\ref{doubleSum2}). The integration of 
the momenta is carried out analogously to earlier examples. Again we limit this demonstration to a small 
part of the numerator by considering only \begin{eqnarray}
f(\Delta_{\mu},k_{1\mu},k_{2\mu},k_{3\mu},p_{\mu},m,\ep)=(\Delta.p)^3~.
\end{eqnarray}
Furthermore we limit ourselves to terms which contain the color factor $C_A$. The following representation is obtained:
\begin{eqnarray}
I_{I_{1}}&=&T_F^2 n_f C_A a_s^3 
\left(\frac {m^2} {\mu^2}\right)^{3/2 \ep} \frac {1} {(4 \pi)^{3/2 D}}
\Gamma\Biggl[\frac[0pt]{2+\ep/2,2+\ep/2,-3/2 \ep} {1-\ep/2,4+\ep}\Biggr] (-1)^N
\N\\
&&
\times
\int_0^1 dx_0 \int_0^1 dx_1 \int_0^1 dx_1 \int_0^1 dx_3 \sum_{j=0}^{N-2} x_0^{\ep} (1-x_1)^{-1-\ep}
(1-x_1x_3)^{N-2-j} (1-x_0x_1)^{3/2 \ep}
\N\\
&&
\times \left[
(1-x_1x_3(1-x_0))^{j} - (x_1x_3(1-x_0))^{j}
\right]
~.
\end{eqnarray}
Here the two terms in the last factor stem from the structure in Eqs. (\ref{doubleSum1}-\ref{doubleSum2}).
Performing the sum $\sum_{j=0}^{N-2}$ in the second term by a geometric sum yields
\begin{eqnarray}
I_{I_{1}}&=&T_F^2 n_f C_A a_s^3 \left(\frac {m^2} {\mu^2}\right)^{3/2 \ep} \frac {1} {(4 \pi)^{3/2 D}}
\Gamma\Biggl[\frac[0pt]{2+\ep/2,2+\ep/2,-3/2 \ep}{1-\ep/2,4+\ep}\Biggr]
(-1)^N
\N\\
&&
\times \int_0^1 dx_0 \int_0^1 dx_1 \int_0^1 dx_1 \int_0^1 dx_3 \sum_{j=0}^{N-2} x_0^{\ep} (1-x_1)^{-1-\ep}
(1-x_0x_1)^{3/2 \ep}
\N\\
&&
\times
\left\{
- (1-x_1x_3)^{N-2-j}(x_1x_3(1-x_0))^{j}
\N \right.\\
&&\left.
+\frac{1} {x_0 x_1 x_3} \left[
(1-x_1x_3(1-x_0))^{N-1}
-(1-x_1x_3)^{N-1}
\right]
\right\}
~.
\end{eqnarray}
Due to the stronger nesting of the Feynman parameters, no simplifications by using the variable 
transformations in Appendix~\ref{App-VarTrans} are possible.
The terms $(1-x_1x_3(1-x_0))^{N-1}$ and $(1-x_1x_3)^{N-1}$ are expressed by binomial series. The remaining integral is then evaluated in terms of the hypergeometric function $_3F_2$ yielding the following representation:
\begin{eqnarray}
I_{I_{1}}&=&T_F^2 n_f C_A a_s^3 \left(\frac {m^2} {\mu^2}\right)^{3/2 \ep} \frac {1} {(4 \pi)^{3/2 D}}
\Gamma\Biggl[\frac[0pt]{-\ep/2,-3/2 \ep,2+\ep/2,2+\ep/2,-\ep}{4+\ep,1-\ep/2}\Biggr]
\N\\
&&
\Biggl\{
-\sum_{j=0}^{N-2} \sum_{j_1=0}^{N-j-2} \binom {N-j-2} {j_1} \Gamma\Biggl[\frac[0pt]{1+\ep,1+j-\ep,N-j_1-1}{2+j,N-j_1-\ep}\Biggr] 
\N\\
&&
\hspace{20mm}\times
_3F_2\Biggl[\frac[0pt]{-3/2 \ep,1+\ep,N-j_1}{2+j,N-j_1-\ep};1\Biggr]
\N\\
&&
+\sum_{j_1=0}^{N-1} \binom {N-1}{j_1} \Gamma\Biggl[\frac[0pt]{1-\ep,\ep,N-j_1-1}{N-j_1-\ep}\Biggr]  
\empty_3F_2\Biggl[\frac[0pt]{-3/2 \ep,\ep,N-j_1}{1,N-j_1-\ep};1\Biggr]
\N\\
&&
-\sum_{j_1=0}^{N-1} \binom {N-1}{j_1} \Gamma\Biggl[\frac[0pt]{\ep,N-j_1-1}{N-j_1}\Biggr]
\empty_2F_1\Biggl[\frac[0pt]{-3/2 \ep,\ep}{N-j_1-\ep};1\Biggr]
\Biggr\}
~.
\end{eqnarray}
The Gau\ss{}--function $_2F_1$ can be mapped to a product of $\Gamma$-functions by applying Gau\ss' 
theorem (\ref{Gauss}). Examples for typical sums, emerging after 
the series expansion in $\ep$ are given in Appendix~G.
 \newpage
\section{The Massive 3--Loop Operator Matrix Elements and 3--Loop Anomalous Dimensions \label{Results}}
In the following we summarize the main results of the calculation. 
We first obtain the unrenormalized massive operator matrix elements $\Ahathat_{ij}$ 
for the various channels. Their analytic structure is known, cf. \cite{Bierenbaum:2009mv}. 
At $3-loop$ order two new quantities can be obtained: {\it i)} the constant part 
of the OMEs $A_{ij}^{(3)}$, $a_{ij}^{(3)}$; {\it ii)} the 
three loop anomalous dimension corresponding to the process considered, $\gamma_{ij}^{(2)}$. 
The latter term is part of the $1/{\ep}$ pole term of \hspace*{1mm} $\Ahathat_{ij}^{(3}$. We consider 
the OMEs $A_{Qg}$, $A_{Qq}^{\sf PS}$, 
$A_{qq,Q}^{\sf PS}$ 
$A_{qq,Q}^{\sf NS}$ 
and $A_{qq,Q}^{\sf NS,TR}$ in the following 
and determine both quantities. 
They are represented in terms of the algebraic basis \cite{Blumlein:2003gb} 
of the harmonic sums, which leads to very essential structural simplifications and avoids redundancies present 
in other representations for e.g. the 3--loop anomalous dimensions \cite{Moch:2004pa,Vogt:2004mw} 
and the massless Wilson 
coefficients if compared to the representations in 
\cite{Vermaseren:2005qc,Blumlein:2009ta,Blumlein:2009fz,Blumlein:2009tj,ABS2010}. The contributions to 
the operator matrix elements calculated below constitute the first contributions for general values of 
$N$ at $3$--loop order and, due to Eqs.~(\ref{eqWIL1}--\ref{eqWIL5}), to the heavy flavor Wilson 
coefficients in the asymptotic region $Q^2\gg m^2$. The corresponding renormalized OMEs 
can be obtained following  Ref.~\cite{Bierenbaum:2009mv}, see also (\ref{AQg3MSren}), 
cf.~Section~\ref{Renorm}.

\subsection{\boldmath The gluonic contribution $\Ahathat_{Qg}^{(3)}$}
The unrenormalized OME \hspace*{1mm} $\Ahathat_{Qg}$ has the following structure, cf. 
\cite{Bierenbaum:2009mv},
   \begin{eqnarray}
   \Ahathat_{Qg}^{(3)}&=&
                  \Bigl(\frac{\hat{m}^2}{\mu^2}\Bigr)^{3\ep/2}
                     \Biggl[
           \frac{\hat{\gamma}_{qg}^{(0)}}{6\ep^3}
             \Biggl(
                   (n_f+1)\gamma_{gq}^{(0)}\hat{\gamma}_{qg}^{(0)}
                 +\gamma_{qq}^{(0)} 
                                \Bigl[
                                        \gamma_{qq}^{(0)}
                                      -2\gamma_{gg}^{(0)}
                                      -6\beta_0
                                      -8\beta_{0,Q}
                                \Bigr]
                 +8\beta_0^2
\N\\ &&
                 +28\beta_{0,Q}\beta_0
                 +24\beta_{0,Q}^2 
                  +\gamma_{gg}^{(0)} 
                                \Bigl[
                                        \gamma_{gg}^{(0)}
                                       +6\beta_0
                                       +14\beta_{0,Q}
                                \Bigr]
             \Biggr)
          +\frac{1}{6\ep^2}
             \Biggl(
                   \hat{\gamma}_{qg}^{(1)}
                      \Bigl[
                              2\gamma_{qq}^{(0)}
                             -2\gamma_{gg}^{(0)}
\N\\ &&
                             -8\beta_0
                             -10\beta_{0,Q} 
                      \Bigr]
                  +\hat{\gamma}_{qg}^{(0)}
                      \Bigl[
                              \hat{\gamma}_{qq}^{(1), {\sf PS}}\{1-2n_f\}
                             +\gamma_{qq}^{(1), {\sf NS}}
                             +\hat{\gamma}_{qq}^{(1), {\sf NS}}
                             +2\hat{\gamma}_{gg}^{(1)}
                             -\gamma_{gg}^{(1)}
                             -2\beta_1
\N\\ &&
                             -2\beta_{1,Q}
                      \Bigr]
                  + 6 \delta m_1^{(-1)} \hat{\gamma}_{qg}^{(0)} 
                      \Bigl[
                              \gamma_{gg}^{(0)}
                             -\gamma_{qq}^{(0)}
                             +3\beta_0
                             +5\beta_{0,Q}
                      \Bigr]
             \Biggr)
          +\frac{1}{\ep}
             \Biggl(
                   \frac{\hat{\gamma}_{qg}^{(2)}}{3}
                  -n_f \frac{\hat{\tilde{\gamma}}_{qg}^{(2)}}{3}
\N\\ &&
                  +\hat{\gamma}_{qg}^{(0)}\Bigl[
                                    a_{gg,Q}^{(2)}
                                   -n_fa_{Qq}^{(2),{\sf PS}}
                                          \Bigr]
                  +a_{Qg}^{(2)}
                      \Bigl[
                              \gamma_{qq}^{(0)}
                             -\gamma_{gg}^{(0)}
                             -4\beta_0
                             -4\beta_{0,Q}
                      \Bigr]
                  +\frac{\hat{\gamma}_{qg}^{(0)}\zeta_2}{16}
                      \Bigl[
                              \gamma_{gg}^{(0)} \Bigl\{
                                                        2\gamma_{qq}^{(0)}
\N\\ &&
                                                       -\gamma_{gg}^{(0)}
                                                       -6\beta_0
                                                       +2\beta_{0,Q}
                                                \Bigr\}
                             -(n_f+1)\gamma_{gq}^{(0)}\hat{\gamma}_{qg}^{(0)}
                             +\gamma_{qq}^{(0)} \Bigl\{
                                                       -\gamma_{qq}^{(0)}
                                                       +6\beta_0
                                                \Bigr\}
                             -8\beta_0^2
\N\\ &&
                             +4\beta_{0,Q}\beta_0
                             +24\beta_{0,Q}^2
                      \Bigr]
                  + \frac{\delta m_1^{(-1)}}{2}
                      \Bigl[
                              -2\hat{\gamma}_{qg}^{(1)}
                              +3\delta m_1^{(-1)}\hat{\gamma}_{qg}^{(0)}
                              +2\delta m_1^{(0)}\hat{\gamma}_{qg}^{(0)}
                      \Bigr]
\N\\ &&
                  + \delta m_1^{(0)}\hat{\gamma}_{qg}^{(0)}
                       \Bigl[
                               \gamma_{gg}^{(0)}
                              -\gamma_{qq}^{(0)}
                              +2\beta_0
                              +4\beta_{0,Q}
                      \Bigr]
                  -\delta m_2^{(-1)}\hat{\gamma}_{qg}^{(0)}
             \Biggr)
                 +a_{Qg}^{(3)}
                  \Biggr]~. \label{AhhhQg3}
\end{eqnarray}
Here and in the following we always project onto the color factors $T_F^2n_f C_F$ and $T_F^2n_f C_A$.
The computation performed in this thesis yields~:
%
%
\begin{eqnarray}
A_{Qg}^{(3)}&=& \frac {T_F^2 n_f} {N (N+1) (N+2)} \Biggl\{
\N\\
&&
C_A \Biggl\{
\frac{1} {\ep^3} \Biggl[
(N^2+N+2) \left(
-\frac{512 (N^2+N+1)}{9 (N-1) N (N+1) (N+2)}
+\frac{256}{9}  S_1
 \right)
\Biggr]
\N\\
&&
+\frac{1} {\ep^2} \Biggl[
(N^2+N+2) \left(
\frac{128}{9}  S_1^2
+\frac{256}{9} S_{-2}
+\frac{128}{9} S_2 
\right)
\N\\
&&
-\frac{128 (5 N^4+20 N^3+59 N^2+76 N+20) S_1}{27 (N+1) (N+2)}
-\frac{32 Q_1(N)}{27 (N-1) N^2 (N+1)^2 (N+2)^2}
\Biggr]
\N\\
&&
+\frac{1} {\ep} \Biggl[
(N^2+N+2) \Biggl(
+\frac{128}{27}  S_1^3
+\frac{32}{3}  \zeta _2 S_1
+\frac{128}{3}  S_{-2} S_1
+\frac{256}{9}  S_2 S_1
+\frac{832}{27} S_3
\N\\
&&
-\frac{128}{3} S_{-2,1}
+\frac{128}{9} S_{2,1} 
+\frac{320}{9} S_{-3}
-\frac{64 (N^2+N+1) }{3 (N-1) N (N+1) (N+2)} \zeta _2
\Biggr)
\N\\
&&
-\frac{128 (5 N^4+14 N^3+53 N^2+82 N+20) }{27 (N+1) (N+2)} S_{-2}
-\frac{32 Q_2(N) }{27 (N-1) N (N+1) (N+2)} S_2
\N\\
&&
-\frac{32 (10 N^4+49 N^3+154 N^2+197 N+58) }{27 (N+1) (N+2)}S_1^2
+\frac{32 Q_3(N)}{81 (N+1)^2 (N+2)^2} S_1
\N\\
&&
+\frac{16 Q_4(N)}{81 (N-1) N^3 (N+1)^3 (N+2)^3}
\Biggr]
\Biggr\}\N
\N\\
&&
+C_F \Biggl\{
\frac{1} {\ep^3} \Biggl[
(N^2+N+2) \left(
\frac{64  Q_{5}(N)}{9 (N-1) N^2(N+1)^2 (N+2)}
-\frac{256}{9} S_1
 \right)
\Biggr]
\N\\
&&
+ \frac {1} {\ep^2} \Biggl[
-\frac{128}{9} (N^2+N+2) S_1^2
+\frac{128 (5 N^3+11 N^2+28 N+12) S_1}{27 N}
\N\\
&&
+\frac{16 (N-2) Q_{6}(N)}{27 (N-1) N^3 (N+1)^3 (N+2)^2}
\Biggr]
+ \frac {1} {\ep} \Biggl[
(N^2+N+2) \Biggl(
-\frac{128}{27}  S_1^3
\N\\
&&
+\frac{704}{27}  S_3
-\frac{32}{3} \zeta _2 S_1
-\frac{128}{9}  S_2 S_1
\Biggr)
+\frac{64 (5 N^3+20 N^2+37 N+12) }{27 N} S_1^2
\N\\
&&
+\frac{8 (N^2+N+2) Q_5(N) }{3 (N-1) N^2 (N+1)^2 (N+2)} \zeta _2
+\frac{32 Q_7(N) }{9 (N-1) N^2 (N+1)^2 (N+2)}S_2 
\N\\
&&
-\frac{64 (10 N^4+86 N^3+483 N^2+341 N+114)}{81 N (N+1)} S_1
\N\\
&&
-\frac{4 Q_8(N)}{81 (N-1) N^4 (N+1)^4 (N+2)^3}
\Biggr]
\Biggr\}
\Biggr\}
+a_{Qg}^{(3)}~.
\end{eqnarray}
The contributions due to the individual diagrams are given in (\ref{DIAGA1}-\ref{DIAGH2}). The constant 
part $a_{Qg}^{(3)}$ reads~:
%
%
\begin{eqnarray}
 a_{Qg}^{(3)}&=& 
    n_fT_F^2C_A\Biggl\{
         \frac{16(N^2+N+2)}{27N(N+1)(N+2)}\Bigl[
             108 S_{-2,1,1}
            -78 S_{2,1,1}
            -90 S_{-3,1}
            +72 S_{2,-2}
            -6 S_{3,1}
\N\\ && \hspace{-12mm}
            -108 S_{-2,1} S_1
            +42 S_{2,1} S_1
            -6 S_{-4}
            +90 S_{-3} S_1
            +118 S_3 S_1
            +120 S_4
            +18 S_{-2} S_2
            +54 S_{-2} S_1^2
\N\\ && \hspace{-12mm}
            +33 S_2 S_1^2
            +15 S_2^2
            +2 S_1^4
            +18S_{-2}\zeta_2
            +9S_2\zeta_2
            +9S_1^2\zeta_2
            -42 S_1\zeta_3
                                    \Bigr]
\N\\ && \hspace{-12mm}
        +32\frac{5N^4+14N^3+53N^2+82N+20}{27N(N+1)^2(N+2)^2}\Bigl[
             6 S_{-2,1}
            -5 S_{-3}
            -6 S_{-2}S_1
                                    \Bigr]
\N\\ && 
\hspace{-12mm}
        -\frac{64(5N^4+11N^3+50N^2+85N+20)}{27N(N+1)^2(N+2)^2}S_{2,1}
\N\\ && \hspace{-12mm}
        -\frac{16(40N^4+151N^3+544N^2+779N+214)}{27N(N+1)^2(N+2)^2}S_2S_1
\N\\ && \hspace{-12mm}
        -\frac{32(65N^6+429N^5+1155N^4+725N^3+370N^2+496N+648)}{81(N-1)N^2(N+1)^2(N+2)^2}S_3
\N\\ && \hspace{-12mm}
        -\frac{16(20N^4+107N^3+344N^2+439N+134)}{81N(N+1)^2(N+2)^2}S_1^3
        +\frac{Q_{9}(N)}{81(N-1)N^3(N+1)^3(N+2)^3} S_2
\N\\ && \hspace{-12mm}
        +\frac{32(47N^6+278N^5+1257N^4+2552N^3+1794N^2+284N+448)}{81N(N+1)^3(N+2)^3}S_{-2}
\N\\ && \hspace{-12mm}
        +\frac{8(22N^6+271N^5+2355N^4+6430N^3+6816N^2+3172N+1256)}{81N(N+1)^3(N+2)^3}S_1^2
\N\\ && \hspace{-12mm}
        +\frac{Q_{10}(N)}{243(N-1)N^2(N+1)^4(N+2)^4} S_1
        +\frac{448(N^2+N+1)(N^2+N+2)}{9(N-1)N^2(N+1)^2(N+2)^2}\zeta_3
\N
\\
&& \hspace{-12mm}
        -\frac{16(5N^4+20N^3+59N^2+76N+20)}{9N(N+1)^2(N+2)^2}S_1\zeta_2
        -\frac{Q_{11}(N)}{9(N-1)N^3(N+1)^3(N+2)^3} \zeta_2
\N\\ && \hspace{-12mm}
        -\frac{Q_{12}(N)}{243(N-1)N^5(N+1)^5(N+2)^5}
       \Biggr\}
\N
\\&&
   +n_fT_F^2C_F\Biggl\{
         \frac{16(N^2+N+2)}{27N(N+1)(N+2)}\Bigl[
             144S_{2,1,1}
            -72S_{3,1} 
            -72S_{2,1}S_1
            +48S_4
            -16S_3S_1
\N\\ && \hspace{-12mm}
            -24S_2^2
            -12S_2S_1^2
            -2S_1^4
            -9 S_1^2\zeta_2
            +42S_1\zeta_3
                                    \Bigr]
        +32\frac{10N^3+49N^2+83N+24}{81N^2(N+1)(N+2)}\Bigl[
             3S_2S_1
             +S_1^3
                                    \Bigr]
\N\\ && \hspace{-12mm}
        -\frac{128(N^2-3N-2)}{3N^2(N+1)(N+2)}S_{2,1}
        -\frac{Q_{13}(N)}{81(N-1)N^3(N+1)^3(N+2)^2} S_3
\N\\ && \hspace{-12mm}
        +\frac{Q_{14}(N)}{27(N-1)N^4(N+1)^4(N+2)^3} S_2
        -\frac{32(10N^4+185N^3+789N^2+521N+141)}{81N^2(N+1)^2(N+2)}S_1^2
\N\\ && \hspace{-12mm}
        -\frac{16(230N^5-924N^4-5165N^3-7454N^2-10217N-2670)}{243N^2(N+1)^3(N+2)}S_1
\N\\ && \hspace{-12mm}
        +\frac{16(5N^3+11N^2+28N+12)}{9N^2(N+1)(N+2)}S_1\zeta_2
        -\frac{Q_{15}(N)}{9(N-1)N^3(N+1)^3(N+2)^2}\zeta_3
\N\\ && \hspace{-12mm}
        +\frac{Q_{16}(N)}{9(N-1)N^4(N+1)^4(N+2)^3}\zeta_2
        +\frac{Q_{17}(N)}{243(N-1)N^6(N+1)^6(N+2)^5} 
        \Biggr\}~,\label{aQg3}
\end{eqnarray}
with the polynomials
\begin{eqnarray} 
Q_{1}(N)&=&15 N^9+90 N^8+146 N^7-32 N^6-501 N^5-610 N^4-244 N^3-48 N^2
\N\\
&&
+224 N+96~,
\\
Q_2(N)&=&10 N^6+105 N^5+231 N^4+109 N^3+143 N^2+122N+144~,
\\
Q_3(N)&=&11 N^6+32 N^5+363 N^4+848 N^3+186 N^2-520 N+88~,
\\
Q_4(N)&=&273 N^{12}+2565 N^{11}+9838 N^{10}+18902 N^9+14303 N^8-8953 N^7
\N\\
&&
-26402 N^6-19930 N^5-1860 N^4+4512 N^3-2560 N^2-2208 N
\N\\
&&
-576~,
\\
Q_{5}(N)&=&3 N^6+9 N^5-N^4-17 N^3-38 N^2-28 N-24~,
\\
Q_{6}(N)&=&45 N^{10}+405 N^9+1606 N^8+3842
   N^7+6717 N^6+9325 N^5+10888 N^4
\N\\
&&
+9804 N^3+6232 N^2+3264 N+864~,
\\
Q_7(N)&=&3 N^8+62 N^7+150 N^6-8 N^5-357 N^4-350 N^3-300 N^2-208 N
\N\\
&&
-144~,
\\
Q_8(N)&=&873 N^{14}+8298 N^{13}+29743 N^{12}+38892 N^{11}-42545 N^{10}-211766 N^9
\N\\
&&
-227439 N^8+95376 N^7+573704 N^6+942576 N^5+995168 N^4
\N\\
&&
+776576 N^3+527232 N^2+232704 N+ 48384~,
\\
 Q_{9}(N)&=&32N^9-936N^8+6448N^7+55208N^6+126160N^5+61760N^4
\N\\
&&
-53152N^3-25024N^2-32256N -13824
~, \\
 Q_{10}(N)&=&+7856N^{10}+84672N^9+377648N^8+985568N^7+1395456N^6
\N\\
&&
+470688N^5-1183712N^4-1180224N^3-182528N^2-42752N
\N\\
&&
+13824
~,
\\
 Q_{11}(N)&=&60N^9+360N^8+584N^7-128N^6-2004N^5-2440N^4-976N^3
\N\\
&&
-192N^2+896N+384
          ~, \\
 Q_{12}(N)&=&28776N^{15}+356112N^{14}+1896088N^{13}+5538320N^{12}+9112264N^{11}
\N\\
&&
+6793968N^{10}-3019528N^9-11879520N^8-11673088N^7 
\N\\
&&
-6450992N^6-3726976N^5-2248128N^4-183296N^3+268032N^2
\N\\
&&
+147456N+27648
          ~, 
\\
 Q_{13}(N)&=&+464N^8-15616N^7-38112N^6+27776N^5+146064N^4+119552N^3
\N\\
&&
+109312N^2+86016N+62208~, \\
 Q_{14}(N)&=&456N^{11}+4376N^{10}+11328N^9-3184N^8-54552N^7-111720N^6
\N\\
&&
-155376N^5-251072N^4-312192N^3-222464N^2-135936N
\N\\
&&
-41472
          ~, \\
 Q_{15}(N)&=&168N^8+672N^7+784N^6-3192N^4-5600N^3-7168N^2-4480N
\N\\
&&
-2688 ~,
\\
 Q_{16}(N)&=&90N^{11}+630N^{10}+1592N^9+1260N^8-1934N^7-8218N^6
\N\\
&&
-15524N^5-23944N^4-26752N^3-18400N^2-11328N-3456
          ~, \\
 Q_{17}(N)&=&15777N^{17}+186525N^{16}+879391N^{15}+1874085N^{14}+575913N^{13}
\N\\
&&
-5568833N^{12}-10465411N^{11}-2970289N^{10}+11884298N^9
\N\\
&&
+12640320N^8-10343664N^7-40750480N^6-55711424N^5
\N\\
&&
-53947712N^{4}-42534912N^3-23256576N^2-7865856N
\N\\
&&
-1244160
          ~.
\end{eqnarray}
We compared $a_{Qg}^{(3)}(N)$, Eq.~(\ref{aQg3}), to the fixed moments (\ref{aQg3MOM2}--\ref{aQg3MOM10}) 
of Ref. \cite{Bierenbaum:2009mv} and obtained agreement.
The anomalous dimension appears in the $1/\ep$ term of (\ref{AhhhQg3}). As all other contributions to 
this term are known, the anomalous dimension can be obtained by comparing with the $1/\ep$ term of the 
present computation. The following expression for $\gamma_{qg}^{(2)}(N)$ is obtained:
\begin{eqnarray} 
{\gamma}_{qg}^{(2)}&=&\frac {T_F^2 n_f} {(N+1) (N+2)} \Biggl\{ 
\N\\
&&
C_A \Biggl[
\left(N^2+N+2\right) \Biggl(
+\frac{128}{3 N} S_{2,1}
+\frac{32}{9 N} S_1^3
+\frac{128}{3 N} S_{-3}
\N\\
&&
+\frac{64}{9 N} S_3
-\frac{32 }{3 N} S_2 S_1
\Biggr)
-\frac{128 (5 N^2+8 N+10) }{9 N} S_{-2}
\N\\
&&
-\frac{64 (5 N^4+26 N^3+47 N^2+43 N+20) }{9 N (N+1) (N+2)} S_2
\N\\
&&
-\frac{64 (5 N^4+20 N^3+41 N^2+49 N+20) }{9 N (N+1) (N+2)} S_1^2
\N\\
&&
+\frac{64 Q_{18}(N)}{27 N (N+1)^2 (N+2)^2} S_1
+\frac{16 Q_{19}(N)}{27 (N-1) N^4 (N+1)^3 (N+2)^3}
\Biggr]
\N
\end{eqnarray}
\begin{eqnarray}&&
+ C_F \Biggl[ 
-\frac{32 (N^2+N+2)}{9 N} S_1^3
+\frac{320 (N^2+N+2)}{9 N} S_3
\N\\
&&
-\frac{32 (N^2+N+2) }{3 N} S_1 S_2
+\frac{32 (5 N^2+3 N+2) }{3 N^2} S_2
\N\\
&&
+\frac{32 (10 N^3+13 N^2+29 N+6) }{9 N^2} S_1^2
\N\\
&&
-\frac{32 (47 N^4+145 N^3+426 N^2+412 N+120)}{27 N^2 (N+1)} S_1
\N\\
&&
+\frac{4 Q_{20}(N)}{27 (N-1) N^5 (N+1)^4 (N+2)^3}
\Biggr]
\Biggr\}\label{gammaqg}~,
\\
\N\\
Q_{18}(N)&=&19 N^6+124 N^5+492 N^4+1153 N^3+1362 N^2+712 N+152~,
\\
Q_{19}(N)&=&165 N^{12}+1485 N^{11}+5194 N^{10}+8534 N^9+3557 N^8-8899 N^7
\N\\
&&
-10364 N^6+6800 N^5+25896 N^4+30864 N^3+19904 N^2
\N\\
&&
+7296 N+1152~,
\\
Q_{20}(N)&=&99 N^{14}+990 N^{13}+4925 N^{12}+17916 N^{11}+46649 N^{10}+72446 N^9
\N\\
&&
+32283 N^8-95592 N^7-267524 N^6-479472 N^5-586928 N^4
\N\\
&&
-455168 N^3-269760 N^2-122112 N-27648~.
\end{eqnarray}
It agrees with the moments (\ref{gaqg2MOM2}-\ref{gaqg2MOM10}), 
\cite{Larin:1993vu,Larin:1996wd,Retey:2000nq,Vogt:2004mw,Bierenbaum:2009mv}. Due to the algebraic 
compactification we obtain 
a lower number of harmonic sums $S_{\vec{a}}(N)$ if compared to Ref. \cite{Vogt:2004mw}, and agree with 
\cite{Blumlein:2009tj}. 
 \newpage
\subsection{\boldmath The pure-singlet contribution $\Ahathat_{Qq}^{\rm{PS,(3)}}$}
The general structure of the pure-singlet contribution $A_{Qq}^{\sf PS,(3)}$ is, \cite{Bierenbaum:2009mv},  
   \begin{eqnarray}
    \Ahathat_{Qq}^{(3),{\sf PS}}&=&
     \Bigl(\frac{\hat{m}^2}{\mu^2}\Bigr)^{3\ep/2}\Biggl[
     \frac{\hat{\gamma}_{qg}^{(0)}\gamma_{gq}^{(0)}}{6\ep^3}
                  \Biggl(
                         \gamma_{gg}^{(0)}
                        -\gamma_{qq}^{(0)}
                        +6\beta_0
                        +16\beta_{0,Q}
                  \Biggr)
   +\frac{1}{\ep^2}\Biggl(     
                        -\frac{4\hat{\gamma}_{qq}^{(1),{\sf PS}}}{3}
                                 \Bigl[
                                        \beta_0
                                       +\beta_{0,Q}
                                 \Bigr]
 \N\\ &&
                        -\frac{\gamma_{gq}^{(0)}\hat{\gamma}_{qg}^{(1)}}{3}
                        +\frac{\hat{\gamma}_{qg}^{(0)}}{6}
                                 \Bigl[
                                        2\hat{\gamma}_{gq}^{(1)}
                                       -\gamma_{gq}^{(1)}
                                 \Bigr]
                        +\delta m_1^{(-1)} \hat{\gamma}_{qg}^{(0)}
                                           \gamma_{gq}^{(0)}
                 \Biggr)
   +\frac{1}{\ep}\Biggl(     
                           \frac{\hat{\gamma}_{qq}^{(2),{\sf PS}}}{3}
                        -n_f\frac{\hat{\tilde{\gamma}}_{qq}^{(2),{\sf PS}}}{3}
         \N\\ &&
                          +\hat{\gamma}_{qg}^{(0)}a_{gq,Q}^{(2)}
                          -\gamma_{gq}^{(0)}a_{Qg}^{(2)}
                          -4(\beta_0+\beta_{0,Q})a_{Qq}^{(2),{\sf PS}}
                   -\frac{\hat{\gamma}_{qg}^{(0)}\gamma_{gq}^{(0)}\zeta_2}{16}
                            \Bigl[
                               \gamma_{gg}^{(0)}
                              -\gamma_{qq}^{(0)}
                              +6\beta_0
                            \Bigr]
         \N\\ &&
                   +\delta m_1^{(0)} \hat{\gamma}_{qg}^{(0)}
                                           \gamma_{gq}^{(0)}
                   -\delta m_1^{(-1)} \hat{\gamma}_{qq}^{(1),{\sf PS}}
                 \Biggr)
   +a_{Qq}^{(3),{\sf PS}}
                              \Biggr]~, \label{AhhhQq3PS}
   \end{eqnarray}
In this process the photon couples to the heavy quark line. 
Summing the individual results of the different diagrams (\ref{AQqA}--\ref{AQqEnd}) weighted by their 
respective multiplicities yields
\begin{eqnarray} 
A_{Qq}^{\rm{PS},(3)}&=&
\frac{T_F^2\,n_f\,C_F} {N^2 (-1+N)  (2+N) (1+N)^2} \Biggl\{
-\frac {1} {\ep^3}
{\frac {256}{9}}\,(N^{2}+N+2)^2
\N\\
&&
+\frac {1} {\ep^2} \Biggl[
-\frac {128}{9}\,(N^{2}+N+2)^2 S_1
+\frac {128}{27}\,\frac {Q_{21}(N)}{N (1+N) (2+N) }
\Biggr]
\N\\
&&
+\frac {1} {\ep} \Biggl[
\left(N^2+N+2\right)^2\Biggl(
-{\frac {416}{9}}\,S_2
-{\frac {32}{9}}\,S_1^2
-{\frac {32}{3}}\zeta_2
\Biggr)
\N\\
&&
+{\frac {64}{27}}\,{\frac {Q_{21}(N)}{N (1+N) (2+N) }} S_1
-{\frac {64}{81}}\,{\frac {Q_{22}(N)} {N^{2} (1+N) ^{2} (2+N)^{2} }}
\Biggr]
\Biggr\}
\N\\
&&
+ a_{Qq}^{\rm PS,(3)}~\label{AQqPS3}.
\end{eqnarray}
The constant part is given by
\begin{eqnarray} 
a_{Qq}^{\rm PS,(3)}&=& \frac{T_F^2\,n_f\,C_F} {N^{2} (1+N) ^{2} (2+N
)  (N-1) } \Biggl\{
(N^2+N+2)^2\Biggl(
-{\frac {1760}{27}}\,S_3
\N\\
&& 
-{\frac {208}{9}}\,S_2 S_1
-{\frac {16}{27}}\, S_1^{3}
-\frac {16}{3}\,S_1\zeta_2
+\frac {224}{9}\,\zeta_3
\Biggr)
\N\\
&&
+{\frac {208}{27}}\,{\frac {Q_{21}(N)}{N (1+N)(2+N) }} S_2
+{\frac {16}{27}}\,\frac {Q_{21}(N)}{N (1+N) (2+N)} S_1^2
\N\\
&&
+{\frac {16}{9}}\,\frac {Q_{21}(N)}{N (1+N) (2+N)} \zeta_2
-{\frac {32}{81}}\,{\frac { Q_{22}(N)}{N^{2} (1+N)^{2} (2+N)^{2}}} S_1
\N\\
&&
+{\frac {32}{243}}\,{\frac {Q_{23}(N)}{N^{3} (1+N) ^{3} (2+N) ^{3}}}
\Biggr\}~,\label{aQqPS3}
\end{eqnarray}
with

{\begin{eqnarray} 
Q_{21}(N)&=& 8\,N^{7}+37\,N^{6}+68\,N^{5}-11\,N^{4}-86\,N^{3}-56\,N^{2}-104\,N-48~,
\\
Q_{22}(N)&=& +25\,N^{10}+176\,N^{9}+417\,N^{8}+30\,N^{7}-20\,N^{6}+1848\,N^{5}
\N\\
&&
+2244\,N^{4}+1648\,N^{3}+3040 N^2+2112 N +576~,
\\
Q_{23}(N)&=& 158\,N^{13}+1663\,N^{12}+7714\,N^{11}+23003\,N^{10}+56186\,N^{9}
\N\\
&&
+89880\,N^{8}+59452\,N^{7}-8896\,N^{6}
-12856\,N^{5}-24944\,N^{4}
\N\\
&&
-84608\,N^{3}-77952\,N^{2}-35712\,N-6912~,
\end{eqnarray}
agreeing with the moments (\ref{aQqMOMstart}--\ref{aQqMOMend}) from \cite{Bierenbaum:2009mv}.
\subsection{\boldmath The pure-singlet contribution $\Ahathat_{qq,Q}^{\rm{PS},(3)}$}
The analytic structure for this OME, \cite{Bierenbaum:2009mv}, is given by

\begin{eqnarray}
    \Ahathat_{qq,Q}^{(3),{\sf PS}}&=&
           n_f\Bigl(\frac{\hat{m}^2}{\mu^2}\Bigr)^{3\ep/2}\Biggl[
            \frac{2\hat{\gamma}_{qg}^{(0)}\gamma_{gq}^{(0)}\beta_{0,Q}}{3\ep^3}
                +\frac{1}{3\ep^2} \Biggl(
                    2\hat{\gamma}_{qq}^{(1),{\sf PS}}\beta_{0,Q}
                   +\hat{\gamma}_{qg}^{(0)}\hat{\gamma}_{gq}^{(1)}
                                  \Biggr)
\N\\ &&
                +\frac{1}{\ep} \Biggl(
                       \frac{\hat{\tilde{\gamma}}_{qq}^{(2),{\sf PS}}}{3}
                      +\hat{\gamma}_{qg}^{(0)}a_{gq,Q}^{(2)}
         -\frac{\hat{\gamma}_{qg}^{(0)}\gamma_{gq}^{(0)}\beta_{0,Q}\zeta_2}{4}
                                \Biggr)
                +\frac{a_{qq,Q}^{(3), {\sf PS}}}{n_f}
                                                     \Biggr]~.
                   \label{Ahhhqq3PSQ}
\end{eqnarray}
For this OME we computed the complete result, since no other color factors than $T_F^2 n_f C_F$ 
contribute in this case.
The matrix element corresponds to the pure-singlet term in which the photon couples to a massless fermion line. One obtains
\begin{eqnarray} 
A_{qq,Q}^{\rm{PS},(3)}&=&\frac{T_F^2\,n_f\,C_F} {N^{2} (1+N) ^{2} (2+N
)  (N-1) } \Biggl\{
-\frac {1} {\ep^3}
\frac {256}{9}\,(N^{2}+N+2)^{2}
\N\\
&&
+\frac {1} {\ep^2} \Biggl[
{\frac {256}{9}}\,(N^{2}+N+2)^{2} S_1
-{\frac {128}{27}}\,{\frac {Q_{24}(N)}{N (2+N) (1+N)}}
\Biggr]
\N\\
&&
+\frac {1} {\ep} \Biggl[
(N^2+N+2)^2 \Biggl(
-{\frac {128}{9}}\,S_2
-{\frac {128}{9}}\,S_1^2
-{\frac {32}{3}}\,\zeta_2
\Biggr)
\N\\
&&
+{\frac {128}{27}}\,{\frac {Q_{24}(N)}{N  (2+N)  (1+N) }} S_1
-{\frac {64}{81}}\,{\frac {Q_{25}(N)}{N^2 (2+N)^2 (1+N)^2}}
\Biggr]
\Biggr\}
\N\\
&&
+ a_{qq,Q}^{\rm PS,(3)}~\label{AqqQ3},
\end{eqnarray}
with the constant part
\begin{eqnarray}
a_{qq,Q}^{\rm PS,(3)}&=&
\frac{T_F^2\,n_f\,C_F} {N^2 (-1+N)  (2+N) (1+N)^2} \Biggl\{
(N^2+N+2)^2\Biggl(
{\frac {256}{27}}\,S_3
+{\frac {128}{9}}\,S_2 S_1
\N\\
&& 
+{\frac {128}{27}}\,S_1^3
+{\frac {32}{3}}\,S_1 \zeta_2
+{\frac {224}{9}}\,\zeta_3
\Biggr)
-{\frac {64}{27}}\,{\frac {Q_{24}(N)}{N (2+N) (1+N)}} S_2
\N\\
&&
-{\frac {64}{27}}\,{\frac { Q_{24}(N)}{N (2+N)(1+N)}} S_1^2
-{\frac {16}{9}}\,{\frac {Q_{24}(N)}{N (2+N) (1+N)}} \zeta_2
\N\\
&&
+{\frac {64}{81}}\,{\frac {Q_{25}(N)}{N^{2} (2+N) ^{2} (1+N)^{2}}} S_1
-{\frac {32}{243}}\,{\frac {Q_{26}(N)}{{N}^{3} (2+N)^{3} (1+N)^{3}}}\Biggr\}~,\label{aqqQPS3}
\end{eqnarray}
and
\begin{eqnarray}
\N\\
Q_{24}(N)&=& 16\,N^{7}+74\,N^{6}+181\,N^{5}+266\,N^{4}+269\,N^{3}+230\,N^{2}
\N\\
&&
+44\,N-24~, \\
Q_{25}(N)&=&  181\,N^{10}+1352\,N^{9}+4737\,N^{8}+10101\,N^{7}
\N\\
&&
+14923\,N^{6}+17085\,N^{5}+14133\,N^{4}+5944\,N^{3}+568\,N^{2}-48\,N
\N\\
&&
+144~,
\\
Q_{26}(N)&=& 2074\,N^{13}+21728\,N^{12}+105173\,N^{11}+311482\,N^{10}+636490\,N^{9}
\N\\
&&
+966828\,N^{8}+1126568\,N^{7}+968818\,N^{6}+550813\,N^{5}
\N\\
&&
+169250\,N^{4}+12104\,N^{3}-3408\,N^{2}-1008\,N-864~,
\end{eqnarray}
agreeing with the fixed moments (\ref{aqqQMOM2}-\ref{aqqQMOM14}) from \cite{Bierenbaum:2009mv}.
From (\ref{AhhhQq3PS}, \ref{AQqPS3}, \ref{Ahhhqq3PSQ}, \ref{AqqQ3}) one may extract the anomalous 
dimension

\begin{eqnarray} 
\hat{\gamma}_{qq}^{(2),PS} &=&C_FT_F^2n_f \frac {1} {(N-1) N^2 (N+1)^2 (N+2)} \Biggl\{
\N\\
&&
-{\frac {32}{3}}\,( {N}^{2}+N+2 )^{2} (S_1^{2}+S_2)
+{\frac {64}{9}}\,{\frac { Q_{27}(N)\,S_1}{{N} ( 1+N ) (2+N )}}
\N\\
&&
-{\frac {64}{27}}\,{\frac {Q_{28}(N)}{{N}^{2} ( 1+N ) ^{2} ( 2+N)^{2}}}
\Biggr\}\label{gammaqqPS}~,
\\
\N\\
Q_{27}(N)&=&68\,{N}^{5}+37\,{N}^{6}+8\,{N}^{7}-11\,{N}^{4}-86\,{N}^{3}-56\,{N}^{2}-104\,N-48~, 
\\
Q_{28}(N)&=&  
+52\,{N}^{10}
+392\,{N}^{9}
+1200\,{N}^{8}
+1353\,{N}^{7}
-317\,{N}^{6}
-1689\,{N}^{5}
\N\\
&&
-2103\,{N}^{4}
-2672\,{N}^{3}
-1496\,{N}^{2}
-48\,N
+144~.
\end{eqnarray}  
Again we obtain agreement with the fixed moments from \cite{Bierenbaum:2009mv}, Eqs. (\ref{gaqqMOM2}-\ref{gaqqMOM12}).
\subsection{\boldmath The non-singlet contribution $\Ahathat_{qq,Q}^{\rm{NS}}$}

The following structure is known for the $\sf {NS}$ contribution:
   \begin{eqnarray}
   \Ahathat_{qq,Q}^{(3),{\sf NS}}&=&
     \Bigl(\frac{\hat{m}^2}{\mu^2}\Bigr)^{3\ep/2}\Biggl\{
            -\frac{4\gamma_{qq}^{(0)}\beta_{0,Q}}{3\ep^3}
                   \Bigl(\beta_0+2\beta_{0,Q}\Bigr)
            +\frac{1}{\ep^2}
              \Biggl(
                      \frac{2\gamma_{qq}^{(1),{\sf NS}}\beta_{0,Q}}{3}
                     -\frac{4\hat{\gamma}_{qq}^{(1),{\sf NS}}}{3}
                             \Bigl[\beta_0+\beta_{0,Q}\Bigr]
\N\\ 
&&
                     +\frac{2\beta_{1,Q}\gamma_{qq}^{(0)}}{3}
                     -2\delta m_1^{(-1)}\beta_{0,Q}\gamma_{qq}^{(0)}
              \Biggr)
            +\frac{1}{\ep} 
              \Biggl(
                      \frac{\hat{\gamma}_{qq}^{(2), {\sf NS}}}{3}
                     -4a_{qq,Q}^{(2),{\sf NS}}\Bigl[\beta_0+\beta_{0,Q}\Bigr]
                     +\beta_{1,Q}^{(1)}\gamma_{qq}^{(0)}\N
\\
&&
                     +\frac{\gamma_{qq}^{(0)}\beta_0\beta_{0,Q}\zeta_2}{2}
                     -2 \delta m_1^{(0)} \beta_{0,Q} \gamma_{qq}^{(0)} 
                     -\delta m_1^{(-1)}\hat{\gamma}_{qq}^{(1),{\sf NS}}
                  \Biggr)
         +a_{qq,Q}^{(3), {\sf NS}}
                              \Biggr\}~. \label{Ahhhqq3NSQ}
   \end{eqnarray}
cf. \cite{Bierenbaum:2009mv}. 
We obtain the following result:
\begin{eqnarray} 
A_{qq,Q}^{\rm{NS,(3)}}&=&
T_F^2\,n_f\,C_F 
\Biggl\{
\frac {1} {\ep^3} \Biggl[
-\frac {512}{27}\,S_1
+\frac {128}{27}\,\frac  {3\,N+3\,N^{2}+2} {(1+N) N}
\Biggr]
\N\\
&&
+\frac {1} {\ep^2} \Biggl[
\frac {256}{27}\,S_2
-\frac {1280}{81}\,S_1
+\frac {32}{81}\, \frac {20\,N+47\,N^{2}+6\,N^{3}+3\,N^{4}-12} {(1+N)^2 N^2}
\Biggr]
\N\\
&&
+\frac {1} {\ep} \Biggl[
-{\frac {128}{27}}\,S_3
-{\frac {64}{9}}\,\zeta_2\,S_1
+{\frac {640}{81}}\,S_2
+{\frac {16}{9}}\,{\frac {3\,N+3\,N^{2}+2}{ (1+N) N}} \zeta_2
-{\frac {1280}{27}}\,S_1
\N\\
&&
+{\frac {8}{81}}\,{\frac {Q_{29}(N)}{ (1+N) ^{3}N^{3}}}
\Biggr]
\Biggr\}+ a_{qq,Q}^{\rm NS,(3)}~,
\end{eqnarray}
with
\begin{eqnarray} 
a_{qq,Q}^{\sf NS,(3)}&=&
T_F^2\,n_f\,C_F 
\Biggl\{
{\frac {64}{27}}\,\,S_4
+{\frac {448}{27}}\,\zeta_3 S_1
+{\frac {32}{9}}\,\zeta_2\,S_2
-{\frac {320}{81}}\,\,S_3
\N\\
&&
-{\frac {160}{27}}\,\zeta_2\,S_1
-{\frac {112}{27}}\,{\frac {3\,N+3\,N^{2}+2}{
    (1+N) N}} \zeta_3
+\frac {640}{27}\,\,S_2
\N\\
&&
+\frac {4}{27}\,\frac {20\,N+47\,N^{2}+6\,N^{3}+3\,N^{4}-12}{(1+N)^2 N^{2}} \zeta_2
-\frac {55552}{729}\,S_1
\N\\
&&
+\frac {2}{729}\,{\frac {Q_{30}(N)}{(1+N) ^{4}N^{4}}}
\Biggr\}~,\label{aqqQNS3}
\end{eqnarray}
and
\begin{eqnarray}
Q_{29}(N)&=&321\,N^{6}+963\,N^{5}+1307\,N^{4}+833\,N^{3}+152\,N^{2}-16\,N+24~,
\\
Q_{30}(N)&=&
+11751\,N^{8}+47004\,N^{7}+93754\,N^{6}+104364\,N^{5}+55287\,N^{4}
\N\\
&&
+6256\,N^{3}-2448\,N^{2}-144\,N-432~.
\end{eqnarray}
Again the anomalous dimension for the non-singlet case is determined by comparing the $1/\ep$ term of our result to the corresponding term of the general structure. We obtain

\begin{eqnarray} 
\hat{\gamma}_{qq}^{(2),NS} &=&C_FT_F^2n_f\Bigl\{
{\frac {128}{9}}\,S_3
-{\frac {640}{27}}\,S_2
-{\frac {128}{27}}\,S_1
+{\frac {8}{27}}\,{\frac {Q_{31}(N)}{{N}^{3} (1+N)^{3}}}\Bigr\}\label{gammaqqNS}~,
\end{eqnarray}
with
\begin{eqnarray}
Q_{31}(N)&=& 51\,{N}^{6}+153\,{N}^{5}+57\,{N}^{4}+35\,{N}^{3}+96\,{N}^{2}+16\,N-24 ~.
\end{eqnarray}  
For fixed moments $a_{qq,Q}^{(3), \sf NS}$ agrees with the values in \cite{Bierenbaum:2009mv}, cf. (\ref{aqqQM1}--\ref{aqqQM14}) and the anomalous dimension $\gamma_{qq}^{\sf (3), NS}$ with (\ref{gqqNSMOM2}--\ref{gqqNSMOM13}), \cite{Retey:2000nq,Moch:2004pa,Bierenbaum:2009mv}.
\subsection{\boldmath
The non-singlet transversity contribution $\Ahathat_{qq,Q}^{\rm{NS,TR}}$}
The non-singlet transversity OME ${A_{qq,Q}^{\rm{NS,TR}}}$, cf. Ref. \cite{Blumlein:2009rg},  obeys the same general structure as the non-singlet OME ${A_{qq,Q}^{\sf {NS}}}$, (\ref{Ahhhqq3NSQ}). One obtains
\begin{eqnarray} 
A_{qq,Q}^{\sf (3),TR} &=& T_F^2n_fC_F \Biggl\{
\frac {1} {\ep^3} 
\Biggl[
\frac {128}{9}
-\frac{512 }{27} S_1
\Biggr]
+\frac {1} {\ep^2} \Biggl[
\frac{256 }{27} S_2
-\frac{1280 }{81} S_1
+\frac{32}{27}
\Biggr]
\N\\
&&
+\frac {1} {\ep} \Biggl[
-\frac{128 }{27} S_3
-\frac{64 }{9} \zeta_2 S_1
+\frac{640 }{81} S_2
-\frac{1280}{27} S_1
+\frac {16} {3} \zeta_2
\N\\
&&
+\frac {8 ( 107 N^2 +107 N +8 )} {27 N (N+1)}
\Biggr]\Biggr\}
+a_{qq,Q}^{(3),TR}~,
\end{eqnarray}
with
\begin{eqnarray}
a_{qq,Q}^{\sf (3),TR} &=&
T_F^2n_fC_F \Biggl\{
\frac{64 }{27} S_4
+\frac{448 }{27} \zeta_3 S_1
+\frac{32 }{9} \zeta_2 S_2
-\frac{320 }{81} S_3
-\frac{160 }{27} \zeta_2 S_1
\N\\
&&
-\frac{112} {9} \zeta_3
+\frac{640}{27} S_2
+\frac {4} {9} \zeta_2
-\frac{55552}{729} S_1
\N\\
&&
+\frac {2 (3917 N^4 + 7834 N^3 +4157 N^2 -48 N -144)} {243 N^2 (1+N)^2}
\Biggr\}\label{aTrans}~,
\end{eqnarray}
agreeing with the moments (\ref{aqqTR3MOM1}-\ref{aqqTR3MOM13}) for fixed values of $N$.
Extracting the anomalous dimension from the $1 /\ep$ term of the OME yields

\begin{eqnarray} 
\hat{\gamma}_{qq}^{(2),TR} &=&C_FT_F^2n_f\Biggl\{
{\frac {128}{9}}\,S_3
-{\frac {640}{27}}\,S_2
-{\frac {128}{27}}\,S_1
+{\frac {8}{9}}\,{\frac {( 17\,{N}^{2}+17\,N-8 )}  {N ( 1+N ) }}
\Biggr\}\label{gammaTrans}~.
\end{eqnarray}  
The results for the anomalous dimensions constitute a first independent check of the result obtained in 
\cite{Gracey:2003yrxGracey:2003mrxGracey:2006zrxGracey:2006ah,Blumlein:2009rg}. It is interesting to 
note that for this color factor the vector- and tensor operators lead to the same structures in the 
harmonic sums for $a_{qq,Q}^{(3)}$ and $\gamma_{qq}^{(3)}$.

\subsection{Harmonic Sums}
\label{sec:HS1}

\vspace*{2mm}
\noindent
The $T_F^2 n_f C_{F,A}$--contributions at $O(a_s^3)$ to the massive operator matrix elements
contain nested harmonic sums up to weight {\sf w = 4}. This also applies to all individual
Feynman diagrams, which we calculated in the Feynman--gauge, cf. Appendix~D. In intermediary results 
generalizations
of harmonic sums occur, see~Section.~\ref{sec:HS2}. As has been observed in the computation of various
other physical quantities before, such as anomalous dimensions and massless Wilson coefficients
to 3-loop order \cite{Moch:2004pa,Vogt:2004mw,Vermaseren:2005qc,Blumlein:2009tj}, 
unpolarized and polarized massive OMEs to 2--loop order
\cite{Bierenbaum:2007qe,Bierenbaum:2008yu,Bierenbaum:2007pn,Bierenbaum:2009zt}, 
the polarized and unpolarized Drell-Yan  and Higgs-boson
production cross section \cite{Blumlein:2005im}, time-like Wilson coefficients \cite{Blumlein:2006rr}, 
and virtual- and soft
corrections to Bhabha-scattering \cite{Blumlein:2007dj}, the classes of contributing harmonic sums are always the
same. For main properties of the nested harmonic sums see Appendix~\ref{App-SpeFunHarm}. They depend on the loop-order and the 
topologies of Feynman diagrams involved. 

In the present case the following harmonic sums emerge~:
\begin{eqnarray}
\label{eq:HSUM1}
& & S_1  \nonumber \\
& & S_2,~S_{-2}  \nonumber \\
& & S_3,~S_{-3},~S_{2,1},~S_{-2,1}  \nonumber \\
& & S_4,~S_{-4},~S_{3,1},~S_{-3,1},~S_{-2,2},~S_{2,1,1},~S_{-2,1,1}~.  
\end{eqnarray}
Note that this class, as for the other processes mentioned above, does not contain the index $\{-1\}$.
Moreover, we used the algebraic relations between the harmonic sums, cf.~\cite{Blumlein:2003gb}. Furthermore,
structural relations exist between harmonic sums, cf.~\cite{Blumlein:2009ta,Blumlein:2009fz,ABS2010}, 
which 
reduce the set 
(\ref{eq:HSUM1})
further. Here  the sums
\begin{eqnarray}
\label{eq:HSUM2}
S_{-2,2},~~S_{3,1} 
\end{eqnarray}
are connected by differential relations w.r.t. their argument $N$ to other sums of (\ref{eq:HSUM1}).
This is also the case for all single harmonic sums $S_{\pm n},~~~n \in \mathbb{N}$, $n > 1,$ using 
both 
the differentiation and argument-duplication relation, cf. \cite{Blumlein:1998if}. Due to this $S_1$ represents the
class of all single harmonic sums. I.e. only the {\it six} basic harmonic sums 
\begin{eqnarray}
\label{eq:HSUM3}
& & S_1  \nonumber \\
& & S_{2,1},~S_{-2,1}  \nonumber \\
& & S_{-3,1},~S_{2,1,1},~S_{-2,1,1}  
\end{eqnarray}
are needed to represent the 3-loop results for the 
$T_F^2 n_f C_{F,A}$--contributions to the OMEs calculated in the present paper. In the final 
representation we refer to the algebraic basis (\ref{eq:HSUM1}) and consider the basis (\ref{eq:HSUM3})
for a later numerical implementation. We sorted the respective expressions keeping a rational function
in $N$ in front of the harmonic sums (\ref{eq:HSUM1}) and $\zeta$--values, like $\zeta_{2}$ and $\zeta_3$.

The harmonic sums emerge from the series--expansion of hypergeometric structures like the
Euler $B$-- and $\Gamma$--functions and the Pochhammer--symbols in the (generalized) hypergeometric
functions $_PF_Q(a_i, b_i;1)$ in the dimensional parameter $\varepsilon$. This leads to single harmonic
sums first, which, through summation, turn into (multiple) zeta values \cite{Blumlein:2009cf} and nested
harmonic sums \cite{Blumlein:1998if,Vermaseren:1998uu}. The principle steps on the way from single--scale
Feynman diagrams to these structures have been described in Ref.~\cite{Blumlein:2009ta}.
\subsection{Generalized Harmonic Sums}
\label{sec:HS2}

\vspace*{2mm}
\noindent
At 3--loop order the expressions for individual Feynman diagrams are rather large. In Section~\ref{Evaluation}
we described how the Feynman parameter integrals which emerge in the present calculation are transformed 
into nested infinite and finite sums. If these sums could be computed analytically as a whole only 
nested harmonic sums would occur in the calculation. However, this is not always possible in practice.
Usually the expressions obtained are split into different parts and the sums are then computed.
In intermediary steps, depending on the summation methods used, more complicated sum--structures emerge.
We will not deal with this aspect here, but only refer to typical structures, which occur in the final 
result separating the expressions to be summed over into tractable terms of an intermediate size.

Here, so-called generalized harmonic sums occur \cite{Moch:2001zr,ABS10A}. They obey the following recursive definition 
\cite{Moch:2001zr}~: 
\begin{eqnarray} 
\label{eq:HSUM4} 
\widetilde{S}_{m_1,...}(x_1,...; N) &=& \sum_{i_1}^N \frac{x_1^{i_1}}{i_1^{m_1}} 
\sum_{i_2=1}^{i_1-1} \frac{x_2^{i_2}}{i_2^{m_2}} \widetilde{S}_{m_3,...}(x_3,...; i_2) 
\nonumber\\ & & + 
\widetilde{S}_{m_1+m_2,m_3,...}( x_1 \cdot x_2,x_3,...; N)~. 
\end{eqnarray} 
The sums 
$\widetilde{S}$ may be reduced to nested harmonic sums for $x_i \in \{-1, 1\}$. In the present calculation the values 
of $x_i$ extend to $\{-1/2, 1/2, -2, 2\}$. These sums occur in ladder like structures, 
cf.~\cite{Vermaseren:2005qc,
ABKSW2010,BHKS}, but may also emerge, if contributions to 3--loop Feynman diagrams containing a 2-point insertion,
are separated into various terms. They were even observed in case of massive 2-loop graphs of 
the type \cite{Bierenbaum:2008yu}
if large expressions are arbitrarily separated, cf.~\cite{BHKS}.
The weight of these sums can reach {\sf w = 5} intermediary, depending on the $\varepsilon$--structure
of the contribution, although only {\sf w = 4} sums will emerge in the final results.
Examples for these sums are~: 
\begin{eqnarray} 
\label{eq:HSUM5} 
&&\widetilde{S}_{1}(1/2,N),~~
\widetilde{S}_2(-2;N),~~
\widetilde{S}_{2,1}(-1,2;N),~~
\widetilde{S}_{3,1}(-2,-1/2;N),~~
\nonumber\\ &&
\widetilde{S}_{1,1,1,2}(-1,1/2,2,-1;N),~~
\widetilde{S}_{2,3}(-2,-1/2;N),~~
\nonumber\\&&
\widetilde{S}_{2,2,1}(-1,-1/2,2;N),~~{\rm etc.}
\end{eqnarray}

The algebraic and structural relations for these sums are worked out in 
Ref.~\cite{ABS10A}. Similar to the case of harmonic sums, corresponding basis representations are obtained. These 
relations were provided in a code used in the present calculation. They finally lead to the reduction of the 
results for the individual diagrams to a representation 
just in terms of nested harmonic sums. In Appendix~G one of the typical lengthy nested sums, 
which also contains the generalized harmonic sums, is shown.

\newpage
\section{Summary}\label{Conclusion}

\vspace*{1mm}
\noindent
In the present work a first contribution to the computation of the massive
Wilson coefficients of the deep--inelastic structure function $F_2(x,Q^2)$ 
for unpolarized charged lepton--nucleon scattering at 3--loop order for general values 
of the Mellin variable $N$ in the region  $Q^2 \gsim 10 \cdot m^2$  has been made. Since the
corresponding massless Wilson coefficients are known, cf. \cite{Vermaseren:2005qc}, only the
massive operator matrix elements remain to be calculated. A series of fixed Mellin moments
for all contributing OMEs has been computed before, Refs.~\cite{Bierenbaum:2009mv,Blumlein:2009rg}.
Here, we extend this work to general values of $N$, calculating a very large class of terms for 
$A_{Qg}^{(3)}(N)$ and
all contributions to 
$A_{Qq}^{{\sf PS}, \rm (3)}(N)$ 
$A_{qq,Q}^{{\sf PS}, \rm (3)}(N)$ 
$A_{qq,Q}^{{\sf NS}, \rm (3)}(N)$, and 
$A_{qq,Q}^{{\sf NS,TR}, \rm (3)}(N)$ for the color factors $T_F^2 n_f C_F$ and $T_F^2 n_f C_A$.
The complete results on $A_{Qg}^{(3)}(N)$ were obtained in a larger team and will be given, including 
$A_{qg,Q}^{(3)}(N)$, in \cite{ABKSW2010} in detail.

The present calculation is a first step within a larger programme to compute all topologies contributing 
to the massive OMEs, including those which define heavy--light quark transitions in the variable flavor 
number scheme. Due to the 
large
heavy flavor contributions of $O(25-30 \%)$ to the nucleon structure functions the experimental precision of the DIS World Data requires these calculations, both to extract the 
different parton distribution functions and to measure the strong coupling constant $\alpha_s(M_Z^2)$ 
at highest possible precision. The knowledge of these quantities is instrumental for precise 
measurements at the {\sf LHC}. $\alpha_s(M_Z^2)$ is furthermore one of the central parameters in 
physics.

The present calculation built technically on a method to compute the Feynman--integrals, containing 
one massive line, directly, i.e. avoiding traditional methods, as integration-by-parts and related 
techniques. Those usually lead to a large proliferation of terms compared to the compact results
being finally obtained. Instead we referred to representations in terms of generalized hypergeometric 
functions. They occur as the analytic result of the Feynman parameter integrals.
Due to this the expansion in the dimensional parameter $\ep$ can be uniquely performed in 
an elegant way. At the same time, various significant simplifications of intermediary results are 
possible choosing particularly this representation.
In the end, a small number of nested finite and infinite sums
over hypergeometric terms equipped with Beta-functions and harmonic sums has to be performed. In the present 
calculation up to triple sums occurred. The main technical work to be performed consisted in finding these 
representation by means of computer 
algebra for the contributing graphs in a partly automated way. Intermediary very large expressions 
had to be handled and 
corresponding codes based on the systems {\sf FORM} and {\sf MAPLE} had to be designed. 

In the present case the final sums, compared to foregoing massless computations of other groups
at 3--loops, cf. e.g. \cite{Moch:2004pa,Vogt:2004mw}, turn out to be much more involved due to the
the increased nesting in the massive case. Moreover, it has  been unclear whether the usual nested 
harmonic sums form a frame to express all the intermediary results. This had to be found out using
strict mathematical methods based on the construction of sum-- and product fields. The corresponding 
software {\sf Sigma}~\cite{sigma1,sigma2} could be used for this purpose. Indeed, it turned out that
in the intermediary results the algebra of harmonic sums is to small to describe the corresponding terms 
and generalized harmonic sums had to be invoked. The methods used would have been pointing to other
more general structures also, if contributing. Thus this method is indispensable in exploring new 
territories in higher order quantum-field theoretic calculations.
The present work triggered mathematical research to find all relations for this new
class of functions, cf.~\cite{ABS10A}, which will help 
to make the codes being used at present even more efficient.

The major new results of the present work are the constant parts of the unrenormalized  massive 
OMEs at 3--loop order for general values of $N$, $a_{ij}^{(3)}(N)$ in $O(T_F^2 n_f C_F,A)$. Furthermore, 
the corresponding 
contributions to the 3-loop anomalous dimensions in the vector-- and transversity case are obtained.
In this way, a first independent recalculation of these important quantities given in 
\cite{Moch:2004pa,Vogt:2004mw,Gracey:2003yrxGracey:2003mrxGracey:2006zrxGracey:2006ah} before, was 
performed using very different methods. Here the anomalous dimensions are computed in the massive case. 
We compared the present results mutually with all the available fixed Mellin moments 
in the literature and find agreement. In case of the anomalous dimensions the comparison could be 
performed for the general expressions. The representation we gave, however, is more compact, since we
applied the algebraic relations between harmonic sums.

The final results for the matrix elements $a_{ij}^{(3)}(N)$ and the 3-loop anomalous 
dimensions $\hat{\gamma}_{ij}^{(2)}(N)$ are given in harmonic sums only. All generalized harmonic sums 
cancel already at the level of the individual diagrams. The complexity of harmonic sums determining the
present results is maximally six for the $a_{ij}^{(3)}(N)$ and two for 
$\hat{\gamma}_{ij}^{(2)}(N)$ at
$O(T_F^2 n_f C_{F,A})$, using also the structural relations, Ref.~\cite{Blumlein:2009ta}. This 
situation is 
comparable to the case of {\it all} massive OMEs
at 2--loop order expanded in the dimensional parameter to $O(\ep)$, cf.~\cite{Bierenbaum:2008yu}.
However, the present calculation, at 3--loop order, has been by far more complex and had to pass  much 
more sophisticated structures intermediary. It is interesting to note that the structure w.r.t. 
harmonic sums for the vector-- and tensor flavor non-singlet operators concerning both the matrix 
elements 
$a_{ij}^{(3)}(N)$ and the contributions to the anomalous dimensions $\hat{\gamma}_{ij}^{(2)}(N)$ are the 
same for the terms given by the harmonic sums in case at the present color factors.

The present calculations unraveled the relevance of hypergeometric structures in computing  Feynman 
integrals in general. They are given in the present by the generalized hypergeometric functions 
emerging. Likewise, the finite and 
infinite sums had to be performed over hypergeometric summands, equipped with products out of
(generalized) harmonic sums. 

The present work constitutes a first step to explore the general structure
of single scale massive observables in QCD at the 3--loop level, which probably bears a rich host of
yet unexplored structures. They form an interesting topic to be studied in the future to understand
the final simplicity of seemingly complex problems at the higher loop level in Quantum Field Theories 
such as Quantum Chromodynamics, and hopefully finally the 
yet unknown reason for that.  

%
%
\newpage
\vspace*{20cm}
\newpage
\newpage
 \thispagestyle{empty}
 \begin{flushleft}
 \end{flushleft}
 \vspace{70mm}
 \begin{center}
  \section{\bf\boldmath Appendix}
 \end{center}
 \newpage
 \begin{flushleft}
 \end{flushleft}
%
%
%

   \begin{appendix} 

%
%
%
    \section{\bf \boldmath Conventions}
     \label{App-Con}
     \renewcommand{\theequation}{\thesection.\arabic{equation}}
     \setcounter{equation}{0}
     We use natural units
     \begin{eqnarray}
      \hbar=1~,\quad c=1~,\quad \ep_0=1~,
     \end{eqnarray}
     where $\hbar$ denotes Planck's constant, $c$ the vacuum speed 
     of light and $\ep_0$ the permittivity of vacuum.
     The electromagnetic fine--structure constant $\alpha$~is
     given by 
     \begin{eqnarray} 
      \alpha=\alpha'(\mu^2=0)=\frac{e^2}{4\pi\ep_0\hbar c}
       =\frac{e^2}{4\pi}\approx 
      \frac{1}{137.03599911(46)}~.
     \end{eqnarray}
     In this convention, energies and momenta are given in the 
     same units, electron volt ($\eV$). 

     The space--time dimension is taken to be $D=4+\ep$ and 
     the metric tensor $g_{\mu\nu}$ in Minkowski--space 
     is defined as
     \begin{eqnarray}
      g_{00}=1~,\quad g_{ii}=-1~,i=1\ldots D-1~,\quad g_{ij}=0~,i\neq j~. 
      \label{metricDdim}
     \end{eqnarray}
     Einstein's summation convention is used, i.e.
     \begin{eqnarray}
      x_{\mu}y^{\mu}:=\sum^{D-1}_{\mu=0}x_{\mu}y^{\mu}~.
     \end{eqnarray}
     Bold--faced symbols represent $(D-1)$--dimensional spatial vectors:
     \begin{eqnarray}
      x=(x_0,{\boldmath x})~.
     \end{eqnarray}
     If not stated otherwise, Greek indices refer to the $D$--component 
     space--time vector and Latin ones to the $D-1$ spatial components 
     only. The dot product of two vectors is defined 
     by 
     \begin{eqnarray}
        p.q=p_0q_0-\sum_{i=1}^{D-1}p_iq_i~. 
     \end{eqnarray}
     The $\gamma$--matrices $\gamma_{\mu}$ are taken to be of dimension
     $D$ and fulfill the anti--commutation relation
     \begin{eqnarray}
      \{\gamma_{\mu},\gamma_{\nu}\}=2g_{\mu\nu} \label{gammaanticom}~.
     \end{eqnarray}
     It follows that
     \begin{eqnarray}
      \gamma_{\mu}\gamma^{\mu}&=&D \\
   Tr \left(\gamma_{\mu}\gamma_{\nu}\right)&=&4g_{\mu\nu} \\
   Tr \left(\gamma_{\mu}\gamma_{\nu}\gamma_{\alpha}\gamma_{\beta}\right)
       &=&4[g_{\mu\nu}g_{\alpha\beta}+
            g_{\mu\beta}g_{\nu\alpha}-
            g_{\mu\alpha}g_{\nu\beta}] \label{gammarelations}~.
     \end{eqnarray}
      The slash--symbol for a $D$-momentum $p$ is defined by
      \begin{eqnarray}
       \adag p:=\gamma_{\mu}p^{\mu} \label{dagger}~.
      \end{eqnarray}
      The conjugate of a bi--spinor $u$ of a particle is given by 
      \begin{eqnarray}
       \overline{u}=u^{\dagger}\gamma_0~,
      \end{eqnarray}
      where $\dagger$ denotes Hermitian and $*$ complex 
      conjugation, respectively. The bi--spinors 
      $u$ and $v$ fulfill the free
      Dirac--equation,
      \begin{eqnarray}
       (\adag p-m )u(p)&=&0~,~\quad \overline{u}(p)(\adag p-m )=0 \\
       (\adag p+m )v(p)&=&0~,~\quad \overline{v}(p)(\adag p+m )=0~. 
      \end{eqnarray}
      Bi--spinors and polarization vectors are normalized to 
      \begin{eqnarray}
       \sum_{\sigma}u(p,\sigma)\overline{u}(p,\sigma)&=&\adag p+m \\   
       \sum_{\sigma}v(p,\sigma)\overline{v}(p,\sigma)&=&\adag p-m \\
       \sum_{\lambda}\epsilon^{\mu}(k,\lambda )\epsilon^{\nu}(k,\lambda)
       &=&-g^{\mu \nu}~,
      \end{eqnarray}
      where $\lambda$ and $\sigma$ represent the spin.

      The commonly used caret~``~$\hat{\empty}~$''~to signify an operator, 
      e.g. $\hat{O}$, is omitted if confusion is not to be expected. 

      The gauge symmetry group of QCD is the Lie--Group $SU(3)_c$. We
      consider the general case of $SU(N_c)$. The 
      non--commutative generators are denoted by $t^a$, where 
      $a$ runs from $1$ to $N_c^2-1$. The generators can 
      be represented by Hermitian, traceless matrices, \cite{Muta:1998vi}.
      The structure constants $f^{abc}$
      and $d^{abc}$ of $SU(N_c)$ are defined via the commutation and 
      anti--commutation relations of its generators, \cite{Yndurain:1999ui},
      \begin{eqnarray}
       [t^a,t^b]&=&if^{abc}t^c \label{structconstf} \\
      \{t^a,t^b\}&=& d^{abc}t^c+\frac{1}{N_c}\delta_{ab} \label{structconstd}~.
      \end{eqnarray}
      The indices of the color matrices, in a certain representation, 
      are denoted by $i,j,k,l,..$. The color invariants
      most commonly encountered are
      \begin{eqnarray}
       \delta_{ab} C_A&=&f^{acd}f^{bcd}   \label{CA} \\
       \delta_{ij} C_F&=&t^a_{il}t^a_{lj} \label{CF} \\
       \delta_{ab} T_F&=&t^a_{ik}t^b_{ki} \label{TR}~.
      \end{eqnarray}
      These constants evaluate to 
      \begin{eqnarray}
       C_A=N_c~,\quad~C_F=\frac{N_c^2-1}{2N_c}~,\quad~T_F=\frac{1}{2}~.\label{Cval}
      \end{eqnarray}
      At higher loops, more color--invariants emerge.
      At $3$--loop order, one additionally obtains
      \begin{eqnarray}
        d^{abd}d_{abc}=(N_c^2-1)(N_c^2-4)/N_c~. \label{dabc2}
      \end{eqnarray}
      In case of $SU(3)_c$, $C_A=3~,~C_F=4/3~,~d^{abc}d_{abc}=40/3$ holds.
%
%
%
\newpage
  \section{\bf \boldmath Feynman Rules}
   \label{App-FeynRules}
   \renewcommand{\theequation}{\thesection.\arabic{equation}}
   \setcounter{equation}{0}
    The Feynman rules can be derived, as in any renormalizable Quantum Field theory, from the path integral representation \cite{Feynman:1965ab,Kleinert:2004ev,KleinertPhi4,ZinnJustin:2002ab}.
    For the QCD Feynman rules, Figure \ref{feynrulesqcd}, we follow Ref.
    \cite{Yndurain:1999ui}, cf. also 
    Refs.~\cite{Veltman:1994wz,'tHooft:1973pz}.
    $D$--dimensional momenta are denoted by $p_i$ and Lorentz-indices by Greek
    letters.
    Color indices are $a,b,...$ and $i,j$ are indices of the color matrices.
    Solid directed lines represent fermions, wavy lines gluons and dashed lines ghosts.
    Arrows denote the direction of the momenta. A factor $(-1)$ has to be
    included for each closed fermion-- or ghost loop.
    \begin{figure}[H]
     \begin{center}
      \includegraphics[angle=0, height=16.5cm]{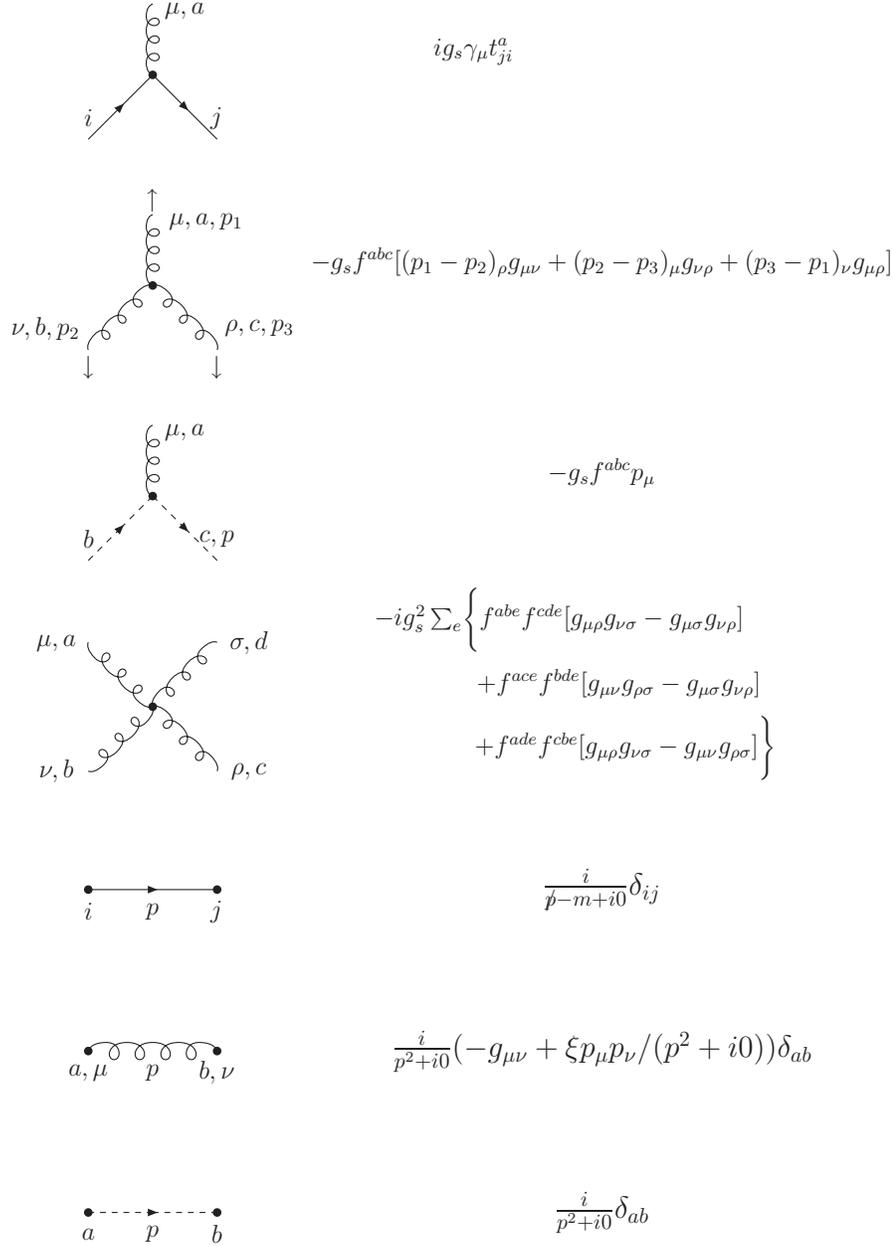}
     \end{center}
     \begin{center} 
      \caption{\sf Feynman rules of QCD.}
        \label{feynrulesqcd}
      \noindent
      \small
     \end{center}
     \normalsize
    \end{figure} 
   \noindent
   The Feynman rules for the quarkonic composite operators are given 
   in Figure \ref{feynrulescompqua}. Up to $O(g^2)$ they can be found 
   in Ref. \cite{Floratos:1977auxFloratos:1977aue1} 
   and also in \cite{Mertig:1995ny}. Note that the 
   $O(g)$ term in the former reference contains a typographical error.
   In Ref. \cite{Bierenbaum:2009mv} these rules were checked and agree up to
   normalization factors, which may be due to different
   conventions. There also the new rule with three external gluons was given.
   The terms $\gamma_{\pm}$ refer to the unpolarized ($+$) and polarized 
   ($-$) case, respectively.
   Gluon momenta are taken to be incoming. 
    \begin{figure}[H]
     \begin{center}
      \includegraphics[angle=0, height=16.5cm]{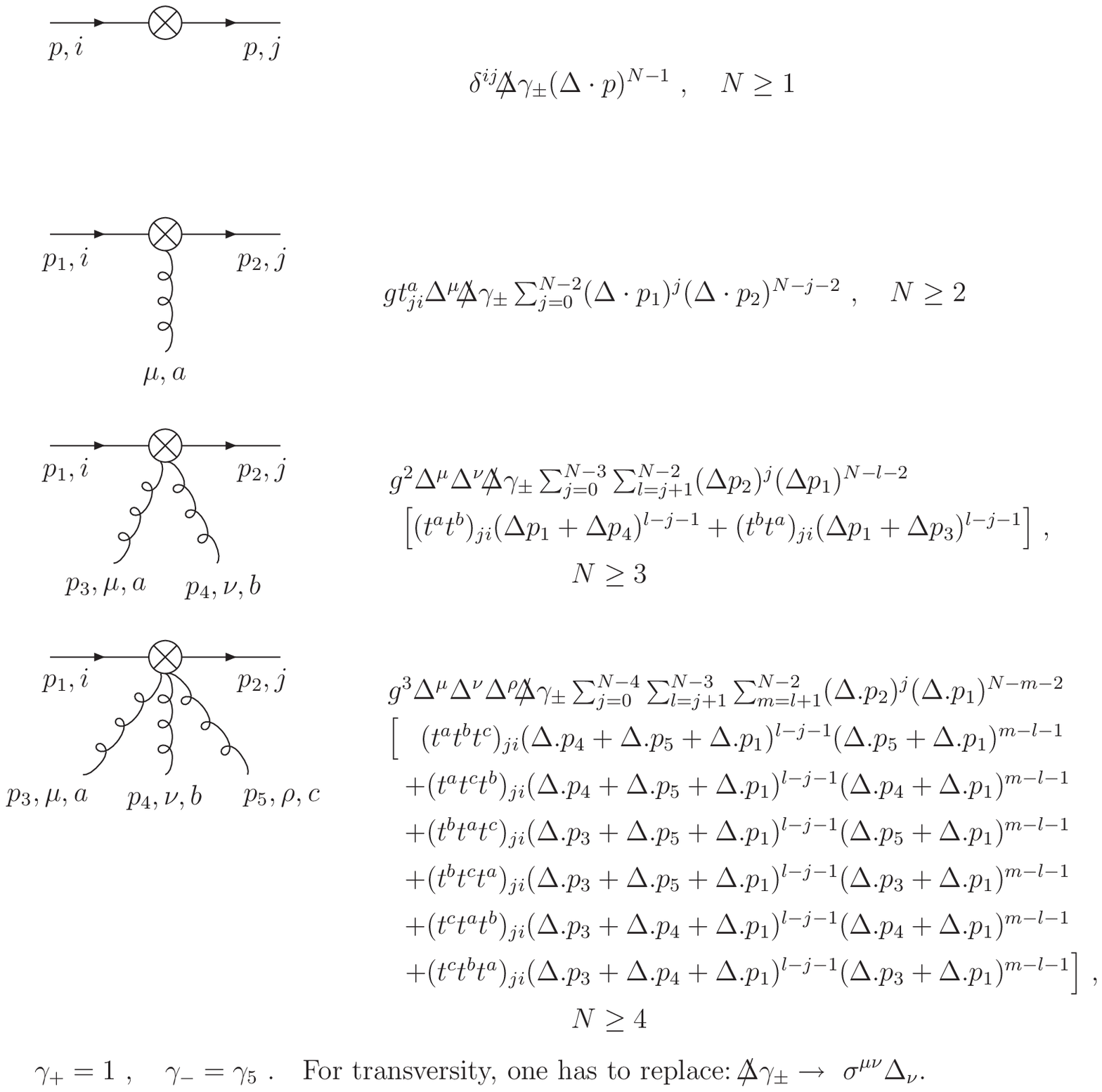}
     \end{center}
     \begin{center} 
      \caption[{\sf Feynman rules for quarkonic composite operators.}]
     {\sf Feynman rules for quarkonic composite operators. $\Delta$
     denotes  a light-like $4$-vector,\\ 
              \phantom{abcdefghijk} $\Delta^2=0$;
              $N$ is a suitably large positive integer, Ref.~\cite{Bierenbaum:2009mv}}
        \label{feynrulescompqua}
      \noindent
      \small
     \end{center}
     \normalsize
    \end{figure} 
\newpage
\section{D-dimensional integrals}
\label{sec:1}

\vspace{1mm}
\noindent
In the calculation of the  $D$-dimensional loop integrals 
\cite{Bollini:1972ui,Speer:1974cz,'tHooft:1972fi}, $D = 4 + 
\varepsilon$,  we perform first 
a Wick-rotation to Euclidean momenta
\begin{eqnarray}
\prod_{i=1}^M \int \frac{d^D k_i}{(2 \pi)^D} f\left(\left. 
k_i\right|_{i=1}^M\right) \prod_{i=1}^M \frac{1}{(k_i^2 - m_i^2)^{a_i}} &=&
(-1)^{-\sum_{j=1}^M a_j} \prod_{i=1}^M \int \frac{d^D k_i}{(2 \pi)^D} 
f\left(\left. 
-k_i\right|_{i=1}^M\right) 
\N\\
&&
\times \prod_{i=1}^M \frac{1}{(-k_i^2 - m_i^2)^{a_i}},
\nonumber
\end{eqnarray}
with $ \forall a_i \in {\mathbb N}$. One obtains the following Euclidean integrals,
where, cf.~\cite{Yndurain:1999ui}, 
\begin{eqnarray}
k_E^2 = k_0^2 + \vec{k}^2, \nonumber
\end{eqnarray}
\begin{eqnarray} 
\int
\frac{d^D k}{(2 \pi)^D} \frac{(k^2)^r}{(k^2 - \varphi)^m} &=& i (-1)^{(r-m)} 
\int
\frac{d^D k_E}{(2 \pi)^D} \frac{(k^2_E)^r}{(k^2_E + \varphi)^m}
\nonumber\\ &=&
i \frac{(-1)^{(r-m)}}{(4\pi)^{D/2}} \frac{\Gamma(r+D/2)}{\Gamma(D/2)} 
\frac{\Gamma(m-r-D/2)}{\Gamma(m)} \varphi^{r-m+D/2},
\N\\
&&
~~m,r \in \mathbb{N}
\nonumber\\
\\
\int d^D k~~k^\alpha_1 k^\alpha_2~~f(k^2) &=& \frac{g^{\alpha_1 \alpha_2}}{D}
\int d^D k~~k^2~~f(k^2) \\
\int d^D k~~k^\alpha_1 k^\alpha_2 k^\alpha_3 k^\alpha_4~~f(k^2) &=& 
\frac{
 g^{\alpha_1 \alpha_2}
 g^{\alpha_3 \alpha_4}
+g^{\alpha_1 \alpha_3}
 g^{\alpha_2 \alpha_4}
+g^{\alpha_1 \alpha_4}
 g^{\alpha_2 \alpha_3}}
{D^2 + 2D}
\N\\
&&
\times \int d^D k~~k^4 f(k^2) \\
\int d^D k~~k^\alpha_1 k^\alpha_2 k^\alpha_3 k^\alpha_4 k^\alpha_5 
k^\alpha_6~~f(k^2) &=& \frac{P_6(g^{\alpha_i \alpha_j})}{D^3 + 6 D^2 +8 D}
\int d^D k~~k^6 f(k^2), 
\end{eqnarray} 
with
\begin{eqnarray} 
P_6(g^{\alpha_i \alpha_j}) &=&~~\hspace{2.2mm} 
                               g^{\alpha_1 \alpha_2} \left[
                               g^{\alpha_3 \alpha_4}  
                               g^{\alpha_5 \alpha_6}  
                              +g^{\alpha_3 \alpha_5}  
                               g^{\alpha_4 \alpha_6}  
                              +g^{\alpha_3 \alpha_6}  
                               g^{\alpha_4 \alpha_5} \right] 
\nonumber\\ &&  
                              +g^{\alpha_1 \alpha_3} \left[
                               g^{\alpha_2 \alpha_4}  
                               g^{\alpha_5 \alpha_6}  
                              +g^{\alpha_2 \alpha_5}  
                               g^{\alpha_4 \alpha_6}  
                              +g^{\alpha_2 \alpha_6}  
                               g^{\alpha_4 \alpha_5} \right]  
\nonumber\\ &&  
                              +g^{\alpha_1 \alpha_4} \left[
                               g^{\alpha_2 \alpha_3}  
                               g^{\alpha_5 \alpha_6}  
                              +g^{\alpha_3 \alpha_5}  
                               g^{\alpha_2 \alpha_6}  
                              +g^{\alpha_3 \alpha_6}  
                               g^{\alpha_2 \alpha_5} \right]  
\nonumber\\ &&  
                              +g^{\alpha_1 \alpha_5} \left[
                               g^{\alpha_3 \alpha_4}  
                               g^{\alpha_2 \alpha_6}  
                              +g^{\alpha_2 \alpha_3}  
                               g^{\alpha_4 \alpha_6}  
                              +g^{\alpha_3 \alpha_6}  
                               g^{\alpha_2 \alpha_4} \right]  
\nonumber\\ &&  
                              +g^{\alpha_1 \alpha_6} \left[
                               g^{\alpha_3 \alpha_4}  
                               g^{\alpha_2 \alpha_5}  
                              +g^{\alpha_3 \alpha_5}  
                               g^{\alpha_2 \alpha_4}  
                              +g^{\alpha_2 \alpha_3}  
                               g^{\alpha_4 \alpha_5} \right]  
\end{eqnarray} 
and
\begin{eqnarray} 
\int d^D k~~\prod_{i=1}^{2M+1} 
k^\alpha_i~~f(k^2)  =  0~.
\end{eqnarray} 
For each loop integral a universal factor
\begin{eqnarray} 
S_\varepsilon = \exp\left[(\gamma_E - \ln(4 \pi))~\frac {\ep} {2}\right]
\end{eqnarray} 
emerges, where $\gamma_E$ denotes the Euler--Mascheroni constant
\begin{eqnarray} 
\gamma_E = \lim_{k \rightarrow \infty} \left[ \sum_{l=1}^k \frac{1}{l} - 
\ln(l) \right]~. \label{gammaE}
\end{eqnarray} 
The factors $S_\varepsilon$ are kept separately and are not expanded in 
$\varepsilon$. In the $\overline{\rm MS}$--scheme \cite{Bardeen:1978yd} they are set to 
$S_\varepsilon = 1$ at the end of the calculation.

The $\Gamma$-function obeys the relation
\begin{eqnarray} 
\Gamma(1+\varepsilon) = \exp \left[ - \gamma_E \varepsilon +\sum_{n=2}^\infty 
\frac{(-\varepsilon)^n}{n} \zeta_n \right]~,
\end{eqnarray} 
with
\begin{eqnarray} 
\label{zetn}
\zeta_n = \sum_{k=1}^\infty \frac{1}{k^n},~~~n \in {\mathbb N},~~n \geq 2~, 
\end{eqnarray} 
the Riemann $\zeta$-function at integer arguments.

We apply the following Feynman parametrization to combine denominators 
\begin{eqnarray} 
\frac{1}{A_1 \ldots A_n} &=& \Gamma(n) \int_0^1 dx_1 \ldots \int_0^1 dx_n
\delta\left(\sum_{k=1}^n x_k -1\right) \frac{1}{(x_1 A_1 +  \ldots + x_n A_n)^n}
\end{eqnarray} 
resp.
\begin{eqnarray} 
\frac{1}{A_1^{a_1} \ldots A_n^{a_n}} &=& \frac{\Gamma\left(\sum_{k=1}^n 
a_k\right)}{\prod_{k=1}^N \Gamma(a_k)}
\int_0^1 dx_1 \ldots \int_0^1 dx_n
\delta\left(\sum_{k=1}^n x_k -1\right) 
\N\\
&&
\times \frac{\prod_{k=1}^n x_k^{a_k - 1}
}{\left(x_1 A_1 +  \ldots + x_n 
A_n\right)^{\left(\sum_{k=1}^n a_k\right)}}~,
\nonumber\\
\label{FeynParam}
\end{eqnarray} 
with $\forall a_i \in \mathbb{N}$.

The integral over the $\delta$--distribution yields
\begin{eqnarray} 
\int_0^1~ dx_l~ \delta\left(\sum_{k=1}^n x_k -1\right) &=& 
\int_{-\infty}^{+\infty}~ dx_l~ \delta\left(\sum_{k=1}^n x_k -1\right)
\theta\left(1- \sum_{k=1}^n x_k\right) \prod_{m=1}^n \theta(x_m) 
\nonumber\\ &=&
\theta\left(1- \sum_{k=1, k \neq l}^n x_k\right) \prod_{m=1, m \neq l}^n 
\theta(x_m)~, 
\end{eqnarray} 
where $\theta(z)$ denotes the Heaviside function
\begin{eqnarray} 
\theta(z) &=& \left\{ \begin{array}{l} 1,~~~z \geq 0 \\ 0,~~~z < 0 \end{array} 
\right.
\end{eqnarray} 
\newpage
  \section{\bf \boldmath Results for the Individual Diagrams}
\label{App-Diag}
 \setcounter{equation}{0}
In this appendix we list the results for the individual diagrams
contributing to the $O\left(T_F^2N_f C_{F,A}\right)$ terms in
the massive operator matrix elements $A_{Qg}^{(3)}$,
$A_{Qq}^{(3),\rm{PS}}$,$A_{qq,Q}^{(3),\rm{PS}}$,
$A_{qq,Q}^{(3),\rm{NS}}$ and $A_{qq,Q}^{(3),\rm{TR}}$. They are all
represented in terms of harmonic sums, despite the fact that in
intermediate results also generalized harmonic sums occur. No sums
beyond $w=4$ contribute. Furthermore, the individual diagrams do not
contain sums, which fully cancel in the final result, similar to
\cite{Bierenbaum:2007qe,Bierenbaum:2008yu}.The individual contributions were compared to the 
corresponding moments in the calculation Ref. \cite{Bierenbaum:2009mv,Blumlein:2009rg}.
The common prefactor 
\begin{eqnarray}
\hat{a}_s^3 S_{\ep}^3 \left(\frac{m^2} {\mu^2}\right)^{3/2 \ep}   \frac{\left(1+(-1)^N\right)} {2}
\end{eqnarray}
 has been taken out.

\subsection{\boldmath ${\Ahathat_{Qg}}$}
The results for the individual diagrams contributing to $\Ahathat_{Qg}^{(3)}$ are:
\begin{eqnarray} 
\Ahathat_{Qg}^{(3),A_{1}}&=&
\frac {T_F^2 n_f C_F} {N^2 (N+1)} \Biggl\{
-\frac {1} {\ep^3}
\Biggl[
\frac{64 (N-1) (N+2)}{9}
\Biggr]
\N\\
&&
-\frac {1} {\ep^2} \Biggl[
\frac{64 (2 N^4-5 N^3+7 N^2+2 N+3)}{27 N (N+1)}
\Biggr]
\N\\
&&
+\frac {1} {\ep} \Biggl[
-\frac{32}{3} (N-1) (N+2) S_2
-\frac{8}{3} (N-1) (N+2) \zeta _2
-\frac{32 P_1(N)}{81 N^2 (N+1)^2 (N+2)^2}
\Biggr]
\N\\
&&
-16 (N-1) (N+2) S_3
+\frac{56}{9} (N-1) (N+2) \zeta _3
\N\\
&&
-\frac{32 (2 N^4-5 N^3+7 N^2+2 N+3) }{9 N (N+1)} S_2
-\frac{8 (2 N^4-5 N^3+7 N^2+2 N+3) }{9 N (N+1)} \zeta _2
\N\\
&&
-\frac{16 P_2(N)}{243 N^3(N+1)^3 (N+2)^3}
\Biggr\}~,\label{DIAGA1}
\end{eqnarray}
with
\begin{eqnarray} 
P_1(N)&=&5 N^8+26 N^7-203 N^6-721 N^5-1079 N^4-460 N^3+269 N^2
\N\\
&&
+444 N+180~,
\\
P_2(N)&=&17 N^{11}+107 N^{10}+824 N^9-927 N^8-8904 N^7-17883
  N^6
\N\\
&&
-19301 N^5-10193 N^4+3973 N^3+15342 N^2+11052 N
\N\\
&&
+2808~.
\end{eqnarray}
%
\begin{eqnarray} 
\Ahathat_{Qg}^{(3),B_{1}}&=&
{T_F^2 n_f C_F} \frac {1} {N} \Biggl\{
\frac {1} {\ep^3}\Biggl[
-\frac{128 }{9} S_1
+\frac{128}{9}
\Biggr]
+\frac {1} {\ep^2} \Biggl[
-\frac{64 }{9} S_1^2
+\frac{128 }{9} S_2
\N\\
&&
+\frac{128 (4 N^2+21 N+8) }{27 (N+1) (N+2)} S_1
-\frac{64 (11 N^2+51 N+22)}{27 (N+1) (N+2)}
\Biggr]
\N
\end{eqnarray}
\begin{eqnarray} 
&& 
+\frac {1} {\ep} \Biggl[
-\frac{64 }{3} S_{2,1}
-\frac{64 }{27} S_1^3
+\frac{736 }{27} S_3
-\frac{64 }{9} S_2 S_1
-\frac{16 }{3}  \zeta _2 S_1 
\N\\
&&
+\frac{64 (4 N^2+21 N+8)} {27 (N+1) (N+2)} S_1^2
+\frac{64 (N^2-15 N+2) }{27 (N+1) (N+2)} S_2
+\frac{16 }{3} \zeta _2
\N\\
&&
-\frac{32 P_3(N)}{81 N (N+1)^2
  (N+2)^2} S_1 
+\frac{32 P_4(N)}{81 (N+1)^3
  (N+2)^2}
\Biggr]
\N\\
&&
-\frac{16 }{27} S_1^4
+\frac{32 }{9} S_2^2
+\frac{352 }{9} S_4
-\frac{128 }{27} S_3 S_1
+\frac{112 }{9} \zeta_3 S_1
-\frac{64 }{3} S_{3,1}
+\frac{32}{3} S_{2,1,1}
\N\\
&&
-\frac{32 }{9} S_2 S_1^2 
-\frac{8 }{3} \zeta_2 S_1^2 
+\frac{16 }{3} \zeta _2 S_2
-\frac{64 }{3} S_{2,1} S_1
+\frac{64 (4 N^2+21 N+8) }{9 (N+1) (N+2)} S_{2,1}
\N\\
&&
+\frac{64 (4 N^2+21 N+8) }{27 (N+1) (N+2)} S_2 S_1
+\frac{16 (4 N^2+21 N+8) }{9 (N+1) (N+2)} \zeta_2 S_1
\N\\
&&
+\frac{64 (4 N^2+21 N+8)}{81 (N+1) (N+2)} S_1^3
-\frac{32 (11 N^2+240 N+22) }{81 (N+1) (N+2)} S_3
\N\\
&&
-\frac{112 }{9} \zeta_3
-\frac{8 (11 N^2+51 N+22) }{9 (N+1) (N+2)} \zeta _2
-\frac{8 P_5(N)}{81 N (N+1)^2 (N+2)^2} S_1^2 
\N\\
&&
-\frac{8 P_6(N)}{81 N (N+1)^2(N+2)^2} S_2
+\frac{16 P_7(N)}{243 N (N+1)^3 (N+2)^3} S_1
\N\\
&&
-\frac{16 P_8(N)}{243 (N+1)^4 (N+2)^3}
\Biggr\}~,
\end{eqnarray}
with
\begin{eqnarray}
P_3(N)&=&88 N^5+780 N^4+1873 N^3+1938 N^2+892 N+216~,
\\
P_4(N)&=&121 N^5+1126 N^4+3280 N^3+3907 N^2+1792 N+484~,
\\
P_5(N)&=&176 N^5+1614 N^4+4097 N^3+4686 N^2+2324 N+648~,
\\
P_6(N)&=&44 N^5+282 N^4+2057 N^3+3966 N^2+1796 N+648~,
\\
P_7(N)&=&824 N^7+10350 N^6+46281 N^5+104256 N^4+124746 N^3
\N\\
&&
+77364 N^2+24952 N+4752~,
\\
P_8(N)&=&1187 N^7+15182 N^6+70743 N^5+161517 N^4+197367 N^3
\N\\
&&
+137022
  N^2+56836 N+9496~.
\end{eqnarray}
\begin{eqnarray} 
\Ahathat_{Qg}^{(3),C_{1}}&=&
{T_F^2 n_f C_F} \frac {1} {N} \Biggl\{
\frac {1} {\ep^2} \Biggl[
\frac{16 (13 N^2+65 N+6)}{3 (N+2) (N+3)}
\Biggr]
\N\\
&&
+\frac {1} {\ep} \Biggl[
64 S_2
-\frac{4 P_9(N)}{3 N (N+1)^2 (N+2)^2 (N+3)}
\Biggr]
\N\\
&&
+96 S_3
+\frac{2 (13 N^2+65 N+6) }{(N+2) (N+3)} \zeta_2
-\frac{8 (13 N^3+86 N^2+191 N+102) }{(N+1) (N+2) (N+3)} S_2
\N\\
&&
+\frac{ P_{10}(N)}{3 N^2 (N+1)^3 (N+2)^3 (N+3)}
\Biggr\}~,
\end{eqnarray}
with
\begin{eqnarray}
P_9(N)&=&91 N^6+855 N^5+2481 N^4+3037 N^3+1436 N^2-28 N-48~,
\\
P_{10}(N)&=&569 N^9+7152 N^8+33604 N^7+79234 N^6+99271 N^5+61206
N^4
\N\\
&&
+13356 N^3+584 N^2+800 N+384~.
\end{eqnarray}
\begin{eqnarray} 
\Ahathat_{Qg}^{(3),D_{1}}&=&
 \frac {{T_F^2 n_f C_F}} {N (N+2) (N+3)} \Biggl\{
\frac {1} {\ep^2} \Biggl[
\frac{16 (N^2+5 N+42)}{3}
\Biggr]
\N\\
&&
+\frac {1} {\ep} \Biggl[
\frac{16 (N+3) (N^4+4 N^3+41 N^2+32 N+12) }{3 N(N+1)^2} S_1
-\frac{4 P_{11}(N)}{3 (N+1)^3 (N+2)}
\Biggr]
\N\\
&&
+2 (N^2+5 N+42) \zeta _2
+\frac{8 P_{12}(N)}{3 N (N+1)^2} S_1^2
+\frac{8 P_{12}(N) }{3 N (N+1)^2} S_2
\N\\
&&
-\frac{4 P_{13}(N) }{3 N (N+1)^3 (N+2)} S_1
+\frac{ P_{14}(N) }{3 (N+1)^4
   (N+2)^2}
\Biggr\}~,
\end{eqnarray}
with
\begin{eqnarray} 
P_{11}(N)&=&7 N^6+70 N^5+748 N^4+2770 N^3+3961 N^2+2320 N+588~,
\\
P_{12}(N)&=&N^5+7 N^4+53 N^3+188 N^2+141 N+54~,
\\
P_{13}(N)&=&7 N^7+70 N^6+730 N^5+3340 N^4+6127
   N^3+5074 N^2+2340 N
\N\\
&&
+504~,
\\
P_{14}(N)&=&53 N^8+665 N^7+7044 N^6+36078 N^5+90789 N^4+122301 N^3
\N\\
&&
+93322 N^2+40780 N+7752~.
\end{eqnarray}
\begin{eqnarray} 
\Ahathat_{Qg}^{(3),E_{1}}&=&
{T_F^2 n_f \left(C_F-\frac {C_A} {2} \right)} \frac {1} {N (N+1)} \Biggl\{
\frac {1} {\ep^3}
\Biggl[
\frac{64 (N-1)}{9 (N+1)}
\Biggr]
\N\\
&&
+\frac {1} {\ep^2} \Biggl[
-\frac{32}{9} (4 N+5) S_1
+\frac{16 (N-1) (15 N^3+47 N^2+44 N+18)}{27 N (N+1)^2}
\Biggr]
\N\\
&&
+\frac {1} {\ep} \Biggl[
-\frac{8 (8 N^3+29 N^2+41 N-66) }{9 (N+2) (N+3)} S_1^2
+\frac{8 (N-1) }{3 (N+1)} \zeta _2
\N\\
&&
-\frac{8 (32 N^4+157 N^3+310 N^2+71 N-66) }{9 (N+1) (N+2) (N+3)} S_2
\N\\
&&
+\frac{8 P_{15}(N) }{27 N (N+1) (N+2) (N+3)} S_1
-\frac{4 P_{16}(N)}{81 N^2 (N+1)^3 (N+2)^2 (N+3)}
\Biggr]
\N\\
&&
-\frac{4 (16 N^3+37 N^2+25 N-258) }{27 (N+2) (N+3)} S_1^3
-\frac{4}{3} (4 N+5) \zeta _2 S_1
\N
\end{eqnarray}
\begin{eqnarray} 
&&
-\frac{4 (16 N^3+37 N^2+25 N-258) }{9 (N+2) (N+3)} S_2 S_1
-\frac{56 (N-1) }{9 (N+1)} \zeta_3
\N\\
&&
-\frac{8 (124 N^4+593 N^3+1142 N^2+199 N-258) }{27 (N+1) (N+2) (N+3)} S_3
-\frac{16}{3} (4 N+5) S_{2,1}
\N\\
&&
+\frac{2 P_{17}(N) }{27 N (N+1) (N+2) (N+3)} S_1^2
+\frac{2 (N-1) (15 N^3+47 N^2+44 N+18) }{9 N (N+1)^2} \zeta_2
\N\\
&&
+\frac{2 P_{18}(N)}{27 N (N+1)^2 (N+2) (N+3)} S_2
-\frac{2 P_{19}(N)}{81 N (N+1)^2 (N+2)^2 (N+3)} S_1 
\N\\
&&
+\frac{ P_{20}(N)}{243 N^3 (N+1)^4 (N+2)^3 (N+3)}
\Biggr\}~,
\end{eqnarray}
with
\begin{eqnarray} 
P_{15}(N)&=&118 N^5+779 N^4+1838 N^3+835 N^2-114 N-432~,
\\
P_{16}(N)&=&231 N^9+1670 N^8+4858 N^7+1994 N^6-12383 N^5
\N\\
&&
-23476 N^4-23234 N^3-9924 N^2+1944 N+1296~,
\\
P_{17}(N)&=&236 N^5+1105 N^4+2428 N^3-2347 N^2-3198 N-1296~,
\\
P_{18}(N)&=&1052 N^6+6813 N^5+18221 N^4+14457 N^3-3865 N^2
\N\\
&&
-8166 N-2592~,
\\
P_{19}(N)&=&2830 N^7+24935 N^6+87817 N^5+134237 N^4+95737 N^3
\N\\
&&
+28280
  N^2-29532 N-19008~,
\\
P_{20}(N)&=&5487 N^{12}+36857 N^{11}+58060 N^{10}-225576 N^9-995433
  N^8
\N\\
&&
-1617615 N^7-1886730 N^6-2199914 N^5-1681960 N^4
\N\\
&&
-544872 N^3-87696
  N^2-51840 N-15552~.
\end{eqnarray}
\begin{eqnarray} 
\Ahathat_{Qg}^{(3),F_{1}}&=&
{T_F^2 n_f \left(C_F-\frac {C_A} {2}\right)} \frac {1} {(N+1) (N+2)} \Biggl\{
\frac {1} {\ep^3} \Biggl[
-\frac{512}{9}
+\frac{512}{9} S_1
\Biggr]
\N\\
&&
+\frac {1} {\ep^2} \Biggl[
+\frac{64}{9} S_1^2
-\frac{64 (4 N^2+9 N+8)}{9 N} S_2
+\frac{128 (3 N^3-N^2+30 N+12)}{27 N^2} S_1
\N\\
&&
-\frac{128 (3 N^3+2 N^2+17 N+6)}{27 N (N+1)}
\Biggr]
\N\\
&&
+\frac {1} {\ep} \Biggl[
-\frac{32}{27} S_1^3
+\frac{64 (3 N^3+20 N^2+3 N+12)}{27 N^2} S_1^2
-\frac{64 P_{21}(N)}{81 N^2
  (N+1) (N+2)} S_1
\N\\
&&
+\frac{64}{3} \zeta _2 S_1
-\frac{32}{9} S_2 S_1
+\frac{64 P_{22}(N)}{81 N (N+1)^2(N+2)}
-\frac{64 \zeta _2}{3}
\N\\
&&
+\frac{64 (4 N^3+87 N^2-85 N-24) }{27 N^2} S_2
-\frac{256 (3 N^2+10 N+6) }{27 N} S_3
\N\\
&&
+\frac{128 (N+1) (N+2) }{3 N} S_{2,1}
\Biggr]
-\frac{20 }{27} S_1^4
+\frac{8 }{3} \zeta_2 S_1^2
-\frac{40 }{9} S_2 S_1^2
-\frac{448 }{9} \zeta _3 S_1
\N
\end{eqnarray}
\begin{eqnarray}
&&
-\frac{160  }{27} S_3 S_1
+\frac{64  }{3} S_{2,1} S_1
-\frac{4 (48 N^2+101 N+96)}{9 N} S_2^2
-\frac{8 (4 N^2+9 N+8) }{3 N} \zeta _2 S_2
\N\\
&&
+\frac{64}{3 } S_{3,1}
+\frac{64 (3 N^2+7 N+6) }{3 N} S_{2,1,1}
-\frac{8 (56 N^2 +169 N + 112)}{9 N} S_4
\N\\
&&
+\frac{32 (6 N^3+61 N^2-21 N+24) }{81 N^2} S_1^3
+\frac{16 (3 N^3-N^2+30 N+12)}{9 N^2} \zeta _2 S_1
\N\\
&&
+\frac{32 (6 N^3+61 N^2-21 N+24)}{27 N^2} S_2 S_1
-\frac{128 (N^3+9 N^2-10 N-6) }{9 N^2} S_{2,1}
\N\\
&&
+\frac{448 }{9} \zeta_3
-\frac{32 (9 N^3-623 N^2+894 N+276)}{81 N^2} S_3
-\frac{16 P_{23}(N)}{81 N^2 (N+1) (N+2)} S_1^2
\N\\
&&
-\frac{16 (3 N^3+2 N^2+17 N+6) }{9 N (N+1)} \zeta _2
-\frac{16 P_{24}(N)}{81 N^2 (N+1) (N+2)} S_2
\N\\
&&
+\frac{32 P_{25}(N)}{243 N^2 (N+1)^2 (N+2)^2} S_1
-\frac{32 P_{26}(N)}{243 N (N+1)^3 (N+2)^2} \Biggr\}~,
\end{eqnarray}
with
\begin{eqnarray}
P_{21}(N)&=&24 N^5+196 N^4+1047 N^3+1748 N^2+984 N+120~,
\\
P_{22}(N)&=&33 N^5+364 N^4+1462 N^3+2543 N^2+1904 N+132~,
\\
P_{23}(N)&=&48 N^5+746 N^4+2697 N^3+2746 N^2+1104 N+240~,
\\
P_{24}(N)&=&124 N^5+198 N^4-2387 N^3-6162 N^2-3632 N-480~,
\\
P_{25}(N)&=&264 N^7+4046 N^6+21591 N^5+52844 N^4+74856 N^3
\N\\
&&
+66812N^2
+30576 N+2640~,
\\
P_{26}(N)&=&363 N^7+6758 N^6+41285 N^5+121235 N^4+190235 N^3
\N\\
&&
+150758 N^2+46964 N+2904~.
\end{eqnarray}
\begin{eqnarray} 
\Ahathat_{Qg}^{(3),G_{1}}&=&
\frac {T_F^2 n_f C_F} {(N+1) (N+2)} \Biggl\{
-\frac{1}{\ep^2}
\frac{224}{3}
+\frac{1}{\ep}
\Biggl[
-64 S_2
\N\\
&&
-\frac{16 \left(N^2+27 N+38\right) }{3 (N+1) (N+2)} S_1
+\frac{8 \left(65 N^4+430 N^3+881 N^2+608 N+44\right)}{3 (N+1)^2 (N+2)^2}
\Biggr]
\N\\
&&
- 96 S_3
-\frac{8 \left(N^3+30 N^2+137 N+144\right) }{3 (N+1) (N+2) (N+3)} S_1^2
-28  \zeta_2
\N\\
&&
+\frac{16 (25 N^3+150 N^2+248 N+105) }{3 (N+1) (N+2) (N+3)} S_2
+\frac{4 P_{27}(N)}{3 N (N+1)^2 (N+2)^2 (N+3)} S_1
\N\\
&&
-\frac{2 P_{28}(N) }{3 (N+1)^3 (N+2)^3 (N+3)}
\Biggr\}~,
\end{eqnarray}
with
\begin{eqnarray}
P_{27}(N)&=&13 N^6+373 N^5+2197 N^4+4907 N^3+4534 N^2+1464 N+144~,
\\
P_{28}(N)&=&455 N^7+5828 N^6+29700 N^5+77486 N^4+111073 N^3
\N\\
&&
+87014 N^2+35364 N+6840~.
\end{eqnarray}
\begin{eqnarray} 
\Ahathat_{Qg}^{(3),H_{1}}&=&
{T_F^2 n_f \left(C_F -\frac {C_A} {2} \right)} \frac {1} {N+1} \Biggl\{
\frac{1} {\ep^2} \Biggl[
\frac{320 }{9 N} S_1
-\frac{128}{3 (N+2)}
\Biggr]
\N\\ 
&&
+\frac{1} {\ep} \Biggl[
\frac{16 \left(N^2+41 N-30\right) }{9 N (N+2) (N+3)} S_1^2
\N\\
&&
-\frac{16 \left(71 N^4+850 N^3+2707 N^2+4160 N+2724\right) }{27 N (N+1) (N+2)^2 (N+3)} S_1
\N\\
&&
-\frac{16 P_{29}(N)}{3 (N+1)^2 (N+2)^3 (N+3)}
+\frac{16 \left(13 N^2+173 N-30\right) }{9 N (N+2) (N+3)} S_2
\Biggr]
\N\\
&&
-\frac{8 \left(7 N^2-73 N+150\right) }{27 N (N+2) (N+3)} S_1^3
+\frac{40 }{3 N} \zeta _2 S_1
\N\\
&&
-\frac{8 \left(7 N^2-73 N+150\right) }{9 N (N+2) (N+3)} S_2 S_1
+\frac{16 \left(47 N^2+667 N-150\right) }{27 N (N+2) (N+3)} S_3
+\frac{160 }{3 N} S_{2,1}
\N\\
&&
+\frac{4 \left(119 N^4-962 N^3-3137 N^2-3784 N-3324\right) }{27 N (N+1) (N+2)^2 (N+3)} S_1^2
-\frac{16 }{(N+2)} \zeta _2
\N\\
&&
-\frac{4 \left(433 N^4+6746 N^3+16769 N^2+13192 N+5052\right) }{27 N (N+1) (N+2)^2 (N+3)} S_2
\N\\
&&
-\frac{4 P_{30}(N) }{81 N (N+1)^2 (N+2)^3 (N+3)} S_1
\N\\
&&
+\frac{8 P_{31}(N)}{9 (N+1)^3 (N+2)^4 (N+3)}
\Biggr\}~,
\end{eqnarray}
with
\begin{eqnarray}
P_{29}(N)&=& 7 N^5+7 N^4-203 N^3-955 N^2-1604 N-948~,
\\
P_{30}(N)&=& 427 N^6-10201 N^5-91517 N^4-319471 N^3-556466 N^2\\
\N
&&
-463084 N
-137784~,
\\
P_{31}(N)&=& 253 N^7+1884 N^6+3444 N^5-9582 N^4-50265 N^3-82446 N^2
\N
\\
&&
-58120 N-13680~.
\end{eqnarray}
\begin{eqnarray} 
\Ahathat_{Qg}^{(3),I_{1}}  
&=&
\frac{T_F^2 n_f}{(N+1)(N+2)} \Biggl\{
\N\\ 
&&
C_F \Biggl\{
\frac{1} {\ep^2} 
\left[
   \frac{32}{3} S_1^2
   -\frac{32 }{3} S_2
   -\frac{128 (N+1)}{3 N} S_1
   +\frac{256}{3}
\right]
\N\\ 
&& 
+\frac {1} {\ep} 
 \Biggl[
   \frac{16}{3} S_1^3
   + 16 S_1 S_2
   -\frac{64}{3} S_3
   +\frac{32 (7 N+3)}{9 N } S_2
   -\frac{32 (13N-3) }{9 N } S_1^2
\N\\  
&& 
   +\frac{64 \left(20N^2+37 N+23\right)}{9 N (N+1)} S_1
     -\frac{2560}{9}
\Biggr]
\N\\
&& 
+\frac{14}{9} S_1^4
+\frac{28}{3} S_2 S_1^2
+\frac{14 }{3} S_2^2
- 64 S_{2,1,1}
+ 32 S_{2,1}  S_1
- 4\zeta_2 S_2
+ 4\zeta_2 S_1^2
+ 32  S_{3,1}
\N 
\end{eqnarray}
\begin{eqnarray} 
&&
+\frac{112}{9} S_3 S_1
-28 S_4
-\frac{16 (11 N-5)}{9 N} S_1^3
-\frac{16 (N+1) }{N} \zeta _2 S_1
-\frac{16 (11 N-5) }{3 N} S_2 S_1
\N\\  
&& 
+\frac{32 (19 N+5)}{9 N} S_3
-\frac{64 (N+1) }{N} S_{2,1}
+\frac{16 \left(179 N^2+200N+75\right)}{27 N (N+1)} S_1^2
+ 32 \zeta _2
\N\\  
&&
-\frac{16 \left(23 N^2+2 N-75\right)}{27 N (N+1)} S_2
-\frac{32 \left(274 N^3+816 N^2+921N+325\right)}{27 N (N+1)^2}
S_1
+\frac{16960}{27} \Biggr\}
\N\\  
&&
+C_A  
\Biggl\{
 \frac{1} {\ep^3} 
 \Biggl[ \frac{64}{9} S_1 -\frac{64 (N+4)}{9
   (N+2)}\Biggr] 
+\frac{1} {\ep^2} \Biggl[
-\frac{16}{9} S_1^2
+\frac{64 (N+1)}{3} S_{-2}
+\frac{16 (6 N+5)}{9} S_2
\N\\  
&&
+\frac{32 \left(N^3+33 N^2+74N+36\right)}{27 N (N+1) (N+2)} S_1
-\frac{32 \left(7N^3+25 N^2+80 N+80\right)}{27 (N+1) (N+2)^2}
\Biggr]
\N\\  
&&
+\frac{1}{\ep} \Biggl[
-\frac{40 }{27} S_1^3
+\frac{8 }{3} \zeta _2 S_1
+\frac{8 (18N+13)}{9} S_2 S_1
+\frac{80 (N+1) }{3} S_{-3}
+\frac{16 (54 N+49) }{27}S_3
\N\\  
&& 
-32 (N+1) S_{-2,1}
-\frac{16(3 N+1) }{3} S_{2,1}
-\frac{32 \left(8N^2+3 N-14\right) }{9 (N+2)} S_{-2}
\N\\  
&&
-\frac{16 \left(24N^4+38 N^3+15 N^2+31 N+18\right) }{27
  N(N+1)(N+2)} S_2
+\frac{16 \left(22 N^3+69 N^2+35 N-18\right)}{27 N (N+1)(N+2)} S_1^2
\N \\  
&& 
-\frac{8 (N+4) }{3(N+2)} \zeta _2
-\frac{16 P_{32}(N)}{81 N(N+1)^2 (N+2)^2} S_1
+32 (N+1) S_{-2} S_1
+\frac{16 P_{33}(N)}{81 (N+1)^2 (N+2)^3}
\Biggr]
\N\\  
&& 
-\frac{13 }{27} S_1^4
-\frac{8 (9 N+11) }{3 }S_{2,1}  S_1
-48 (N+1) S_{-2,1} S_1
+\frac{8 (162 N+149) }{27} S_3  S_1 
\N\\  
&&
+ 40 (N+1) S_{-3} S_1
-\frac{56}{9} \zeta _3 S_1
-\frac{ (36 N+73)}{9} S_2^2
+8 (N+1) \zeta _2 S_{-2}
\N\\
&& 
+\frac{2 (6 N+5)}{3}  \zeta _2 S_2
+8(N+1) S_{-2} S_2
+24 (N+1) S_{-2} S_1^2
-\frac{16 (9 N+10) }{3} S_{3,1}
\N\\  
&& 
+ 48 (N+1) S_{-2,1,1}
+\frac{8 (9 N+19)}{3} S_{2,1,1}
+\frac{2 (228N+203) }{9} S_4
-40 (N+1)S_{-3,1}
\N \\
&& 
+ 32 (N+1)S_{2,-2}
+\frac{8 \left(65 N^3+174 N^2+31 N-90\right)}{81 N (N+1) (N+2)} S_1^3
\N\\
&& 
-\frac{8\left(72 N^4+34 N^3-273 N^2-157 N+90\right) }{27 N
  (N+1)(N+2)} S_2 S_1
+\frac{4 \left(N^3+33N^2+74 N+36\right) }{9 N (N+1)(N+2)} \zeta_2
S_1
\N\\  
&& 
-\frac{16\left(8 N^2+3 N-14\right) }{3(N+2)} S_{-2} S_1
-\frac{8(N+1) }{3} S_{-4}
-\frac{40 \left(8 N^2+3N-14\right) }{9(N+2)} S_{-3}
\N\\  
&& 
+\frac{56 (N+4) }{9 (N+2)} \zeta _3
+\frac{2 (54 N+41) }{9}  S_2 S_1^2
-\frac{8 \left(432 N^4+707 N^3+111N^2+154 N+180\right) }{81 N
  (N+1) (N+2)} S_3
\N\\  
&& 
- \frac {2}{3} \zeta _2 S_1^2
+\frac{16\left(8 N^2+3 N-14\right) }{3
  (N+2)} S_{-2,1}
+\frac{8 \left(24 N^4+35 N^3+33 N^2+106 N+72\right)}{9 N
  (N+1) (N+2)} S_{2,1}
\N\\  
&& 
-\frac{4 P_{34}(N)}{81 N(N+1)^2 (N+2)^2} S_1^2
-\frac{4 \left(7 N^3+25 N^2+80 N+80\right) }{9 (N+1) (N+2)^2}
\zeta_2
\N
\end{eqnarray}
\begin{eqnarray} && 
+\frac{64 \left(13N^4+36 N^3+73 N^2+153 N+130\right)}{27 (N+1)
   (N+2)^2} S_{-2}
+\frac{4 P_{35}(N)}{81 N(N+1)^2 (N+2)^2} S_2
\N\\
&&
+\frac{8 P_{36}(N)}{243 N (N+1)^3 (N+2)^3} S_1
-\frac{8 P_{37}(N)}{243
  (N+1)^3(N+2)^4}
\Biggr\}
\Biggr\}~,
\end{eqnarray}
with
\begin{eqnarray} 
P_{32}(N)&=&173N^5+1494 N^4+4706 N^3+7191 N^2+5588
      N+1872~,
\\
P_{33}(N)&=&320 N^5+2720 N^4+9797 N^3+17507N^2+15272N+5216~,
\\
P_{34}(N)&=&700N^5+4572 N^4+10909 N^3+12447 N^2+7624
    N+2448~,
\\
P_{35}(N)&=&312 N^6+962N^5+456 N^4-2947 N^3-7203 N^2-6880
      N
\N\\
&&-2448~,
\\
P_{36}(N)&=&3019 N^7+33159 N^6+152637
   N^5+388497 N^4+593193 N^3
\N\\
&&
+540399 N^2+267080 N+52848~,
\\
P_{37}(N)&=&5041 N^7+58633 N^6+293478 N^5+808539
    N^4+1321572N^3
\N\\ &&
+1281873 N^2+683576 N
    +154784~.
\end{eqnarray}
\begin{eqnarray} 
\Ahathat_{Qg}^{(3),J_{1a}}&=&
\frac {T_F^2 n_f C_A} {N^2 (N+1)^2} \Biggl\{
\frac {1} {\ep^3}\Biggl[
\frac{32 (N-3) (2 N^2+N-4)}{9}
\Biggr]
\N\\
&&
-\frac {1} {\ep^2} \Biggl[
\frac{16 P_{38}(N)}{27 (N-1) N (N+1) (N+2)}
\Biggr]
\N\\
&&
+\frac {1} {\ep} \Biggl[
\frac{16}{3} (N-3) (2 N^2+N-4) S_2
+\frac{4}{3} (N-3) (2 N^2+N-4) \zeta _2
\N\\
&&
+\frac{8 P_{39}(N)}{81 (N-1)^2 N^2 (N+1)^2 (N+2)^2}
\Biggr]
+8 (N-3) (2 N^2+N-4) S_3
\N\\
&&
-\frac{28}{9} (N-3) (2 N^2+N-4) \zeta_3
-\frac{8 P_{38}(N)}{9 (N-1) N (N+1) (N+2)} S_2
\N\\
&&
-\frac{2 P_{38}(N)}{9 (N-1) N (N+1) (N+2)} \zeta_2
\N\\
&&
-\frac{4 P_{40}(N)}{243 (N-1)^3 N^3 (N+1)^3 (N+2)^3}
\Biggr\}~,
\end{eqnarray}
with
\begin{eqnarray} 
P_{38}(N)&=&16 N^7-17 N^6-209 N^5+142 N^4-65 N^3-629 N^2+186 N
\N\\
&&
+216~,
\\
P_{39}(N)&=& 68 N^{11}+75 N^{10}-1715 N^9-2112 N^8-3360 N^7+24 N^6
\N\\
&&
+24544 N^5+11469 N^4-24037 N^3-6432 N^2+4932 N+3024~,
\\
P_{40}(N)&=&460 N^{15}+1799 N^{14}-11123 N^{13}-40963 N^{12}-38795
N^{11}
\N\\
&&
+155048 N^{10}+356978 N^9-365210 N^8-953263 N^7+458455 N^6
\N\\
&&
+924557 N^5-305717 N^4-358086 N^3-66708 N^2+87048
   N
\N\\
&&
+38880~.
\end{eqnarray}
\begin{eqnarray} 
\Ahathat_{Qg}^{(3),J_{1b}}&=&
{T_F^2 n_f C_A} \frac {1} {N^2 (N+1)^2} \Biggl\{
-\frac {1} {\ep^3}
\Biggl[
\frac{64 (4 N^2+4N-5)}{9}
\Biggr]
\N\\
&&
+\frac {1} {\ep^2} \Biggl[
-\frac{32}{9} (4 N^2+4 N-5) S_1
+\frac{32 P_{41}(N)}{27 N (N+1) (N+2)}
\Biggr]
\N\\
&&
+\frac {1} {\ep} \Biggl[
-\frac{8}{9} (4 N^2+4 N-5) S_1^2
-\frac{104}{9} (4 N^2+4 N-5) S_2
\N\\
&&
-\frac{8}{3} (4 N^2+4 N-5) \zeta_2
+\frac{16 P_{41}(N) }{27 N (N+1) (N+2)} S_1
\N\\
&&
-\frac{16 P_{42}(N)}{81 N^2 (N+1)^2 (N+2)^2}
\Biggr]
-\frac{4}{27} (4 N^2+4 N-5) S_1^3
\N\\
&&
-\frac{440}{27} (4 N^2+4 N-5) S_3
+\frac{56}{9} (4 N^2+4 N-5) \zeta _3
\N\\
&&
+\frac{4 P_{41}(N) }{27 N (N+1) (N+2)} S_1^2
+\frac{52 P_{41}(N) }{27 N (N+1) (N+2)} S_2
\N\\
&&
+\frac{4 P_{41}(N) }{9 N (N+1) (N+2)} \zeta_2
-\frac{52}{9} (4 N^2+4 N-5) S_2 S_1
\N\\
&&
-\frac{4}{3} (4 N^2+4 N-5) \zeta _2 S_1
-\frac{8 P_{42}(N)}{81 N^2 (N+1)^2 (N+2)^2} S_1
\N\\
&&
+\frac{8 P_{43}(N)}{243 N^3 (N+1)^3 (N+2)^3}
\Biggr\}~,
 \end{eqnarray}
with
 \begin{eqnarray}
P_{41}(N)&=&44 N^5+191 N^4+99 N^3-17 N^2+136 N+60~,
\\
P_{42}(N)&=&232 N^8+1978 N^7+4292 N^6+4447 N^5+3446 N^4-1058 N^3
\N\\
&&
-5180 N^2-2712 N-720~,
\\
P_{43}(N)&=&1328 N^{11}+16796 N^{10}+66200 N^9+134952 N^8+148833
N^7
\N\\
&&
+40020 N^6-52496 N^5+66940 N^4+180160 N^3+115296 N^2
\N\\
&&
+45504 N+8640~.
\end{eqnarray}
\begin{eqnarray} 
\Ahathat_{Qg}^{(3),K_{1a}}&=&
\frac {T_F^2 n_f C_A} {(N-1) N (N+1)^2 (N+2)} \Biggl\{
-\frac {1} {\ep^3} \frac{32 (N^3-10 N^2+43 N+46)}{9}
\N\\
&&
+\frac {1} {\ep^2} \Biggl[
+\frac{16 P_{44}(N)}{27 (N-1) N (N+1) (N+2)}
\Biggr]
\N\\
&&
+\frac {1} {\ep} \Biggl[
-\frac{16}{3} (N^3-10N^2+43 N+46) S_2
-\frac{4}{3} (N^3-10 N^2+43 N+46) \zeta _2
\N\\
&&
-\frac{8 P_{45}(N)}{81 (N-1)^2 N^2 (N+1)^2 (N+2)^2}
\Biggr]
\N
\end{eqnarray}
\begin{eqnarray}&& 
 -8 (N^3-10 N^2+43 N+46) S_3
+\frac{28}{9} (N^3-10 N^2+43 N+46) \zeta _3
\N\\
&&
+\frac{8 P_{44}(N)}{9 (N-1) N (N+1) (N+2)} S_2
+\frac{2 P_{44}(N)} {9 (N-1) N (N+1) (N+2)} \zeta _2
\N\\
&&
+\frac{4 P_{46}(N)}{243 (N-1)^3 N^3 (N+1)^3 (N+2)^3}
\Biggr\}~,
\end{eqnarray}
with
\begin{eqnarray}
P_{44}(N)&=&11 N^7-100 N^6+317 N^5-59 N^4-2184 N^3-765 N^2+1064 N
\N\\
&&
+276~,
\\
P_{45}(N)&=&67 N^{11}-606 N^{10}+221 N^9-5084 N^8-7700 N^7+41990
N^6
\N\\
&&
+48735 N^5-45550 N^4-33703 N^3+18682 N^2+7212 N+1656~,
\\
P_{46}(N)&=&431 N^{15}-3320 N^{14}-7624 N^{13}-6288 N^{12}+161483
N^{11}
\N\\
&&
+372478 N^{10}-690814 N^9-1786393 N^8+805264 N^7+2328180 N^6
\N\\
&&
-851672 N^5-1227601 N^4+257348 N^3+123792 N^2+48240 N
\N\\
&&
+9936~.
\end{eqnarray}
\begin{eqnarray} 
\Ahathat_{Qg}^{(3),K_{1b}}&=&
 \frac {T_F^2 n_f C_A} {(N-1) N (N+1)^2 (N+2)} \Biggl\{
\frac {1} {\ep^3}\Biggl[
\frac{64 (3 N^2-23 N-20)}{9 }
\Biggr]
\N\\
&&
+\frac {1} {\ep^2} \Biggl[
\frac{32 (3 N^2-23 N-20) }{9 } S_1
\N\\
&&
-\frac{32 (54 N^4-127 N^3-289 N^2+322 N+376)}{27 (N+1) (N+2)}
\Biggr]
\N\\
&&
+\frac {1} {\ep} \Biggl[
\frac{8 (3 N^2-23 N-20) }{9} S_1^2
+\frac{104 (3 N^2-23 N-20) }{9} S_2
\N\\
&&
+\frac{8 (3 N^2-23 N-20) }{3} \zeta _2
\N\\
&&
-\frac{16 (54 N^4-127 N^3-289 N^2+322 N+376) }{27 (N+1) (N+2)} S_1
+\frac{16 P_{47}(N)}{81 (N+1)^2 (N+2)^2}
\Biggr]
\N\\
&&
+\frac{4 (3 N^2-23 N-20) }{27} S_1^3
\N\\
&&
+\frac{440 (3 N^2-23 N-20) }{27} S_3
-\frac{56 (3 N^2-23 N-20) }{9} \zeta_3
\N\\
&& 
+\frac{52 (3 N^2-23 N-20) }{9 }  S_2 S_1
+\frac{4 (3 N^2-23 N-20) }{3} \zeta_2 S_1 
\N
\end{eqnarray}
\begin{eqnarray} && 
-\frac{4 (54 N^4-127 N^3-289 N^2+322 N+376)}{27 (N+1) (N+2)}
S_1^2
\N\\
&&
-\frac{52 (54 N^4-127 N^3-289 N^2+322 N+376) }{27 (N+1)(N+2)} S_2
\N\\
&&
-\frac{4 (54 N^4-127 N^3-289 N^2+322 N+376) }{9 (N+1) (N+2)} \zeta_2 
\N\\
&&
+\frac{8 P_{48}(N)}{81 (N+1)^2 (N+2)^2} S_1 
\N\\
&&
-\frac{8 P_{49}(N)}{243 (N+1)^3 (N+2)^3}
\Biggr\}~,
\end{eqnarray}
with
\begin{eqnarray}
P_{47}(N)&=&468 N^6+574 N^5+199 N^4+1615 N^3-5564 N^2-18092 N
\N\\
&&
-10304~,
\\
P_{48}(N)&=&468 N^6+574 N^5+199 N^4+1615 N^3-5564 N^2-18092 N
\N\\
&&-10304~,
\\
P_{49}(N)&=&3096 N^8+11084 N^7+7658 N^6-72687 N^5-301881 N^4
\N\\
&&
-412434 N^3+10260 N^2+433240 N+231808~.
\end{eqnarray}
\begin{eqnarray} 
\Ahathat_{Qg}^{(3),L_{1a}}&=&
 \frac {T_F^2 n_f C_A} {N (N+1)} \Biggl\{
\frac {1} {\ep^3}
\Biggl[
+\frac{16}{9} (4 N+5) S_1
+\frac{16 (2 N^3+5 N^2+6 N+4)}{9 N (N+1)}
\Biggr]
\N\\
&&
+\frac {1} {\ep^2} \Biggl[
\frac{4}{3} (4 N+5) S_1^2
+\frac{4}{3} (4 N+5) S_2
-\frac{8 (32 N^3+96 N^2+49 N-12) }{27 N (N+1)} S_1
\N\\
&&
-\frac{8 P_{50}(N)}{27 N^2 (N+1)^2 (N+2)}
\Biggr]
+\frac {1} {\ep} \Biggl[
\frac{14}{27} (4 N+5) S_1^3
+\frac{2}{3} (4 N+5) \zeta _2 S_1
\N\\
&&
+\frac{14}{9} (4 N+5) S_2 S_1 
+\frac{28}{27} (4 N+5) S_3
+\frac{8}{3} (4 N+5) S_{2,1}
\N\\
&&
-\frac{2 (36 N^3+106 N^2+61 N-4) }{9 N (N+1)} S_1^2
+\frac{2 (2 N^3+5 N^2+6 N+4) }{3 N (N+1)} \zeta_2
\N\\
&&
-\frac{2 (84 N^3+130 N^2+13 N-52) }{9 N (N+1)} S_2
+\frac{4 P_{51}(N)}{81 N^2 (N+1)^2 (N+2)} S_1
\N\\
&&
+\frac{2 P_{52}(N)}{81 N^3 (N+1)^3 (N+2)^2}
\Biggr]
+\frac{5}{36} (4 N+5) S_1^4
+\frac{4 N+5} {2} \zeta _2 S_1^2
\N\\
&&
+\frac{5}{6} (4 N+5) S_2 S_1^2
+\frac{10}{9} (4 N+5) S_3 S_1
+4 (4 N+5) S_{2,1} S_1
\N
\end{eqnarray}
\begin{eqnarray} && 
+\frac{29}{12} (4 N+5) S_2^2
+\frac{4 N+5} {2} \zeta _2 S_2
+\frac{17}{6} (4 N+5) S_4
\N\\
&&
+4 (4 N+5) S_{3,1}
-4 (4 N+5) S_{2,1,1}
-\frac{ (260 N^3+762 N^2+451 N-12) }{81 N (N+1)} S_1^3
\N\\
&&
-\frac{ (32 N^3+96 N^2+49 N-12) }{9 N (N+1)} \zeta_2 S_1
-\frac{14}{9} (4 N+5) \zeta_3 S_1
\N\\
&&
-\frac{ (404 N^3+834 N^2+307 N-156) }{27 N (N+1)} S_2 S_1
-\frac{14 (2 N^3+5 N^2+6 N+4) }{9 N (N+1)} \zeta_3
\N\\
&&
-\frac{2 (908 N^3+1086 N^2-197 N-660) }{81 N (N+1)} S_3
-\frac{4 (20 N^2+90 N+61) }{9 (N+1)} S_{2,1}
\N\\
&&
+\frac{ P_{53}(N)}{27 N^2 (N+1)^2 (N+2)} S_1^2
-\frac{ P_{50}(N) }{9 N^2 (N+1)^2 (N+2)} \zeta_2
\N\\
&&
+\frac{ P_{54}(N)}{27 N^2 (N+1)^2 (N+2)} S_2
-\frac{2 P_{55}(N)}{243 N^3 (N+1)^3 (N+2)^2} S_1
\N\\
&&
-\frac{P_{56}(N)}{486 N^4 (N+1)^4 (N+2)^3}
\Biggr\}~,
\end{eqnarray}
with
\begin{eqnarray} 
P_{50}(N)&=&43 N^6+301 N^5+587 N^4+484 N^3+132 N^2-80 N-48~,
\\
P_{51}(N)&=&190 N^6+870 N^5+802 N^4-543 N^3-938 N^2-192 N+144~,
\\
P_{52}(N)&=&991 N^9+9415 N^8+32215 N^7+52113 N^6+43252 N^5
\N\\
&&
+17484 N^4+2928
N^3+3104 N^2+3648 N+1152~,
\\
P_{53}(N)&=&276 N^6+1472 N^5+1976 N^4+425 N^3-674 N^2-352 N
\N\\
&&
+48~,
\\
P_{54}(N)&=&948 N^6+4328 N^5+6176 N^4+2945 N^3-2 N^2+608 N
\N\\
&&
+624~,
\\
P_{55}(N)&=&1379 N^9+9765 N^8+20193 N^7+10203 N^6-15291 N^5
\N\\
&&
-30216 N^4-27620 N^3-14160 N^2-5472 N-1728~,
\\
P_{56}(N)&=&21055 N^{12}+262474 N^{11}+1324746 N^{10}+3564860 N^9
\N\\
&&
+5666307
N^8+5550126 N^7+3387956 N^6+1293128 N^5
\N\\
&&
+273984 N^4-120320 N^3-219648 N^2-129024 N-27648~.
\end{eqnarray}
\begin{eqnarray} 
\Ahathat_{Qg}^{(3),L_{1b}}&=&
{T_F^2 n_f C_A} \frac {1} {N} \Biggl\{
\frac {1} {\ep^3}
\Biggl[
\frac{64}{9} S_1 
+\frac{16 (2 N^3+7 N^2+6 N+3)}{9 N (N+1)^2}
\Biggr]
\N\\
&&
+\frac {1} {\ep^2} \Biggl[
\frac{64 S_2}{9}
+\frac{32 }{9} S_1^2
-\frac{32 (5 N+14) }{27 (N+1)} S_1
-\frac{8 P_{57}(N)}{27 N^2
  (N+1)^3 (N+2)}
\Biggr]
\N
\end{eqnarray}
\begin{eqnarray}
&&
+\frac {1} {\ep} \Biggl[
\frac{32 }{27} S_1^3
+\frac{8 }{3} \zeta _2 S_1
+\frac{64 }{9} S_2 S_1
+\frac{208 }{27} S_3
+\frac{64 }{9} S_{2,1}
-\frac{16 (5 N+14) }{27 (N+1)} S_1^2
\N\\
&&
+\frac{2 (2 N^3+7 N^2+6 N+3) }{3 N (N+1)^2} \zeta_2
-\frac{8 (74 N^3+121 N^2+38 N-27) }{27 N (N+1)^2} S_2
\N\\
&&
+\frac{8 (47 N^3+13 N^2-196 N-108)}{81 N (N+1)^2} S_1
+\frac{2 P_{58}(N)}{81 N^3 (N+1)^4 (N+2)^2}
\Biggr]
\N\\
&&
+\frac{8 S_1^4}{27}
-\frac{56 }{9} \zeta_3 S_1
+\frac{4 }{3} \zeta _2 S_1^2
+\frac{32 }{9} S_2 S_1^2
+\frac{208 }{27}  S_3 S_1
+\frac{64 }{9} S_{2,1} S_1 
\N\\
&&
+\frac{80 }{9} S_2^2
+\frac{8 }{3} \zeta_2 S_2
+\frac{128 }{9} S_4
+\frac{32} {3} S_{3,1}
-\frac{64} {9} S_{2,1,1}
\N\\
&&
-\frac{16 (5 N+14) }{81 (N+1)} S_1^3
-\frac{4 (5 N+14) }{9 (N+1)} \zeta_2 S_1
\N\\
&&
-\frac{32 (5 N+14) }{27 (N+1)} S_2 S_1
-\frac{14 (2 N^3+7 N^2+6 N+3) }{9 N (N+1)^2} \zeta_3
\N\\
&&
-\frac{4 (616 N^3+899 N^2+202 N-243) }{81 N (N+1)^2} S_3
-\frac{32 (5 N+14)}{27 (N+1)}  S_{2,1}
\N\\
&&
+\frac{4 (47 N^3+13 N^2-196 N-108)}{81 N (N+1)^2} S_1^2
-\frac{ P_{57}(N) }{9 N^2 (N+1)^3 (N+2)} \zeta_2
\N\\
&&
+\frac{4 P_{59}(N)}{81 N^2 (N+1)^3 (N+2)} S_2
-\frac{2 P_{60}(N)}{243 N^2 (N+1)^3} S_1
\N\\
&&
-\frac{P_{61}(N)}{486 N^4 (N+1)^5 (N+2)^3}
\Biggr\}~,
\end{eqnarray}
with
\begin{eqnarray} 
P_{57}(N)&=&43 N^6+302 N^5+566 N^4+445 N^3+129 N^2-81 N-54~,
\\
P_{58}(N)&=&1207 N^9+11543 N^8+40819 N^7+70919 N^6+65266 N^5
\N\\
&&
+27608
N^4+312 N^3+162 N^2+3672 N+1512~,
\\
P_{59}(N)&=&742 N^6+3260 N^5+5157 N^4+3025 N^3+251 N^2+513 N
\N\\
&&
+486~,
\\
P_{60}(N)&=&323 N^5-3972 N^4-9291 N^3-4456 N^2-1080 N+648~,
\\
P_{61}(N)&=&25807 N^{12}+324932 N^{11}+1659342 N^{10}+4520784 N^9
\N\\
&&
+7180599
  N^8+6692496 N^7+3387488 N^6+763528 N^5
\N\\
&&
+119892 N^4+17388 N^3-147960 N^2-143856 N-38880~.
\end{eqnarray}
\begin{eqnarray} 
\Ahathat_{Qg}^{(3),M_{1}}&=&
 \frac {T_F^2 n_f C_A} {N^2 (N+1)^2} \Biggl\{
\frac {1} {\ep^3}
\Biggl[
\frac{16 (4 N^2-8 N-9)}{9}
\Biggr]
\N\\
&&
+\frac {1} {\ep^2} \Biggl[
\frac{16}{9} (N^2-2 N-2) S_1
+\frac{8 P_{62}(N)}{27 N (N+1) (N+2)}
\Biggr]
\N
\end{eqnarray}
\begin{eqnarray} 
&&
+\frac {1} {\ep} \Biggl[
\frac{4}{9} (N^2-2 N-2) S_1^2
+\frac{4}{9} (5 N-14) (5 N+4) S_2
\N\\
&&
+\frac{2}{3} (4 N^2-8 N-9) \zeta _2
-\frac{8 P_{63}(N) }{27 N (N+1) (N+2)} S_1
\N\\
&&
-\frac{4 P_{64}(N)}{81 N^2 (N+1)^2 (N+2)}
\Biggr]
+\frac{2}{27} (N^2-2 N-2) S_1^3
\N\\
&&
+\frac{4}{27} (109 N^2-218 N-245) S_3
-\frac{14}{9} (4 N^2-8 N-9) \zeta _3
\N\\
&&
+\frac{26}{9} (N^2-2 N-2) S_2 S_1
+\frac{2}{3} (N^2-2 N-2) \zeta _2 S_1
\N\\
&&
-\frac{2 P_{63}(N) }{27 N (N+1) (N+2)} S_1^2
+\frac{2 P_{65}(N) }{27 N(N+1) (N+2)} S_2
\N\\
&&
+\frac{ P_{62}(N) }{9 N (N+1) (N+2)} \zeta_2
+\frac{4 P_{66}(N)}{81 N^2 (N+1)^2 (N+2)^2} S_1
\N\\
&&
+\frac{2 P_{67}(N)}{243 N^3 (N+1)^3 (N+2)^3}
\Biggr\}~,
\end{eqnarray}
with
\begin{eqnarray}
P_{62}(N)&=&12 N^6-17 N^5-155 N^4-43 N^3+83 N^2-207 N-138~,
\\
P_{63}(N)&=&17 N^5+59 N^4+21 N^3-38 N^2+16 N+24~,
\\
P_{64}(N)&=&24 N^8-91 N^7-713 N^6-558 N^5-424 N^4-1202 N^3+912 N^2
\N\\
&&
+2055 N+918~,
\\
P_{65}(N)&=&72 N^6-119 N^5-989 N^4-279 N^3+536 N^2-1258 N-852~,
\\
P_{66}(N)&=&88 N^8+553 N^7+977 N^6+241 N^5-436 N^4-26 N^3-296 N^2
\N\\
&&
-624 N-288~,
\\
P_{67}(N)&=&264 N^{12}+1381 N^{11}-4166 N^{10}-47256 N^9-152187
  N^8
\N\\
&&
-251073 N^7-189276 N^6+16418 N^5+68699 N^4-79947 N^3
\N\\
&&
-148962
  N^2-95364 N-23112~.
\end{eqnarray}
\begin{eqnarray} 
\Ahathat_{Qg}^{(3),N_{1a}}&=&
 \frac {T_F^2 n_f C_A} {N (N+1) (N+2)} \Biggl\{ 
\frac {1} {\ep^3} \Biggl[
\frac{32}{9} (2 N^2+3 N+2) S_1
\N\\
&&
-\frac{32 (N^3+2 N^2+1)}{9 (N+1)}
\Biggr]
+\frac {1} {\ep^2} \Biggl[
\frac{8}{9} (6 N^2+11N+8) S_1^2
\N\\\N
&&
-\frac{64}{9} (N^2-4) S_{-2}
-\frac{8}{9} (14 N^2+33 N+12) S_2
\\\N
&&
-\frac{16 P_{68}(N)}{27N (N+1) (N+2)} S_1
+\frac{16 P_{69}(N)}{27 N (N+1)^2
  (N+2)}
\Biggr]\N
\end{eqnarray}
\begin{eqnarray}&& 
+\frac {1} {\ep} \Biggl[
\frac{4}{27} (14 N^2+27N+20) S_1^3
+\frac{4}{3} (2 N^2+3 N+2) \zeta _2 S_1
\N\\\N
&&
-\frac{32}{3} (N^2-4) S_{-2} S_1
-\frac{4}{9} (6 N^2+N-44) S_2 S_1
\\\N
&&
-\frac{80}{9} (N^2-4) S_{-3}
+\frac{32}{3} (N^2-4) S_{-2,1}
+\frac{16}{9} N (11 N+16) S_{2,1}
\\\N
&&
-\frac{8}{27} (94 N^2+171 N-2) S_3
-\frac{4 P_{70}(N)}{27 N (N+1) (N+2)} S_1^2
-\frac{4 (N^3+2 N^2+1)}{3 (N+1)} \zeta _2
\N\\\N
&&
+\frac{64 (7 N^4-17 N^3-43 N^2+5 N+6) }{27 N (N+1)} S_{-2}
+\frac{4 P_{73}(N) }{27 N (N+1) (N+2)} S_2
\N\\
&&
+\frac{8 P_{71}(N)}{81 N^2 (N+1)^2 (N+2)^2} S_1
-\frac{8 P_{72}(N)}{81 N^2 (N+1)^3 (N+2)^2}
\Biggr]
\N\\
&&
+\frac{1}{54} (30 N^2+59 N+44) S_1^4
+\frac{1}{3} (6 N^2+11 N+8) \zeta _2 S_1^2
\N\\
&&
-8 (N^2-4) S_{-2} S_1^2
+\frac{1}{9} (2 N^2+63 N+188) S_2 S_1^2
\N\\
&&
-\frac{28}{9} (2 N^2+3 N+2) \zeta _3 S_1
-\frac{40}{3} (N^2-4) S_{-3} S_1
\N\\
&&
-\frac{4}{27} (114 N^2+67 N-368) S_3 S_1
+16 (N^2-4) S_{-2,1} S_1
\N\\\N
&&
+\frac{8}{9} (25 N^2+32 N-12) S_{2,1} S_1
+\frac{1}{18} (6 N^2-205 N-292) S_2^2
\\\N
&&
+\frac{8}{9} (N^2-4) S_{-4}
-\frac{8}{3} (N^2-4) \zeta _2 S_{-2}
\N\\\N
&&
-\frac{1}{3} (14 N^2+33 N+12) \zeta _2 S_2
-\frac{8}{3} (N^2-4) S_{-2} S_2
\\\N
&&
-\frac{7}{9} (54 N^2+99 N-4) S_4
+\frac{40}{3} (N^2-4) S_{-3,1}
\\\N
&&
-\frac{32}{3} (N^2-4) S_{2,-2}
+\frac{16}{3} (7 N^2+8 N-6) S_{3,1}
\\\N
&&
-16 (N^2-4) S_{-2,1,1}
-\frac{8}{9} (25 N^2+32 N-12) S_{2,1,1}
\\\N
&&
-\frac{2 P_{74}(N)}{81 N (N+1) (N+2)} S_1^3
-\frac{2 P_{68}(N)}{9 N (N+1) (N+2)} \zeta _2 S_1
\\\N
&&
+\frac{32 (7 N^4-17 N^3-43 N^2+5 N+6)}{9 N (N+1)} S_{-2} S_1
+\frac{2 P_{75}(N)}{27 N (N+1)(N+2)} S_2 S_1
\\
\N&&
+\frac{28 (N^3+2 N^2+1)}{9 (N+1)} \zeta_3
+\frac{80 (7 N^4-17 N^3-43 N^2+5 N+6) }{27 N (N+1)} S_{-3}
\\\N
&&
+\frac{4 P_{76}(N)}{81 N (N+1) (N+2)} S_3
-\frac{32 (7 N^4-17 N^3-43 N^2+5 N+6)}{9 N (N+1)} S_{-2,1}
\N\\
&&
-\frac{16 P_{77}(N)}{27 N (N+1) (N+2)} S_{2,1}
+\frac{2 P_{78}(N)}{81 N^2 (N+1)^2 (N+2)^2} S_1^2 
\N\\
&&
+\frac{2 P_{69}(N)}{9 N (N+1)^2 (N+2)} \zeta_2
-\frac{32 P_{79}(N)}{81 N^2 (N+1)^2 (N+2)} S_{-2}
\N
\end{eqnarray}
\begin{eqnarray} &&
-\frac{2 P_{80}(N)}{81 N^2 (N+1)^2 (N+2)^2} S_2
-\frac{4 P_{81}(N)}{243 N^3 (N+1)^3 (N+2)^3} S_1
\N\\
&&
+\frac{4 P_{82}(N)}{243 N^3 (N+1)^4 (N+2)^3}
\Biggr\}~,
\end{eqnarray}
with
\begin{eqnarray}
P_{68}(N)&=&16 N^5+72 N^4+75 N^3-138 N^2-232 N-48~,
\\
P_{69}(N)&=&14 N^6+67 N^5+85 N^4+37 N^3-18 N^2-47 N-6~,
\\
P_{70}(N)&=&42 N^5+217 N^4+316 N^3-190 N^2-532 N-144~,
\\
P_{71}(N)&=&140 N^8+1131 N^7+2545 N^6-258 N^5-6077 N^4-4596 N^3
\N\\
&&
+476
N^2+816 N+288~,
\\
P_{72}(N)&=&154 N^9+1208 N^8+3481 N^7+4866 N^6+3091 N^5+612 N^4
\N\\
&&
+975 N^3+1549 N^2+336 N+36~,
\\
P_{73}(N)&=&82 N^5+417 N^4+686 N^3+210 N^2-348 N-144~,
\\
P_{74}(N)&=&94 N^5+507 N^4+798 N^3-294 N^2-1132 N-336~,
\\
P_{75}(N)&=&90 N^5-283 N^4-1618 N^3-410 N^2+1900 N+432~,
\\
P_{76}(N)&=&716 N^5+2238 N^4+939 N^3-2532 N^2-2252 N-528~,
\\
P_{77}(N)&=&47 N^5+136 N^4+10 N^3-295 N^2-252 N-60~,
\\
P_{78}(N)&=&336 N^8+2909 N^7+7295 N^6+2096 N^5-11749 N^4-9748 N^3
\N\\
&&
+1580
N^2+2448 N+864~,
\\
P_{79}(N)&=&50 N^7-N^6-70 N^5+530 N^4+716 N^3-97 N^2-84 N-36~,
\\
P_{80}(N)&=& 344 N^8+2775 N^7+9197 N^6+13188 N^5+6341 N^4+1560 N^3
\N\\
&&
+4404
  N^2+2880 N+864~,
\\
P_{81}(N)&=& 1252 N^{11}+14118 N^{10}+58795 N^9+122580 N^8+153807
  N^7
\N\\
&&
+131190 N^6+43487 N^5-64116 N^4-66172 N^3-14880 N^2
\N\\
&&
-7488
  N-1728~,
\\
P_{82}(N)&=&1550 N^{12}+16786 N^{11}+74226 N^{10}+174527 N^9+224721
  N^8
\N\\
&&
+133572 N^7-9824 N^6-69685 N^5-68916N^4-46499 N^3
\N\\
&&
-12426 N^2-2340
  N-216~.
\end{eqnarray}
\begin{eqnarray} 
\Ahathat_{Qg}^{(3),N_{1b}}&=&
 \frac {T_F^2 n_f C_A} {N (N+1) (N+2)} \Biggl\{
\frac {1} {\ep^3} \Biggl[
\frac{16}{9} (4 N^2+5 N+2) S_1
\N\\
&&
-\frac{32 N (N^2+2 N-2)}{9 (N+1)}
\Biggr]
+\frac {1} {\ep^2} \Biggl[
\frac{4}{9} (8 N^2+5 N-2) S_1^2
-\frac{128}{9} (N-1) (N+2) S_{-2}
\N\\
&&
-\frac{4}{9} (32 N^2+71 N+18) S_2
-\frac{8 P_{83}(N)}{27 N (N+1) (N+2)} S_1
+\frac{16 P_{84}(N)}{27 (N+1)^2 (N+2)}
\Biggr]
\N
\end{eqnarray}
\begin{eqnarray}
&&
+\frac {1} {\ep} \Biggl[
\frac{2}{27} (16 N^2+5 N-10) S_1^3
+\frac{2}{3} (4 N^2+5 N+2) \zeta _2  S_1
\N\\
&&
-\frac{256}{9} (N-1) (N+2) S_{-2}  S_1
-\frac{2}{9} (48 N^2+59 N-118) S_2  S_1
\N\\
&&
-\frac{160}{9} (N-1) (N+2) S_{-3}
+\frac{64}{3} (N-1) (N+2) S_{-2,1}
\N\\
&&
+\frac{8}{9} (28 N^2+31 N-26) S_{2,1}
-\frac{4}{27} (284 N^2+457 N-158) S_3
\N\\
&&
-\frac{2 P_{85}(N) }{27 N (N+1) (N+2)} S_1^2
-\frac{4 N (N^2+2 N-2) }{3 (N+1)} \zeta_2
\N\\
&&
+\frac{64 (11 N^3-14 N^2-59 N-10)}{27 (N+1)} S_{-2}
+\frac{2 P_{86}(N)}{27 N (N+1) (N+2)} S_2
\N\\
&&
+\frac{4 P_{87}(N)}{81 N (N+1)^2 (N+2)^2}  S_1
-\frac{8 P_{88}(N)}{81 (N+1)^3 (N+2)^2}
\Biggr]
\N\\
&&
+\frac{1}{108} (32 N^2+5 N-26) S_1^4
+\frac{1}{6} (8 N^2+5 N-2) \zeta _2  S_1^2
\N\\
&&
-\frac{256}{9} (N-1) (N+2) S_{-2}  S_1^2
+\frac{1}{18} (-224 N^2-251 N+486) S_2  S_1^2
\N\\
&&
-\frac{14}{9} (4 N^2+5 N+2) \zeta _3  S_1
-\frac{320}{9} (N-1) (N+2) S_{-3}  S_1
\N\\
&&
-\frac{2}{27} (736 N^2+763 N-1510) S_3  S_1
+\frac{128}{3} (N-1) (N+2) S_{-2,1}  S_1
\N\\
&&
+\frac{4}{9} (88 N^2+79 N-134) S_{2,1}  S_1
+\frac{1}{36} (128 N^2-403 N-1034) S_2^2
\N\\
&&
+\frac{16}{9} (N-1) (N+2) S_{-4}
-\frac{16}{3} (N-1) (N+2)  \zeta _2 S_{-2}
\N\\
&&
+\frac{1}{6} (-32 N^2-71 N-18) \zeta _2  S_2
-\frac{64}{9} (N-1) (N+2) S_{-2}  S_2
\N\\
&&
-\frac{64}{3} (N-1) (N+2) S_{2,-2}
+\frac{1}{18} (-1120 N^2-1903 N+526) S_4
\N\\
&&
+\frac{80}{3} (N-1) (N+2) S_{-3,1}
+\frac{4}{3} (48 N^2+45 N-82) S_{3,1}
\N\\
&&
-32 (N-1) (N+2) S_{-2,1,1}
-\frac{4}{9} (68 N^2+41 N-142) S_{2,1,1}
\N\\
&&
-\frac{ P_{89}(N)} {81 N (N+1) (N+2)} S_1^3
-\frac{ P_{83}(N) }{9 N (N+1) (N+2)}  \zeta _2 S_1
\N\\
&&
+\frac{128 (11 N^3-14 N^2-59 N-10) }{27 (N+1)}  S_{-2} S_1
+\frac{ P_{90}(N) }{27 N (N+1) (N+2)}  S_2 S_1
\N\\
&&
+\frac{28 N (N^2+2 N-2) }{9 (N+1)} \zeta _3
+\frac{2 P_{91}(N)}{81 N (N+1) (N+2)} S_3
\N\\
&&
+\frac{80 (11 N^3-14 N^2-59 N-10) }{27 (N+1)} S_{-3}
-\frac{4 P_{92}(N)}{27 N (N+1) (N+2)} S_{2,1}
\N\\
&&
-\frac{32 (11 N^3-14 N^2-59 N-10) }{9 (N+1)} S_{-2,1}
+\frac{ P_{93}(N)}{81 N (N+1)^2 (N+2)^2}  S_1^2
\N
\end{eqnarray}
\begin{eqnarray} &&
+\frac{2 P_{84}(N) }{9 (N+1)^2 (N+2)} \zeta _2
-\frac{128 P_{94}(N)}{81 (N+1)^2 (N+2)} S_{-2}
\N\\
&&
-\frac{P_{95}(N)}{81 N (N+1)^2 (N+2)^2} S_2
-\frac{2 P_{96}(N)}{243 N (N+1)^3 (N+2)^3}  S_1
\N\\
&&
+\frac{4 P_{97}(N)}{243 (N+1)^4 (N+2)^3}
\Biggl\}~,
\end{eqnarray}
with
\begin{eqnarray}
P_{83}(N)&=&20 N^5+76 N^4+35 N^3-256 N^2-316 N-96~,
\\
P_{84}(N)&=&8 N^5+43 N^4+45 N^3+29 N^2+22 N-24~,
\\
P_{85}(N)&=&40 N^5+94 N^4-155 N^3-856 N^2-836 N-192~,
\\
P_{86}(N)&=&184 N^5+622 N^4+251 N^3-1240 N^2-1596 N-384~,
\\
P_{87}(N)&=&184 N^7+1388 N^6+2586 N^5-2671 N^4-12540 N^3
\N\\
&&
-13444 N^2-6160 N-960~,
\\
P_{88}(N)&=&88 N^7+530 N^6+227 N^5-3946 N^4-11473 N^3-14065 N^2
\N\\
&&
-7202 N-528~,
\\
P_{89}(N)&=&80 N^5+130 N^4-535 N^3-2056 N^2-1876 N-384~,
\\
P_{90}(N)&=&624 N^5+454 N^4-4961 N^3-7000 N^2-844 N+384~,
\\
P_{91}(N)&=&2248 N^5+4052 N^4-9473 N^3-22736 N^2-11468 N-2208~,
\\
P_{92}(N)&=&236 N^5+374 N^4-1269 N^3-3032 N^2-1628 N-288~,
\\
P_{93}(N)&=&368 N^7+2498 N^6+3042 N^5-12661 N^4-37056 N^3
\N\\
&&
-35356 N^2-14144 N-1920~,
\\
P_{94}(N)&=&19 N^5+23 N^4+N^3+131 N^2+178 N-28~,
\\
P_{95}(N)&=&1376 N^7+7298 N^6+11630 N^5-8683 N^4-47392 N^3
\N\\
&&
-46660 N^2-15168 N-3840~,
\\
P_{96}(N)&=&1448 N^9+14464 N^8+45664 N^7+27660 N^6-157311 N^5
\N\\
&&
-465972 N^4-638594 N^3-483184 N^2-176200 N-21120~,
\\
P_{97}(N)&=&824 N^9+7474 N^8+20726 N^7-2277 N^6-140187 N^5
\N\\
&&
-335448 N^4-348968 N^3-155743 N^2-25874 N-11616~.
\end{eqnarray}
\begin{eqnarray} 
\Ahathat_{Qg}^{(3),O_1}&=&
\frac {T_F^2 n_f C_A} {(N+1) (N+2)} \Biggl\{
\frac {1} {\ep^3}
\Biggl[
-\frac{16 (9 N+10) }{9 N} S_1
-\frac{128 (2 N+3)}{9 (N+1) (N+2)}
\Biggr]
\N\\
&&
+\frac {1} {\ep^2}
\Biggl[
-\frac{4 (23 N+26)}{9 N} S_1^2
-\frac{4 (31 N+34) }{9 N} S_2
\N\\
&&
+\frac{8 (96 N^3+383 N^2+552 N+268) }{27 N (N+1) (N+2)} S_1
\N\\
&&
+\frac{16 (99 N^4+562 N^3+1035 N^2+656 N+48)}{27 (N+1)^2 (N+2)^2}
\Biggr]
+\frac {1} {\ep}
\Biggl[
-\frac{2 (51 N+58) }{27 N} S_1^3
\N
\end{eqnarray}
\begin{eqnarray}
&&
-\frac{2 (9 N+10) }{3 N} \zeta_2 S_1
-\frac{2 (67 N+74) }{9 N} S_2 S_1
-\frac{4 (87 N+94) }{27 N} S_3
\N\\
&&
-\frac{8 (23 N+26) }{9 N} S_{2,1}
-\frac{16 (2 N+3) }{3 (N+1) (N+2)} \zeta_2
\N\\
&&
+\frac{2 (644 N^3+2135 N^2+2692 N+1516) }{27 N (N+1) (N+2)} S_2
\N\\
&&
+\frac{2 (244 N^3+1003 N^2+1436 N+668) }{27 N (N+1) (N+2)} S_1^2
+\frac{4 P_{102}(N)}{81 N (N+1)^2 (N+2)^2} S_1
\N\\
&&
-\frac{4 P_{103}(N)}{81 (N+1)^3(N+2)^3}
\Biggr]
-\frac{(107 N+122)}{108 N} S_1^4
\N\\
&&
+\frac{(540 N^3+2243 N^2+3204 N+1468)}{81 N (N+1) (N+2)} S_1^3
-\frac{4 (23 N+26) }{3 N} S_{3,1}
\N\\
&&
+\frac{4 (61 N+70) }{9 N} S_{2,1,1}
-\frac{(139 N+154)}{18 N} S_2 S_1^2
-\frac{(23 N+26) }{6 N} \zeta _2 S_1^2
\N\\
&&
-\frac{(755 N+842) }{36 N} S_2^2
-\frac{(31 N+34) }{6 N} \zeta _2 S_2
\N\\
&&
-\frac{2 (179 N+194)}{27 N} S_3 S_1
-\frac{4 (61 N+70) }{9 N} S_{2,1} S_1
+\frac{14 (9 N+10) }{9 N} \zeta_3 S_1
\N\\
&&
-\frac{(511 N+562) }{18 N} S_4
+\frac{2 (2340 N^3+7337 N^2+8856 N+5284) }{81 N (N+1) (N+2)} S_3
\N\\
&&
+\frac{ 836 N^3+3283 N^2+4636 N+2300 }{27 N (N+1) (N+2)} S_2 S_1
+\frac{112 (2 N+3) }{9 (N+1) (N+2)} \zeta _3
\N\\
&&
+\frac{4 (214 N^3+889 N^2+1298 N+596) }{27 N (N+1) (N+2)} S_{2,1}
\N\\
&&
+\frac{ (96 N^3+383 N^2+552 N+268)}{9 N (N+1) (N+2)} \zeta _2 S_1
-\frac{ P_{98}(N)}{81 N (N+1)^2 (N+2)^2} S_1^2
\N\\
&&
-\frac{ P_{101}(N)}{81 N (N+1)^2 (N+2)^2} S_2
\N\\
&&
+\frac{2 (99 N^4+562 N^3+1035 N^2+656 N+48) }{9 (N+1)^2
  (N+2)^2} \zeta _2
\N\\
&&
-\frac{2 P_{99}(N)}{243 N (N+1)^3 (N+2)^3} S_1
+\frac{ P_{100}(N)}{243 (N+1)^4 (N+2)^4}
\Biggr\}~,
\end{eqnarray}
with
\begin{eqnarray}
P_{98}(N)&=&560 N^5+4964 N^4+24041 N^3+55706 N^2+58136 N+22256~,
\\
P_{99}(N)&=&5646 N^7+27934 N^6-28488 N^5-426969 N^4-1064262 N^3
\N\\
&&
-1202880
N^2
-648456 N
-133672~,
\\
P_{100}(N)&=&66465 N^8+706420 N^7+3202098 N^6+8134152 N^5
\N\\
&&
+12726405
N^4+12559980 N^3+7477056 N^2+2214080 N
\N\\
&&
+120576~,
\\
P_{101}(N)&=&4996 N^5+40516 N^4+140065 N^3+238954 N^2+191872
  N
\N\\
&&
+54736~,
\end{eqnarray}
\begin{eqnarray}
P_{102}(N)&=&42 N^5-388 N^4-6171 N^3-18898 N^2-21564 N-8416~,
\\
P_{103}(N)&=&2745 N^6+20977 N^5+59601 N^4+76519 N^3+40266 N^2
\N\\
&&
+3700 N-480~.
\end{eqnarray}
\begin{eqnarray} 
\Ahathat_{Qg}^{(3),P_{1a}}&=&
 \frac {T_F^2 n_f C_A} {(N+1) (N+2)} \Biggl\{
\frac {1} {\ep^3} \Biggl[
-\frac{64 (N+4)}{9 (N+2)}
-\frac{64 (N-4) }{9 N} S_1
\Biggr]
\N\\
&&
+\frac {1} {\ep^2} \Biggl[
-\frac{16 (3 N-8) }{9 N} S_1^2
+\frac{16 (5 N+16)}{9 N} S_2
\N\\
&&
+\frac{32 \left(8 N^3-98 N^2-251 N-112\right) }{27 N (N+1) (N+2)} S_1
+\frac{32 \left(23 N^3+149 N^2+277 N+160\right)}{27 (N+1)(N+2)^2}
\Biggr]
\N\\
&&
+\frac {1} {\ep} \Biggl[
-\frac{8 (7 N-16) }{27 N}S_1^3
-\frac{8 (N-4) }{3 N} \zeta _2 S_1
-\frac{8 (3 N-32) }{9 N} S_2 S_1
\N\\
&& 
+\frac{16 (29 N+52) }{27 N}S_3
-\frac{128 (N-2)}{9 N}S_{2,1}
\N\\
&&
+\frac{8 \left(30 N^3-184 N^2-525 N-224\right) }{27 N(N+1) (N+2)}
S_1^2
-\frac{8 (N+4) }{3 (N+2)}\zeta _2
\N\\
&&
-\frac{8 \left(94 N^3+500 N^2+797 N+448\right) }{27 N (N+1) (N+2)} S_2
+\frac{16 P_{104}(N)}{81 N (N+1)^2 (N+2)^2}  S_1
\N\\
&&
-\frac{16 P_{105}(N)}{81(N+1)^2 (N+2)^3}
\Biggr]
-\frac{ (15 N-32) }{27 N} S_1^4
-\frac{2 (3 N-8) }{3 N} \zeta _2 S_1^2
\N\\
&&
-\frac{2 (19 N-64) }{9 N} S_2 S_1^2
+\frac{56 (N-4) }{9 N} \zeta_3 S_1
+\frac{8 (3 N+104)}{27 N} S_3 S_1
\N\\
&&
-\frac{128 (N-2) }{9 N} S_{2,1} S_1
+\frac{ (9 N+320)}{9 N} S_2^2
+\frac{2 (5 N+16) }{3 N} \zeta _2 S_2
\N\\
&&
-\frac{64 (N-2) }{3 N} S_{3,1}
+\frac{128 (N-2) }{9 N} S_{2,1,1}
+\frac{2 (113 N+256) }{9 N} S_4
\N\\
&&
+\frac{4 \left(74 N^3-356 N^2-1073 N-448\right) }{81 N (N+1) (N+2)} S_1^3
\N\\
&&
+\frac{4 \left(8 N^3-98 N^2-251 N-112\right) }{9 N (N+1) (N+2)} \zeta _2 S_1
\N\\
&&
-\frac{4 \left(6 N^3+844 N^2+1893 N+896\right) }{27 N (N+1) (N+2)} S_2 S_1
+\frac{56 (N+4) }{9 (N+2)}\zeta _3
\N\\
&&
-\frac{8 \left(484 N^3+1778 N^2+2297 N+1456\right) }{81 N (N+1) (N+2)} S_3
\N\\
&&
+\frac{64 \left(11 N^3-43 N^2-137 N-56\right) }{27 N
  (N+1)(N+2)} S_{2,1}
\N\\
&&
-\frac{4 P_{106}(N)}{81 N (N+1)^2 (N+2)^2}S_1^2
+\frac{4 \left(23 N^3+149N^2+277 N+160\right) }{9 (N+1) (N+2)^2}\zeta_2
\N\\
&&
+\frac{4 P_{107}(N)}{81 N (N+1)^2 (N+2)^2} S_2
-\frac{8 P_{108}(N)}{243 N (N+1)^3 (N+2)^3} S_1
\N
\end{eqnarray}
\begin{eqnarray}
&&
+\frac{8 P_{109}(N)}{243 (N+1)^3 (N+2)^4}
\Biggr\}~,
\end{eqnarray}
with
\begin{eqnarray}
P_{104}(N)&=&38 N^5+1660 N^4+6647 N^3+10051 N^2+7013
    N+2464~,
\\
P_{105}(N)&=&307 N^5+2794 N^4+9631 N^3+16150 N^2+13219 N +4240~,
\\
P_{106}(N)&=&24 N^5-3140 N^4-13401 N^3-20075 N^2-13587
    N-4928~,
\\
P_{107}(N)&=&1052 N^5+8260 N^4+25625 N^3+40447
    N^2+32003N
\N\\
&&
+9856~,
\\
P_{108}(N)&=&1210 N^7+22502 N^6+119904 N^5+319074
    N^4+502572N^3
    \N\\&&+486540 N^2+259511 N+49600~,
\\
P_{109}(N)&=&3395 N^7+40451 N^6+201441 N^5+547869 N^4+879225
    N^3
    \N\\&&+832767 N^2+431665 N+94720~.
\end{eqnarray}
\begin{eqnarray} 
\Ahathat_{Qg}^{(3),P_{1b}}&=& 
\frac {T_F^2 n_f C_A} {N (N+1) (N+2)} \Biggl\{
-\frac{1} {\ep^3} \Biggl[
\frac{32}{9} (N-10) S_1
+\frac{64 (2 N^2+7 N+2)}{9 (N+2)}
\Biggr]
\N\\
&&
+\frac{1} {\ep^2} \Biggl[
\frac{8}{9} (11 N+30) S_2
-\frac{8}{9} (N-30) S_1^2
+\frac{16(2 N^3-257 N^2-542 N-172) }{27 (N+1) (N+2)} S_1
\N\\
&&
+\frac{32 P_{110}(N)}{27 N (N+1) (N+2)^2}
\Biggr]
+\frac{1} {\ep} \Biggl[
-\frac{4}{27} (N-70) S_1^3
-\frac{4}{3} (N-10) \zeta _2  S_1
\N\\
&&
-\frac{4}{9} (N-70) S_2  S_1
+\frac{8}{27} (53 N+70) S_3
-\frac{16}{3} (N-10) S_{2,1}
\N\\
&&
+\frac{4 (8 N^3-663 N^2-1454 N-516) }{27 (N+1) (N+2)} S_1^2
-\frac{8 (2 N^2+7 N+2) \zeta _2}{3 (N+2)}
\N\\
&&
-\frac{4 (124 N^3+663 N^2+1070 N+660) S_2}{27 (N+1) (N+2)}
+\frac{8 P_{111}(N)}{81 (N+1)^2 (N+2)^2}  S_1
\N\\
&&
-\frac{16 P_{112}(N)}{81 N^2 (N+1)^2 (N+2)^3}
\Biggr]
+\frac{1}{54} (150-N) S_1^4
+\frac{1}{9} (227 N+510) S_4
\N\\
&&
-\frac{4}{27} (N-150) S_3 S_1
-\frac{8}{3} (N-30) S_{2,1} S_1
+\frac{1}{3} (30-N) \zeta _2 S_1^2
\N\\
&&
+\frac{1}{9} (150-N) S_2 S_1^2
+\frac{1}{18} (119 N+870) S_2^2
+\frac{1}{3} (11 N+30) \zeta _2 S_2
\N\\
&&
-\frac{8}{3} (N-30) S_{3,1}
-\frac{8}{3} (N+30) S_{2,1,1}
+\frac{2 (20 N^3-1475 N^2-3278 N-1204) }{81 (N+1) (N+2)} S_1^3
\N\\
&&
+\frac{2 (2 N^3-257 N^2-542 N-172) }{9 (N+1) (N+2)} \zeta _2 S_1
+\frac{28}{9} (N-10) \zeta _3 S_1
\N
\end{eqnarray}
\begin{eqnarray}
&&
+\frac{2 (20 N^3-1475 N^2-3278 N-1204) }{27 (N+1) (N+2)} S_2 S_1
+\frac{56 (2 N^2+7 N+2) }{9 (N+2)}  \zeta _3
\N\\
&&
-\frac{4 (574 N^3+1475 N^2+1550 N+1852) }{81 (N+1) (N+2)} S_3
+\frac{8 (2 N^3-257 N^2-542 N-172) }{9 (N+1) (N+2)} S_{2,1}
\N\\
&&
+\frac{2 P_{113}(N)}{81 (N+1)^2 (N+2)^2} S_1^2
+\frac{4 P_{110}(N) }{9 N (N+1) (N+2)^2}  \zeta _2
+\frac{2 P_{114}(N)}{81 N (N+1)^2 (N+2)^2} S_2
\N\\
&&
-\frac{4 P_{115}(N)}{243 (N+1)^3 (N+2)^3} S_1
+\frac{8 P_{116}(N)}{243 N^3 (N+1)^3 (N+2)^4}
\Biggr\}~,
\end{eqnarray}
with
\begin{eqnarray} 
P_{110}(N)&=&22 N^5+134 N^4+204 N^3+25 N^2-88 N-12~,
\\
P_{111}(N)&=&140 N^5+3958 N^4+15101 N^3+22156 N^2+15566 N+6088~,
\\
P_{112}(N)&=&458 N^8+4306 N^7+15574 N^6+28027 N^5+26233 N^4
\N\\
&&
+12193 N^3+2798 N^2+636 N+72~,
\\
P_{113}(N)&=&290 N^5+9918 N^4+39839 N^3+62088 N^2+46622 N
\N\\
&&
+18264~,
\\
P_{114}(N)&=&1094 N^6+9414 N^5+31391 N^4+52296 N^3+41630 N^2
\N\\
&&
+11064 N-864~,
\\
P_{115}(N)&=&4276 N^7+68264 N^6+351618 N^5+915849 N^4+1394436 N^3
\N\\
&&
+1270848 N^2+610766 N+89584~,
\\
P_{116}(N)&=&4426 N^{11}+52400 N^{10}+256348 N^9+673350 N^8+1012896 N^7
\N\\
&&
+836430 N^6+281102 N^5-71327 N^4-82696 N^3-22728 N^2
\N\\
&&
-4464 N-432~.
\end{eqnarray}
\begin{eqnarray} 
\Ahathat_{Qg}^{(3),Q_{1}}&=& 0~.
\end{eqnarray}
\begin{eqnarray} 
\Ahathat_{Qg}^{(3),R_{1}}&=& 0~.
\end{eqnarray}
\begin{eqnarray} 
\Ahathat_{Qg}^{(3),R'_{1a}}&=& 0~.
\end{eqnarray}
\begin{eqnarray} 
\Ahathat_{Qg}^{(3),R'_{1b}}&=& 0~.
\end{eqnarray}
\begin{eqnarray} 
\Ahathat_{Qg}^{(3),S_{1a}}&=& 
\frac {T_F^2 n_f C_A} {N^2 (N+1)^2} \Biggl\{
-\frac {1} {\ep^3}
\Biggl[
\frac{16 (N+2)}{9}
\Biggr]
\N\\
&&
+\frac {1} {\ep^2} \Biggl[
-\frac{8}{9} (N+2) S_1
+\frac{8 (32 N^4+73 N^3+10 N^2-64 N-24)}{27 N (N+1) (N+2)}
\Biggr]
\N\\
&&
+\frac {1} {\ep} \Biggl[
-\frac{2}{9} (N+2) S_1^2
-\frac{26}{9} (N+2) S_2
-\frac{2}{3} (N+2) \zeta _2
\N\\
&&
+\frac{4 (32 N^4+73 N^3+10 N^2-64 N-24) }{27 N (N+1) (N+2)} S_1
\N\\
&&
-\frac{4 P_{117}(N)}{81 N^2 (N+1)^2 (N+2)^2}
\Biggr]
-\frac{1}{27} (N+2) S_1^3
-\frac{110}{27} (N+2) S_3
\N\\
&&
+\frac{14}{9} (N+2) \zeta_3
-\frac{13}{9} (N+2) S_2 S_1
-\frac{1}{3} (N+2) \zeta_2 S_1
\N\\
&&
+\frac{1 (32 N^4+73 N^3+10 N^2-64 N-24) }{27 N (N+1) (N+2)} S_1^2
\N\\
&&
+\frac{13 (32 N^4+73 N^3+10 N^2-64 N-24) }{27 N (N+1) (N+2)} S_2
\N\\
&&
+\frac{ (32 N^4+73 N^3+10 N^2-64 N-24) }{9 N (N+1) (N+2)} \zeta_2
\N\\
&&
-\frac{2 P_{117}(N)}{81 N^2 (N+1)^2 (N+2)^2} S_1
+\frac{2 P_{118}(N)}{243 N^3 (N+1)^3 (N+2)^3}
\Biggr\}~,
\end{eqnarray}
with
\begin{eqnarray} 
P_{117}(N)&=&352 N^7+1166 N^6+775 N^5-550 N^4+436 N^3+1976 N^2
\N\\
&&
+1200 N+288~,
\\
P_{118}(N)&=&2432 N^{10}+12004 N^9+24210 N^8+36105 N^7+56184 N^6
\N\\
&&
+49188 N^5-15320 N^4-62080 N^3-47040 N^2-19584 N
\N\\
&&
-3456~.
\end{eqnarray}
\begin{eqnarray} 
\Ahathat_{Qg}^{(3),S_{1b}}&=&
\frac {T_F^2 n_f C_A} {N^2 (N+1)^2} \Biggl\{
\frac {1} {\ep^3}
\Biggl[
\frac{16 (N-2)}{9}
\Biggr]
\N\\
&&
+\frac {1} {\ep^2} \Biggl[
\frac{8}{9} (N-2) S_1
+\frac{8 (4 N^4+43 N^3+26 N^2-64 N-24)}{27 N (N+1) (N+2)}
\Biggr]
\N\\
&&
+\frac {1} {\ep} \Biggl[
\frac{2}{9} (N-2) S_1^2
+\frac{26}{9} (N-2) S_2
+\frac{2}{3} (N-2) \zeta _2
\N
\end{eqnarray}
\begin{eqnarray}
&&
+\frac{4 (4 N^4+43 N^3+26 N^2-64 N-24) }{27 N (N+1) (N+2)} S_1
-\frac{4 P_{119}(N)}{81 N^2 (N+1)^2 (N+2)^2}
\Biggr]
\N\\
&&
+\frac{1}{27} (N-2) S_1^3
+\frac{110}{27} (N-2) S_3
-\frac{14}{9} (N-2) \zeta _3
+\frac{13}{9} (N-2) S_2 S_1
\N\\
&&
+\frac{1}{3} (N-2) \zeta_2 S_1
+\frac{(4 N^4+43 N^3+26 N^2-64 N-24) }{27 N (N+1) (N+2)} S_1^2
\N\\
&&
+\frac{13 (4 N^4+43 N^3+26 N^2-64 N-24) }{27 N (N+1) (N+2)} S_2
\N\\
&&
+\frac{(4 N^4+43 N^3+26 N^2-64 N-24) }{9 N (N+1) (N+2)} \zeta_2
\N\\
&&
-\frac{2 P_{119}(N)}{81 N^2 (N+1)^2 (N+2)^2} S_1
+\frac{2 P_{120}(N)}{243 N^3 (N+1)^3 (N+2)^3}
\Biggr\}~,
\end{eqnarray}
with
\begin{eqnarray} 
P_{119}(N)&=&152 N^7+722 N^6+461 N^5-1186 N^4-364 N^3+1976 N^2
\N\\
&&
+1200 N+288~,
\\
P_{120}(N)&=&1168 N^{10}+6028 N^9+8238 N^8+5523 N^7+27480 N^6
\N\\
&&
+51132 N^5-1240 N^4-62080 N^3-47040 N^2-19584 N
\N\\
&&
-3456~.
\end{eqnarray}
\begin{eqnarray} 
\Ahathat_{Qg}^{(3),T_{1a}}&=&
 \frac {T_F^2 n_f C_A} {(N-1) N (N+2)} \Biggl\{
\frac {1} {\ep^3}
\frac{64}{9}
+\frac {1} {\ep^2} \Biggl[
\frac{32 }{9} S_1
\N\\
&&
-\frac{64 (7 N^2+6 N-10)}{27 (N+1) (N+2)}
\Biggr]
+\frac {1} {\ep} \Biggl[
\frac{8 }{9} S_1^2
+\frac{104 }{9} S_2
+\frac{8 }{3} \zeta_2
\N\\
&&
-\frac{32 (7 N^2+6 N-10) }{27 (N+1) (N+2)} S_1
\N\\
&&
+\frac{64 (25 N^4+45 N^3-14 N^2+36 N+124)}{81 (N+1)^2 (N+2)^2}
\Biggr]
+\frac{4 }{27} S_1^3
+\frac{440 }{27} S_3
\N\\
&&
-\frac{56}{9} \zeta_3
+\frac{52 }{9} S_2 S_1
+\frac{4 }{3} \zeta_2 S_1
-\frac{8 (7 N^2+6 N-10) }{27 (N+1)(N+2)} S_1^2
\N\\
&&
-\frac{104 (7 N^2+6 N-10) }{27 (N+1) (N+2)} S_2
-\frac{8 (7 N^2+6 N-10) }{9 (N+1) (N+2)} \zeta_2
\N\\
&&
+\frac{32 (25 N^4+45 N^3-14 N^2+36 N+124)}{81 (N+1)^2 (N+2)^2} S_1
\N\\
&&
-\frac{64 P_{121}(N)}{243 (N+1)^3 (N+2)^3}
\Biggr\}~,
\end{eqnarray}
with
\begin{eqnarray} 
P_{121}(N)&=&79 N^6+336 N^5+765 N^4+1488 N^3+1392 N^2-744 N
\N\\
&&
-1480~.
\end{eqnarray}
\begin{eqnarray} 
\Ahathat_{Qg}^{(3),T_{1b}}&=&
\frac {T_F^2 n_f C_A} {(N-1) N (N+1)^2 (N+2)} \Biggl\{
\frac {1} {\ep^3}
\Biggl[
\frac{64 (N+3)}{9}
\Biggr]
\N\\
&&
+\frac {1} {\ep^2} \Biggl[
\frac{32}{9} (N+3) S_1
-\frac{32 (11 N^3+39 N^2+10 N-36)}{27 (N+1) (N+2)}
\Biggr]
\N\\
&&
+\frac {1} {\ep} \Biggl[
\frac{8}{9} (N+3) S_1^2
+\frac{104}{9} (N+3) S_2
+\frac{8}{3} (N+3) \zeta _2
\N\\
&&
-\frac{16 (11 N^3+39 N^2+10 N-36) }{27 (N+1) (N+2)} S_1
+\frac{16 P_{122}(N)}{81 (N+1)^2 (N+2)^2}
\Biggr]
\N\\
&&
+\frac{4}{27} (N+3) S_1^3
+\frac{440}{27} (N+3) S_3
-\frac{56}{9} (N+3) \zeta_3
\N\\
&&
+\frac{52}{9} (N+3) S_2 S_1
+\frac{4}{3} (N+3) \zeta_2 S_1
-\frac{4 (11 N^3+39 N^2+10 N-36) }{27 (N+1) (N+2)} S_1^2
\N\\
&&
-\frac{52 (11 N^3+39 N^2+10 N-36) }{27 (N+1) (N+2)} S_2
-\frac{4 (11 N^3+39 N^2+10 N-36) }{9 (N+1) (N+2)} \zeta_2
\N\\
&&
+\frac{8 P_{122}(N)}{81 (N+1)^2 (N+2)^2} S_1
-\frac{8 P_{123}(N)}{243 (N+1)^3 (N+2)^3}
\Biggr\}~,
\end{eqnarray}
with
\begin{eqnarray} 
P_{122}(N)&=&58 N^5+225 N^4+181 N^3+324 N^2+1372 N+1296~,
\\
P_{123}(N)&=&332 N^7+2418 N^6+9531 N^5+27393 N^4+45426 N^3
\N\\
&&
+24108 N^2-24536 N-25920~.
\end{eqnarray}
\begin{eqnarray} 
\Ahathat_{Qg}^{(3),T_{1c}}&=&
 \frac {T_F^2 n_f C_A} {(N-1) N (N+1)^2 (N+2)} \Biggl\{
\frac {1} {\ep^3}
\Biggl[
\frac{32 (N^2+3 N+4)}{9}
\Biggr]
\N\\
&&
+\frac {1} {\ep^2} \Biggl[
\frac{16}{9} (N^2+3 N+4) S_1
-\frac{16 (14 N^4+51 N^3+57 N^2-18 N-56)}{27 (N+1)(N+2)}
\Biggr]
\N\\
&&
+\frac {1} {\ep} \Biggl[
\frac{4}{9} (N^2+3 N+4) S_1^2
+\frac{52}{9} (N^2+3 N+4) S_2
+\frac{4}{3} (N^2+3 N+4) \zeta _2
\N\\
&&
-\frac{8 (14 N^4+51 N^3+57 N^2-18 N-56) }{27 (N+1) (N+2)} S_1
+\frac{8 P_{124}(N)}{81 (N+1)^2 (N+2)^2}
\Biggr]
\N
\\&&
+\frac{2}{27} (N^2+3 N+4) S_1^3
+\frac{220}{27} (N^2+3 N+4) S_3
-\frac{28}{9} (N^2+3 N+4) \zeta _3
\N
\end{eqnarray}
\begin {eqnarray}
&&
+\frac{26}{9} (N^2+3 N+4) S_2 S_1
+\frac{2}{3} (N^2+3 N+4) \zeta _2 S_1
\N\\
&&
-\frac{2 (14 N^4+51 N^3+57 N^2-18 N-56) }{27 (N+1) (N+2)} S_1^2
\N\\
&&
-\frac{26 (14 N^4+51 N^3+57 N^2-18 N-56) }{27 (N+1)(N+2)}  S_2
\N\\
&&
-\frac{2 (14 N^4+51 N^3+57 N^2-18 N-56) }{9 (N+1) (N+2)} \zeta_2
+\frac{4 P_{124}(N)}{81 (N+1)^2 (N+2)^2} S_1
\N\\
&&
-\frac{4 P_{125}(N)}{243 (N+1)^3 (N+2)^3}
\Biggr\}~,
\end{eqnarray}
with
\begin{eqnarray} 
P_{124}(N)&=&100 N^6+438 N^5+629 N^4+393 N^3+1052 N^2+2508 N
\N\\
&&
+1792~,
\\
P_{125}(N)&=&632 N^8+4284 N^7+14546 N^6+36363 N^5+68457 N^4
\N\\
&&
+73650 N^3+11500 N^2-54168 N-37760~.
\end{eqnarray}
\begin{eqnarray} 
\Ahathat_{Qg}^{(3),A_{2}}&=&
\frac {T_F^2 n_f C_F} {(N-1) N^2 (N+1) (N+2)} \Biggl\{
-\frac {1} {\ep^3}
\frac{512}{3}
\N\\
&&
+\frac {1} {\ep^2} \Biggl[
+\frac{128 (3 N^4-14 N^3-21 N^2+16 N+4)}{3 (N-1) N (N+1) (N+2)}
\Biggr]
\N\\
&&
+\frac {1} {\ep} \Biggl[
-256 S_2
-64 \zeta _2
+\frac{64 P_{126}(N)}{3 (N-1)^2 N^2 (N+1)^2 (N+2)^2}
\Biggr]
\N\\
&&
-384 S_3
+\frac{448 }{3} \zeta_3
+\frac{64 (3 N^4-14 N^3-21 N^2+16 N+4) }{(N-1) N (N+1) (N+2)} S_2
\N\\
&&
+\frac{16 (3 N^4-14 N^3-21 N^2+16 N+4) }{(N-1) N (N+1) (N+2)} \zeta_2
\N\\
&&
+\frac{32 P_{127}(N)}{3 (N-1)^3 N^3 (N+1)^3 (N+2)^3}
\Biggr\}~,
\end{eqnarray}
with
\begin{eqnarray}
P_{126}(N)&=&N^8+34 N^7-45 N^6-239 N^5+38 N^4+269 N^3-86 N^2
\N\\
&&
-36 N-8~,
\\
P_{127}(N)&=& 4 N^{12}+34 N^{11}+286 N^{10}-20 N^9-1845 N^8-826 N^7
\N\\
&&
+3276 N^6+632 N^5-2657 N^4+396 N^3+192 N^2+80 N
\N\\
&&
+16~.
\end{eqnarray}
\begin{eqnarray} 
\Ahathat_{Qg}^{(3),B_2}&=&
\frac {T_F^2 n_f C_F} {N^3 (N+1)^2} \Biggl\{
-\frac {1} {\ep^3}
\Biggl[
\frac{64 (N-1) (N+2)}{3}
\Biggr]
\N\\
&&
+\frac {1} {\ep^2} \Biggl[
\frac{32 P_{128}(N)}{3 (N-1) N (N+1) (N+2)}
\Biggr]
\N\\
&&
+\frac {1} {\ep} \Biggl[
-32 (N-1) (N+2) S_2
-8 (N-1) (N+2) \zeta _2
\N\\
&&
-\frac{16 P_{129}(N)}{3 (N-1)^2N^2 (N+1)^2 (N+2)^2}
\Biggr]
-48 (N-1) (N+2) S_3
\N\\
&&
+\frac{56}{3} (N-1) (N+2) \zeta _3
+\frac{16 P_{128}(N) }{(N-1) N (N+1) (N+2)} S_2
\N\\
&&
+\frac{4 P_{128}(N) }{(N-1) N (N+1) (N+2)} \zeta_2
\N\\
&&
+\frac{8 P_{130}(N)}{3 (N-1)^3 N^3 (N+1)^3 (N+2)^3}
\Biggr\}~,
\end{eqnarray}
with
\begin{eqnarray}
P_{128}(N)&=&2 N^6+4 N^5-13 N^4+14 N^3+41 N^2-12 N-12~,
\\
P_{129}(N)&=&N^{10}+N^9-13 N^8+100 N^7+46 N^6-494 N^5-212 N^4
\N\\
&&
+473 N^3+110
N^2-100 N-56~,
\\
P_{130}(N)&=& 4 N^{14}+26 N^{13}+67 N^{12}+134 N^{11}-719 N^{10}-1462
  N^9
\N\\
&&
+3010 N^8+5074 N^7-4187 N^6-5412 N^5+2609 N^4+
\N\\
&&
2240 N^3+296 N^2-576 N-240~.
\end{eqnarray}
\begin{eqnarray} 
\Ahathat_{Qg}^{(3),C_{2}}&=&
{T_F^2 n_f \left(C_F-\frac {C_A} {2} \right)} \frac {1} {(N-1) N (N+1)^2 (N+2)^2} \Biggl\{
\frac {1} {\ep^3} \Biggl[
\frac{2048}{3}
\Biggr]
\N\\
&&
+\frac {1} {\ep^2} \Biggl[
-\frac{256 P_{131}}{3 (N-1) N (N+1) (N+2)}
\Biggr]
\N\\
&&
+\frac {1} {\ep} \Biggl[
1024 S_2
+256 \zeta _2
+\frac{128 P_{132}(N)}{3 (N-1)^2 N^2 (N+1)^2 (N+2)^2}
\Biggr]
\N\\
&&
+1536 S_3
-\frac{1792 \zeta _3}{3}
-\frac{128 P_{131}(N) }{(N-1) N (N+1) (N+2)} S_2
\N\\
&&
-\frac{32 P_{131}(N)}{(N-1) N (N+1) (N+2)} \zeta _2
\N\\
&&
-\frac{64 P_{133}(N)}{3 (N-1)^3 N^3 (N+1)^3 (N+2)^3}
\Biggr\}~,
\end{eqnarray}
with
\begin{eqnarray} 
P_{131}(N)&=&N^5+9 N^4-25 N^3-53 N^2-4~,
\\
P_{132}(N)&=& 3 N^9+5N^8-71 N^7+66 N^6+531 N^5+243 N^4-267 N^3
\N\\
&&
+14 N^2-84 N-8~,
\\
P_{133}(N)&=&4 N^{13}+10N^{12}+38 N^{11}+568 N^{10}+201 N^9-3687 N^8
\N\\
&&
-3872 N^7+3330 N^6+2677 N^5-1573 N^4+648 N^3
\N\\
&&
-744 N^2-176 N-16~.
\end{eqnarray}
\begin{eqnarray} 
\Ahathat_{Qg}^{(3),D2}&=&
{T_F^2 n_f \left(C_F - \frac{C_A} {2} \right)} \frac {1} {N^2 (N+1)^3} \Biggl\{
\frac {1} {\ep^3}
\frac{256 (N-1)}{3}
\N\\
&&
-\frac {1} {\ep^2} \Biggl[
\frac{128 P_{134}(N)}{3 N (N+1) (N+2)^2 (N-1)}
\Biggr]
+\frac {1} {\ep} \Biggl[
128 (N-1) S_2
\N\\
&&
+32 (N-1) \zeta _2
-\frac{32 P_{135}(N)}{3 (N-1)^2 N^2 (N+1)^2 (N+2)^3}
\Biggr]
+192 (N-1) S_3
\N\\
&&
-\frac{224}{3} (N-1) \zeta _3
-\frac{64 P_{134}(N) S_2}{N (N+1) (N+2)^2 (N-1)}
\N\\
&&
-\frac{16 P_{134}(N) }{N (N+1) (N+2)^2 (N-1)} \zeta _2
\N\\
&&
+\frac{16 P_{136}(N)}{3 N^3 (N+1)^3 (N+2)^4 (N-1)^3}
\Biggr\}~,
\end{eqnarray}
with
\begin{eqnarray}
P_{134}(N)&=&2 N^6+6 N^5-9 N^4+16 N^3+53 N^2+8 N-4~,
\\
P_{135}(N)&=&N^{11}+4 N^{10}+17 N^9+58 N^8-237 N^7-278 N^6+1057 N^5
\N\\
&&
+1156
  N^4-358 N^3-420 N^2-152 N+16~,
\\
P_{136}(N)&=&4 N^{15}+22 N^{14}+79 N^{13}+95 N^{12}-575 N^{11}+335
N^{10}
\N\\
&&
+3743 N^9-3193 N^8-13937 N^7-1437 N^6+10542 N^5
\N\\
&&
+3490 N^4-832 N^3-2752 N^2-800 N+32~.
\end{eqnarray}
\begin{eqnarray} 
\Ahathat_{Qg}^{(3),E_2}&=&
\frac {T_F^2 n_f C_A} {(N+1)^2 (N+2)} \Biggl\{
-\frac {1} {\ep^3} \Biggl[
\frac{32 (2 N^2-13 N-33)}{9 (N-1) N}
\Biggr]
\N\\
&&
+\frac {1} {\ep^2} \Biggl[
-\frac{32 }{9} S_1
+\frac{16 P_{137}(N)}{27 (N-1)^2 N^2 (N+1) (N+2)}
\Biggr]
\N\\
&&
+\frac {1} {\ep} \Biggl[
-\frac{8 (13 N^2-79 N-198) }{9 (N-1) N} S_2
-\frac{4 (2 N^2-13 N-33) }{3 (N-1) N} \zeta _2
-\frac{8 }{9} S_1^2
\N\\
&&
+\frac{16 (26 N^4+103 N^3+253 N^2+338 N+144) }{27 (N-1) N (N+1) (N+2)} S_1
\N
\end{eqnarray}
\begin{eqnarray}
&&
-\frac{8 P_{138}(N)}{81 (N-1)^3 N^3 (N+1)^2 (N+2)^2}
\Biggr]
-\frac{88 (5 N^2-32 N-81) }{27 (N-1) N} S_3
\N
\\
&&
+\frac{28 (2 N^2-13 N-33) }{9 (N-1) N} \zeta_3
-\frac{52 }{9} S_2 S_1
-\frac{4 }{3} \zeta _2 S_1
-\frac{4 }{27} S_1^3
\N\\
&&
+\frac{4 P_{139}(N) }{27 (N-1)^2 N^2 (N+1) (N+2)} S_2
+\frac{2 P_{140}(N)}{9 (N-1)^2 N^2 (N+1) (N+2)} \zeta _2
\N\\
&&
+\frac{4 (26 N^4+103 N^3+253 N^2+338 N+144) }{27 (N-1) N (N+1) (N+2)} S_1^2
\N\\
&&
-\frac{8 P_{141}(N)}{81 (N-1) N (N+1)^2 (N+2)^2} S_1
\N\\
&&
+\frac{4 P_{142}(N)}{243 (N-1)^4 N^4 (N+1)^3 (N+2)^3}
\Biggl\}~,
\end{eqnarray}
with
\begin{eqnarray}
P_{137}(N)&=&52 N^6+24 N^5+28 N^4+765 N^3+532 N^2-555 N-54~,
\\
P_{138}(N)&=&572 N^{10}+1696 N^9+2460 N^8+207 N^7-18024 N^6-24537 N^5
\N\\
&&
+16705 N^4+20618 N^3-12729 N^2-900 N-324~,
\\
P_{139}(N)&=&338 N^6+221 N^5+318 N^4+4675 N^3+2998 N^2-3474 N
\N\\
&&
-324~,
\\
P_{140}(N)&=&52 N^6+24 N^5+28 N^4+765 N^3+532 N^2-555 N-54~,
\\
P_{141}(N)&=&286 N^6+1856 N^5+5575 N^4+8375 N^3+4624 N^2-1276 N
\N\\
&&
-1296~,
\\
P_{142}(N)&=&4276 N^{14}+16804 N^{13}+10992 N^{12}-81178 N^{11}-181276 N^{10}
\N\\
&&
+209316 N^9+783074 N^8-83677 N^7-1043730 N^6+235534 N^5
\N\\
&&
+660610 N^4-261267 N^3-4554 N^2-6372 N-1944~.
\end{eqnarray}
\begin{eqnarray} 
\Ahathat_{Qg}^{(3),F2}&=&
 \frac {T_F^2 n_f C_A} {N (N+1)} \Biggl\{
\frac {1} {\ep^3}\Biggl[
\frac{16 (5 N^3+15 N^2+5 N-17)}{9 N (N+1)^2}
\Biggr]
\N\\
&&
+\frac {1} {\ep^2} \Biggl[
\frac{16 }{9} S_1
-\frac{8 P_{143}(N)}{27 (N-1) N^2 (N+1)^3 (N+2)}
\Biggr]
\N\\
&&
+\frac {1} {\ep} \Biggl[
+\frac{4 (31 N^3+92 N^2+31 N-102) }{9 N (N+1)^2} S_2
\N\\
&&
+\frac{2 (5 N^3+15 N^2+5 N-17) }{3 N (N+1)^2} \zeta _2
+\frac{4 }{9} S_1^2
\N\\
&&
-\frac{8 (5 N^3+30 N^2+34 N+60) }{27 N (N+1) (N+2)} S_1
+\frac{4 P_{144}(N)}{81 (N-1)^2 N^3 (N+1)^4 (N+2)^2}
\Biggr]
\N
\\
&&
+\frac{4 (136 N^3+407 N^2+136 N-459) }{27 N (N+1)^2} S_3
-\frac{14 (5 N^3+15 N^2+5 N-17) }{9 N (N+1)^2} \zeta_3
\N\\
&&
+\frac{26 }{9} S_2 S_1
+\frac{2 }{3} \zeta_2 S_1
+\frac{2 }{27} S_1^3
-\frac{2 (5 N^3+30 N^2+34 N+60) }{27 N (N+1) (N+2)} S_1^2
\N
\end{eqnarray}
\begin{eqnarray}
&&
-\frac{2 P_{145}(N) }{27 (N-1) N^2 (N+1)^3 (N+2)} S_2
-\frac{ P_{146}(N)}{9 (N-1) N^2 (N+1)^3 (N+2)}  \zeta_2
\N\\
&&
-\frac{4 P_{147}(N)}{81 N^2 (N+1)^2 (N+2)^2} S_1
\N\\
&&
-\frac{2 P_{148}(N)}{243 (N-1)^3 N^4 (N+1)^5 (N+2)^3}
\Biggr\}~,
\end{eqnarray}
with
\begin{eqnarray}
P_{143}(N)&=&46 N^7+251 N^6+255 N^5-250 N^4+234 N^3+737 N^2
\N\\
&&
-247 N-234~,
\\
P_{144}(N)&=&146 N^{11}+1928 N^{10}+4491 N^9+869 N^8-3135 N^7-10143
  N^6
\N\\
&&
-26309 N^5-3731 N^4+25311 N^3+4129 N^2-4824 N-2988~,
\\
P_{145}(N)&=&281 N^7+1541 N^6+1589 N^5-1441 N^4+1400 N^3+4328 N^2
\N\\
&&
-1542 N-1404~,
\\
P_{146}(N)&=&46 N^7+251 N^6+255 N^5-250 N^4+234 N^3+737 N^2-247 N
\N\\
&&
-234~,
\\
P_{147}(N)&=&35 N^6-243 N^5-1303 N^4-2082 N^3-136 N^2+2136 N
\N\\
&&
+720~,
\\
P_{148}(N)&=&454 N^{15}+15564 N^{14}+57172 N^{13}+34999 N^{12}-143229
  N^{11}
\N\\
&&
-317074 N^{10}
-15876 N^9
+850116 N^8
+667430 N^7
\N\\
&&
-871408 N^6-712308 N^5+468109 N^4+287657 N^3
\N\\
&&
+51030 N^2-79092 N-36936~.
\end{eqnarray}
\begin{eqnarray} 
\Ahathat_{Qg}^{(3),G_{2}}&=&
\frac {T_F^2 n_f C_A} {N (N+1) (N+2)} \Biggl\{
\frac {1} {\ep^3} \Biggl[
\frac{32 (N-4)}{9} S_1
+\frac{32 \left(N^2+2 N+4\right)}{9 (N+2)}
\Biggr]
\N\\
&&
+\frac {1} {\ep^2} \Biggl[
\frac{8 (N-4) }{3} S_1^2
+\frac{8 (N-4) }{3} S_2
\N\\
&&
-\frac{16 \left(26 N^4-16 N^3-87 N^2-83 N-128\right)}{27 (N-1) (N+1)
  (N+2)}  S_1
\N\\
&&
+\frac{16 P_{150}(N)}{27 (N-1) N (N+1) (N+2)^2}
\Biggr]
\N\\
&&
+\frac {1} {\ep} \Biggl[
+\frac{28 (N-4) }{27} S_1^3
+\frac{4 (N-4) }{3}  \zeta _2 S_1
+\frac{4 (31 N-28) }{9}  S_2 S_1
\N\\
&&
-\frac{16 (7 N+12) }{9} S_{2,1}
-\frac{8 (17 N+28) }{27} S_3
\N
\\
&&
-\frac{4 \left(28 N^4-12 N^3-35 N^2+19 N-96\right) }{9 (N-1) (N+1) (N+2)} S_1^2
+\frac{4 \left(N^2+2 N+4\right) }{3 (N+2)} \zeta _2 
\N
\end{eqnarray}
\begin{eqnarray}
&&
+\frac{4 \left(12 N^4+172 N^3+215 N^2-159 N-144\right) }{9 (N-1) (N+1) (N+2)} S_2
\N\\
&&
+\frac{8 P_{151}(N)}{81 (N-1) (N+1)^2 (N+2)^2}  S_1
-\frac{8 P_{152}(N)}{81 (N-1) N^2 (N+1)^2 (N+2)^3}
\Biggr]
\N
\\
&&
+\frac{5 (N-4)}{18} S_1^4
+ (N-4)  \zeta_2 S_1^2
+\frac{ (29 N-20) }{3} S_2 S_1^2
-\frac{28 (N-4) }{9} \zeta_3 S_1
\N\\
&&
+\frac{20 (5 N-4) }{9}  S_3 S_1
-\frac{8 (17 N+36) }{9} S_{2,1} S_1
+\frac{29 (N-4) }{6} S_2^2
\N\\
&&
-\frac{ (181 N+204) }{9} S_4
-\frac{8 (5 N+36) }{9} S_{3,1}
\N\\
&&
-\frac{2 (200 N^4-76 N^3-141 N^2+337 N-608) }{81 (N-1) (N+1) (N+2)}
S_1^3
\N\\
&&
-\frac{2 (26 N^4-16 N^3-87 N^2-83 N-128) }{9 (N-1) (N+1) (N+2)} \zeta_2 S_1
+\frac{8 (17 N+36) }{9} S_{2,1,1}
\N\\
&&
-\frac{2 (296 N^4-940 N^3-2133 N^2-359 N-608) }{27 (N-1) (N+1) (N+2)} S_2 S_1
-\frac{28 \left(N^2+2 N+4\right) }{9 (N+2)} \zeta _3
\N\\
&&
+\frac{4 (676 N^4+2332 N^3+1947 N^2-2035 N-2632) }{81 (N-1) (N+1)
  (N+2)} S_3
\N\\
&&
+\frac{8 (2 N^3-150 N^2-491 N-384) }{27 (N+1) (N+2)} S_{2,1}
+(N-4) \zeta_2 S_2
\N\\
&&
+\frac{2 P_{149}(N)}{27 (N-1) (N+1)^2 (N+2)^2} S_1^2
+\frac{2 P_{150}(N) }{9 (N-1) N (N+1) (N+2)^2} \zeta_2
\N\\
&&
-\frac{2 P_{153}(N)}{27 (N-1) N (N+1)^2 (N+2)^2} S_2
-\frac{4 P_{154}(N)}{243 (N-1) (N+1)^3 (N+2)^3} S_1
\N
\\
&&
+\frac{4 P_{155}(N)}{243 (N-1) N^3 (N+1)^3 (N+2)^4}
\Biggr\}~,
\end{eqnarray}
with
\begin{eqnarray}
P_{149}(N)&=&242 N^6+246 N^5-495 N^4-958 N^3-2384 N^2-1979 N
\N\\
&&
+1296~,
\\
P_{150}(N)&=&13 N^6+82 N^5+16 N^4-353 N^3-446 N^2-152 N-24~,
\\
P_{151}(N)&=&268 N^6+412 N^5-1147 N^4-4532 N^3-6520 N^2-2513 N
\N\\
&&
+1936~,
\\
P_{152}(N)&=&53 N^9+341 N^8-693 N^7-6717 N^6-13305 N^5-8433 N^4\\\N&&+2422 N^3+4324 N^2+1128 N+144~,
\\
P_{153}(N)&=&342 N^7+3314 N^6+8767 N^5+8398 N^4+2732 N^3+879 N^2
\N\\
&&
+48 N+288~,
\\
P_{154}(N)&=&1940 N^8+8156 N^7+9782 N^6-20076 N^5-94500 N^4
\N\\
&&-153048 N^3-142667 N^2-123635 N-76928~,
\\
P_{155}(N)&=&2833 N^{12}+32644 N^{11}+146294 N^{10}+316196 N^9
\N\\
&&
+266628 N^8-211206 N^7-715870 N^6-709393 N^5
\N\\
&&
-362414 N^4-119936 N^3
-36528 N^2
-8064 N
-864~,
\end{eqnarray}
\begin{eqnarray} 
\Ahathat_{Qg}^{(3),H_{2}}&=&
 \frac {T_F^2 n_f C_A} {(N+1)^2 (N+2)} \Biggl\{
-\frac {1} {\ep^3} \Biggl[
\frac{32 (2 N^2-13 N-33)}{9 (N-1) N}
\Biggr]
\N\\
&&
+\frac {1} {\ep^2} \Biggl[
-\frac{32}{9} S_1
+\frac{16 P_{156}(N)}{27 (N-1)^2 N^2 (N+1) (N+2)}
\Biggr]
\N\\
&&
+\frac {1} {\ep} \Biggl[
-\frac{8 }{9} S_1^2
-\frac{8 (13 N^2-79 N-198) }{9 (N-1) N} S_2
-\frac{4 (2 N^2-13 N-33) }{3 (N-1) N} \zeta _2
\N\\
&&
+\frac{16 (26 N^4+103 N^3+253 N^2+338 N+144) }{27 (N-1) N (N+1) (N+2)} S_1
\N\\
&&
-\frac{8 P_{157}(N)}{81 (N-1)^3 N^3 (N+1)^2 (N+2)^2}
\Biggr]
-\frac{88 (5 N^2-32 N-81) }{27 (N-1) N} S_3
\N\\
&&
+\frac{28 (2 N^2-13 N-33) }{9 (N-1) N} \zeta_3
-\frac{52 }{9} S_2 S_1
-\frac{4 }{3} \zeta_2 S_1
-\frac{4 }{27} S_1^3
\N\\
&&
+\frac{4 (26 N^4+103 N^3+253 N^2+338 N+144) }{27 (N-1) N (N+1) (N+2)} S_1^2
\N\\
&&
+\frac{4 P_{158}(N)}{27 (N-1)^2 N^2 (N+1) (N+2)} S_2
\N\\
&&
+\frac{2 P_{159}(N) }{9 (N-1)^2 N^2 (N+1) (N+2)} \zeta _2
\N\\
&&
-\frac{8 P_{160}(N)}{81 (N-1) N (N+1)^2 (N+2)^2} S_1
\N\\
&&
+\frac{4 P_{161}(N)}{243 (N-1)^4 N^4 (N+1)^3 (N+2)^3}
\Biggl\}~,
\end{eqnarray}
with
\begin{eqnarray}
P_{156}(N)&=&52 N^6+6 N^5-26 N^4+783 N^3+658 N^2-555 N-126~,
\\
P_{157}(N)&=&572 N^{10}+1480 N^9+1920 N^8+1233 N^7-15432 N^6
\N\\
&&
-26697 N^5+12385 N^4+23048 N^3-10029 N^2-1980 N
\N\\
&&
-756~,
\\
P_{158}(N)&=&338 N^6+113 N^5-6 N^4+4783 N^3+3754 N^2-3474 N
\N\\
&&
-756~,
\\
P_{159}(N)&=&52 N^6+6 N^5-26 N^4+783 N^3+658 N^2-555 N-126~,
\\
P_{160}(N)&=&286 N^6+1856 N^5+5575 N^4+8375 N^3+4624 N^2-1276 N
\N\\
&&
-1296~,
\\
P_{161}(N)&=&4276 N^{14}+15508 N^{13}+8400 N^{12}-72916 N^{11}-173500 N^{10}
\N\\
&&
+169140 N^9+778700 N^8+23729 N^7-1060254 N^6
\N\\
&&
+98482 N^5+694468 N^4-190635 N^3-20106 N^2-14148 N
\N\\
&&
-4536~.\label{DIAGH2}
\end{eqnarray}
\vspace*{1cm}
\subsection{\boldmath ${\Ahathat_{Qq}^{PS}}$}
The individual contributions to ~$\Ahathat_{Qq}^{\sf PS}$~ are:
\begin{eqnarray} 
\Ahathat_{Qq}^{(3),\rm{PS},a}&=& \frac {T_F^2\,n_f\,C_F} {N^2 (N+1)^2} \Biggl\{
-\frac {1} {\ep^3} \Biggl[
{\frac {64}{9}}\,(2+N)  (-1+N)
\Biggr]
\N\\
&&
+\frac{1} {\ep^2} \Biggl[ 
-{\frac {32}{9}}\,(2+N)  (-1+N) S_1
+{\frac {32}{27}}\,{\frac {\, P_{162}(N)\,}{N (1+N) (2+N)}}
\Biggr]
\N\\
&&
+\frac {1} {\ep} \Biggl[
-{\frac {104}{9}}\,(2+N) (-1+N)  S_2
-{\frac {8}{9}}\, (2+N)  (-1+N)  S_1^2
\N\\
&&
-\frac {8} {3}\,(2+N) (-1+N) \zeta_2
+{\frac{16}{27}}\,\frac {P_{162}(N)} {N (1+N) (2+N) } S_1
\N\\
&&
-{\frac {16}{81}}\,\frac {P_{163}(N)}{N^{2} (1+N) ^{2} (2+N)^{2}}
\Biggr]
-{\frac {440}{27}}\,(2+N)  (-1+N) S_3
\N\\
&&
-{\frac {52}{9}}\,(2+N)  (-1+N) S_2 S_1
-{\frac {4}{27}}\,(2+N)  (-1+N)  S_1^3
\N\\
&&
+{\frac {56}{9}}\, (2+N)  (-1+N) \zeta_3
-\frac {4} {3}\, (2+N)  (-1+N) S_1 \zeta_2
\N\\
&&
+{\frac {52}{27}}\,{\frac {P_{162}(N)}{N (1+N) (2+N) }}  S_2
+{\frac {4}{27}} \frac {P_{162}(N)}{N (1+N) (2+N)} S_1^2
\N\\
&&
+{\frac{4}{9}}\,{\frac {P_{162}(N)}{N (1+N) (2+N) }} \zeta_2
-{\frac {8}{81}}\,{\frac {P_{163}(N)}{N^{2} (1+N) ^{2} (2+N) ^{2}}}
S_1
\N\\
&&
+{\frac {8}{243}}\,{\frac {P_{164}(N)}{N^{3} (1+N) ^{3} (2+N) ^{3}}}
\Biggr\}~, \label{AQqA}
\end{eqnarray}
with
\begin{eqnarray}
P_{162}(N)&=& 8\,N^{5}+29\,N^{4}-9\,N^{3}-8\,N^{2}+64\,N+24~,\\
P_{163}(N)&=& 25\,N^{8}+151\,N^{7}+116\,N^{6}+352\,N^{5}+1052\,N^{4}
\\\N
&&
-428\,N^{3}-2264
\,N^{2}-1200\,N-288~,\\
P_{164}(N)&=& 158\,N^{11}+1505\,N^{10}+5261\,N^{9}+12912\,N^{8}+13860\,N^{7}\N\\&&\N
-16140\,N^{6}
-33344\,N^{5}+27880\,N^{4}+77344\,N^{3}
\N\\
&&
+50496\,N^{2}
+19584\,N+3456~.
\end{eqnarray}
\begin{eqnarray} 
\Ahathat_{Qq}^{(3),\rm{PS},b}&=&
 \frac {T_F^2\,n_f\,C_F} {(N-1) N (N+1) (N+2)}
\Biggl\{
- \frac {1} {\ep^3} 
{\frac {512}{9}}
\N\\
&&
+ \frac {1} {\ep^2} \Biggl[
-{\frac {256}{9}}\,S_1
+{\frac {1024}{27}}\,{\frac { (2\,N^{2}-5) }{ (2+N) (1+N)}}
\Biggr]
\N
\end{eqnarray}
\begin{eqnarray}
&&
+ \frac {1} {\ep} \Biggl[
-{\frac {832}{9}}\,S_2
-{\frac {64}{9}}\,S_1^2
-{\frac {64}{3}}\,\zeta_2
+{\frac {512}{27}}\,\frac { (2\,N^{2}-5)}{ (2+N) (1+N)} S_1
\N\\
&&
-{\frac {128}{81}}\,\frac { (25\,N^{4}-42\,N^{3}-107\,N^{2}+348\,N+532) }{(2+N) ^{2} (1+N) ^{2} }
\Biggr]
\N
\\&&
-{\frac {3520}{27}}\,S_3
-{\frac {416}{9}}\,S_2 S_1
-{\frac {32}{27}}\,S_1^3
-{\frac {32}{3}}\,S_1\zeta_2
+{\frac {448}{9}}\,\zeta_3
\N\\
&&
+{\frac {1664}{27}}\,\frac { (2\,N^{2}-5)}{(2+N) (1+N)} \,S_2
+{\frac {128}{27}}\,\frac { (2\,N^{2}-5)}{ (2+N)^1 (1+N) ^1 } \,
S_1^2
\N\\
&&
+{\frac {128}{9}}\,\frac { (2\,N^{2}-5)}{(2+N) (1+N)} \,\zeta_2
\N\\
&&
-{\frac {64}{81}}\,\frac {(25\,N^{4}-42\,N^{3}-107\,N^{2}+348\,N+532)}{(2+N)^{2} (1+N)^{2} } \,S_1
\N\\
&&
+{\frac {128}{243}}\,\frac {P_{165}(N)}{(2+N) ^{3} (1+N)^{3}}
\Biggr\}~,
\end{eqnarray}
with
\begin{eqnarray}
P_{165}(N)&=&79\,N^{6}+411\,N^{5}+2085\,N^{4}+5289\,N^{3}+3252\,N^{2}
\N\\
&&
-5484\,N-6064~.\label{AQqEnd}
\end{eqnarray}
\subsection{\boldmath ${\Ahathat_{qq,Q}^{\rm PS}}$}
The contributions to $\Ahathat_{qq,Q}^{\sf PS}$ read:
\begin{eqnarray} 
\Ahathat_{qq,Q}^{(3),\rm{PS},a}&=&
 \frac {T_F^2\,n_f\,C_F} {(N-1) N (N+1) (N+2)} \Biggl\{
-\frac {1} {\ep^3}\Biggl[
\frac{512}{9}
\Biggr]
+\frac {1} {\ep^2} \Biggl[
\frac{512 }{9} S_1
-\frac{512 (2 N+1) (4 N+7)}{27 (N+1) (N+2)}
\Biggr]
\N\\
&&
+\frac {1} {\ep} \Biggl[
-\frac{256 }{9} S_2
-\frac{64 }{3} \zeta_2
-\frac{256 }{9} S_1^2
+\frac{512 (2 N+1) (4 N+7) }{27 (N+1) (N+2)} S_1
\N\\
&&
-\frac{128 (181 N^4+894 N^3+1597 N^2+1248 N+400)}{81 (N+1)^2 (N+2)^2}
\Biggr]
+\frac{256 }{9} S_2 S_1
+\frac{64 }{3} \zeta_2 S_1
\N
\\
&&
+\frac{256 }{27} S_1^3
+\frac{512 }{27} S_3
+\frac{448 }{9} \zeta_3
+\frac {(2 N +1) (4 N +7)} {(N+1) (N+2)} \Bigl[
-\frac{256}{27} S_1^2
-\frac{256}{27} S_2
-\frac{64}{9} \zeta_2
\Bigr]
\N\\
&&
+\frac{128 (181 N^4+894 N^3+1597 N^2+1248 N+400)}{81 (N+1)^2(N+2)^2} S_1
\N\\
&&
-\frac{128 P_{166}(N)}{243 (N+1)^3 (N+2)^3}
\Biggr\}~,
\end{eqnarray}
with
\begin{eqnarray}
P_{166}(N)&=&1037 N^6+8247 N^5+26940 N^4+46191 N^3+43809 N^2
\N\\
&&
+21648 N+4192~.
\end{eqnarray}
\begin{eqnarray} 
\Ahathat_{qq,Q}^{(3),\rm{PS},b}&=& \frac {T_F^2\,n_f\,C_F} {N^2 (N+1)^2} \Biggl\{
- \frac {1} {\ep^3} \frac {64}{9} \,(2+N)  (-1+N)
\N\\
&&
+\frac {1} {\ep^2} \Biggl[
\frac {64}{9}\,(2+N)  (-1+N) S_1
\N\\
&&
-\frac {32}{27}\,\frac {16\,N^{4}+26\,N^{3}-25\,N^{2}-11\,N+6}{N (1+N)}
\Biggr]
\N\\
&&
+\frac {1} {\ep} \Biggl[
(2+N)  (-1+N)\Biggl(
-\frac {32}{9}\, S_2
-\frac {32}{9}\,S_1^2
-\frac {8} {3}\,\zeta_2
\Biggr)
\N
\\
&&
+\frac {32}{27}\,\frac {(16\,N^{4}+26\,N^{3}-25\,N^{2}-11\,N+6)}{N (1+N)} S_1
-\frac {16}{81}\,\frac {P_{167}(N)}{N^{2} (1+N) ^{2}}
\Biggr]
\N\\
&&
+(N+2)(N-1)\Biggl(
+\frac {64}{27}\,S_3
+\frac {32}{9}\,S_2 S_1
+\frac {32}{27}\,S_1^3
+\frac{8}{3} \, S_1\zeta_2
\N\\
&&
+\frac {56}{9}\,\zeta_3
\Biggr)
-\frac {16}{27}\,\frac {(16\,N^{4}+26\,N^{3}-25\,N^{2}-11\,N+6)}{N (1+N)} S_2
\N\\
&&
-\frac {16}{27}\,\frac {(16\,N^{4}+26\,N^{3}-25\,N^{2}-11\,N+6)}{N (1+N)} S_1^2
\N\\
&&
-\frac {4}{9}\,\frac {
    (16\,N^{4}+26\,N^{3}-25\,N^{2}-11\,N+6)}{N (1+N)} \zeta_2
+\frac {16}{81}\,\frac { P_{167}(N)}{N^{2} (1+N) ^{2}} S_1
\N\\
&&
-\frac {8}{243}\,\frac {P_{168}(N)}{N^{3} (1+N) ^{3}}
\Biggr\}~,
\end{eqnarray}
with
\begin{eqnarray}
P_{167}(N)&=&181\,N^{6}+447\,N^{5}-32\,N^{4}-297\,N^{3}-92\,N^{2}+15\,N-18~,
\\
P_{168}(N)&=&2074\,N^{8}+7210\,N^{7}+4927\,N^{6}-3503\,N^{5}-5309\,N^{4}-929\,N^{3}
\N\\
&&
+231\,N^{2}+9 N+54~.
\end{eqnarray}
\subsection{\boldmath ${\Ahathat^{\rm{NS}}_{qq,Q}}$}
The contributions to ~$\Ahathat_{qq,Q}^{\sf NS}$ are given by~:
\begin{eqnarray} 
\Ahathat_{qq,Q}^{(3),\rm{NS},a}
&=&
\frac {T_F^2\,n_f\,C_F} {N (N+1)}
\Biggl\{
- \frac {1} {\ep^3}
\Biggl[
\frac {64}{27}\,\,(N+2)  (N-1) 
\Biggr]
\N\\
&&
- \frac {1} {\ep^2}
\Biggl[
\frac {64}{81} \,\frac { (N^{4}+2\,N^{3}-10\,N^{2}-5\,N+3)}{N (1+N)}
\Biggr]
\N\\
&&
+\frac{1} {\ep}
\Biggl[
-\frac {8}{9}\,(N+2)(N-1) \zeta_2
-\frac {32}{81}\,\frac {P_{169}(N)}{ (1+N)^{2} N^{2}}
\Biggr]
\N
\end{eqnarray}
\begin{eqnarray}
&&
+\frac {56}{27}\,(N+2)  (-1+N) \zeta_3
-\frac {8}{27}\,\frac {(-5\,N-10\,N^{2}+2\,N^{3}+N^{4}+3) }{(1+N) ^{1}N^{1}} \zeta_2
\N\\
&&
-\frac {16}{729}\,\frac {P_{170}(N)}{ (1+N) ^{3}N^{3}}
\Biggr\}~,
\end{eqnarray}
with
\begin{eqnarray}
P_{169}(N)&=& 
+14\,N^{6}
+42\,{N}^{5}
-N^{4}
-50\,N^{3}
-19\,N^{2}
+2\,N-3
~,
\\
P_{170}(N)&=&
+308\,N^{8}
+1232\,N^{7}
+395\,N^{6}
-2353\,N^{5}
-2413\,N^{4}
\N\\
&&
-391\,N^{3}
+153\,N^{2}
+9\,N
+27~.
\end{eqnarray}
\begin{eqnarray}
\Ahathat_{qq,Q}^{(3),\rm{NS},b}
&=&
T_F^2\,n_f\,C_F \Biggl\{
\N\\
&&
\frac {1} {\ep^3}
\Biggl[
-\frac {128}{27}\,S_1
+\frac {128}{27}
\Biggr]
+\frac {1} {\ep^2}
\Biggl[
\frac {128}{81}\,
+\frac {64}{27}\,\,S_2
-\frac {320}{81}\,\,S_1
\Biggr]
\N\\
&&
+\frac {1} {\ep}
\Biggl[
-\frac {32}{27}\,\,S_3
-\frac {16}{9}\,\zeta_2\,S_1
+\frac {160}{81}\,\,S_2
+\frac {16}{9}\,\zeta_2
-\frac {320}{27}\,\,S_1
+\frac {896}{81}\,
\Biggr]
\N\\
&&
+\frac {16}{27}\,\,S_4
+\frac {112}{27}\,\,\zeta_3 S_1
+\frac {8}{9}\,\zeta_2\,S_2
-\frac {80}{81}\,S_3
-\frac {112}{27}\,\,\zeta_3
-\frac {40}{27}\,\zeta_2\,S_1
\N\\
&&
+\frac {16}{27}\,\zeta_2
+\frac {160}{27}\,\,S_2
-\frac {13888}{729}\,\,S_1
+\frac {9856}{729}\,
\Biggr\}~.
\end{eqnarray}
\begin{eqnarray}
\Ahathat_{qq,Q}^{(3),\rm{NS},c}&=&T_F^2\,n_f\,C_F
\Biggl\{
- \frac {1} {\ep^2} {\frac {16}{9}}
- \frac {1} {\ep} {\frac {20}{27}}
-\frac {2} {3} \,\,\zeta_2
-\frac {337}{81}
\Biggr\}~.
\end{eqnarray}
\subsection{\boldmath ${A^{\rm{NS,TR}}_{qq,Q}}$}
The contributions to~$\Ahathat_{qq,Q}^{\sf NS,TR}$ read:
\begin{eqnarray} 
\Ahathat_{qq,Q}^{(3),\rm{TR},a}&=& T_F^2 n_f C_F \Biggl\{
-\frac {1} {\ep^3} \frac{64}{27}
-\frac {1} {\ep^2} \frac{64}{81}
-\frac {1} {\ep} \Biggl[
\frac{32 (14 N^2+14 N-3)}{81 N (N+1)}
+\frac{8 }{9} \zeta_2
\Biggr]
\N\\
&&
-\frac{16 (308 N^4+616 N^3+263 N^2+9
   N+27)}{729 N^2 (N+1)^2}
-\frac{8 }{27} \zeta_2
\N\\
&&
+\frac{56 }{27} \zeta_3
\Biggr\}~.
\end{eqnarray}
\begin{eqnarray} 
\Ahathat_{qq,Q}^{(3),\rm{TR},b}&=& T_F^2 n_f C_F \Biggl\{
\frac {1} {\ep^3} \Biggl[
-\frac{128 }{27} S_1
+\frac{128}{27}
\Biggr]
+\frac {1} {\ep^2} \Biggl[
\frac{64 }{27} S_2
-\frac{320 }{81} S_1
+\frac{128}{81}
\Biggr]
\N\\
&&
+\frac {1} {\ep} \Biggl[
-\frac{32 }{27} S_3
-\frac{16 }{9} \zeta_2 S_1 
+\frac{160 }{81} S_2
+\frac{16 }{9} \zeta_2
-\frac{320}{27} S_1 
+\frac{896}{81}
\Biggr]
\N\\
&&
+\frac{16 }{27} S_4
+\frac{112 }{27} \zeta_3 S_1
+\frac{8 }{9} \zeta_2 S_2
-\frac{80 }{81} S_3
-\frac{40 }{27} \zeta_2 S_1
-\frac{112 }{27} \zeta_3
+\frac{160}{27} S_2
\N\\
&&
+\frac{16 }{27} \zeta_2
-\frac{13888}{729} S_1
+\frac{9856}{729}
\Biggr\}~.
\end{eqnarray}
\begin{eqnarray}
\Ahathat_{qq,Q}^{(3),\rm{TR},c}&=& T_F^2 n_f C_F \Biggl\{
-\frac {1} {\ep^2} \frac{16}{9}
-\frac {1} {\ep} \frac{20}{27}
-\frac{2}{3} \zeta_2
-\frac{337}{81}
\Biggr\}~.
\end{eqnarray}

\newpage
  \section{\boldmath Variable Transformations}
   \label{App-VarTrans}
   \setcounter{equation}{0}
In many cases the structure of the emerging Feynman parameter integrals can be simplified by applying 
transformations to the integration variables, which were given in Ref.~\cite{Hamberg:thesis}~.
      \begin{itemize} 
      \item
        To define the product $x'=xy$ as the new integration variable, 
        one maps:
        \begin{eqnarray}
         x'&:=&xy~, \quad \hspace{16.5mm} y':=\frac{x(1-y)}{1-xy}~, \N\\
         x&=&x'+y'-x'y'~,\quad y=\frac{x'}{y'+x'-x'y'}~,\N\\
         &&\frac{\partial(x,y)}{\partial(x',y')}=\frac{1-x'}{x'+y'-x'y'}~.
         \label{trafo1}
       \end{eqnarray}
       One obtains
       \begin{eqnarray}
        \int_0^1\int_0^1 dx dy ~~f(x,y)(xy)^N &=&
        \int_0^1\int_0^1 dx' dy'~~\frac{(1-x')(x')^N}{x'+y'-x'y'}
\N\\
&&                                   
f\left(y'+x'-x'y',
                                  \frac{x'}{x'+y'-x'y'}\right)~.
       \end{eqnarray}
      \item
       Terms of the form $(x-y)^N$ can be combined by \\ 
       %
       \begin{alignat}{3}
          \underline{x>y}:&                & \qquad     \underline{x<y}:&    \N\\
                   x'&:=x-y~,             & \qquad               x'&:=y-x~,    \N\\
                   y'&:=\frac{y}{1-x+y}~, & \qquad               y'&:=\frac{1-y}{1+x-y}~, \N\\
                    x&=x'+y'-x'y'~,        & \qquad                x&=(1-x')(1-y')~,\N\\
                    y&=(1-x')y'~,         & \qquad                y&=1-(1-x')y'~,\N\\
         \frac{\partial(x,y)}{\partial(x',y')}&=1-x'~. & \qquad   \frac{\partial(x,y)}{\partial(x',y')}&=1-x'~. \label{trafo2}
       \end{alignat} \\ 
%
%
       Thus, one obtains
       \begin{eqnarray}
\hspace{-25mm}
        \int_0^1\int_0^1 dx dy~~f(x,y)(x-y)^N\!& =&\!
        \int_0^1\int_0^1 dx' dy'~~
\N\\&&\hspace{5mm}
{x'}^N(1-x')\left[
                                f(y'+x'-x'y',(1-x')y')\N \right.
\N\\&&\hspace{5mm}
                                +(-1)^Nf((1-y')(1-x'),1-(1-x')y')\left.\right]~.
       \end{eqnarray}
      \end{itemize}
      If one applies Eq. (\ref{trafo2}) to factors $(x-y)$ 
      in the denominator, special care is needed to avoid
      possible divergences. The above transformations allow to simplify the Feynman parameter integrals analytically. In this way higher transcendental functions like generalized hypergeometric functions, which obey single-sum representations are obtained, cf. Eq. (\ref{fpq}).
\newpage
  \section{\bf \boldmath Special Functions}
   \label{App-SpeFun}
   \renewcommand{\theequation}{\thesection.\arabic{equation}}
   \setcounter{equation}{0}
    In the following we summarize for convenience some relations for special
    functions 
    which occur repeatedly in quantum field theory calculations and are mutually used within the present work.
   \subsection{The Euler Integrals}
   \label{App-SpeFunGA}
    The $\Gamma$-function, cf. \cite{stegun,Nielsen:1906}, is analytic 
    in the whole complex plane except at the non-positive
    integers, where it possesses single poles. Euler's infinite product defines 
    \begin{eqnarray}
     \frac{1}{\Gamma(z)}=z\exp(\gamma_Ez)
               \prod_{i=1}^{\infty} 
               \Biggl[\Bigl(1+\frac{z}{i}\Bigr)\exp\left(-\frac {z} {i}\right)\Biggr]~. 
               \label{eulerprod}
    \end{eqnarray}
   The residues of the $\Gamma$-function at its poles are given 
   by 
   \begin{eqnarray}
   {\sf Res}[\Gamma(z)]_{z=-N}=\frac{(-1)^N}{N!}~,\quad N \in {\mathbb{N}}\cup 0~.
       \label{gammares} 
   \end{eqnarray}
   In case of ${\sf Re}(z) > 0$, the $\Gamma$-function can be expressed by Euler's
   integral
   \begin{eqnarray}
    \Gamma(z)=\int_0^{\infty}dt \exp(-t)~t^{z-1}~,
   \end{eqnarray}
   from which one infers the well known
   functional equation of the $\Gamma$-function
   \begin{eqnarray}
    \Gamma(z+1)=z\Gamma(z)~,\label{funcrelgam}
   \end{eqnarray}
   which may be used for its analytic continuation.
   Around $z=1$, the following series expansion is obtained
   \begin{eqnarray}
    \Gamma(1-\ep)
      &=&\exp(\ep \gamma_E)
         \exp\Biggl\{\sum_{i=2}^{\infty}\zeta_i\frac{\ep^i}{i}\Biggr\}~,
      |\ep|<1~.
          \label{gammaser}
   \end{eqnarray}
   Here and in (\ref{eulerprod}), $\gamma_E$ denotes the Euler-Mascheroni 
   constant, cf. (\ref{gammaE}), and $\zeta_k$  
Riemann's $\zeta$--function for integer arguments $k$, cf. (\ref{zetn}).  
   A shorthand notation for rational functions of $\Gamma$--functions is
   \begin{eqnarray}
    \Gamma\Biggl[\frac[0pt]{a_1,...,a_i}{b_1,...,b_j}\Biggr]:=
    \frac{\Gamma(a_1)...\Gamma(a_i)}{\Gamma(b_1)...\Gamma(b_j)}~.
    \label{gammashort}
   \end{eqnarray}
   Functions closely related to the $\Gamma$-function 
   are  the Euler Beta-function $B(A,C)$ function, the $\psi(x)-$, 
   and the $\beta(x)$--function.

   The Beta-function can be defined by Eq.~(\ref{gammashort})
   \begin{eqnarray}
    B(A,C)=\Gamma\Biggl[\frac[0pt]{A,C}{A+C}\Biggr]~. \label{betafun1}
   \end{eqnarray}
   If ${\sf Re}(A),{\sf Re}(C) > 0 $, the following 
   integral representation is valid 
   \begin{eqnarray}
    B(A,C)=\int_0^1 dx~x^{A-1}(1-x)^{C-1}~. \label{betafun2}
   \end{eqnarray}
   For arbitrary values of $A$ and $C$, (\ref{betafun2})
   can be continued analytically outside of the respective singularities using Eqs. (\ref{eulerprod},~\ref{betafun1}).
   Its expansion around singularities can be performed 
   via Eqs. (\ref{gammares},~\ref{gammaser}).
   The $\psi$-function and $\beta(x)$ are defined as logarithmic derivatives of the 
   $\Gamma$-function via
   \begin{eqnarray}
    \psi(x)  &=& \frac{1}{\Gamma(x)} \frac{d}{dx} \Gamma(x)~, \label{psifun}\\
    \beta(x) &=& \frac{1}{2} \left[
                  \psi\Bigl(\frac{x+1}{2}\Bigr)
                -  \psi\Bigl(\frac{x}{2}\Bigr)\right]~. \label{smallbeta}
   \end{eqnarray}
   \subsection{\boldmath The Generalized Hypergeometric Functions}
   \label{App-SpeFunFPQ}
   The generalized hypergeometric function $\empty_{P}F_Q$ is defined by,
   cf. \cite{Slater,Bailey,Roy:2001}, 
   \begin{eqnarray}
    \empty_{P}F_Q\Biggl[\frac[0pt]{a_1,...,a_P}
                                  {b_1,...,b_Q}
                                  ;z\Biggr]
    =\sum_{i=0}^{\infty}
     \frac{(a_1)_i...(a_P)_i}
          {(b_1)_i...(b_Q)_i}
          \frac{z^i}{\Gamma(i+1)}~.
     \label{fpq}
   \end{eqnarray}
   Here $(c)_n$ denotes Pochhammer's symbol
   \begin{eqnarray}
    (c)_n=\frac{\Gamma(c+n)}{\Gamma(c)} \label{pochhammer}~, 
   \end{eqnarray} 
   for which the following relation holds
   \begin{eqnarray}
    (N+1)_{-i}&=&\frac{(-1)^i}{(-N)_i}~,~N\in~\mathbb{N}~.  \label{reflect}
   \end{eqnarray}
   In (\ref{fpq}), there are $P$ numerator parameters $a_1...a_P$, 
   $Q$ denominator parameters $b_1...b_Q$, and one variable 
   $z$, all of which may be real or complex. Additionally, the 
   denominator parameters must not be negative integers, since in 
   that case (\ref{fpq}) is not defined. The generalized 
   hypergeometric series $\empty_{P}F_Q$ are evaluated 
   at a certain value of $z$, which is always $z=1$ for the
   final expressions in this thesis.  \\
   Gau\ss{} was the first to study this 
   kind of functions, introducing the function $\empty_2F_1$, and proving 
   the theorem, cf. \cite{Slater},
   \begin{eqnarray} 
    \empty_{2}F_1[a,b;c;1]=
    \Gamma\Biggl[\frac[0pt]{c,c-a-b}{c-a,c-b}\Biggr]
                 \label{Gauss}~, \quad {\sf Re}(c-a-b)>0~, 
   \end{eqnarray}
  which is called Gau\ss' theorem.
  An integral representation for the hypergeometric Gau\ss function 
  is given by
  \begin{eqnarray}
   \empty_2F_1\Biggl[\frac[0pt]{a,b+1}{c+b+2};z\Biggl]=
   \Gamma\Biggl[\frac[0pt]{c+b+2}{c+1,b+1}\Biggr]
   \int_0^1 dx~x^{b}(1-x)^c(1-zx)^{-a}~,\label{pochint}
  \end{eqnarray}
cf. \cite{Slater}, provided that the conditions 
  \begin{eqnarray}
    |z|< 1~,\quad  {\sf Re}(c+1),~{\sf Re}(b+1)~> 0~,\label{condpoch}
  \end{eqnarray}
  hold. 
  Applying Eq. (\ref{pochint}) recursively, one obtains the following 
  integral representation for the general hypergeometric function $\empty_{P+1}F_P$:
  \begin{eqnarray}
   &&\empty_{P+1}F_P\Biggl[\frac[0pt]{a_0,a_1,\ldots ,a_P}
                                  {b_1,\ldots ,b_P}
                                  ;z\Biggr]
    =
        \Gamma\Biggl[\frac[0pt]{b_1,\ldots ,b_P}
                             {a_1,\ldots ,a_P,b_1-a_1,\ldots ,b_P-a_P}\Biggr] 
        \times
\N\\ &&~
                \int_0^1dx_1\ldots \int_0^1dx_P~~
              x_1^{a_1-1}(1-x_1)^{b_1-a_1-1}\ldots x_P^{a_P-1}(1-x_P)^{b_P-a_P-1} 
                 (1-zx_1\ldots x_P)^{-a_0}~,\N \\ \label{FPQint}
  \end{eqnarray}
  under similar conditions as in Eq. (\ref{condpoch}).
As generalized hypergeometric functions appeared frequently during this computation, taking advantage of their properties is of importance.

If one considers the fraction 
\begin{eqnarray}
\frac {\left(m\right)_i} {\left(n\right)_i}~,
\end{eqnarray}
with integers $m>n>1$ one may transform the arguments in the following way by using the definition of Pochhammer's symbol~:
\begin{eqnarray}
\frac {\left( m \right)_i} {\left( n \right)_i}&=&\frac{\Gamma(n)} {\Gamma(m)} ~\frac {\Gamma(m+i-1)} {\Gamma(n+i-1)} ~\frac {m+i-1} {n+i-1}
\N\\
&=& \frac{\Gamma(n)} {\Gamma(m)} ~ \frac {\Gamma(m+i-1)} {\Gamma(n+i-1)} \left(1+\frac {m-n} {n-1+i} \right)
\N\\
&=& \frac{\Gamma(n) \Gamma(m-1)} {\Gamma(m)} \left(m-1\right)_i 
\left[ \frac {1} {\left(n-1\right)_i \Gamma(n-1)} 
+ \frac {(m-n) \Gamma(N)} {\left(n \right)_i}\right]
\N\\
&=& \frac {m-1} {n-1} ~ \frac {\left(m-1\right)_i} {\left(m-1\right)_i} + \frac {m-n} {m-1} ~ \frac {\left(m-1\right)_i} {\left(n\right)_i}
\end{eqnarray}
This relation can be applied to an arbitrary function $_PF_Q$ and proves to be especially useful if one considers the generalized hypergeometric function $_3F_2$ of the form 
$\empty_{3}F_2\Biggl[\frac[0pt]{a_1,a_2,m}
                                  {b_1,n}
                                  ;1\Biggr]~, $ with $m,n$ integers, $n>m>0$.
In this case repeated application of 
\begin{eqnarray}
\empty_{3}F_2\Biggl[\frac[0pt]{a_1,a_2,m}
                                  {b_1,n}
                                  ;z\Biggr]&=&\frac {n-1} {m-1} ~\empty_{3}F_2\Biggl[\frac[0pt]{a_1,a_2,m-1}
                                  {b_1,n-1};1\Biggr]
\N\\
&&+\frac {m-n} {m-1} ~\empty_{3}F_2\Biggl[\frac[0pt]{a_1,a_2,m-1}
                                  {b_1,n}
                                  ;1\Biggr] \label{3F2mn}
\end{eqnarray}
leads to a linear combination of terms of the form 
\begin{eqnarray}
_3F_2\Biggl[\frac[0pt]{a_1,a_{2},1} {b_1,n};1\Biggr]
\label{3F2_1UP}
\end{eqnarray} 
and terms that do not contain any sum any longer. These relations belong to the class of contiguous 
relations \cite{Slater,Roy:2001,PAULE,Vidunas:2003,Kalmykov:2009tw}.
If $a_1,a_2,b_1$ are non--integers or $a_1,a_2,b_1>n-1$ (\ref{3F2_1UP}) can be simplified by considering the following relations:
\begin{eqnarray}
\frac {1} {\left(n\right)_i}&=&\frac {\Gamma(n)} {\Gamma(n+i)}
\label{rel1}
=\frac {\Gamma(n)} {(n+i-1)!}~,\\
\left(A\right)&=&\frac {\Gamma(A+i)} {\Gamma(A)}\N\\
&=& \frac {\Gamma(A+i-(n-1)+(n-1))} {\Gamma(A)}\N\\
&=&\frac {\Gamma(A-(n-1))} {\Gamma(A)} \left(A+1-n\right)_{i+n-1} \label{rel2}~.
\end{eqnarray}
After applying (\ref{rel1}) and (\ref{rel2}) one obtains
\begin{eqnarray}
_3F_2\Biggl[\frac[0pt]{a_1,a_{2},1} {b_1,n};1\Biggr]&=&\sum_{i=0}^{\infty}
     \frac{(a_1)_i(a_{2})_i}
          {(b_1)_i}
          \frac{1}{(n)_i}
\N\\
&=&
\frac {\Gamma(a_1+1-n) \Gamma(a_2+1-n) \Gamma(b_1)} {\Gamma(a_1) \Gamma(a_2) \Gamma(b_1+1-n)}
\N\\
&&
\sum_{i=0}^{\infty}
     \frac{(a_1+1-n)_{i}(a_{2}+1-n)_i}
          {(b_1+1-n)_i}
          \frac{1}{(n+i-1)!}
\N\\
&=&
\frac {\Gamma(a_1+1-n) \Gamma(a_2+1-n) \Gamma(b_1)} {\Gamma(a_1) \Gamma(a_2) \Gamma(b_1+1-n)} 
\N\\
&&
\times
\Biggl\{
~_2F_1\Biggl[\frac[0pt]{a_1+1-n,a_{2}+1-n} {b_1+1-n};1\Biggr]
\N\\
&&
- \sum_{i=0}^{n-2}
     \frac{(a_1+1-n)_{i}(a_{2}+1-n)_i}
          {(b_1+1-n)_i}
          \frac{1}{(i)!} \label{3F21n}
\Biggr\}~.
\end{eqnarray}
The infinite sum contained inside the hypergeometric function in (\ref{3F21n}) can now be evaluated by applying Gau\ss' theorem, (\ref{Gauss}), such that no infinite sum remains. The relations (\ref{3F2mn}) and (\ref{3F21n}) have been implemented into {\sf FORM}-algorithms and in many cases made it possible to perform the infinite sums before expanding in the dimensional regularization parameter $\ep$.
A similar useful relation is the following, which has been given in Ref. \cite{Coffey:2005} and holds under the condition, that ${\sf Re}(e-a-b-1-k)>0~:$
\begin{eqnarray}
     \empty_{3}F_2\Biggl[\frac[0pt]{a,b,d+k}{d,e};1\Biggr]
               =\Gamma\Biggl[\frac[0pt]{e}{e-a,e-b}\Biggr]
          \Biggl[\sum_{i=0}^{k}\Gamma(e-a-b-i)\binom{k}{i}\frac{(a)_i(b)_i}
           {(d)_i}\Biggr] \label{corollary1}~.
\end{eqnarray}
The following relations are restricted to special cases, but prove often to be useful. 
One defines the parametric excess of the series by $s:=d+e-a-b-c$. 
  Saalsch\"utz's theorem, cf. \cite{Slater}, states that 
  \begin{eqnarray}
   \empty_3F_2\Biggl[\frac[0pt]{a,b,c}{d,e};1\Biggr]
   =\Gamma\Biggl[\frac[0pt]{d,1+a-e,1+b-e,1+c-e}{1-e,d-a,d-b,d-c}\Biggr]~,
   \label{saalschutz}
  \end{eqnarray}
  provided that $s=1$, i.e. the series is Saalsch\"utzian, and one 
  of the numerator parameters is equal to a negative integer. Another theorem is a generalization of Dixon's theorem, \cite{Slater},
  \begin{eqnarray}
   \empty_3F_2\Biggl[\frac[0pt]{a,b,c}{d,e};1\Biggr]
   =\Gamma\Biggl[\frac[0pt]{d,e,s}{a,b+s,c+s}\Biggr]~
    \empty_{3}F_2\Biggl[\frac[0pt]{d-a,e-a,s}{s+b,s+c};1\Biggr]~.
    \label{dixonth}
  \end{eqnarray}   
   \subsection{Harmonic Sums and Nielsen--Integrals}
   \label{App-SpeFunHarm}
    Expanding the $\Gamma$--function in $\ep$, 
    its logarithmic derivatives, the $\psi^{(k)}$-functions, emerge. 
    In many applications of perturbative QCD and 
    QED, harmonic sums occur, cf. \cite{Blumlein:1998if,Vermaseren:1998uu},
    which can be considered as generalization 
    of the $\psi$-function and the $\beta$-function. 
    These are defined by
    \begin{eqnarray}
      S_{a_1, \ldots, a_m}(N)&=&
        \sum_{n_1=1}^N  \sum_{n_2=1}^{n_1} \ldots \sum_{n_m=1}^{n_{m-1}} 
        \frac{({\rm sign}(a_1))^{n_1}}{n_1^{|a_1|}}
        \frac{({\rm sign}(a_2))^{n_2}}{n_2^{|a_2|}} \ldots
        \frac{({\rm sign}(a_m))^{n_m}}{n_m^{|a_m|}}~, \N\\
        && N~\in~{\mathbb{N}},~\forall~l~a_l~\in {\mathbb{Z}}\setminus 0~, 
      \label{harmdef} \\
     S_{\emptyset}&=&1~.
    \end{eqnarray}
    We adopt the convention
    \begin{eqnarray}
     S_{a_1, \ldots ,a_m} \equiv S_{a_1, \ldots ,a_m}(N)~, 
    \end{eqnarray}
    i.e. harmonic sums are taken at argument $N$, if no argument is 
    indicated.
    Related quantities are the $Z$--sums defined by 
    \begin{eqnarray}
        \label{SZSums}
     Z_{m_1, \ldots, m_k}(N) &= \sum_{N \geq i_1 > i_2 \ldots > i_k > 0}
     \D{\frac {\prod_{l=1}^k [{\rm sign}(m_l)]^{i_l}}{i_l^{|m_l|}}}~.
    \end{eqnarray}
    The depth $d$ and the weight $w$ of a harmonic sum are
    defined by
    \begin{eqnarray}
     d&:=&m~,\label{depth} \\
     w&:=&\sum_{i=1}^m |a_i| ~.\label{weight}
    \end{eqnarray}
    Harmonic sums of depth $d=1$ are referred to as single harmonic sums. 
    The complete set of algebraic relations connecting harmonic 
    sums to other harmonic sums of the same or lower weight 
    is known~\cite{Blumlein:2003gb}. Thus the number of 
    independent harmonic sums can be reduced significantly,
    e.g., up to $w=4$ the $80$ possible harmonic 
    sums can be expressed algebraically in terms of $31$ basic harmonic 
    sums only \cite{Blumlein:2009ta}.
    One introduces a product for the harmonic sums, 
    the shuffle product \SH, cf.~\cite{Blumlein:2003gb}. 
    For the product of a single and
    a general finite harmonic sum it is given by
    \begin{eqnarray}
      S_{a_1}(N) \SH S_{b_1, \ldots, b_m}(N)
       = S_{a_1, b_1, \ldots, b_m}(N) 
        + S_{b_1, a_1, b_2, \ldots, b_m}(N)
         + \ldots + S_{b_1, b_2,  \ldots, b_m, a_1}(N)~. \N\\
         \label{shuffle}  
    \end{eqnarray}
    For sums $S_{a_1, \ldots, a_n}(N)$ and 
    $S_{b_1, \ldots,b_m}(N)$ of arbitrary depth, the shuffle 
    product is then
     the sum of all harmonic sums of depth $m+n$ in the index set of 
     which $a_i$ occurs left of $a_j$ for $i < j$, likewise for
     $b_k$ and $b_l$ for $k < l$. Note that the shuffle 
     product is symmetric. \\
     One shows that the following relation holds, cf.~\cite{Blumlein:2003gb}, 
     \begin{eqnarray}
      S_{a_1}(N) \cdot S_{b_1, \ldots, b_m}(N)
      &=& S_{a_1}(N) \SH S_{b_1, \ldots, b_m}(N) \N\\
      & & -S_{a_1 \wedge b_1, b_2, \ldots, b_m}(N) - \ldots -
          S_{b_1, b_2, \ldots, a_1 \wedge b_m}(N)~,
          \label{genshuff}
     \end{eqnarray}
     where the $\wedge$ symbol is defined as
     \begin{eqnarray}
      a \wedge b = \si(a) \si(b) \left(|a| + |b|\right)~.
      \label{wedge}
     \end{eqnarray}
     Due to the additional terms containing wedges ($\wedge$) between 
     indices, harmonic sums form a quasi--shuffle 
     algebra, \cite{Hoffman:1997,Hoffman:2004bf}.
     By summing (\ref{genshuff}) over permutations, one 
     obtains the symmetric algebraic relations between harmonic sums.  
     At depth $2$ and $3$ these read,~\cite{Blumlein:1998if},
     \begin{eqnarray}
      S_{m,n} + S_{n,m} &=&  S_m S_n + S_{m \wedge n}~, \label{algrel1}\\
      \sum_{{\rm perm}\{l,m,n\}} S_{l,m,n} &=& S_l S_m S_n 
      + \sum_{{\rm inv~perm}\{l,m,n\}} S_{l}
      S_{m \wedge n}
      + 2~S_{l \wedge m \wedge n}~, \label{algrel2}
     \end{eqnarray}
     \normalsize
     which we used extensively to simplify our expressions. 
     In (\ref{algrel1},~\ref{algrel2}), 
     ``{\sf perm}'' denotes all permutations and ``{\sf inv perm}''
     invariant ones.

     The limit $N \rightarrow \infty$ of finite harmonic sums exists
     only if $a_1\neq 1$ in (\ref{harmdef}). Additionally, one defines
     all $\sigma$-values symbolically as
     \begin{eqnarray}
      \sigma_{k_l, \ldots, k_1} = \lim_{N \rightarrow \infty}
      S_{a_1, \ldots, a_l}(N)~. \label{sigmaval}
     \end{eqnarray}
     The finite 
     $\sigma$-values are related to multiple 
$\zeta$-values,~\cite{Blumlein:1998if,Vermaseren:1998uu,Euler:1775,Zagier:1994,Borwein:1999js,Blumlein:2009cf,
ABS2010}. Further we define the symbols
     \begin{eqnarray}
      \sigma_0&:=&\sum_{i=1}^{\infty}\frac {1} {i}~,\\
      \sigma_1&:=&\sum_{i=1}^{\infty}1~.
     \end{eqnarray}
     It is useful to include these $\sigma$-values into the algebra, since they
     allow to treat parts of sums individually, accounting 
     for the respective divergences, cf. also \cite{Blumlein:1998if,Vermaseren:1998uu,ABS2010}.
     These divergent pieces
     cancel in the end if the overall sum is finite. 

     The relation of single harmonic sums with positive 
     or negative indices to the $\psi^{(k)}$--functions is then given by
     \begin{eqnarray}
      S_1(N)  &=&\psi(N+1)+\gamma_E~,\label{s1psi}    \\
      S_a(N)  &=&\frac{(-1)^{a-1}}{\Gamma(a)}\psi^{(a-1)}(N+1)
                 +\zeta_a~, k \ge 2~,\label{sapsi}    \\
      S_{-1}(N)&=&(-1)^N\beta(N+1)-\ln(2)~,\label{sm1beta}\\
      S_{-a}(N)&=&-\frac{(-1)^{N+a}}{\Gamma(a)}\beta^{(a-1)}(N+1)-
                  \left(1-2^{1-a}\right)\zeta_a~, 
                 \label{smabeta}k \ge 2~.
     \end{eqnarray}
     Single harmonic sums can be analytically continued 
     to complex values of $N$ by these relations. 
     At higher depths, harmonic sums can be expressed in terms 
     of Mellin--transforms of polylogarithms and the more general 
     Nielsen-integrals,~\cite{Nielsen:1909,Kolbig:1983qt,Devoto:1983tc}. 
     The latter are defined by
     \begin{eqnarray}
      {\rm S}_{n,p}(z) = \frac{(-1)^{n+p-1}}{(n-1)! p!} \int_0^1 \frac{dx}{x}
      \log^{n-1}(x) \log^p(1-zx)~ \label{nielsenint}
     \end{eqnarray}
     and fulfill the relation
     \begin{eqnarray}
      \frac{d \SN_{n,p}(x)}{d \log(x)} = \SN_{n-1,p}(x)~.
     \end{eqnarray}
     If $p=1$, one obtains the polylogarithms
     \begin{eqnarray}
      \Li_n(x) =  \SN_{n-1,1}(x)~,
     \end{eqnarray}
     where 
     \begin{eqnarray}
      \Li_0(x) =  \frac{x}{1-x}~.
     \end{eqnarray}
     These functions
     do not suffice for arbitrary harmonic sums, in which case 
     the harmonic polylogarithms have to be considered, \cite{Remiddi:1999ew}.
     The latter functions obey a direct shuffle algebra, cf. 
     \cite{Borwein:1999js,Blumlein:2003gb,Blumlein:2009ta,Blumlein:2009fz,ABS2010}.
     The representation in terms of Mellin--transforms then allows 
     an analytic continuation of arbitrary harmonic sums to complex $N$, 
     cf. \cite{Carlson:thesis,Titchmarsh:1939,Blumlein:2000hw,Blumlein:2005jg}.
     Equivalently, one may express harmonic sums by factorial 
     series, \cite{Nielsen:1906,Knopp:1947,Landau:1906}, up to polynomials 
     of $S_1(N)$ and harmonic sums of lower degree, and use this 
     representation for the analytic continuation to $N \in \mathbb{C}$, cf. 
     \cite{Blumlein:2009ta,Blumlein:2009fz,ABS2010}.
\newpage
  \section{\bf Examples for Sums}
\label{SEC:SUMS}
  \renewcommand{\theequation}{\thesection.\arabic{equation}}
   \setcounter{equation}{0}

\vspace{1mm}\noindent
In the present calculation numerous single-- to triple finite and infinite
sums of an extension of the hypergeometric type had to be calculated. For these sums, depending
on various summation parameters, $n_i$, the ratio of the summands, except the part containing harmonic 
sums,
\begin{eqnarray}
\frac{a(...,n_i +1, ...)}
     {a(...,n_i, ...)},~~~~~\forall i
\end{eqnarray}
is a rational function in all variables $n_i$. Sums of this type can be represented
by basic sums of a certain type, which are transcendental to each other and form
sum-- and product--fields, cf.~\cite{Refined,Schneider:2007,sigma1,sigma2} and 
references therein.
The general form of these sums is
\begin{eqnarray}
\sum_{k_1 = 1}^{N_1(N)}
\sum_{k_2 = 1}^{N_2(k_1,N)}
\sum_{k_3 = 1}^{N_3(k_2,k_1,N)}
R(k_1, k_2, k_3, N) \prod_{l=1}^4 S_{\vec{a_l}}(s(k_i,N)) \Gamma\left[\frac[0pt]
{s_1(k_i,N) ... s_p(k_i,N)}
{s_1(k_i,N) ... s_q(k_i,N)} \right]~,
\end{eqnarray}
with $R$ a rational function, $s(k_i,N)$ a linear combination of the arguments
with weight ${\pm 1}$, $\vec{a_l}$ an index set, $p, q \in \mathbb{N}$,~$N_i \in
\mathbb{N} \cup \infty$. The generalized $\Gamma$--function usually includes both Beta--functions and
binomials.

In the present calculation one faces more complicated sums than occurring in earlier two--loop
calculations up to $O(\varepsilon)$, \cite{Bierenbaum:2007qe,Bierenbaum:2008yu}. 
We present a few examples.

\begin{eqnarray}
&& \sum_{j_1=1}^{N-2} \sum_{n=1}^{\infty }
(-1)^{j_1} B(n,N-j_1) \binom{N-2}{j_1} \frac{S_2(-j_1+n+N)}{n^2 (j_1-N-2)} =
\N\\
&&
\Biggl\{(-1)^N \frac{6-23 N+9 N^2+2 N^3}{2 (N-1)^2 N^2 (1+N) (2+N)}
+\Biggl[\frac{1}{N+2}-\frac{27 (-1)^N}{(N-1) N (N+1) (N+2)}\Biggr] S_1
\N\\
&&
-\frac{1}{N (N+2)}\Biggr\} S_2^2
\N\\
&&
+\Biggl[\frac{1}{N+2}-\frac{48 (-1)^N}{(N-1) N (N+1) (N+2)}\Biggr] S_3 S_2-\frac{2 S_{-2}^2}{N (N+2)}
\N\\
&&
+\Biggl\{-(-1)^N \frac{7 \big(12+6 N-37 N^2+6 N^3+N^4\big)}{20 (-1+N)^2 N^2 (1+N) (2+N)}
+\Biggl[-(-1)^N \frac{21}{5 (-1+N) N (1+N) (2+N)}
\N\\
&&
-\frac{7}{10 (N+2)}\Biggr] S_1+\frac{7}{10 N (N+2)}\Biggr\} \zeta_2^2+\Biggl\{(-1)^N \frac{6-23 N+9
N^2+2 N^3}{2 (-1+N)^2 N^2 (1+N) (2+N)}
\N\\
&&
+\Biggl[\frac{3 (-1)^N}{(N-1) N (N+1) (N+2)}+\frac{3}{N+2}\Biggr] S_1-\frac{3}{N (N+2)}\Biggr\} S_4
\N\\
&&
+ \Biggl[\frac{3}{N+2}-\frac{18 (-1)^N}{(N-1) N (N+1) (N+2)}\Biggr] S_5
+ \Biggl[\frac{2 S_{-2}^2}{N+2}+\frac{(-1)^N (3 N-1)}{(N-1)^3 N^3}\Biggr] S_1
\N\\
&&
+\frac{2}{2+N} S_{-2} S_{-3} + \Biggl\{-(-1)^N \frac{3 \big(12-6 N-14 N^2+7 N^3+12
N^4+N^5\big)}{(-1+N)^3 N^3 (1+N) (2+N)}
\N
\end{eqnarray}
\begin{eqnarray}
&&
+(-1)^N \frac{9 S_1^2}{(-1+N) N (1+N) (2+N)}+(-1)^N
\frac{3 \big(6-23 N+9 N^2+2 N^3\big)}{(-1+N)^2 N^2 (1+N) (2+N)} S_1
\N\\
&&
+\Biggl[\frac{3 (-1)^N}{(N-1) N (N+1) (N+2)}-\frac{1}{N+2}\Biggr] S_2\Biggr\} S_{2,1}
\N\\
&&
+\Biggl[(-1)^N \frac{2 \big(12-37 N+9 N^2+4 N^3\big)}{(-1+N)^2 N^2 (1+N) (2+N)}+(-1)^N \frac{24}{(-1+N)
N (1+N) (2+N)} S_1\Biggr] S_{3,1}
\N\\
&&
+\frac{2 S_{3,2}}{N+2}+\Biggl[-\frac{12 (-1)^N}{(N-1) N (N+1) (N+2)}-\frac{3}{N+2}\Biggr]
S_{4,1}
\N\\
&&
- \frac{2 S_{-2} S_{-2,1}}{2+N}
+ \frac{4 S_{-3,-2}}{N+2}+\Biggl[-(-1)^N \frac{2 \big(6-23 N+9 N^2+2 N^3\big)}{(-1+N)^2 N^2 (1+N)
(2+N)}
\N\\
&&
-(-1)^N \frac{12}{(-1+N) N (1+N) (2+N)} S_1\Biggr] S_{2,1,1}
-\frac{2 S_{-2,1,-2}}{N+2}
\N\\
&&
+(-1)^N \frac{1}{(-1+N) N (1+N) (2+N)} \Biggl[ 42 S_{2,2,1}
- 24 S_{3,1,1} + 54 S_{2,1,1,1} \Biggr]
\N\\
&&
-(-1)^N \frac{30}{(-1+N) N (1+N) (2+N)} S_3 \tilde{S}_1\left(\frac{1}{2}\right) \tilde{S}_1(2)
\N\\
&&
+(-1)^N \frac{30}{(-1+N) N (1+N) (2+N)} S_1 \tilde{S}_1\left(\frac{1}{2}\right) \tilde{S}_3(2)
+\Biggl\{\frac{(-1)^N 2^{N+2}}{(-1+N)^3 N}
\N\\
&&
+\Biggl[-(-1)^N \frac{2 \big(6-12 N+7 N^2+N^3\big)}{(-1+N)^3 N^3}+(-1)^N \frac{6 S_1^2}{(-1+N) N (1+N)
(2+N)}
\N\\
&&
+(-1)^N \frac{2 \big(6-23 N+9 N^2+2 N^3\big)}{(-1+N)^2 N^2 (1+N) (2+N)} S_1\Biggr] \tilde{S}_1(2)
\N\\
&&
+\Biggl[(-1)^N \frac{2 \big(6-23 N+9 N^2+2 N^3\big)}
{(-1+N)^2 N^2 (1+N) (2+N)}+(-1)^N \frac{12}{(-1+N)
N (1+N) (2+N)} S_1\Biggr] \tilde{S}_2(2)
\N\\
&&
+(-1)^N \frac{6}{(-1+N) N (1+N) (2+N)} \tilde{S}_3(2)\Biggr\} \tilde{S}_{1,1}\left(\frac{1}{2},1\right)
\N\\
&&
+\Biggl\{\Biggl[(-1)^N \frac{12 S_1^2}{(-1+N) N (1+N) (2+N)}
+(-1)^N \frac{2 \big(6-23 N+9 N^2+2 N^3\big)}{(-1+N)^2 N^2 (1+N) (2+N)} S_1\Biggr]
\tilde{S}_1\left(\frac{1}{2}\right)
\N\\
&&
+\Biggl[-(-1)^N \frac{2 \big(6-23 N+9 N^2+2 N^3\big)}{(-1+N)^2 N^2 (1+N) (2+N)}
\N\\
&&
-(-1)^N \frac{12}{(-1+N) N (1+N) (2+N)} S_1\Biggr] \tilde{S}_{1,1}\left(\frac{1}{2},1\right)\Biggr\}
\tilde{S}_{1,1}(2,1)
\N\\
&&
+\Biggl[-(-1)^N \frac{2 \big(6-12 N+7 N^2+N^3\big)}{(-1+N)^3 N^3}
+(-1)^N \frac{6 S_1^2}{(-1+N) N (1+N) (2+N)}
\N\\
&&
+(-1)^N \frac{2 \big(6-23 N+9 N^2+2 N^3\big)}{(-1+N)^2 N^2 (1+N) (2+N)} S_1
\N
\end{eqnarray}
\begin{eqnarray}
&&
+(-1)^N \frac{30}{(-1+N) N (1+N) (2+N)} S_2\Biggr] \tilde{S}_{1,2}\left(\frac{1}{2},2\right)
\N\\
&&
-(-1)^N \frac{6}{(-1+N) N (1+N) (2+N)} \tilde{S}_{1,1}\left(\frac{1}{2},1\right) \tilde{S}_{1,2}(2,1)
\N\\
&&
+\Biggl[(-1)^N \frac{2 \big(6-23 N+9 N^2+2 N^3\big)}{(-1+N)^2 N^2 (1+N) (2+N)}
+(-1)^N \frac{12}{(-1+N) N (1+N) (2+N)} S_1\Biggr]
\N\\ &&
\times \left[
\tilde{S}_{1,3}\left(\frac{1}{2},2\right)
- \tilde{S}_{1,3}\left(2,\frac{1}{2}\right) \right]
\N\\
&&
+ \frac{(-1)^N}{(-1+N) N (1+N) (2+N)} \Biggl\{ 30 \tilde{S}_1\left(\frac{1}{2}\right)
\tilde{S}_{1,3}(2,1)
                                           + 36 \tilde{S}_{1,4}\left(\frac{1}{2},2\right)
\N\\
&&
                                 +  \Biggl[  - 30 S_1 \tilde{S}_1\left(\frac{1}{2}\right)
                                           - 30 \tilde{S}_2\left(\frac{1}{2}\right)
                                           + 30 \tilde{S}_{1,1}\left(\frac{1}{2},1\right)\Biggr]
\tilde{S}_{2,1}(1,2)
\Biggr\}
\N
\\&&
+\Biggl\{\Biggl[(-1)^N \frac{2 \big(6-23 N+9 N^2+2 N^3\big)}{(-1+N)^2 N^2 (1+N) (2+N)}
\N\\
&&
+(-1)^N \frac{24}{(-1+N) N (1+N) (2+N)} S_1\Biggr] 
\tilde{S}_1\left(\frac{1}{2}\right) \Biggr\} \tilde{S}_{2,1}(2,1)
\N\\
&&
+
\frac{(-1)^N}{(-1+N) N (1+N) (2+N)} \Biggl[
- 12 \tilde{S}_{1,1}\left(\frac{1}{2},1\right) \tilde{S}_{2,1}(2,1)
+ 30 \tilde{S}_{2,3}\left(\frac{1}{2},2\right)
\N\\
&&
- 12 \tilde{S}_{2,3}\left(2,\frac{1}{2}\right)
- 18 \tilde{S}_1\left(\frac{1}{2}\right) \tilde{S}_{3,1}(2,1)
+ 30 \tilde{S}_{3,2}\left(\frac{1}{2},2\right)
- 30 \tilde{S}_{4,1}\left(\frac{1}{2},2\right) \Biggr]
\N\\
&&
+\Biggl[(-1)^N \frac{2 \big(6-12 N+7 N^2+N^3\big)}{(-1+N)^3 N^3}
-(-1)^N \frac{6 S_1^2}{(-1+N) N (1+N) (2+N)}
\N
\\&&
-(-1)^N \frac{2 \big(6-23 N+9 N^2+2 N^3\big)}{(-1+N)^2 N^2 (1+N) (2+N)} S_1
\N\\
&&
-(-1)^N \frac{30}{(-1+N) N (1+N) (2+N)} S_2\Biggr] \Biggl[
\tilde{S}_{1,1,1}\left(\frac{1}{2},1,2\right)
+
\tilde{S}_{1,1,1}\left(\frac{1}{2},2,1\right) \Biggr]
\N\\
&&
+\Biggl[-(-1)^N \frac{2 \big(6-23 N+9 N^2+2 N^3\big)}{(-1+N)^2 N^2 (1+N) (2+N)}
\N\\
&&
-(-1)^N \frac{12}{(-1+N) N (1+N) (2+N)} S_1\Biggr] \tilde{S}_1\left(\frac{1}{2}\right)
\tilde{S}_{1,1,1}(1,2,1)
\N\\
&&
+(-1)^N \frac{12}{(-1+N) N (1+N) (2+N)} \tilde{S}_{1,1}\left(\frac{1}{2},1\right)
\tilde{S}_{1,1,1}(2,1,1)
\N\\
&&
+\Biggl[-(-1)^N \frac{2 \big(6-23 N+9 N^2+2 N^3\big)}{(-1+N)^2 N^2 (1+N) (2+N)}
-(-1)^N \frac{12}{(-1+N) N (1+N) (2+N)} S_1\Biggr]
\N\\
&& \times
\Biggl[
     \tilde{S}_{1,1,2}\left(\frac{1}{2},1,2\right)
   - \tilde{S}_{1,1,2}\left(2,\frac{1}{2},1\right)
 - 2 \tilde{S}_{1,1,2}\left(2,1,\frac{1}{2}\right)
\Biggr]
\N
\end{eqnarray}
\begin{eqnarray}
&&
-\frac{(-1)^N}{(-1+N) N (1+N) (2+N)}
\Biggl[
  66 \tilde{S}_{1,1,3}\left(\frac{1}{2},1,2\right)
+ 36 \tilde{S}_{1,1,3}\left(\frac{1}{2},2,1\right)
\N\\
&&
+ 30 \tilde{S}_{1,1,3}\left(1,\frac{1}{2},2\right)
+ 30 \tilde{S}_{1,1,3}\left(1,2,\frac{1}{2}\right)
\Biggr]
\N\\
&&
+\Biggl[-(-1)^N \frac{4 \big(6-23 N+9 N^2+2 N^3\big)}{(-1+N)^2 N^2 (1+N) (2+N)}
         -(-1)^N \frac{24}{(-1+N) N (1+N) (2+N)} S_1\Biggr]
\N\\
&&
 \times \Biggl[ \tilde{S}_{1,2,1}\left(\frac{1}{2},2,1\right)
-\tilde{S}_{1,2,1}\left(2,\frac{1}{2},1\right)
- \frac{1}{2}
 \tilde{S}_{1,2,1}\left(2,1,\frac{1}{2}\right)
\Biggr]
\N\\
&&
-(-1)^N \frac{12}{(-1+N) N (1+N) (2+N)} \tilde{S}_1\left(\frac{1}{2}\right) \tilde{S}_{1,2,1}(1,2,1)
\N\\ &&
-(-1)^N \frac{1}{(-1+N) N (1+N) (2+N)} \Biggl[
  30 \tilde{S}_{1,2,2}\left(\frac{1}{2},1,2\right)
 +36 \tilde{S}_{1,2,2}\left(\frac{1}{2},2,1\right)
\nonumber\\
&&
 +48 \tilde{S}_{1,3,1}\left(\frac{1}{2},2,1\right)
 +30 \tilde{S}_{1,3,1}\left(1,\frac{1}{2},2\right)
+ 30 \tilde{S}_{1,3,1}\left(1,2,\frac{1}{2}\right)
+ 30 \tilde{S}_{2,1,2}\left(\frac{1}{2},2,1\right)
\N\\
&&
+ 30 \tilde{S}_{2,1,2}\left(1,\frac{1}{2},2\right)
- 12 \tilde{S}_{2,1,2}\left(2,\frac{1}{2},1\right)
- 24 \tilde{S}_{2,1,2}\left(2,1,\frac{1}{2}\right)
+ 30 \tilde{S}_{2,2,1}\left(1,2,\frac{1}{2}\right)
\N\\
&&
- 24 \tilde{S}_{2,2,1}\left(2,\frac{1}{2},1\right)
- 12 \tilde{S}_{2,2,1}\left(2,1,\frac{1}{2}\right)
+ 30 \tilde{S}_{3,1,1}\left(\frac{1}{2},1,2\right)
+ 30 \tilde{S}_{3,1,1}\left(\frac{1}{2},2,1\right) \Biggr]
\N\\ &&
+\Biggl[(-1)^N \frac{2 \big(6-23 N+9 N^2+2 N^3\big)}{(-1+N)^2 N^2 (1+N) (2+N)}
\N\\ &&
+(-1)^N \frac{12}{(-1+N) N (1+N) (2+N)} S_1\Biggr]
\Biggl[
   \tilde{S}_{1,1,1,1}\left(\frac{1}{2},1,2,1\right)
\N\\ &&
-2 \tilde{S}_{1,1,1,1}\left(2,\frac{1}{2},1,1\right)
-2 \tilde{S}_{1,1,1,1}\left(2,1,\frac{1}{2},1\right)
-2 \tilde{S}_{1,1,1,1}\left(2,1,1,\frac{1}{2}\right)
\Biggr]
\N\\
&&
+\frac{(-1)^N}{(-1+N) N (1+N) (2+N)}
\Biggl[ 66 \tilde{S}_{1,1,1,2}\left(\frac{1}{2},1,2,1\right)
      + 48 \tilde{S}_{1,1,1,2}\left(\frac{1}{2},2,1,1\right)
\N\\
&&
      + 30 \tilde{S}_{1,1,1,2}\left(1,\frac{1}{2},2,1\right)
      + 30 \tilde{S}_{1,1,1,2}\left(1,2,\frac{1}{2},1\right)
      + 30 \tilde{S}_{1,1,2,1}\left(\frac{1}{2},1,1,2\right)
\N\\
&&
      + 12 \tilde{S}_{1,1,2,1}\left(\frac{1}{2},1,2,1\right)
      + 48 \tilde{S}_{1,1,2,1}\left(\frac{1}{2},2,1,1\right)
      + 30 \tilde{S}_{1,1,2,1}\left(1,\frac{1}{2},1,2\right)
\N\\
&&
      + 30 \tilde{S}_{1,1,2,1}\left(1,2,1,\frac{1}{2}\right)
      + 30 \tilde{S}_{1,2,1,1}\left(\frac{1}{2},1,1,2\right)
      + 30 \tilde{S}_{1,2,1,1}\left(\frac{1}{2},1,2,1\right)
\N\\
&&
      + 12 \tilde{S}_{1,2,1,1}\left(\frac{1}{2},2,1,1\right)
      + 30 \tilde{S}_{1,2,1,1}\left(1,1,\frac{1}{2},2\right)
      + 30 \tilde{S}_{1,2,1,1}\left(1,1,2,\frac{1}{2}\right)
\N\\
&&
      + 30 \tilde{S}_{2,1,1,1}\left(1,\frac{1}{2},1,2\right)
      + 30 \tilde{S}_{2,1,1,1}\left(1,\frac{1}{2},2,1\right)
      + 30 \tilde{S}_{2,1,1,1}\left(1,1,\frac{1}{2},2\right)
\N\\
&&
      + 30 \tilde{S}_{2,1,1,1}\left(1,1,2,\frac{1}{2}\right)
      + 30 \tilde{S}_{2,1,1,1}\left(1,2,\frac{1}{2},1\right)
      + 30 \tilde{S}_{2,1,1,1}\left(1,2,1,\frac{1}{2}\right)
\N
\end{eqnarray}
\begin{eqnarray}
&&
- 24 \tilde{S}_{2,1,1,1}\left(2,\frac{1}{2},1,1\right)
- 24 \tilde{S}_{2,1,1,1}\left(2,1,\frac{1}{2},1\right)
- 24 \tilde{S}_{2,1,1,1}\left(2,1,1,\frac{1}{2}\right)
\N\\
&&
- 12 \tilde{S}_{1,1,1,1,1}\left(\frac{1}{2},1,2,1,1\right)
- 36 \tilde{S}_{1,1,1,1,1}\left(\frac{1}{2},2,1,1,1\right)
\Biggr]
\N\\
&&
+\Biggl\{(-1)^N \frac{3 S_1^2}{(-1+N) N (1+N) (2+N)}
+(-1)^N \frac{2+9 N-5 N^2}{(-1+N)^2 N (1+N) (2+N)}
\N\\ &&
+(-1)^N \frac{6-11 N+2 N^2}{(-1+N)^2 N^2 (2+N)} S_1
\N\\ &&
+\Biggl[\frac{3 (-1)^N}{(N-1) N (N+1) (N+2)}
+\frac{1}{N+2}\Biggr] S_2\Biggr\} \zeta_3
\N\\ &&
+\zeta_2 \Biggl\{
(-1)^N \frac{9 S_1^2}{2 (-1+N) N (1+N) (2+N)}
\N\\ &&
+(-1)^N \frac{2+3 N-2^{2+N} N-2 N^2-3 \cdot 2^{1+N} N^2+6 N^3-2^{1+N} N^3-3 N^4}{(-1+N)^3 N^2 (1+N)
(2+N)}
\N\\ &&
+\Biggl\{
(-1)^N \frac{-12+N+27 N^2-4 N^3}{2 (-1+N)^2 N^2 (1+N) (2+N)}
\N\\ &&
-\Biggl[\frac{1}{N+2}+\frac{6 (-1)^N}{(N-1) N (N+1) (N+2)}\Biggr] S_1
\N\\ &&
+\frac{1}{N (N+2)}\Biggr\} S_2
+\Biggl[-\frac{6 (-1)^N}{(N-1) N (N+1) (N+2)}
-\frac{2}{N+2}\Biggr] S_3
\N\\ &&
-\frac{2 S_{-2}}{N (N+2)}
+ \Biggl[(-1)^N \frac{-6+3 N+18 N^2-20 N^3-3 N^4+2 N^5}{(-1+N)^3 N^3 (1+N) (2+N)}
\N\\ &&
+\frac{2 S_{-2}}{N+2}\Biggr]
 S_1
+\frac{3 S_{-3}}{N+2}
+(-1)^N \frac{12}{(-1+N) N (1+N) (2+N)} S_{2,1}
\N\\ &&
-\frac{2 S_{-2,1}}{N+2}+\bigg[-(-1)^N \frac{3 S_1^2}{(-1+N) N (1+N) (2+N)}
\N\\ &&
+(-1)^N \frac{6-12 N+7 N^2+N^3}{(-1+N)^3 N^3}+(-1)^N \frac{-6+23 N-9 N^2-2 N^3}{(-1+N)^2 N^2 (1+N) (2+N)}
S_1\bigg] \tilde{S}_1(2)
\N\\
&&
+\bigg[(-1)^N \frac{-6+23 N-9 N^2-2 N^3}{(-1+N)^2 N^2 (1+N) (2+N)}
\N\\ &&
-(-1)^N \frac{6}{(-1+N) N (1+N) (2+N)} S_1\bigg] \tilde{S}_2(2)
\N\\ &&
+\bigg[(-1)^N \frac{6-23 N+9 N^2+2 N^3}{(-1+N)^2 N^2 (1+N) (2+N)}
\N\\ &&
+(-1)^N \frac{6}{(-1+N) N (1+N) (2+N)} S_1
\bigg]
\tilde{S}_{1,1}(2,1)
\N\\ &&
-(-1)^N \frac{3}{(-1+N) N (1+N) (2+N)} \tilde{S}_3(2)
\N\\
&&
+\frac{(-1)^N}{(-1+N) N (1+N) (2+N)}\Biggl[
  3 \tilde{S}_{1,2}(2,1)
-15 \tilde{S}_{2,1}(1,2)
+ 6 \tilde{S}_{2,1}(2,1)
- 6 \tilde{S}_{1,1,1}(2,1,1) \Biggr]
\N
\end{eqnarray}
\begin{eqnarray}
&&
+\Biggl[\frac{3}{N+2}-\frac{18 (-1)^N}{(N-1) N (N+1) (N+2)}\Biggr] \zeta_3
\Biggr\}
\N\\ &&
+\Biggl[\frac{27 (-1)^N}{(N-1) N (N+1) (N+2)}-\frac{9}{2 (N+2)}\Biggr] \zeta_5
\N\\ &&
- 30 (-1)^N \frac{1}{(N-1) N (N+1)(N+2)} \tilde{S}_1\left(\frac{1}{2}\right) \tilde{S}_{1,1,2}(1,2,1)~,
\label{SUM:main}
\\
\N\\
\N\\
\N\\
 &&\sum_{j=1}^{N-2} \sum_{j_1=1}^{-j+N-2} \sum_{n=1}^{\infty } \frac{(-1)^{j_1} j B(j,n) \binom{-j+N-2}{j_1} 
S_1(j) S_1(n)}{(j+n) (j+n+1) (j+n+2) (j+n+3) (j_1-N-2)} =
\N\\
&&
 (-1)^N \frac{-1+2 N-3 N^2}{N^3 (1+N) (2+N)}
+\frac{-16+12 N+10 N^2-17 N^3-31 N^4}{8 N^3 (1+N) (2+N)}
\N\\
&&
+
\Bigg[(-1)^N\Big[
-\frac{1}{N (2+N)} S_1
+\frac{1}{(N+1) (N+2)}
\Big]
+\frac{1}{4 (N+2)}\Bigg] S_3-\frac{S_4}{2 (N+2)}
\N\\
&&
-(-1)^N \frac{4}{(1+N) (2+N)} S_{-2}
+S_2 \Biggl\{\frac{-8-2 N+N^2+5 N^3+5 N^4}{4 N^3 (1+N) (2+N)}
\N\\
&&
+\Bigg[\frac{(-1)^N (N-1)}{N^2 (N+2)}
+\frac{1}{2 N^2 (N+1)}\Bigg] S_1
+(-1)^N \frac{2}{N (2+N)} S_{-2}
\N\\
&&
-\frac{2 (-1)^N}{(N+1) (N+2)}\Biggr\}
+(-1)^N \frac{1+2 N^2}{N^2 (1+N) (2+N)} S_{-3}
+S_1 \Biggl\{
\frac{2+N+N^2}{2 N^2 (1+N) (2+N)}
\N\\
&&
+(-1)^N \Big[
\frac{2 (-1+N)}{N^2 (2+N)} S_{-2}
-\frac{3}{N (2+N)} S_{-3}\Big] \Biggr\}
+ \frac{3 (-1)^N}{N (2+N)} S_{-4}
\N\\
&&
+\Big[\frac{(-1)^N (1-N)}{N^2 (N+2)}
+\frac{1}{2 (N+2)}\Big] S_{2,1}
-(-1)^N \frac{2}{N (2+N)} S_{2,-2}
\N\\
&&
+\Big[\frac{(-1)^N}{N (N+2)}
+\frac{1}{2 (N+2)}\Big] S_{3,1}
+\Big[(-1)^N \frac{2}{N (2+N)} S_1
\N\\
&&
-(-1)^N\frac{2 \big(-1+2 N^2\big)}{N^2 (1+N) (2+N)}\Big] S_{-2,1}
+(-1)^N\Big[
\frac{2}{N (2+N)} S_{-3,1}
\N\\
&&
-\frac{4}{N (2+N)} S_{-2,1,1}
\Big]
+\Big[\frac{16+4 N-2 N^2+N^3+N^4}{8 N^3 (1+N) (2+N)}
-\frac{S_1}{2 N^2 (N+1)}
\N\\
&&
+\frac{(-1)^N}{N^3 (N+1) (N+2)}\Big] \zeta_2~,
\\
\N\\
&& \sum_{j=1}^{N-2} \sum_{j_1=1}^{-j+N-2} \sum_{n=1}^{\infty } \frac{(-1)^{j_1} j B(j,n) \binom{-j+N-2}{j_1} S_2(n)}{(j+n) (j+n+1) (j+n+2) (j+n+3) (j_1-N-2)} =
\N\\
&&
-\frac{S_2^2}{4 (N+2)}
+\Bigg[\frac{(-1)^N S_1^2}{2 N (N+2)}
+\Big[-\frac{(-1)^N}{N (N+2)}
-\frac{1}{2 N^2 (N+1)}\Big] S_1
\N
\end{eqnarray}
\begin{eqnarray}
&&
-\frac{N^2+1}{2 N^2 (N+1)}
+(-1)^N\Big[ \frac{1}{N (2+N)} S_{-2}
+\frac{1}{(N+1) (N+2)}\Big] \Bigg]S_2
\N
\\
&&
+(-1)^N \frac{8-19 N+24 N^2}{8 (-1+N) N^2 (1+N) (2+N)}
+\frac{-48-24 N+71 N^2+95 N^3}{48 N^2 (1+N) (2+N)}
\N\\
&&
-\frac{S_3}{2 N^2 (N+1)}
+\Bigg[-\frac{(-1)^N}{2 N (N+2)}
-\frac{1}{4 (N+2)}\Bigg] S_4
\N\\
&&
+(-1)^N \Bigg[\frac{2}{(1+N) (2+N)} S_{-2}
+\frac{1}{N (2+N)} S_1^2 S_{-2}
+\frac{1}{N (2+N)} S_{-3}
\N
\\
&&
+S_1 \Big[- \frac{2}{N (2+N)} S_{-2}
-\frac{1}{N (2+N)} S_{-3}\Big]
\N\\
&&
-\frac{1}{N (2+N)} S_{-4}
\Bigg]
+\Bigg[-(-1)^N \frac{1}{N (2+N)} S_1
+\frac{1}{2 N^2 (N+1)}
\N\\
&&
+\frac{(-1)^N}{N (N+2)}\Bigg] S_{2,1}
+\Bigg[\frac{2 (-1)^N}{N (N+2)}
-(-1)^N \frac{2}{N (2+N)} S_1\Bigg] S_{-2,1}
\N\\
&&
+(-1)^N
\Bigg[
\frac{1}{N (2+N)} S_{-3,1}
+\frac{1}{N (2+N)} S_{2,1,1}
+\frac{2}{N (2+N)} S_{-2,1,1}
\Bigg]
\N\\
&&
+\Bigg[(-1)^N \frac{-2+N-2 N^2}{2 (-1+N) N^2 (1+N) (2+N)}
+\frac{2+N-N^2-2 N^3}{2 N^2 (1+N) (2+N)}
+\frac{S_1}{2 N^2 (N+1)}
\N\\
&&
+\Big[\frac{1}{2 (N+2)}
-\frac{(-1)^N}{N (N+2)}\Big] S_2
-(-1)^N \frac{2}{N (2+N)} S_{-2}
\Bigg] \zeta_2
\N\\
&&
+\Biggl[\frac{-12-6 N+N^2+N^3}{12 N^2 (1+N) (2+N)}
+\frac{(-1)^N}{(N-1) N^2 (N+1) (N+2)} \Biggr]\zeta_3~,
\\
\N\\
&& \sum_{j=1}^{N-2} \sum_{j_1=1}^{-j+N-2} \frac{(-1)^{j_1} \binom{-j+N-2}{j_1} 
S_1(j) S_2(-j_1+N)}{(j+2) (j_1-N-2)} =
\N\\
&&
\Bigg[\frac{(-1)^N}{2 (N+1) (N+2)}
-\frac{1}{2 (N+2)}\Bigg] S_2^2
+\Bigg[\frac{S_1^2}{2 (N+2)}
\N\\
&&
+\Biggl[\frac{-3-3 N-N^2}{(1+N)^2 (2+N)^2}
+(-1)^N \frac{-4-5 N-3 N^2-N^3}{N (1+N)^2 (2+N)^2}\Bigg] S_1
\N\\
&&
+(-1)^N \frac{8+28 N+37 N^2-42 N^4-38 N^5-14 N^6-2 N^7}{2 N^2 (1+N)^3 (2+N)^3}
\N\\
&&
+\frac{8+28 N+49 N^2+39 N^3+6 N^4-10 N^5-6 N^6-N^7}{N^2 (1+N)^3 (2+N)^3}\Biggr] S_2
\N
\\
&&
+(-1)^N \Bigg[-\frac{2}{(1+N) (2+N)} S_{-2} S_2
+\frac{\big(-8-28 N-27 N^2-8 N^3\big) S_1^2}{2 N^2 (1+N)^3 (2+N)^3}\Bigg]
\N\\
&&
+\frac{4+5 N+3 N^2+N^3}{N (1+N)^3 (2+N)^2}
+(-1)^N \frac{6+2 N-8 N^2-6 N^3-N^4}{(1+N)^3 (2+N)^3}
\N
\end{eqnarray}
\begin{eqnarray}
&&
+(-1)^N\Bigg[ \frac{-4-3 N+2 N^2+3 N^3+N^4}{N (1+N)^2 (2+N)^2}
+ \frac{1}{(1+N) (2+N)} S_1\Bigg] S_3
\N\\
&&
+(-1)^N \Bigg[\frac{3}{2 (1+N) (2+N)} S_4
- \frac{2 \big(-4+2 N^2+N^3\big)}{N^2 (2+N)^2} S_{-2}
\N\\
&&
+ \frac{3 \big(-4-3 N+2 N^2+3 N^3+N^4\big)}{N (1+N)^2 (2+N)^2} S_{-3}
\Bigg]
+S_1 \Bigg[(-1)^N \frac{16+8 N-4 N^2-N^3}{N^3 (2+N)^3}
\N
\\
&&
-(-1)^N \frac{2 \big(4+5 N+3 N^2+N^3\big)}{N (1+N)^2 (2+N)^2} S_{-2}
+(-1)^N \frac{3}{(1+N) (2+N)} S_{-3}
\N\\
&&
+\frac{1}{(N+1)^2 (N+2)}\Bigg]
+ (-1)^N \Bigg[
+\frac{2}{(1+N) (2+N)} S_{-4}
+\frac{4}{(1+N) (2+N)} S_{2,-2}
\N\\
&&
+\Big[- \frac{2 \big(-4-3 N+2 N^2+3 N^3+N^4\big)}{N (1+N)^2 (2+N)^2}
- \frac{2}{(1+N) (2+N)} S_1\Big] S_{-2,1}
\N\\
&&
-\frac{6}{(1+N) (2+N)} S_{-3,1}
+\frac{4}{(1+N) (2+N)} S_{-2,1,1}
\Bigg]
\end{eqnarray}

The nested sums emerging in this work,  which were not given before in
Refs.~\cite{Bierenbaum:2007qe,Bierenbaum:2008yu} and those being closer related to 
the structure of harmonic sums \cite{Vermaseren:1998uu}, are 
of the type illustrated 
above. The latter have been calculated using C. Schneider's
packages {\sf Sigma}~\cite{Refined,Schneider:2007,sigma1,sigma2}, {\sf EvaluateInfiniteSums} 
\cite{CS:PUB1}
and J. Ablinger's package {\sf HarmonicSums} \cite{JA:DIPL,ABS2010}. The involved structure
of Eq.~(\ref{SUM:main}) requires simplifications due to explicit algebraic
and structural relations for the generalized harmonic sums. These will be given in
an upcoming publication Ref.~\cite{ABS10A}.
\newpage
  \section{\bf \boldmath Fixed Moments}
   \label{App-Moments}
   \renewcommand{\theequation}{\thesection.\arabic{equation}}
   \setcounter{equation}{0}
\subsection{3-loop Moments to the anomalous dimensions   \label{App-AnDim}}
In the following we list the $O\left(T_F^2n_fC_{F,A}\right)$ contributions to the 
fixed moments of the anomalous dimensions related to the present calculation. We 
have used them for comparison in recalculating the corresponding contributions 
to the anomalous dimensions at general values of $N$, cf. Eqs. (\ref{gammaqg},\ref{gammaqqPS},\ref{gammaqqNS}).
\vspace*{2mm}\noindent

\underline{$(i)$~~~\large $\hat{\gamma}_{qg}^{(2)}$}~:

%
%
\begin{eqnarray}
  \hat{\gamma}_{qg}^{(2)}(2)&=&T_F^2n_f \Bigl(
                             \frac{16928}{243}C_A
                            -\frac{2768}{243}C_F
                                        \Bigr)\label{gaqg2MOM2}
                         \\\N\\
  \hat{\gamma}_{qg}^{(2)}(4)&=&T_F^2n_f \Bigl(
                             \frac{4481539}{151875}C_A
                            +\frac{9613841}{1518750}C_F
                                        \Bigr)
                      \\\N\\
  \hat{\gamma}_{qg}^{(2)}(6)&=&T_F^2n_f \Bigl(
                             \frac{86617163}{5834430}C_A
                            +\frac{1539874183}{170170875}C_F
                                        \Bigr)
                                            \\  \N \\
  \hat{\gamma}_{qg}^{(2)}(8)&=&T_F^2n_f \Bigl(
                          \frac{10379424541}{1377810000}C_A
                         +\frac{7903297846481}{810152280000}C_F
                                        \Bigr)
             \\ \N \\
  \hat{\gamma}_{qg}^{(2)}(10)&=&T_F^2n_f \Bigl(
                              \frac{1669885489}{494133750}C_A
                             +\frac{1584713325754369}{161800390434375}C_F
               \Bigr)\label{gaqg2MOM10}
\end{eqnarray}


%
\vspace*{2mm}\noindent
\underline{$(ii)$~~~\large $\hat{\gamma}_{qq}^{(2), {\sf PS}}$}~:
%
%
\begin{eqnarray}
 \hat{\gamma}_{qq}^{(2),{\sf PS}}(2)&=&-T_F^2n_fC_F~\frac{10048}{243}\label{gaqqMOM2}
                     \\\N\\
 \hat{\gamma}_{qq}^{(2),{\sf PS}}(4)&=&-T_F^2n_fC_F~\frac{123734}{151875}
                        \\\N\\
 \hat{\gamma}_{qq}^{(2),{\sf PS}}(6)&=&-T_F^2n_fC_F~\frac{252446104}{72930375}
                         \\\\\
 \hat{\gamma}_{qq}^{(2),{\sf PS}}(8)&=&-T_F^2n_fC_F~
                                     \frac{13131081443}{6751269000}
                                                 \\ \N \\
 \hat{\gamma}_{qq}^{(2),{\sf PS}}(10)&=&-T_F^2n_fC_F~
                                      \frac{531694610144}{420260754375}
                       \\ \N \\
 \hat{\gamma}_{qq}^{(2),{\sf PS}}(12)&=&-T_F^2n_fC_F~
                                      \frac{2566080055386457}{2851637832143100} \label{gaqqMOM12}
                        \end{eqnarray}
\vspace*{2mm}\noindent
\underline{$(iii)$~~~\large $\hat{\gamma}_{qq}^{(2), {\sf NS,+}}$}~:
%
%
\begin{eqnarray}
 \hat{\gamma}_{qq}^{(2),{\sf NS},+}(2)&=&-T_F^2n_fC_F~
                                     \frac{3584}{243} \label{gqqNSMOM2}
                        \end{eqnarray}
\begin{eqnarray}
 \hat{\gamma}_{qq}^{(2),{\sf NS},+}(4)&=&-T_F^2n_fC_F~
                                     \frac{768554}{30375}
                        \\\N\\
 \hat{\gamma}_{qq}^{(2),{\sf NS},+}(6)&=&-T_F^2n_fC_F~
                                     \frac{321390284}{10418625}
                        \\\N\\
 \hat{\gamma}_{qq}^{(2),{\sf NS},+}(8)&=&-T_F^2n_fC_F~
                                     \frac{38920977797}{1125211500}
                        \\ \N \\
 \hat{\gamma}_{qq}^{(2),{\sf NS},+}(10)&=&-T_F^2n_fC_F~
                                     \frac{27995901056887}{748828253250}
                        \\ \N \\
 \hat{\gamma}_{qq}^{(2),{\sf NS},+}(12)&=&-T_F^2n_fC_F~
                                     \frac{65155853387858071}{1645175672390250}
                         \\\N \\ 
 \hat{\gamma}_{qq}^{(2),{\sf NS},+}(14)&=&-T_F^2n_fC_F~
                       \frac{68167166257767019}{1645175672390250}
                       \\\N\\
\end{eqnarray}

%
\vspace*{2mm}\noindent
\underline{$(iv)$~~~\large $\hat{\gamma}_{qq}^{(2), {\sf NS,-}}$}~:
%
%
\begin{eqnarray}
 {\hat{\gamma}_{qq}^{(2),{\sf NS},-}(1)}&=&  0~, \\
 \hat{\gamma}_{qq}^{(2),{\sf NS},-}(3)&=&-T_F^2n_fC_F~
                         \frac{5138}{243}
\\\N\\
 \hat{\gamma}_{qq}^{(2),{\sf NS},-}(5)&=&-T_F^2n_fC_F~
                         \frac{862484}{30375}
\\\N\\
 \hat{\gamma}_{qq}^{(2),{\sf NS},-}(7)&=&-T_F^2n_fC_F~
                         \frac{1369936511}{41674500}
\\\N\\
 \hat{\gamma}_{qq}^{(2),{\sf NS},-}(9)&=&-T_F^2n_fC_F~
                         \frac{20297329837}{562605750}
\\\N\\
 \hat{\gamma}_{qq}^{(2),{\sf NS},-}(11)&=&-T_F^2n_fC_F~
                         \frac{28869611542843}{748828253250}
\\ \N \\
 \hat{\gamma}_{qq}^{(2),{\sf NS},-}(13)&=&-T_F^2n_fC_F~
                         \frac{66727681292862571}{1645175672390250}  \label{gqqNSMOM13}
\end{eqnarray}

%
%
%
\newpage
  \subsection{\bf \boldmath 
            The $O(\ep^0)$ Contributions to $\Ahathat_{ij}^{(3)}$}
   \label{App-OMEs}
   \renewcommand{\theequation}{\thesection.\arabic{equation}}
We list the contributions $O\left(T_F^2 n_f C_{F,A}\right)$ to the fixed moments of the constant part of the unrenormalized massive OMEs, $\hat{a}_{ij}^{(3,K)}$ from Ref. \cite{Bierenbaum:2009mv}. We used these values for comparisons to the general $N$-results computed in the present paper, Eqs. (\ref{aQg3},\ref{aQqPS3},\ref{aqqQPS3},\ref{aqqQNS3}).   

\vspace*{2mm}\noindent
\underline{$(v)$~\large $a_{Qg}^{\rm (3)}$}~: 
\begin{eqnarray}
a_{Qg}^{(3)}(2)&=&
n_fT_F^2C_A
      \Biggl(
                -\frac{6706}{2187}
                -\frac{616}{81}\zeta_3
                -\frac{250}{81}\zeta_2
      \Biggr)
\N \\ \N \\ &&
+T_F^2n_FC_F
      \Biggl(
                 \frac{158}{243}
                +\frac{896}{81}\zeta_3
                +\frac{40}{9}\zeta_2
      \Biggr)~, \label{aQg3MOM2}\\
a_{Qg}^{(3)}(4)&=&
n_fT_F^2C_A
      \Biggl(
                 \frac{947836283}{72900000}
                -\frac{18172}{2025}\zeta_3
                -\frac{11369}{13500}\zeta_2
      \Biggr)
\N \\ \N \\ &&
+T_F^2n_FC_F
      \Biggl(
                 \frac{8164734347}{4374000000}
                +\frac{130207}{20250}\zeta_3
                +\frac{1694939}{810000}\zeta_2
      \Biggr)~, \\
a_{Qg}^{(3)}(6)&=&
n_fT_F^2C_A
      \Biggl(
                 \frac{12648331693}{735138180}
                -\frac{4433}{567}\zeta_3
                +\frac{23311}{111132}\zeta_2
      \Biggr)
\N \\ \N \\ &&
+T_F^2n_FC_F
      \Biggl(
                -\frac{8963002169173}{1715322420000}
                +\frac{111848}{19845}\zeta_3
                +\frac{11873563}{19448100}\zeta_2
      \Biggr)~, \\
a_{Qg}^{(3)}(8)&=&
n_fT_F^2C_A
      \Biggl(
                 \frac{24718362393463}{1322697600000}
                -\frac{125356}{18225}\zeta_3
                +\frac{2118187}{2916000}\zeta_2
      \Biggr)
\N \\ \N \\ &&
+T_F^2n_FC_F
      \Biggl(
                -\frac{291376419801571603}{32665339929600000}
                +\frac{887741}{174960}\zeta_3
\N\\ &&
                -\frac{139731073}{1143072000}\zeta_2
      \Biggr)~, \\
a_{Qg}^{(3)}(10)&=&
n_fT_F^2C_A
      \Biggl(
                 \frac{297277185134077151}{15532837481700000}
                -\frac{1505896}{245025}\zeta_3
\N \\ \N \\ &&
                +\frac{189965849}{188669250}\zeta_2
      \Biggr)
+T_F^2n_FC_F
      \Biggl(
                -\frac{1178560772273339822317}{107642563748181000000}
\N \\ \N \\ &&
                +\frac{62292104}{13476375}\zeta_3
                -\frac{49652772817}{93391278750}\zeta_2
      \Biggr)~.\label{aQg3MOM10}
\end{eqnarray}

\vspace*{2mm}\noindent
\underline{$(vi)$~\large $a_{Qq}^{(3), \sf PS}$}~:
\begin{eqnarray}
a_{Qq}^{(3), {\sf PS}}(2)&=&
T_F^2n_FC_F
      \Biggl(
                -\frac{76408}{2187}
                +\frac{896}{81}\zeta_3
                -\frac{112}{81}\zeta_2
      \Biggr)~, \label{aQqMOMstart}
\\
a_{Qq}^{(3), {\sf PS}}(4)&=&
T_F^2n_FC_F
      \Biggl(
                -\frac{474827503}{109350000}
                +\frac{3388}{2025}\zeta_3
                -\frac{851}{20250}\zeta_2
      \Biggr)~,
\\
a_{Qq}^{(3), {\sf PS}}(6)&=&
T_F^2n_FC_F
      \Biggl(
                -\frac{82616977}{45378900}
                +\frac{1936}{2835}\zeta_3
                -\frac{16778}{694575}\zeta_2
      \Biggr)~,
\\
a_{Qq}^{(3), {\sf PS}}(8)&=&
T_F^2n_FC_F
      \Biggl(
                -\frac{16194572439593}{15122842560000}
                +\frac{1369}{3645}\zeta_3
                -\frac{343781}{14288400}\zeta_2
      \Biggr)~,
\\
a_{Qq}^{(3), {\sf PS}}(10)&=&
T_F^2n_FC_F
      \Biggl(
                -\frac{454721266324013}{624087220246875}
                +\frac{175616}{735075}\zeta_3
                -\frac{547424}{24257475}\zeta_2
      \Biggr)~,
\\
a_{Qq}^{(3), {\sf PS}}(12)&=&
T_F^2n_FC_F
      \Biggl(
                -\frac{6621557709293056160177}{12331394510293050192000}
                +\frac{24964}{150579}\zeta_3
\N \\ \N \\ &&
                -\frac{1291174013}{63306423180}\zeta_2
      \Biggr)~.\label{aQqMOMend}
\end{eqnarray} 


\vspace*{2mm}\noindent
\underline{$(vii)$~\large $a_{qq,Q}^{(3), \sf PS}$}~: 
\begin{eqnarray}
a_{qq,Q}^{(3), {\sf PS}}(2)&=&
n_fT_F^2C_F
      \Biggl( 
                -\frac{100096}{2187}
                +\frac{896}{81}\zeta_3
                -\frac{256}{81}\zeta_2
      \Biggr)~, \label{aqqQMOM2}\\
a_{qq,Q}^{(3), {\sf PS}}(4)&=&
n_fT_F^2C_F
      \Biggl( 
                -\frac{118992563}{21870000}
                +\frac{3388}{2025}\zeta_3
                -\frac{4739}{20250}\zeta_2
      \Biggr)~, \\
a_{qq,Q}^{(3), {\sf PS}}(6)&=&
n_fT_F^2C_F
      \Biggl( 
                -\frac{17732294117}{10210252500}
                +\frac{1936}{2835}\zeta_3
                -\frac{9794}{694575}\zeta_2
      \Biggr)~,\\
a_{qq,Q}^{(3), {\sf PS}}(8)&=&
n_fT_F^2C_F
      \Biggl( 
                -\frac{20110404913057}{27221116608000}
                +\frac{1369}{3645}\zeta_3
                +\frac{135077}{4762800}\zeta_2
      \Biggr)~,\\
a_{qq,Q}^{(3), {\sf PS}}(10)&=&
n_fT_F^2C_F
      \Biggl( 
                -\frac{308802524517334}{873722108345625}
                +\frac{175616}{735075}\zeta_3
                +\frac{4492016}{121287375}\zeta_2
      \Biggr)~,
\\
a_{qq,Q}^{(3), {\sf PS}}(12)&=&
n_fT_F^2C_F
      \Biggl( 
                -\frac{6724380801633998071}{38535607844665781850}
                +\frac{24964}{150579}\zeta_3
\\
&&
                +\frac{583767694}{15826605795}\zeta_2
      \Biggr)~, \N 
\end{eqnarray}
\begin{eqnarray}
a_{qq,Q}^{(3), {\sf PS}}(14)&=&
n_fT_F^2C_F
      \Biggl( 
                -\frac{616164615443256347333}{7545433703850642600000}
                +\frac{22472}{184275}\zeta_3
\N\\ \N \\ && \hspace{-15mm}
                +\frac{189601441}{5533778250}\zeta_2
      \Biggr)~.\label{aqqQMOM14}
\N\\
\end{eqnarray}

\vspace*{2mm}\noindent
\underline{$(viii)$~\large $a_{qq,Q}^{(3), \sf NS}$}~: 
\begin{eqnarray}
a_{qq,Q}^{(3), {\sf NS}}(1)&=& 0~,\label{aqqQM1}\\
a_{qq,Q}^{(3), {\sf NS}}(2)&=&
T_F^2n_FC_F
      \Biggl(
                -\frac{100096}{2187}
                +\frac{896}{81}\zeta_3
                -\frac{256}{81}\zeta_2
      \Biggr)~,\\
a_{qq,Q}^{(3), {\sf NS}}(3)&=&
T_F^2n_FC_F
      \Biggl(
                -\frac{1271507}{17496}
                +\frac{1400}{81}\zeta_3
                -\frac{415}{81}\zeta_2
      \Biggr)~,\\
a_{qq,Q}^{(3), {\sf NS}}(4)&=&
T_F^2n_FC_F
      \Biggl(
                -\frac{1006358899}{10935000}
                +\frac{8792}{405}\zeta_3
                -\frac{13271}{2025}\zeta_2
      \Biggr)~,\\
a_{qq,Q}^{(3), {\sf NS}}(5)&=&
T_F^2n_FC_F
      \Biggl(
                -\frac{195474809}{1822500}
                +\frac{10192}{405}\zeta_3
                -\frac{15566}{2025}\zeta_2
      \Biggr)~, \\
a_{qq,Q}^{(3), {\sf NS}}(6)&=&
T_F^2n_FC_F
      \Biggl(
                -\frac{524427335513}{4375822500}
                +\frac{11344}{405}\zeta_3
                -\frac{856238}{99225}\zeta_2
      \Biggr)~,
\\
a_{qq,Q}^{(3), {\sf NS}}(7)&=&
T_F^2n_FC_F
      \Biggl(
                -\frac{54861581223623}{420078960000}
                +\frac{4108}{135}\zeta_3
                -\frac{3745727}{396900}\zeta_2
      \Biggr)~,\\
a_{qq,Q}^{(3), {\sf NS}}(8)&=&
T_F^2n_FC_F
      \Biggl(
                -\frac{4763338626853463}{34026395760000}
                +\frac{39532}{1215}\zeta_3
                -\frac{36241943}{3572100}\zeta_2
      \Biggr)~,\\
a_{qq,Q}^{(3), {\sf NS}}(9)&=&
T_F^2n_FC_F
      \Biggl(
                -\frac{2523586499054071}{17013197880000}
                +\frac{8360}{243}\zeta_3
                -\frac{19247947}{1786050}\zeta_2
      \Biggr)~,\\
a_{qq,Q}^{(3), {\sf NS}}(10)&=&
T_F^2n_FC_F
      \Biggl(
                -\frac{38817494524177585991}{249090230161080000}
                +\frac{96440}{2673}\zeta_3
\N \\ \N \\ &&
                -\frac{2451995507}{216112050}\zeta_2
      \Biggr)~,
\\
a_{qq,Q}^{(3), {\sf NS}}(11)&=&
T_F^2n_FC_F
      \Biggl(
                -\frac{40517373495580091423}{249090230161080000}
                +\frac{502528}{13365}\zeta_3
\N \\ \N \\ &&
                -\frac{512808781}{43222410}\zeta_2
      \Biggr)~,
\end{eqnarray}
\begin{eqnarray}
a_{qq,Q}^{(3), {\sf NS}}(12)&=&
T_F^2n_FC_F
      \Biggl(
                -\frac{1201733391177720469772303}{7114266063630605880000}
                +\frac{6774784}{173745}\zeta_3
\N \\ \N \\ &&
                -\frac{90143221429}{7304587290}\zeta_2
      \Biggr)~,\\
\end{eqnarray}\begin{eqnarray}
a_{qq,Q}^{(3), {\sf NS}}(13)&=&
T_F^2n_FC_F
      \Biggl(
                -\frac{1242840812874342588467303}{7114266063630605880000}\N
                +\frac{6997864}{173745}\zeta_3
\N\\
&&
                -\frac{93360116539}{7304587290}\zeta_2
      \Biggr)~,\\
a_{qq,Q}^{(3), {\sf NS}}(14)&=&
T_F^2n_FC_F
      \Biggl(
                -\frac{256205552272074402170491}{1422853212726121176000}
                +\frac{1440968}{34749}\zeta_3
\N \\ \N \\ &&
                -\frac{481761665447}{36522936450}\zeta_2
      \Biggr)~.\label{aqqQM14}
\end{eqnarray}
%
%
%
\newpage
  \subsection{\bf \boldmath $3$--loop Moments for Transversity}
   \label{App-Trans}
   \renewcommand{\theequation}{\thesection.\arabic{equation}}
We list below the contributions $O\left(T_F^2n_fC_{F,A}\right)$ to the fixed moments of the 
transversity anomalous dimension, cf.~\cite{Blumlein:2009rg}, to which we compared the result for 
general values of $N$ 
calculated in the present paper, Eqs. (\ref{gammaTrans}), cf. also \cite{Gracey:2003yrxGracey:2003mrxGracey:2006zrxGracey:2006ah}.
\begin{eqnarray}
\gamma_{qq}^{(2),\rm NS, TR}(1) &=&
 -\frac{16}{3}\,T_F^2 n_f C_F \\
\gamma_{qq}^{(2),\rm NS, TR}(2) &=&
-{\frac {368}{27}}\,T_F^2 n_f C_F     \\
\gamma_{qq}^{(2),\rm NS, TR}(3) &=&
-{\frac {4816}{243}}\,T_F^2 n_f C_F   \\
\gamma_{qq}^{(2),\rm NS, TR}(4) &=&
-{\frac {29444}{1215}}\,T_F^2 n_f C_F   \\
\gamma_{qq}^{(2),\rm NS, TR}(5) &=&
-{\frac {837188}{30375}}\,T_F^2 n_f C_F  \\
\gamma_{qq}^{(2),\rm NS, TR}(6) &=&
-{\frac {6419516}{212625}}\,T_F^2 n_f C_F \\
\gamma_{qq}^{(2),\rm NS, TR}(7) &=&
-{\frac {337002284}{10418625}}\,T_F^2 n_f C_F  \\
\gamma_{qq}^{(2),\rm NS, TR}(8) &=&
-{\frac {711801943}{20837250}}\,T_F^2 n_f C_F  \\
\gamma_{qq}^{(2),\rm NS, TR}(9) &=&
-{\frac {20096458061}{562605750}}\,T_F^2 n_f C_F  \\
\gamma_{qq}^{(2),\rm NS, TR}(10) &=&
-{\frac {229508848783}{6188663250}}\,T_F^2 n_f C_F \\
\gamma_{qq}^{(2),\rm NS, TR}(11) &=&
-{\frac {28677274464343}{748828253250}}\,T_F^2 n_f C_F \\
\gamma_{qq}^{(2),\rm NS, TR}(12) &=&
-{\frac {383379490933459}{9734767292250}}\,T_F^2 n_f C_F      \\
\gamma_{qq}^{(2),\rm NS, TR}(13) &=&
-{\frac {66409807459266571}{1645175672390250}}\,T_F^2 n_f C_F 
\end{eqnarray}
%
%
Here we list the $O\left(T_F^2n_f C_{F,A}\right)$ moments of the constant part of the unrenormalized 
massive OMEs $a_{qq,Q}^{(3),\rm{TR}}$ from Ref.~\cite{Blumlein:2009rg}. We used these values for 
comparisons 
to the general $N$-result computed in the present paper, Eq. (\ref{aTrans}). 
\begin{eqnarray}
a_{qq,Q}^{(3), {\rm TR}}(1)&=&
T_F^2n_FC_F
      \Biggl(
                -\frac{15850}{729}
                +\frac{112}{27}\zeta_3
                -\frac{52}{27}\zeta_2
      \Biggr)\label{aqqTR3MOM1}
~, \\
a_{qq,Q}^{(3), {\rm TR}}(2)&=&
T_F^2n_FC_F
      \Biggl(
                -\frac{4390}{81}
                +\frac{112}{9}\zeta_3-4\zeta_2
      \Biggr)
~, \\
a_{qq,Q}^{(3), {\rm TR}}(3)&=&
T_F^2n_FC_F
      \Biggl(
                -\frac{168704}{2187}
                +\frac{1456}{81}\zeta_3
                -\frac{452}{81}\zeta_2
      \Biggr)
~, \\
\N\end{eqnarray}\begin{eqnarray}
a_{qq,Q}^{(3), {\rm TR}}(4)&=&
T_F^2n_FC_F
      \Biggl(
                -\frac{20731907}{218700}
                +\frac{1792}{81}\zeta_3
                -\frac{554}{81}\zeta_2
      \Biggr)
~, \\
a_{qq,Q}^{(3), {\rm TR}}(5)&=&
T_F^2n_FC_F
      \Biggl(
                -\frac{596707139}{5467500}
                +\frac{10304}{405}\zeta_3
                -\frac{15962}{2025}\zeta_2
      \Biggr)
~, \\
a_{qq,Q}^{(3), {\rm TR}}(6)&=&
T_F^2n_FC_F
      \Biggl(
                -\frac{32472719011}{267907500}
                +\frac{3808}{135}\zeta_3
                -\frac{17762}{2025}\zeta_2
      \Biggr)
~, \\
a_{qq,Q}^{(3), {\rm TR}}(7)&=&
T_F^2n_FC_F
      \Biggl(
                -\frac{1727972700289}{13127467500}
                +\frac{1376}{45}\zeta_3
                -\frac{947138}{99225}\zeta_2
      \Biggr)
~, \\
a_{qq,Q}^{(3), {\rm TR}}(8)&=&
T_F^2n_FC_F
      \Biggl(
                -\frac{29573247248999}{210039480000}
                +\frac{4408}{135}\zeta_3
                -\frac{2030251}{198450}\zeta_2
      \Biggr)
~, \N \\ \\
a_{qq,Q}^{(3), {\rm TR}}(9)&=&
T_F^2n_FC_F
      \Biggl(
                -\frac{2534665670688119}{17013197880000}
                +\frac{41912}{1215}\zeta_3
                -\frac{19369859}{1786050}\zeta_2
      \Biggr)
~, \\
a_{qq,Q}^{(3), {\rm TR}}(10)&=&
T_F^2n_FC_F
      \Biggl(
                -\frac{321908083399769663}{2058596943480000}
                +\frac{43928}{1215}\zeta_3
\N\\
&&
                -\frac{4072951}{357210}\zeta_2
      \Biggr)
~, \\
a_{qq,Q}^{(3), {\rm TR}}(11)&=&
T_F^2n_FC_F
      \Biggl(
                -\frac{40628987857774916423}{249090230161080000}
                +\frac{503368}{13365}\zeta_3
\N\\
&&
                -\frac{514841791}{43222410}\zeta_2
      \Biggr)
~, \\
a_{qq,Q}^{(3), {\rm TR}}(12)&=&
T_F^2n_FC_F
      \Biggl(
                -\frac{7126865031281296825487}{42096248897222520000}
\N \\ \N \\ &&
                +\frac{521848}{13365}\zeta_3
                -\frac{535118971}{43222410}\zeta_2
      \Biggr)
~,\\
a_{qq,Q}^{(3), {\rm TR}}(13)&=&
T_F^2n_FC_F
      \Biggl(
                -\frac{1245167831299024242467303}{7114266063630605880000}
\N \\ \N \\ &&
                +\frac{7005784}{173745}\zeta_3
                -\frac{93611152819}{7304587290}\zeta_2
      \Biggr)\label{aqqTR3MOM13}
~.
\end{eqnarray}

\end{appendix}

\newpage

   \newpage
 \thispagestyle{empty}
 \setcounter{page}{0}
\begin{center}
{\bf Acknowledgement}
\end{center}
I am deeply grateful to J. Bl\"umlein for devoting very much time and 
effort in supervising me and invoking my interest in particle physics. \\
I would like to thank S. Klein, for advice and numerous comparisions to earlier results.\\
Additionally I would like to thank C. Schneider and J. Ablinger for providing excellent
support for their packages {\SigmaP} and {\sf Harmonic Sums}. \\
Further thanks go to A. Hasselhuhn for many helpful discussions. \\
\newpage
 \thispagestyle{empty}
 \begin{center}
{\Large\bf Selbstst\"andigkeitserkl\"arung}
 \end{center}
 \vspace{5mm}
Hiermit erkl\"are ich, dass ich diese Arbeit im Rahmen der Betreuung am 
Deutschen Elektronen-Synchrotron in Zeuthen ohne unzul\"assige Hilfe Dritter 
verfasst und alle Quellen als solche gekennzeichnet habe. \\
  \begin{flushleft}
  \vspace{20mm}
  {\small Fabian Wi\ss brock, Berlin den $31$. Mai $2010$.}
  \end{flushleft}
\end{document}